\documentclass[12pt]{article}

\usepackage{moreverb}

\usepackage[colorlinks,bookmarksopen,bookmarksnumbered,citecolor=red,urlcolor=red]{hyperref}

\usepackage{natbib,gentium}

\usepackage{float,amsmath,amsthm,verbatim}
\usepackage{graphicx}
\usepackage{subfigure}
\evensidemargin \oddsidemargin

\theoremstyle{definition}

\newcommand\BibTeX{{\rmfamily B\kern-.05em \textsc{i\kern-.025em b}\kern-.08em
		T\kern-.1667em\lower.7ex\hbox{E}\kern-.125emX}}

\newcommand{\tr}{\mbox{tr}}

\newcommand{\balpha}{\boldsymbol{\alpha}}

\newcommand{\bLambda}{\boldsymbol{\Lambda}}

\newcommand{\bseta}{\boldsymbol{\eta}}
\newcommand{\bdelta}{\boldsymbol{\delta}}

\newcommand{\bSigma}{\boldsymbol{\Sigma}}
\newcommand{\bsigma}{\boldsymbol{\sigma}}

\newcommand{\bepsilon}{\boldsymbol{\epsilon}}

\newcommand{\bB}{\boldsymbol{B}}

\newcommand{\bC}{\boldsymbol{C}}

\newcommand{\bg}{\boldsymbol{g}}
\newcommand{\bD}{\boldsymbol{D}}
\newcommand{\bff}{\boldsymbol{f}}
\newcommand{\bV}{\boldsymbol{V}}

\newcommand{\bh}{\boldsymbol{h}}

\newcommand{\bI}{\boldsymbol{I}}

\newcommand{\bK}{\boldsymbol{K}}

\newcommand{\bA}{\boldsymbol{A}}
\newcommand{\ba}{\boldsymbol{a}}
\newcommand{\bS}{\boldsymbol{S}}

\newcommand{\bP}{\boldsymbol{P}}

\newcommand{\bQ}{\boldsymbol{Q}}
\newcommand{\bR}{\boldsymbol{R}}

\newcommand{\bu}{\boldsymbol{u}}

\newcommand{\bs}{\boldsymbol{s}}

\newcommand{\bx}{\boldsymbol{x}}
\newcommand{\bX}{\boldsymbol{X}}
\newcommand{\by}{\boldsymbol{y}}

\newcommand{\bzero}{\boldsymbol{0}}

\newcommand{\mC}{\mathcal C}

\newcommand{\apprasym}{
	\mathrel{\ooalign{$\sim$\cr\kern+1.25pt\large $\colon$}}}

\begin{document}

\title{Reduced Basis Kriging for Big Spatial Fields}

\author{Karl Pazdernik\thanks{Department of Statistics, Pacific
    Northwest National Laboratory, Richland,
    Washington, USA}. Ranjan Maitra\thanks{Department of  Statistics, Iowa State University, Ames, Iowa, USA},
  Douglas Nychka\thanks{National Center for Atmospheric Research, Boulder, Colorado, USA}, Stephan Sain\thanks{The Climate Corporation, San Francisco, California, USA}
  }

\date{}
\maketitle
\renewcommand{\baselinestretch}{1.5}
\normalsize

\begin{abstract}
In spatial statistics, a common method for prediction over a Gaussian
random field (GRF) is maximum likelihood estimation combined with
kriging.  For massive data sets, kriging is computationally intensive,
both in terms of CPU time and memory, and so fixed rank kriging has
been proposed as a solution. The method however still involves
operations on large matrices, so we develop an alteration to this
method by utilizing the approximations made in fixed rank kriging combined
with restricted maximum likelihood estimation and sparse matrix
methodology.  Experiments show that our methodology can provide
additional gains in computational efficiency over fixed-rank kriging
without loss of accuracy in prediction.  The methodology is applied to
climate data archived by the United States National Climate Data
Center, with very good results.
\end{abstract}

\maketitle

\section{Introduction}
Data collected in a spatial domain are often incomplete. In a 
geostatistical setup, the spatial domain is defined as having an 
infinite set of locations from which a finite subset have been observed. 
Unobserved locations can be the result of data collected at irregularly 
spaced locations (such as temperature recorded only at locations with 
weather stations) or they may be due to a desired spatial resolution 
that is finer than that of the data (such as satellite images that record 
observations in intervals of every kilometer when the desired resolution 
is in terms of a few meters). In either case, these unobserved locations 
may need to be imputed: this task is generally accomplished by a form of 
interpolation known as spatial prediction.

To fix ideas,  consider two sets from a spatial domain $\mathcal D$: one
containing all the observed locations ($\bs =
\{\bs_1,\bs_2,\ldots,\bs_n\}\in \mathcal D$) and the other having the desired
locations given by q($\bs_0 = \{\bs_{01},\bs_{02},\ldots,\bs_{0N}\} \in \mathcal
D$),
  which may (or may not) include the observed set
  $\{\bs_i:i=1,2,\ldots,n\}$.  
A random process $\by(\cdot)$ is defined on $\mathcal D$ by a linear combination
consisting of three parts: a mean structure $\bx(\cdot)$, a zero-mean spatial
process $\bff(\cdot)$, and a zero-mean measurement error process
$\bepsilon(\cdot)$, formulated as 
\begin{equation}
\label{model1}
	\by(\bs) = \bx(\bs) + \bff(\bs) + \bepsilon(\bs).
\end{equation}

Kriging, one of the most common methods of spatial prediction, interpolates 
the missing values by utilizing spatial dependence in the prediction
method. It was
developed by the French mathematician Georges Matheron
\citet{matheron1962} and named by him in honor of the
empirical work by D. G. Krige, a South African mining engineer, for
his work \citep{krige1951} which originated as his Masters'
thesis. Specifically, in the setup of \eqref{model1}, the best linear unbiased
predictions (BLUP) for the random variable at the desired 
locations ($\by(\bs_0)$) can be obtained under squared error loss
using the first and second  moments of $\by(\cdot)$ -- see, {\em
    e.g.} \citet{cressie1993}: 
\begin{equation}
\label{krige}
	\hat\by(\bs_0) = \mbox{E}(\by(\bs_0)) +\mC_{\by}(\bs_0,\bs) \mC^{-1}_{\by}(\bs,\bs)  [\by(\bs_0) - \mbox{E}(\by(\bs_0))].
\end{equation}
where $\mC_{\by}(\bs_0,\bs) = \mbox{Cov}(\by(\bs_0), \by(\bs))$.
Equation~\eqref{krige} is also called the kriging predictor: the
methodology is extensively used in many applications, such as in
mining \citep{richmond2003}, 
hydrogeology  \citep{chilesdelfiner1999}, natural sciences
\citep{goovaerts1997}, environmental sciences
\citep{bayraktarturalioglu2005}, remote sensing
\citep{steinvandermeergorte2002}, or  black-box modeling in computer  
experiments \citep{sackswelchmitchellwynn1989}.
The kriging predictor involves solving an $n \times n$ linear system,
which is an $O(n^3)$ operation: thus for datasets with large
observations, kriging can be computationally impractical to implement
in terms of both CPU time and memory. 





There has been significant work in overcoming the computational limitations
associated with kriging massive spatial fields. Most approaches can
broadly be grouped into two categories. The first kind of methods use
approximations to the kriging equations of \eqref{krige}
\citep{vecchia1988, nychkaetal1996, higdon1998, nychka2000,nychkawikleroyle2002, 
  fuentes2002, billingsbeatsonnewsam2002a, billingsbeatsonnewsam2002b, 
  steinetal2004, quinoerocandelarasmussen2005, furrergentonnychka2006, 
  kaufmanschervishnychka2008}. 
These approximations include the use of
orthogonal bases, covariance tapering, approximate iterative methods,
inducing variables, and a reduced set of space-filling locations. 

The other set of approaches in this area are concerned with choosing a class
of covariance functions, within the framework of which kriging can be
done exactly, regardless of the size of the data. 
In particular, multiresolution spatial processes can be
formulated to produce covariance functions that allow for exact
kriging \citep{huangcressiegabrosek2002, johannessoncressie2004, 
  johannessoncressiehuang2007, cressiejohannesson2008, banerjeeetal2008, 
  finleyetal2008, lindgren2011, nychka2015}. 
One advantage to exact kriging is that
there are no concerns of how close approximate predictions and
prediction errors are to their exact counterparts.  However,
performing estimation within the framework of a general $\bff(\cdot)$ may
still be computationally burdensome, so \citet{cressiejohannesson2008}
suggested ``fixed rank kriging'' which models the
underlying spatial process $\bff(\cdot)$ through basis functions, but
allows for exact kriging. 
Because the methods proposed in this paper use fixed-rank
kriging as a starting point, we describe it in somewhat more detail.

We continue with the setup of~\eqref{model1}, but also assume without
loss of generality that the field for $\by(\cdot)$ has mean zero.
The Spatial Random Effects model~\citep{cressiejohannesson2008} uses a
reduced set of independent locations ($m < n$) coupled with a linear
combination of basis functions to approximate the spatial process at
the original set of locations. 
Specifically, let $\bseta(\cdot)$ define the spatial process at locations or
{\em knots} $\bu = \{\bu_1,...,\bu_m\}$ and let
$\bS_k(\cdot)$ be the basis functions 
corresponding to the $k$th knot. The spatial process is then
approximated by $\bff(\bs) \approx
\sum_{k=1}^{m}\bS_k(\bs)\bseta(\bu_k) + \bdelta(\bs)$ and the model reduces to

\begin{equation}
\label{model2}
	\by(\bs) = \sum_{k=1}^{m}\bS_k(\bs)\bseta(\bu_k) + \bdelta(\bs) + \bepsilon(\bs),
\end{equation}
where $\bseta(\bu)\sim N(\bzero,\bK)$ is mutually independent of
$\bdelta(\bs) \sim N(\bzero,\sigma_{\delta}^2\bV_{\delta})$
and $\bepsilon(\bs) \sim N(\bzero,\sigma_{\epsilon}^2\bV_{\epsilon})$ for known 
$\bV_{\delta}$ and $\bV_{\epsilon}$, both diagonal matrices with entries corresponding 
to the weights for the fine-scale and measurement error variances, respectively. 
Writing $\bS$ as the $n\times m$-matrix with $k$th column given by $\bS_k(\bs)$, we
get that the variance-covariance matrix of $\by(\bs)$ is given by
$\bSigma = \bS\bK\bS'+\bD$, where 
$\bD = \sigma_{\delta}^2\bV_{\delta}+\sigma_{\epsilon}^2\bV_{\epsilon}$. 
With this model and by defining
the particular set of basis functions between the unobserved locations
and the knots as $\bA = \bS_k(\bs_0)$ for $1 \le k \le m$, the
log-likelihood is $\ell(\bK, \sigma_{\delta}^2, \sigma_{\epsilon}^2;\by) =
-\frac{1}{2}\by'(\bS\bK\bS'+\bD)^{-1}\by -
\frac{1}{2}\log(\mid\bS\bK\bS'+\bD\mid) -
\frac{n}{2}\log(2\pi)$. Assuming $\bs \cap \bs_0 = \emptyset$, it follows 
that the kriging predictor is then 
$\hat\bff(\bs_0) = \bA\hat\bK\bS' (\bS\hat\bK\bS' + \hat\bD)^{-1}\by$, 
with kriging standard error
$\hat\bsigma_k(\bs_0) = [\bA\hat\bK\bA' + \hat\sigma_{\delta}^2\bV_{\delta}(\bs_0) 
- \bA\hat\bK\bS'(\bS\hat\bK\bS' + \hat\bD)^{-1}\bS\hat\bK\bA']^{\frac{1}{2}}$, 
where $\hat\bK$ and $\hat\sigma_{\delta}^2$ being the ML estimates of $\bK$ 
and $\sigma_{\delta}^2$. The measurement error variance, $\sigma_{\epsilon}^2$, is 
assumed known or estimated \textit{a priori}, thus, $\hat\bD$ is completely 
defined given $\sigma_{\delta}^2$.

\citet{cressiejohannesson2008} used the Sherman-Morrison-Woodbury
formula \citep{hendersonsearle1981} to reduce the dimension of
fixed-rank kriging by using the identity 
$	\bSigma^{-1} = \bD^{-1} - \bD^{-1}\bS[\bK^{-1} + \bS'\bD^{-1}\bS]^{-1}S'\bD^{-1},
$
whereupon calculating $\bSigma^{-1}$ reduces to inverting two $m
\times m$ matrices as opposed to one $n \times n$ matrix. 
The Method of Moments was then used to obtain parameter estimates.

A likelihood-based approach was developed by
\citet{katzfusscressie2011} who adopted the expectation-maximization (EM)
algorithm~\citep{dempsteretal77} to maintain a positive definite covariance 
matrix throughout the estimation procedure. Their approach considers
 $\by$ as the observed data while $\bseta$ and  $\bdelta$ are
missing. The parameter estimates  are then obtained iteratively
starting from appropriate initial values and updating, till
convergence, via equations which update from the $t$th
to the $(t+1)$th iteration: 
\begin{eqnarray}
\label{M}
  \bK^{(t+1)} &=& \bK^{(t)} -
                  \bK^{(t)}\bS'{\bSigma^{(t)}}^{-1}\bS\bK^{(t)}
                  \nonumber \\&&+   (\bK^{(t)}\bS'{\bSigma^{(t)}}^{-1}\by)(\bK^{(t)}\bS'{\bSigma^{(t)}}^{-1}\by)' \\
	\sigma_{\delta}^{2(t+1)} &=& \sigma_{\delta}^{2(t)} + \frac{(\sigma_{\delta}^{2(t)})^2}{n}\tr[{\bSigma^{(t)}}^{-1}(\by\by'{\bSigma^{(t)}}^{-1} - \bI)\bV_{\delta}]. \nonumber
\end{eqnarray}

Despite improvements in CPU time, implementation of this approach
still requires matrix algebra involving operations on $n \times n$
matrices which can be computationally burdensome, and in some cases,
prohibitive. In this paper, therefore, we propose alterations to the
predetermined specifications of this method called ``reduced basis
kriging'' that take advantage of sparse matrix operations, require
less computer memory, result in faster estimation, and  maintain or
reduce mean square prediction error.  

This paper is organized as follows. Section~\ref{methodology}
illustrates how a QR-decomposition 
\citep{higham2002} of the matrix of basis functions $\bS$ can be used
to reduce the linear system solved in ML estimation,
and also outlines how a version of the Sherman-Morrison-Woodbury
formula can be utilized to reduce the linear system solved during
prediction. 
Section~\ref{experiments} evaluates our methodology via a simulation
study, and compares performance to the E-M approach. 
Section~\ref{application} demonstrates the applicability of our
reduced-basis kriging methodology to predicting  temperatures across
the US. We conclude with some discussion 
and pointers to future work. 
We also have a supplement providing some additional illustrations. 
Sections and figures in the supplement referred to in this paper are
labeled with the prefix ``S-''.

\section{Methodology}
\label{methodology}
\subsection{Reduced Model Reparametrization for increased computational efficiency}
\label{eff.model}
Consider writing $\bS$ in terms of its
QR-decomposition~\citep{golubandvanloan96}, {\em i.e.},
$\bS=\bQ_1\bR_1$, where $\bR_1$ is an $m \times m$ upper triangular matrix 
and $\bQ_1$ is an $n \times m$ matrix being such that $\bQ_1'\bQ_1 = \bI_m$,
the $m\times m$ identity matrix. Let $\by^\ast = \bQ_1'\by$, write
$\bepsilon^\ast = \bQ_1'(\bdelta + \bepsilon)$, and define $\bD^\ast = \bQ_1'\bD\bQ_1$. 
Then equation~\eqref{model2} reduces
to $\by^\ast = \bR_1\bseta + \bepsilon^\ast$, and the distribution of
$\by^\ast \sim N_m(\bzero,\bSigma^\ast)$ where $\bSigma^\ast$ is
defined as $\bSigma^\ast = \bR_1\bK\bR_1' + \bD^\ast$. The likelihood
function of the parameters of interest, given the transformed
observations $\by^\ast$ is then given by
\begin{equation}
\label{llnew}
\ell(\bK,\sigma^2;\by^{\ast}) = -\frac{1}{2}{\by^\ast}' (\bR_1\bK\bR_1'+\bD^\ast)^{-1}\by^\ast - \frac{1}{2}\log(|\bR_1\bK\bR_1'+\bD^\ast|) - \frac{n}{2}\log(2\pi).
\end{equation}
The transformation from $\by$ to $\by^\ast$ concentrates the original
dataset into a smaller dataset of length $m$, located at the knots. 
It also reduces the size 
of the variance-covariance matrix of the dataset (from the original $n\times n$
matrix $\bSigma$ to the smaller $m \times m$ matrix
$\bSigma^\ast$). A few computational benefits accrue with this
reduction. For one, computations involving the determinant and the
inverse of the matrices in the calculation of \eqref{llnew} are
reduced from $O(n^3)$ to $O(m^3)$. A further speedup is obtained if we
combine the matrix determinant and inverse calculations. We propose 
using a Cholesky decomposition \citep{higham2002} for its versatility in  
evaluating determinants as well as solving linear systems. The use of
a Cholesky method coupled with our transformation eliminates the need
to solve an additional $n 
\times n$ linear system to obtain the inverse of the variance-covariance
matrix beyond the determinant: in its stead, we are reduced to solving
a linear system involving an upper-triangular coefficient matrix,
whose diagonal elements when multiplied lead to the determinant. 

A further efficiency is obtained because of the reduction of
computational burden 
from $O(n^2m)$ to $O(m^3)$ with regard to matrix multiplication because
this approach avoids the  matrix multiplications involving
$\Sigma^{-1}$ necessary with the E-M algorithm (\ref{M}), using
instead the $m\times m$ matrix ${\Sigma^\ast}^{-1}$. 

\subsection{Computationally Efficient Kriging Predictions}
\label{eff.predict}
We now turn our attention to making kriging predictions and
calculating the corresponding standard errors. Using the identities
$\bR_1'\bR_1=\bS'\bS$ and $\bR_1'\by^* = \bS'\by$, we 
note that the ML-estimated kriging predictor and standard errors are unchanged: 
$\hat\bff(\bs_0) = \bA\hat\bK\bS' (\bS\hat\bK\bS' + \hat\bD)^{-1}\by$ and 
$\hat\bsigma_k(\bs_0) = [\bA\hat\bK\bA' + \hat\sigma_{\delta}^2\bV_{\delta}(\bs_0) 
 - \bA\hat\bK\bS'(\bS\hat\bK\bS' + \hat\bD)^{-1}\bS\hat\bK\bA']^{\frac{1}{2}}$.
Note that both equations need the product
$\hat\bK\bS' (\bS\hat\bK\bS' + \hat\bD)^{-1}$. To simplify
calculations, we 
use the following matrix identity provided in~\citet{searle2006} or in
\citet{matrixcookbook2008}:
Let $\bB$ be a $n\times n$ invertible matrix and let $\bC$ be any
$n\times m$ matrix such that $\bB +\bC\bC'$ is invertible. Then 
$\bC'(\bB + \bC\bC')^{-1} = (\bI + \bC'\bB^{-1}\bC)^{-1} \bC'\bB^{-1}$.
This result can then be used to obtain a special case of the Woodbury formula, 
known for its use in Kalman filtering as an alternative algorithm 
\citep{matrixcookbook2008, brownhwang92}.
Let $\bS$ be a $n\times m$ matrix ($m < n$, of rank $m$). Further, let 
$\bK$ and $\bD$ be positive definite matrices of real values of rank
$m$ and $n$, respectively, with unique square root matrix $\bK^{\frac12}$. 
Then
  $\bK\bS'(\bS\bK\bS' + \bD)^{-1} = (\bK^{-1} 
+ \bS'\bD^{-1}\bS)^{-1} \bS'\bD^{-1}$.	 
(For a full proof, refer to the supplementary material.)
This means that the kriging predictor reduces to 
\begin{equation}
\label{ridgereg}
\hat\bff(\bs_0) =
\bA(\bS'\hat\bD^{-1}\bS + \hat\bK^{-1})^{-1}\bS'\hat\bD^{-1}\by,
\end{equation}
while the kriging standard error becomes
\begin{equation}
\label{krigse2}
\begin{split}
\hat\bsigma_k(\bs_0) = \left[\bA\hat\bK\bA' +\right.
  \hat\sigma_{\delta}^2\bV_{\delta}&(\bs_0) -  \\ \bA(\bS'\hat\bD^{-1}\bS + 
& \left. \hat\bK^{-1})^{-1}\bS'\hat\bD^{-1}\bS\hat\bK\bA'\right]^{\frac12}.\\
\end{split}
\end{equation}
thus bringing down the computational effort needed in the matrix
multiplication in the kriging predictor. 
Specifically, we note that in~\eqref{ridgereg} and \eqref{krigse2}, we invert two $m
\times m$ matrices, namely $\hat\bK$ and $\bS'\bD^{-1}\bS + \hat\bK^{-1}$,
with each an $O(m^3)$ operation -- instead of the $O(n^3)$ operation of
inverting one $n \times n$ matrix.
Both \eqref{ridgereg} and \eqref{krigse2} also require significantly fewer matrix operations: 
thus for large $n$ (and especially when $m<<n$), there are substantial 
gains in computational efficiency in using these approaches. 

The material developed in Sections~\ref{eff.model} and
Section~\ref{eff.predict} have presented the wherewithal to further
reduce the CPU time and memory required to perform fixed-rank
kriging. Because our simplifications are built on a setup with a
reduced set of basis functions, we call our developed methodology
``Reduced Basis Kriging''. In general, while reduced basis kriging has
the ability to bring down computational overhead, it should be noted
that the exact form of the basis functions and the covariance
structure also play a significant role on computational efficiency.
We next discuss specific 
implementations of reduced basis kriging that address these issues. 

\subsection{Issues in Implementation for Computational Efficiency} 
\subsubsection{Choice of Basis Functions}
\label{basis}
In general, the specific choice of basis functions is subjective since
it is difficult to estimate from data, however it is common to use a
multi-resolution smoothing function \citep{cressiejohannesson2008} 
which defines the $l$th resolution basis function as 
\begin{equation}
\label{B}
  S_{k(l)}(\bx) = \Psi \left( \frac{\|\bx - \bu_{k(l)}\|}{r_l} \right)
  \quad \forall\mbox{ }\bx \in D,
\end{equation}
where $\Psi(d)$ is the smoothing function. A commonly used $\Psi(d)$ 
is the local bisquare function defined as
\begin{equation}
  \label{bisquare}
	\Psi(d) = \left\{ 
	\begin{array}{l l}
		\left(1 - d^2 \right)^2  & \quad 0 \le d \le 1 \\
		0 & \quad d > 1 \\
	\end{array} \right. \\
\end{equation}
and $	r_l = b \hspace{1 mm} \mbox{min}\{\|\bu_{i(l)}-\bu_{j(l)}\| : j \neq i, 1 \le i,j \le m\}$.
We follow \citet{cressiejohannesson2008} and use the
multi-resolution local bisquare function (\ref{B}) together with (\ref{bisquare}) 
in our illustrations. In our paper, we have assumed that all distances 
are defined in terms of the Euclidean norm. Further, $r_l$ is a 
standardizing quantity that is known as the ``bandwidth'' of the basis
function where $b$ is some constant. 

The local bisquare function (\ref{bisquare}) sets any value equal to
zero where the distance between the location and the knot is greater
than $r_l$. This means that the matrices $\bS$ and $\bA$ are sparse as
long as the $r_l$ for each resolution is not too large. Recognition of
this aspect of $\bS$ and $\bA$ presents some computational advantages
because we can now utilize matrix operations and algorithms~\citep{davis2006} 
that have been specifically designed to exploit sparsity and can therefore 
handle larger matrices. Such algorithms are computationally advantageous
because they provide memory savings and allow for fast matrix
manipulations. This is because they can be compressed into
memory-reduced forms because only non-zero values and their
corresponding indexes require storage. There are savings in computer
processing time also because manipulations involving sparse matrices
utilize the sparsity to reduce the number of calculations
performed. In this paper, we used the \texttt{R} package
\texttt{Matrix} \citep{matrix2015} which uses functions from
the \texttt{LAPACK}~\citep{andersonetal99} and
\texttt{SuiteSparse}~\citep{davis2006} packages   
for fast matrix manipulations such as in efficient computation of  
QR-decomposition algorithms.  

Much of the computational advantage of using a sparse matrix lies in
the degree of sparsity that exists. Given that a local bisquare
function is used for the basis functions, the bandwidth constant ($b$)
is a crucial factor in determining the amount of computational
reduction we can expect. An exact estimate of $b$ may not be
necessary, but a suitable range for $b$ should be determined to ensure
fast computations and precise
predictions. In their work, \citet{cressiejohannesson2008} suggested a
bandwidth constant of 1.5. In this paper, we investigate the effects of
a range of values of $b$ through a series of simulation experiments.

\subsubsection{Choice of $\bK$}
\label{K}

The other a priori specification that can greatly affect prediction is
the form of $\bK$ which defines the spatial covariance between
knots. The Spatial Random Effects model~\citep{cressiejohannesson2008}
does not restrict $\bK$ to any particular form, 
however it is important to understand what effect a smoothing
function, in particular the local bisquare function, will have on
the covariance of the reduced spatial process. We discuss the effect
of $\bK$ by first specifying the covariance structure for the original
spatial process.   

There are several choices for the spatial covariance
structure~\citep{cressie1993}. For this particular paper, we will
assume that the original spatial 
covariance, $\bSigma_f$, can be modeled in terms of a covariance
matrix from the 
Mat\'{e}rn family \citep{stein1999}, which is a common choice for
spatial data. The Mat\'{e}rn covariance between $\bff(\bs_i)$ and
$\bff(\bs_j)$ is defined by 
$Cov[\bff(\bs_i), \bff(\bs_j)] = 	\frac{\rho}{2^{\nu - 1} \Gamma(\nu)}
\left(\frac{\|\bs_i -
          \bs_j\|}{\theta}\right)^{\nu}
        K_{\nu}\left(\frac{\|\bs_i -
          \bs_j\|}{\theta}\right)$, 
where $K_{\nu}$ is the
modified Bessel function of the second kind of order $\nu$, $\nu$ is
the smoothness parameter, $\rho$ is the sill parameter, and $\theta$
is the range parameter, with $\nu,\theta \in (0,\infty)$ and 
$\rho \in [0,\infty)$. 
Note that given the assumption of the
Euclidean norm, the Mat\'ern covariance function is second-order
stationary in our setup. 

It is paramount to understand how the fixed rank kriging
approximations model common forms of spatial variance. Thus, the form
of $\bK$ when $\bSigma_f$ is Mat\'{e}rn is of significant
importance. Although the exact form of $\bK$ is not easily recovered,
an empirical covariance structure can be obtained by
combining the QR-decomposition of $\bS$ with an eigenvalue decomposition
of $\bSigma$ \citep{higham2002}. Specifically, writing $\bSigma_f$ in
terms of its spectral decomposition $\bSigma_f  = \bP\bLambda
\bP'$. Write $\bP = [\bP_1\vdots\bP_2]$, where
$\bP_1$ contains the first $m$ columns of $\bP$,
and let $\bLambda_1$ be the diagonal matrix containing the $m$
eigenvalues corresponding to the eigenvectors in $\bP_1$. Then,
since $\bff(\bs) = \sum_{k=1}^{m}\bS_k(\bs)\ba(\bu_k)$, we have
that $\bSigma_f = \bS\bK\bS'$ and
\begin{equation}
\label{theoryK}
\bK = \bR_1^{-1} \bQ_1'
\bP_1 \bLambda_1 \bP_1' \bQ_1 (\bR_1^{-1})'. 
\end{equation}
\citet{katzfusscressie2011} noted how the usual form of
maximum likelihood estimation of $\bK$ was problematic because it
required maintaining a positive definite matrix through a numerical
optimizing technique. Thus, in addition to identifying a form of $\bK$
that will be computationally efficient and produce accurate
predictions, a numerically stable form is also important.  In the next
section, we investigate the behavior of $\bK$ to conclude a
reasonable covariance structure. We will also explore, through
simulation, the loss in accuracy that may be caused by using an
approximate but simpler model ($\bK\propto\bI$), and outline methods
for minimizing this loss by properly constructing the basis functions. 

We conclude this section briefly reiterating that our setup here (and in
our experiments) has assumed second order stationarity. In cases when
the assumption can not be sustained, tactics such as detrending and
median-polishing \citep{hoaglinmostellertukey1983} may be adopted to transform
the observations into a second order stationary process.

\section{Experimental Evaluations}
\label{experiments}
\subsection{Simulation Details}
We investigated the behavior of our reduced basis kriging methodology
through a series of simulation experiments using the local bisquare
basis function of (\ref{B}) and a Mat\'{e}rn
covariance structure for the original spatial process. 
For simplicity, we assume a Gaussian random spatial field with second 
order stationarity and no measurement error, $\sigma_{\epsilon}^2 = 0$.

We investigated the effect of fine-scale variation in our simulation
experiments by setting $\sigma_{\delta}^2$ to be equal to 0, 0.1, 0.25, and
0.4, while holding $\rho$ constant at unity and set $\bV_{\delta} = \bI$ for 
simplicity. The smoothness parameter
$\nu$, is most commonly set to be between 0.5 and 2, so we let $\nu$
be over the set of values in  $\{0.5, 1, 1.5, 2\}$. Note
that the range parameter, $\theta$, is dependent on $\nu$. Therefore
the value of $\theta$ was paired to $\nu$ so that the correlation
would be 0.2 between observations at a distance of 1/3 of the spatial
domain. Specifically, $\theta$ was set to be equal to 0.205, 0.137, 0.110,
and 0.095, respectively. 

The domain for the simulation experiments consisted of two grids ($50
\times 50$ or $200 \times 200$) with locations ranging from 0 to
1. Three hundred randomly selected locations out of the possible 2500 or 40000
represented the locations of the observations with varying degrees of
missing data. The selection of basis functions depends on a variety of
factors: these are the number of knots ($m$), the location of the knots,
the number of resolutions ($l$), and the bandwidth constant
($b$). These variables all need specification prior to implementation of
fixed rank kriging. Since the simulated spatial field is relatively small 
and without much complexity, using one resolution always produced 
predicted values with the lowest corresponding mean square prediction error.
Thus, the focus of the results in this paper will be on estimation and 
prediction using a single resolution, $l=1$. The values for $m$ and $b$, 
in contrast, are presented varied to test the effect on time and accuracy. 
   
A regular, triangular grid was chosen for the knot location after
preliminary results suggested that different space-filling designs for
the locations of the knots had an insignificant effect on prediction
when the observations were not clustered. The number of knots used
were $m = 23$, $m=77$ or $m=  175$, which represented a reduction of roughly
5\% to 50\% of $n$. \citet{cressiejohannesson2008} suggested a
bandwidth of 1.5 multiplied by the shortest distance between
knots. Thus, in our experiments, we tested our method for bandwidths
ranging from 0.5 to 2.5 in increments of 0.1 and  multiplied by the
shortest distance between knots.

\subsection{Covariance of the Reduced Spatial Process}
\begin{figure}
	\begin{center}
		\includegraphics[width=0.5\textwidth]{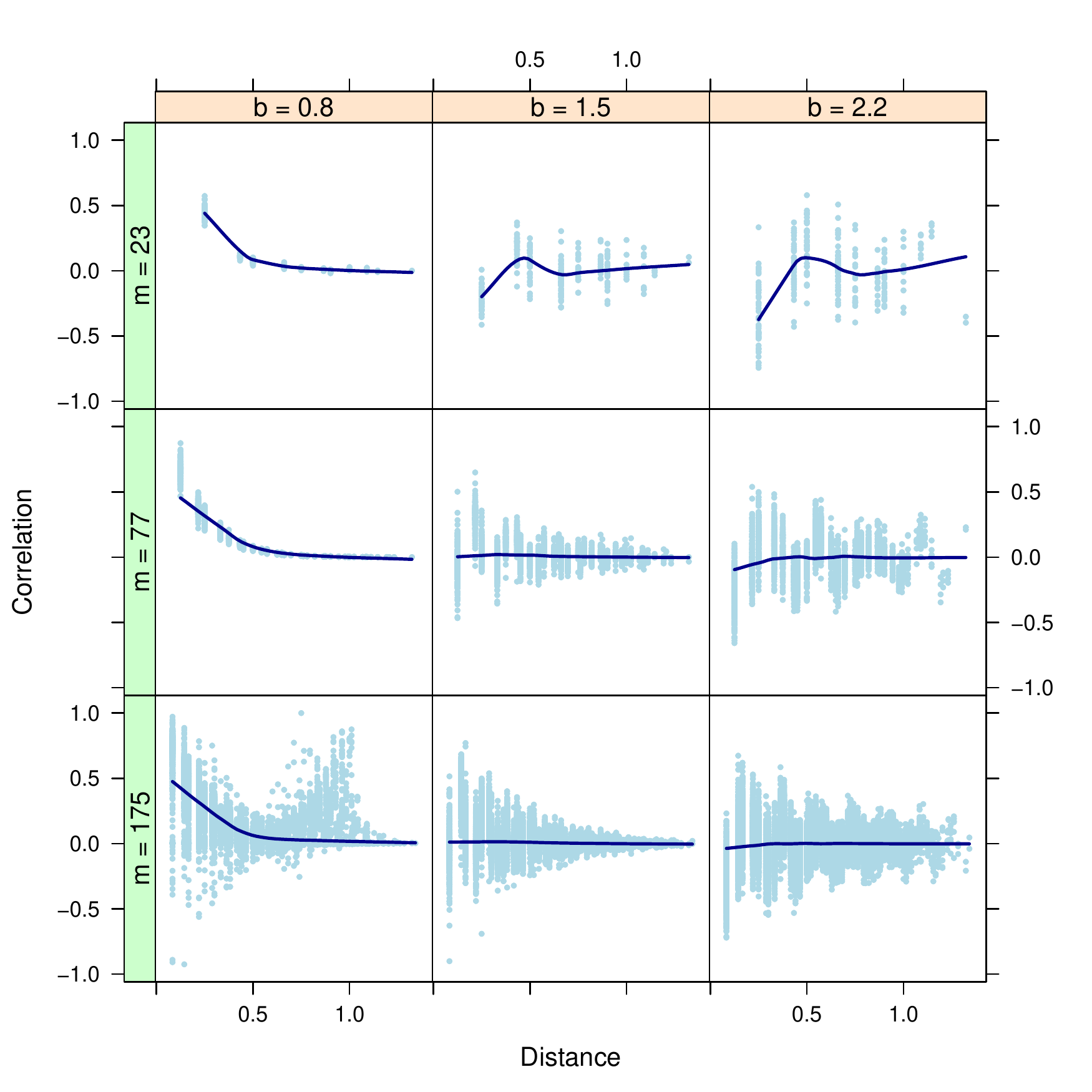}
	\end{center}
	\caption{Correlation between knots from a $50 \times 50$ grid plotted against distance for $\nu = 1$, $\theta = 0.137$, $\sigma^2=0.25$.}
	\label{Kform}
\end{figure}

Using the setup from the previous subsection and the method outlined
in (\ref{theoryK}), we simulated $\bK$ for
every combination of parameter values, number of knots, and bandwidth
constants and converted the covariance into a correlation. Figure \ref{Kform} 
provides an example of that correlation relative to distance (excluding the 
superfluous case where distance = 0) for all $m$ at select bandwidth 
constants on a $50 \times 50$ grid when $\nu = 1$, $\theta = 0.137$,
$\sigma^2=0.25$. (All other combinations of grid size, bandwidth
constant, and parameter values yielded figures of the same kind and
are omitted for brevity.)

The stationarity commonly visible in a Mat\'{e}rn covariance is
clearly weakened by the use of basis functions to describe the large
scale spatial structure. Some of the Mat\'{e}rn structure is
maintained when using lower bandwidth constants, particularly for
lower $m$ and a wave pattern is often visible for bandwidth constants
larger than one. The majority of the plots, however, suggest that
covariance structures such as a Mat\'{e}rn may be somewhat needlessly
and overly complex for 
describing the correlation between knots at varying distances and
could be simplified to a matrix proportional to the identity matrix. 
Choosing $\bK$ to be proportional to the identity matrix is particularly attractive because its
simplicity can provide huge computational gains while also increasing
the stability in the numerical optimization steps.

\subsection{Simulation Results}
One hundred Gaussian random fields were simulated using the \texttt{R}
package {\tt fields} \citep{fields2015} with a single resolution for
each set of parameter values. For each field, fine-scale errors were added
to each response at the 300 randomly selected locations, resulting in values
that represented the dataset. Then, we acquired  ML estimates and the
corresponding predicted values for each simulated field using three
different methods: by optimizing (\ref{llnew}) 
using an identity covariance for $\bK=\rho \bI$ and utilizing
(\ref{ridgereg}), by the E-M approach (\ref{M}) using the full form of
$\bK$, and by the E-M approach approximating $\bK$ with the identity
matrix, $\rho \bI$. As mentioned in Section~\ref{methodology}, these
computations were done using sparse matrix 
methods \citep{davis2006} implemented as per the \texttt{R} package
``Matrix'' \citep{matrix2015}. 
 
\begin{figure*}
	\mbox{
	\subfigure[$m=23$]{\includegraphics[width=0.5\textwidth]{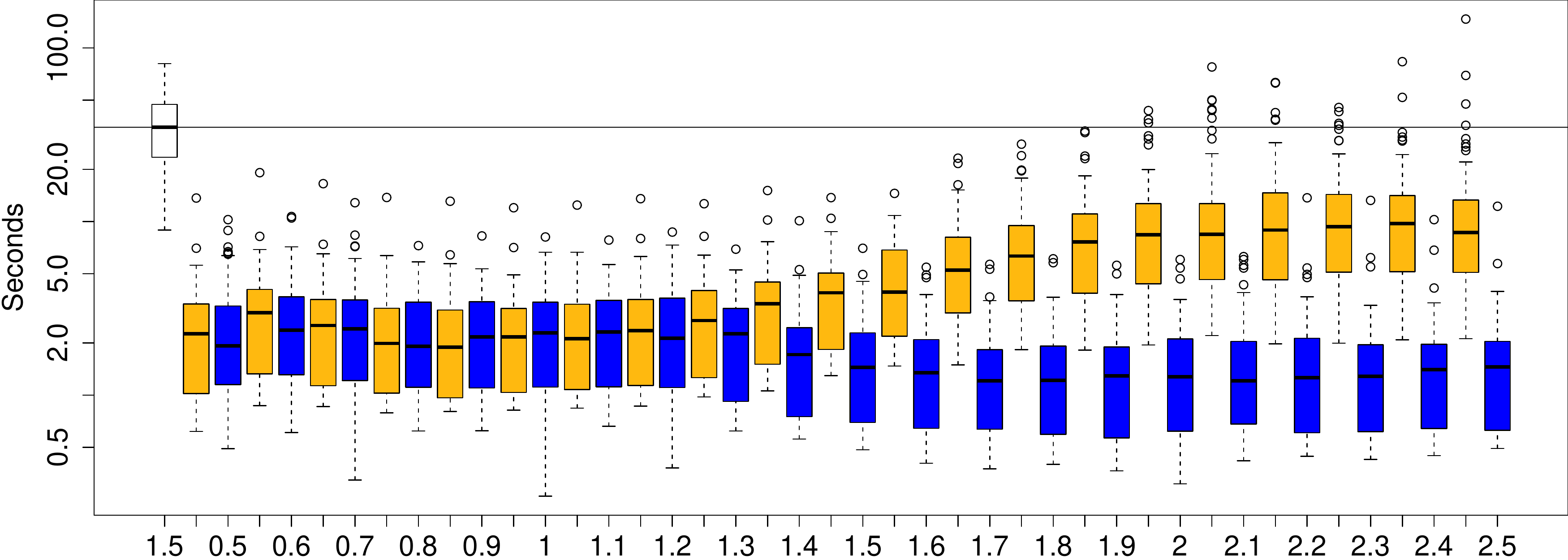}}
		\subfigure[$m=23$]{\includegraphics[width=.5\textwidth]{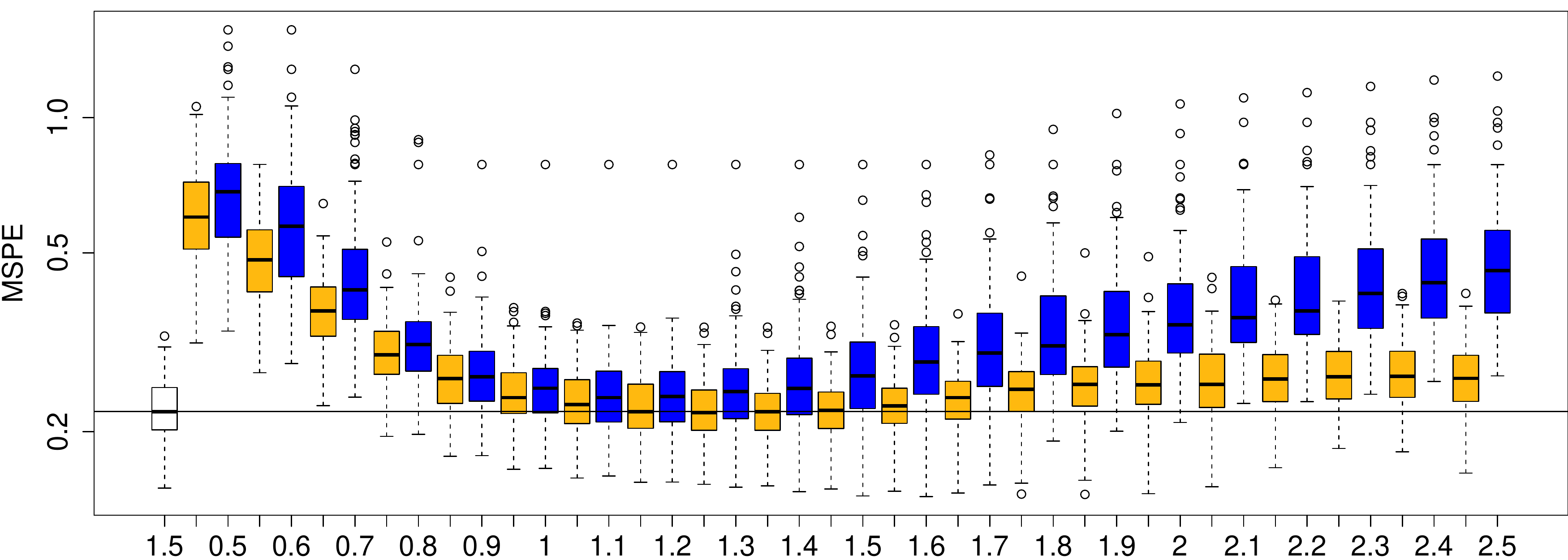}}}
		\mbox{
		\subfigure[$m=77$]{\label{time77} \includegraphics[width=.5\textwidth]{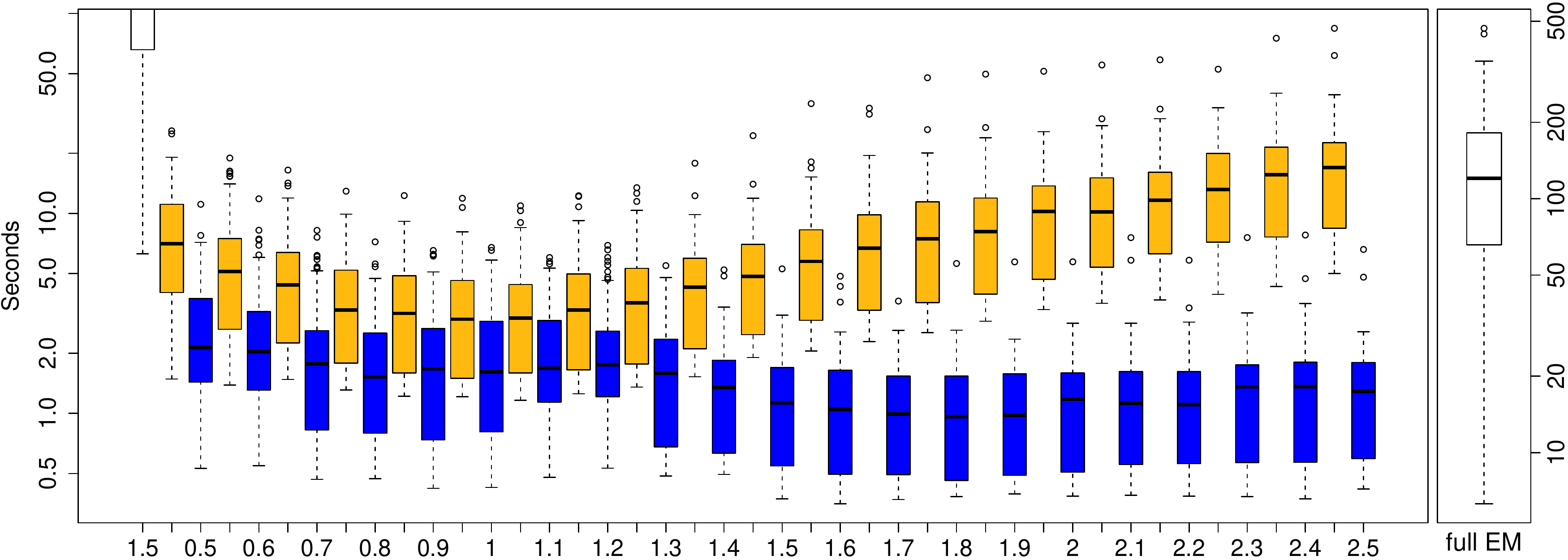}}
		\subfigure[$m=77$]{\includegraphics[width=.5\textwidth=0.5\textwidth]{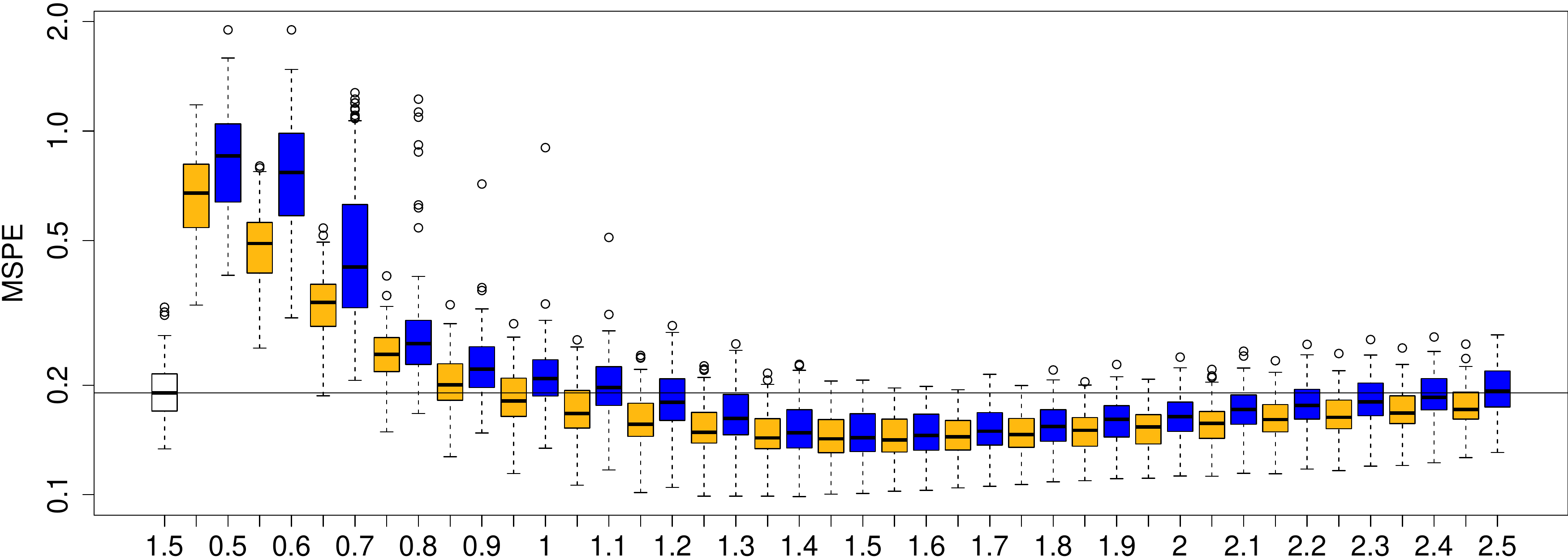}}}
		\mbox{
		\subfigure[$m=175$]{\includegraphics[width=.5\textwidth]{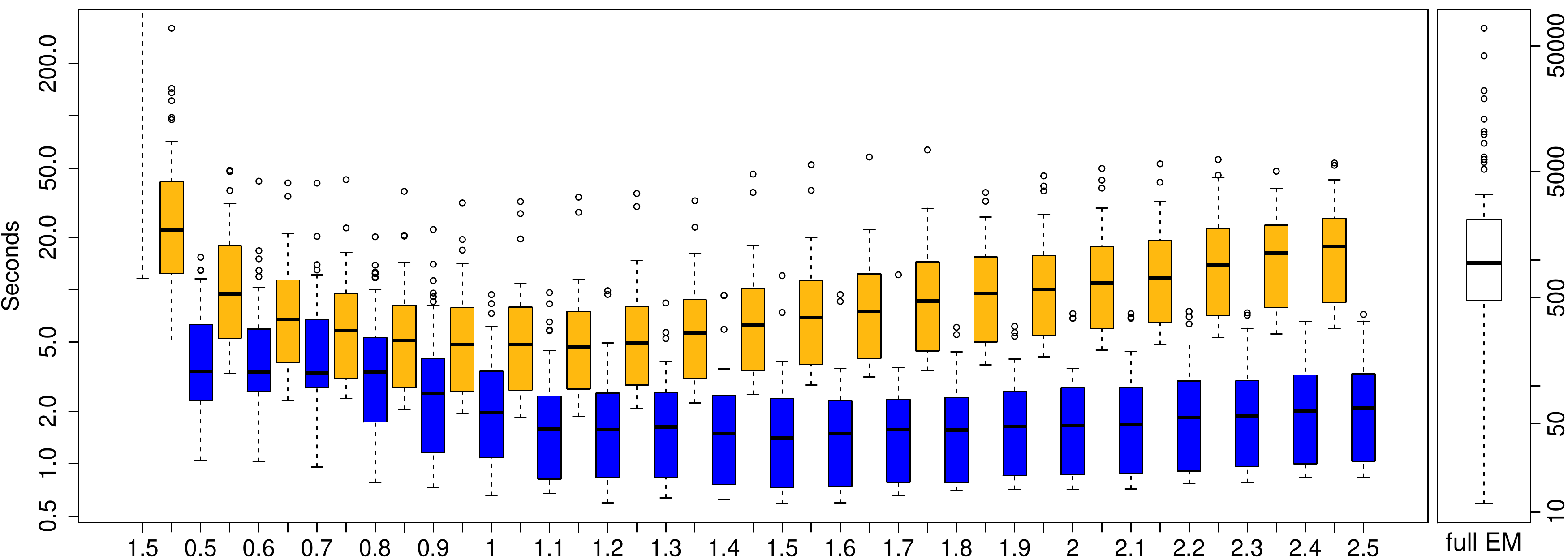}}		
		\subfigure[$m=175$]{\includegraphics[width=.5\textwidth]{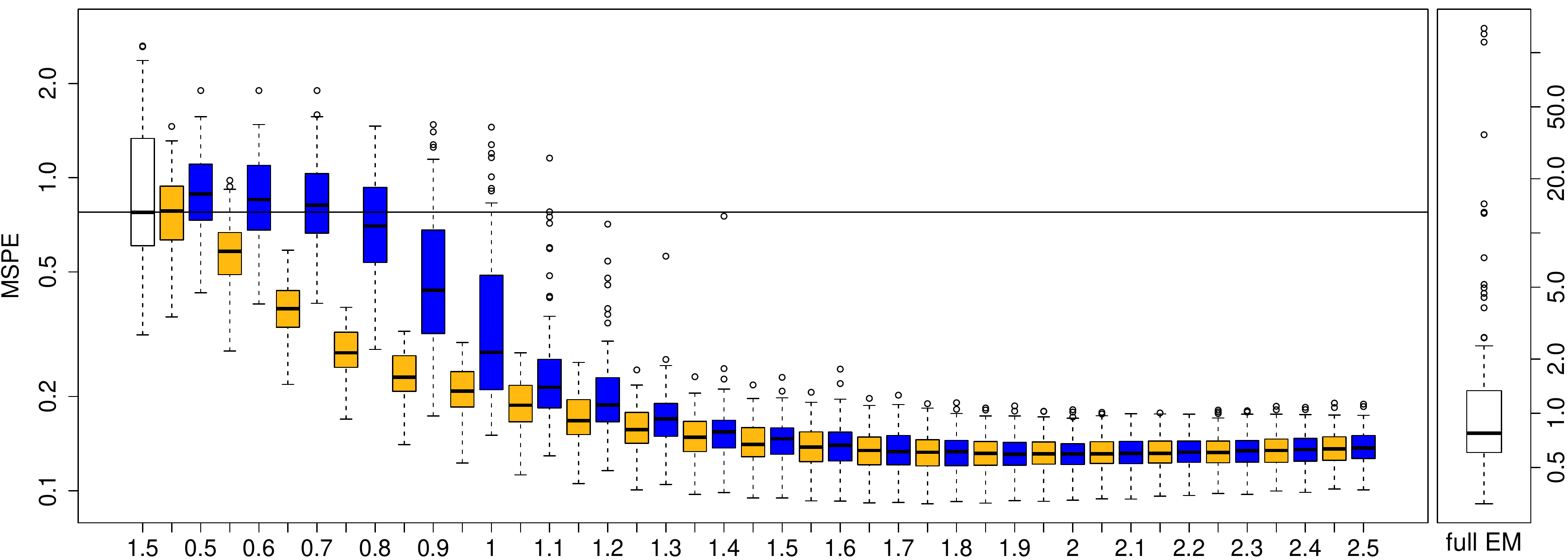}}}
	\caption{Seconds (left panel) and Mean Squared Prediction
          Errors (MSPEs, right panel) for varying bandwidths constants
          (x-axis) and number of knots ($m$) for all three estimation
          methods on a $50 \times 50$ grid when $\nu=1$ and
          $\sigma^2=.25$. The white boxplot is the E-M method with a
          full covariance for $K$ for fixed $b=1.5$.} 
	\label{time}
\end{figure*}

The results of our experiments indicated that the simplicity of the
identity covariance for $\bK$ resulted in an  expected gain in
computational efficiency. To illustrate this gain, the time required to 
iterate to convergence and obtain predicted values for each combination of
inputs was recorded. This value combines the cost of inverting a
matrix with a penalty for covariance structures that require
additional iterations to meet convergence. Example plots of the
distributions of seconds on the log scale for one resolution and
varying number of knots, bandwidth constants, and estimation methods
on a $50 \times 50$ grid when $\nu=1$ and $\sigma_{\delta}^2=.25$ are provided
in Figure \ref{time} (left panel). These boxplots display all three estimation
methods: reduced basis kriging (blue), the E-M with an identity
covariance for $\bK$ (gold), and the E-M with a full covariance for $\bK$
for fixed $b=1.5$ (white). The horizontal line represents the median
seconds for the E-M approach using a full $\bK$. For clarity of presentation, 
the distribution of log seconds for the E-M approach using the full form of 
$\bK$ are presented separately when the maximum is significantly beyond the 
range of the corresponding results for the identity form of $\bK$, such as in 
Figure \ref{time77}. These example plots represent the patterns seen 
throughout all input combinations. Additional figures (see Figures S-1 
through S-16) in the supplementary material also support the following 
conclusions.

Fewer knots resulted in less computation time for the E-M approach and
a bandwidth constant near unity provided the optimal level of sparsity in
$S$ in terms of CPU time. An interesting difference between the E-M
approach and reduced basis kriging is the robustness of reduced basis
kriging to increasing $m$ or a poor choice of bandwidth
constant. Reduced basis kriging is efficient regardless of the
specifics of the basis functions used and is at least as efficient as
the E-M approach with the identity covariance for any choice of
$b$. The E-M approach is consistently more computationally expensive
for larger values of $b$. 

Although reduced basis kriging improves computational efficiency, the 
additional compression of the data could increase prediction error. To
assess accuracy the mean square prediction error (MSPE) was computed relative 
the true simulated field. Figure~\ref{time} (right panel) also provides the
distributions of MSPE on a $50 \times 50$ grid when $\nu=1$ and 
$\sigma_{\delta}^2=.25$. These boxplots are organized as before, with the
horizontal line representing the median MSPE for the E-M approach
using a full $\bK$. Once again, these example plots represent 
the patterns seen throughout all input combinations. 

The distributions of MSPE show that for a reasonable $b$, the identity
covariance structure is at least as accurate in terms of prediction as
the full covariance structure. In fact, as $m$ approaches $n$, the
identity covariance performs substantially better than the full
covariance with regards to both median and maximum MSPE. This suggests
that added complexity to $\bK$ is unnecessary when our focus is on
prediction. Considering its simplicity and the speed of estimation,
proceeding sections will focus only on spatial prediction using the
identity covariance model. 

Using the identity covariance for $\bK$, reduced basis kriging and the
E-M approach are comparable for choices of bandwidth constant where
MSPE is minimized, however an unreasonable $b$ favors the E-M approach
in terms of MSPE. This is opposite of computational efficiency, in
which reduced basis kriging was more robust to poor choices of
$b$. Fortunately, the substantial reduction in computation achieved by
using the identity covariance coupled with parallel computing allow
for the estimation of multiple models (varying $b$ and the number of
resolutions). A model selection approach can thus be implemented by 
minimizing MSPE through cross-validation. However, as a continuous variable, 
the domain of $b$ is infinite and thus a model selection approach is less 
than ideal. Selection of an appropriate value for $b$ and the correct number 
of resolutions are paramount to producing accurate predictions and will 
be explored in further work.

Provided a reasonable bandwidth and number of resolutions can be
obtained, reduced basis kriging and the E-M approach using the
identity covariance produce very comparable distributions of MSPE. The
major advantage to reduced basis kriging is that the memory necessary
to perform each iteration of the estimation process is greatly reduced
by concentrating the data from length $n$ to length $m$. Distributions
of seconds for bandwidth constants that minimize MSPE ($b \approx 1.3$
for $m=23$, $b \approx 1.5$ for $m=77$, and $b \approx 2$ for $m=175$)
illustrate a distinct advantage to using reduced basis kriging. All
patterns visible in Figure \ref{time} were also consistent for all other 
combinations of parameter and grid values, which can
be found in the supplementary materials. 

In summary, the results of our experiments show 
that the identity form of $\bK$ is optimal in terms of minimizing 
computational cost without sacrificing accurate prediction, provided 
that a reasonable bandwidth constant and resolution are selected. The massive 
computational gains achieved by simplifying the covariance between knots 
allows for the estimation of numerous ``models'', leading us to a model 
selection approach that is achieved by minimizing the mean kriging standard 
error, $\hat\bsigma_k(\bs_0)$. Additional and robust computational gains 
are also achieved by reduced basis kriging with a minimal increase in MSPE. 
We now apply our methodology to the National Climate Data Center (NCDC) 
data of monthly temperatures recorded across the continental United States 
of America.

\section{Predicting temperatures over the continental US} 
\label{application}
The Cooperative Observer Program (COOP), formally established in 1890,
is the nation's largest and oldest 
weather and climate observing network \citep{coop2000}, consisting of 
over 11,700 volunteer citizens and institutions observing and 
reporting weather information on a 24-hour basis. The data is archived 
at the US National Climate Data Center (NCDC) and available online at 
{\em http://www.image.ucar.edu/Data/US.monthly.met/}. 

A basic summary in climate science is provided by mean temperature and
precipitation fields on a regular grid 
\citep{johnsnychkakitteldaly2003}. One important application is to
compare these fields from observational data to those simulated by
climate derived models. For our example, we consider mean temperatures in
April in 1990 observed over the entire contiguous United States of America. 
Thus kriging the observed field is important in this setting.

For our example, we use mean temperatures in April in 1990 observed
over the entire continental United States. The daily minimum and
maximum temperatures were observed at 5030 locations across the United
States and the mean monthly minimum and maximum temperatures were
calculated. To obtain an overall monthly average, the mean monthly
minimum and maximum temperatures were averaged together. This average
will be our ``monthly mean'' temperature. 
 
Under the normal 
kriging method, prediction using the entire U.S. observational record
would require a Cholesky decomposition on a matrix of dimension
$5030\times 5030$, which can be computationally prohibitive, both in
terms of CPU time and memory.  Additionally, obtaining kriging
predictions generally requires the use of an iterative process to
estimate the parameter 
values of the model, meaning that this expensive calculation will need
to be repeated.  Thus, the normal kriging approach clearly becomes 
impractical for large data sets, and we are led to 
approaches such as reduced basis kriging.  

\subsection{Data Preprocessing}

Intrinsic stationarity is not a reasonable assumption for this
data. Temperature is directly affected by latitude, longitude, 
and elevation and so the spatial field does not have a constant mean. A
simple approach to adjust for the non-stationarity due to these
factors is to use the additive model $\by(\bs,\bh,\bu) = \bg_{Elev}(\bh(\bs)) +
\bg_{Lat}(\bs) + \bg_{Lon}(\bs) + \bS(\bs,\bu)\bseta(\bu) + \bdelta(\bs) + \bepsilon(\bs)$ where $\bh(\bs)$ is the elevation
at location $\bs$, $\bu$ are the knots, $\bseta(\bu) \sim N(\bzero, \rho\bI)$, 
$\bdelta(\bs) \sim N(\bzero, \sigma_{\delta}^2\bI)$ and 
$\bepsilon(\bs) \sim N(\bzero, \sigma_{\epsilon}^2\bI)$, 
and $\bseta(\bu)$, $\bdelta(\bs)$, and $\bepsilon(\bs)$ are mutually independent.

\begin{figure*}
\mbox{
\subfigure[Cubic regression spline]{\includegraphics[width=.33\textwidth]{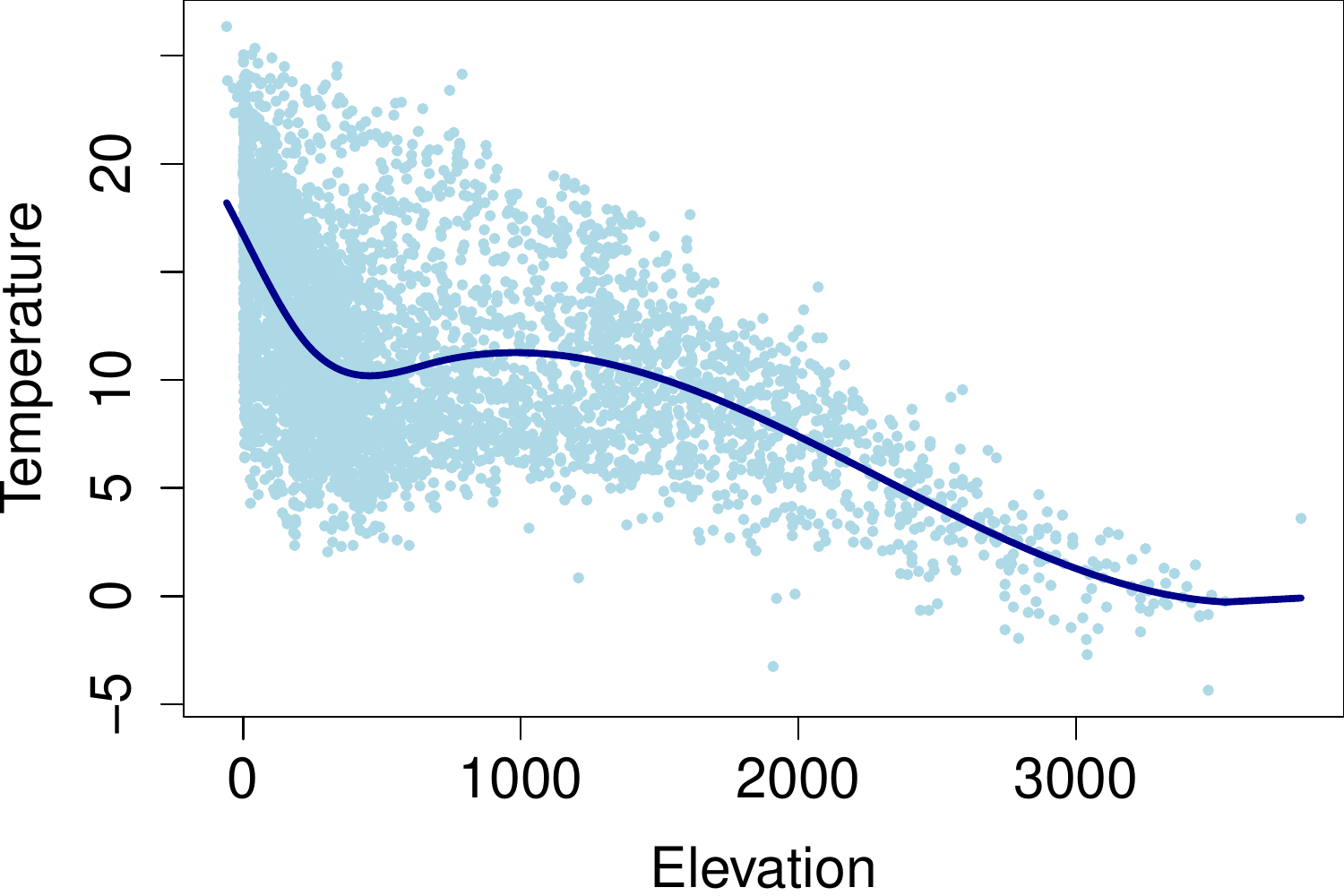}}
		\subfigure[Quadratic regression spline]{\includegraphics[width=.33\textwidth]{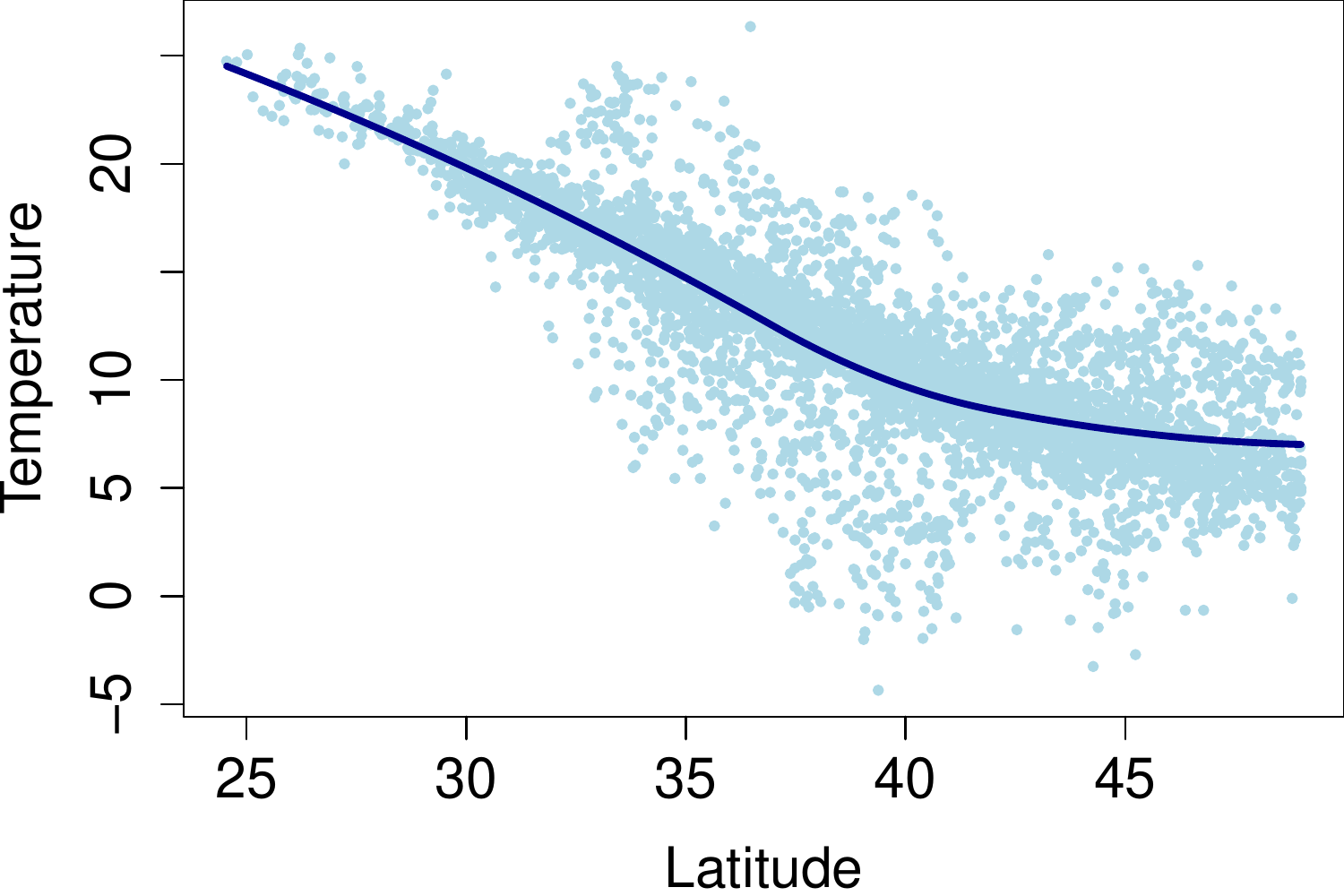}}
		\subfigure[Cubic regression spline]{\includegraphics[width=.33\textwidth]{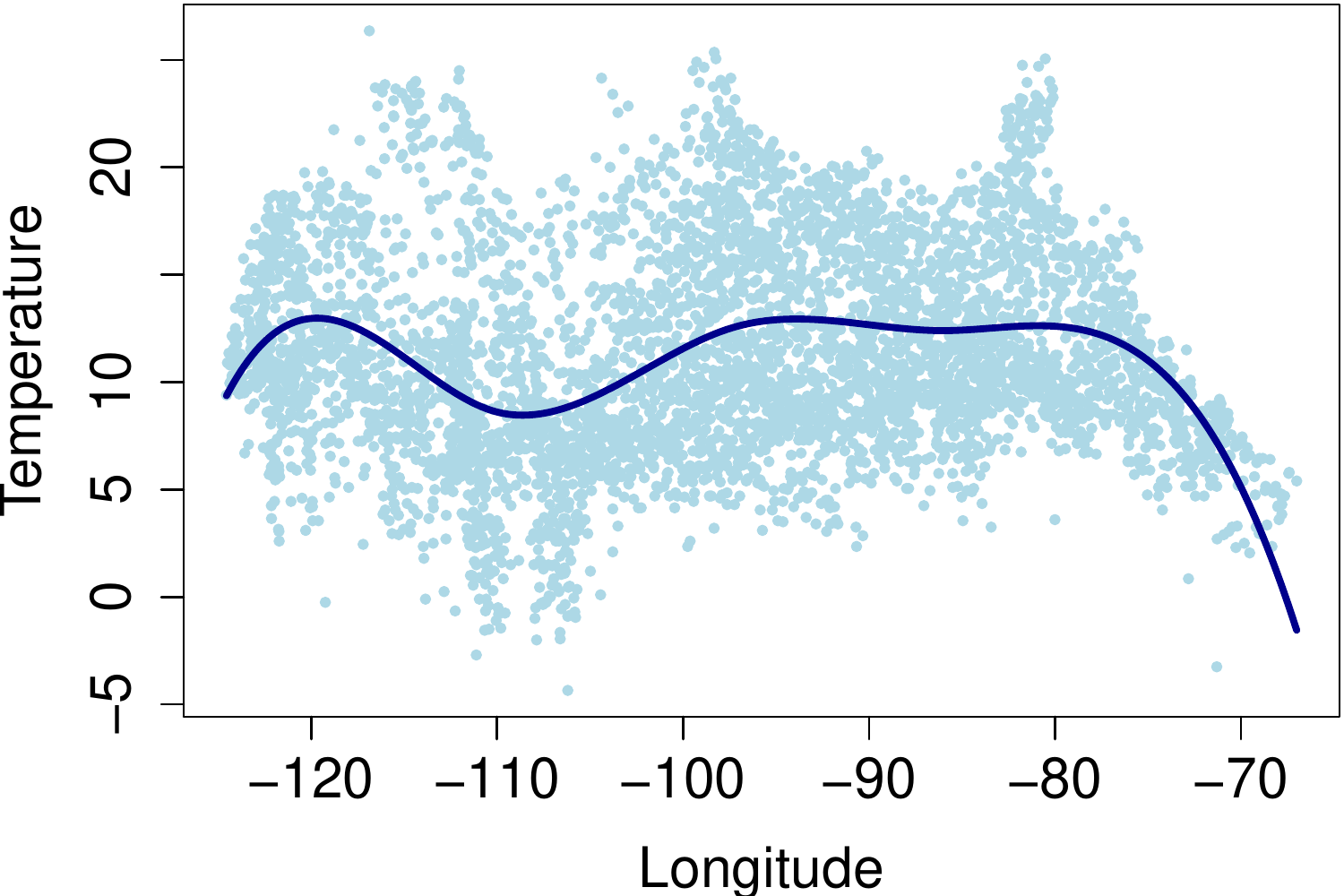}}}
\caption{Monthly mean temperature plotted against covariates with regression spline.}
	\label{splines}
\end{figure*}

We estimated $\bg_{Elev}(\bh(\bs))$ by a cubic regression spline
with five degrees of freedom to the data, $\bg_{Lat}(\bs)$ in terms of a 
quadratic regression spline with 4 degrees of freedom, and $\bg_{Lon}(\bs)$ 
using a cubic regression spline with 6 degrees of freedom, all fit to the 
dataset by AIC. These splines are shown in Figure \ref{splines} 
and demonstrate that the majority of the mean structure is captured.

\subsection{Results and Analysis}
We begin with the linear equation, $\by = \bX\balpha + \bS\bseta + \bdelta + \bepsilon$ 
and define the QR-decompositions $\bX = [\bQ_{\bX1}, \bQ_{\bX2}][\bR_{\bX},\bzero]'$ 
and $\tilde{\bS} = \bQ_1 \bR_1$.
Given this additive model, a QR-decomposition of the basis matrix for
the polynomial splines was used to remove the trend in the data by
multiplying through by the portion of $\bQ_{\bX}$ that is orthogonal to $\bX$,
as shown in equation (\ref{REML}). Thus, the effects of location and 
elevation are removed from the observations. Reduced basis kriging was
subsequently performed on the detrended data set, $\tilde{\by}$, and
least-squares estimates of the regression splines were used to add the
effects of elevation and latitude back into the response following 
reduced basis kriging.

\begin{eqnarray*}
	\by &=& \bX\balpha + \bS\bseta + \bdelta + \bepsilon \nonumber \\
	\bX &=& \left[ \begin{array}{cc} \bQ_{\bX1} & \bQ_{\bX2} \end{array} \right]
	\left[ \begin{array}{c} \bR_{\bX} \\ \bzero \end{array} \right] \nonumber \\
	\end{eqnarray*}	
Then, 
\begin{eqnarray}
\bQ_{\bX2}' \by & = \bQ_{\bX2}' \bX\balpha + \bQ_{\bX2}' \bS\bseta + \bQ_{\bX2}' (\bdelta + \bepsilon)\quad\qquad\nonumber\\
	    \Rightarrow \tilde{\by} &= \tilde{\bS}\bseta + \tilde{\bepsilon},\hfill \quad \mbox{ where }
	\tilde{\bS} = \bQ_1 \bR_1, \qquad\quad\qquad\nonumber\\
	\mbox{and } \bQ_1' \tilde{\by} &= \bQ_1'\tilde{\bS}\bseta + \bQ_1'\tilde{\bepsilon} 
\quad	       \Rightarrow\quad \by^\ast  = \bR_1\bseta + \bepsilon^\ast \qquad\label{REML}
\end{eqnarray}

\begin{figure*}
\mbox{
\subfigure[Observations] 		{\includegraphics[width=.5\textwidth]{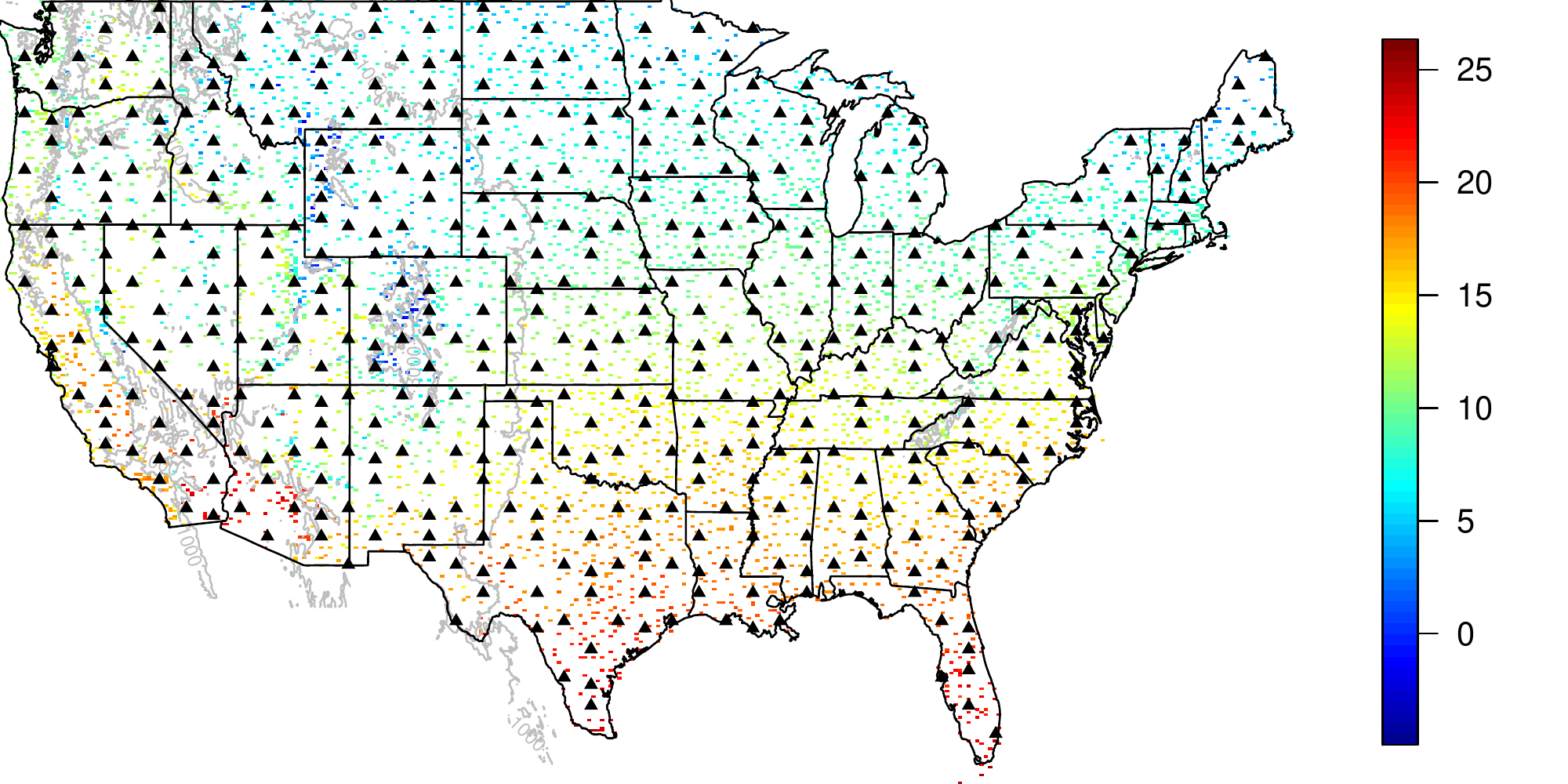}\label{USdata}}
		\subfigure[Kriging estimates only] {\includegraphics[width=.5\textwidth]{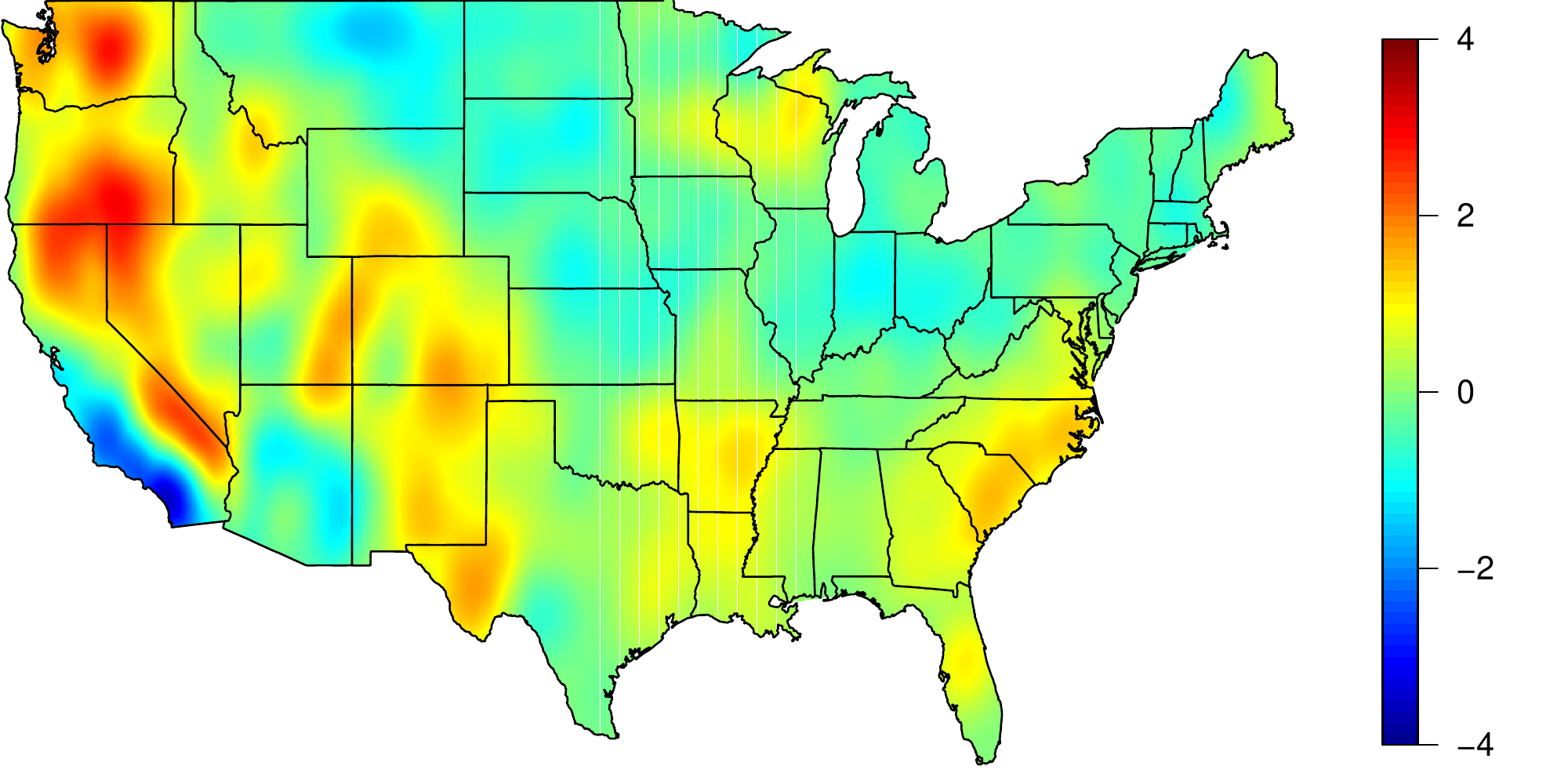}\label{USrr}}}
		\mbox{
		\subfigure[Kriging estimates plus covariates] {\includegraphics[width=.5\textwidth]{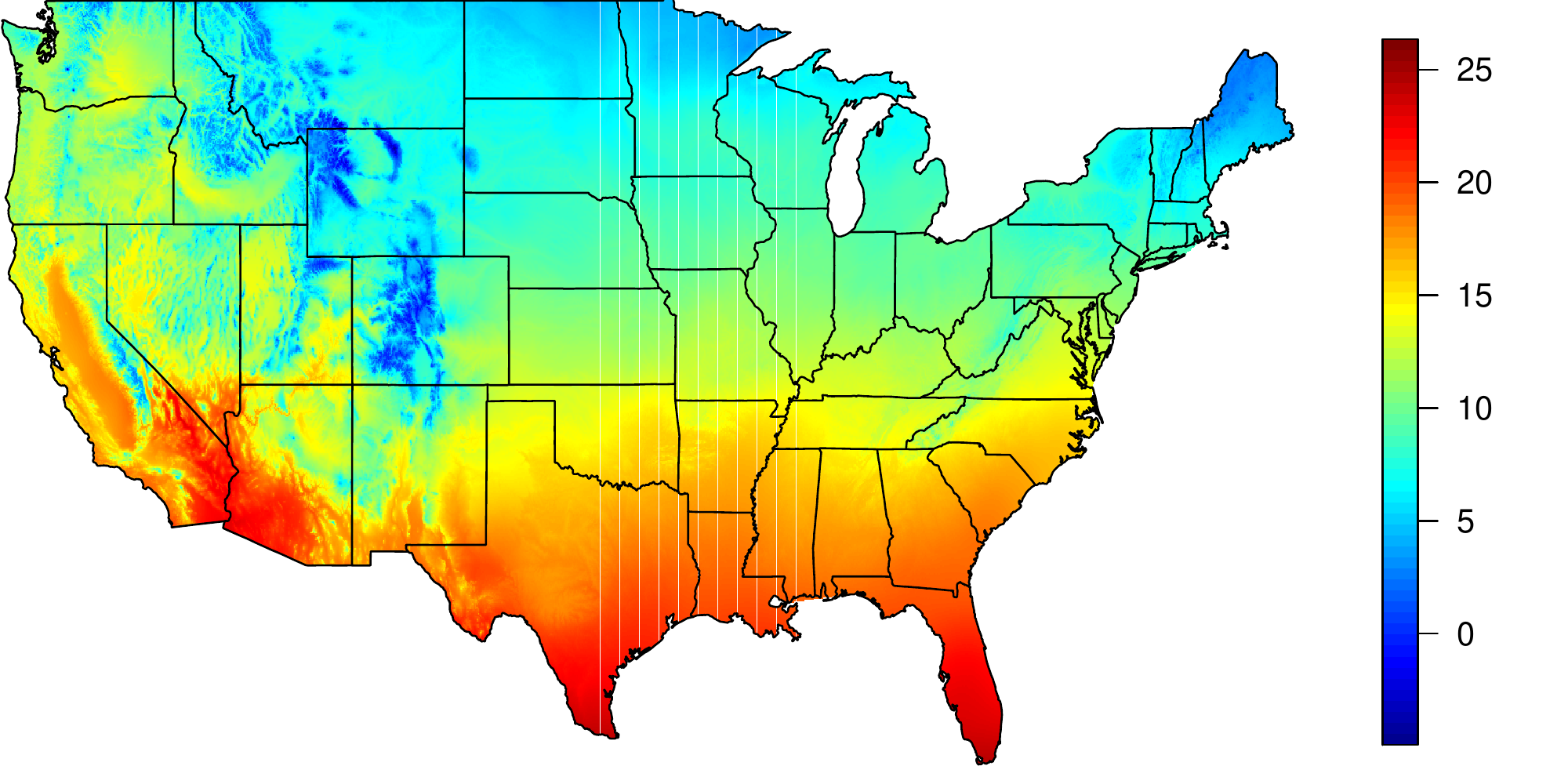}\label{USadd}}
		\subfigure[Kriging standard errors] {\includegraphics[width=.5\textwidth]{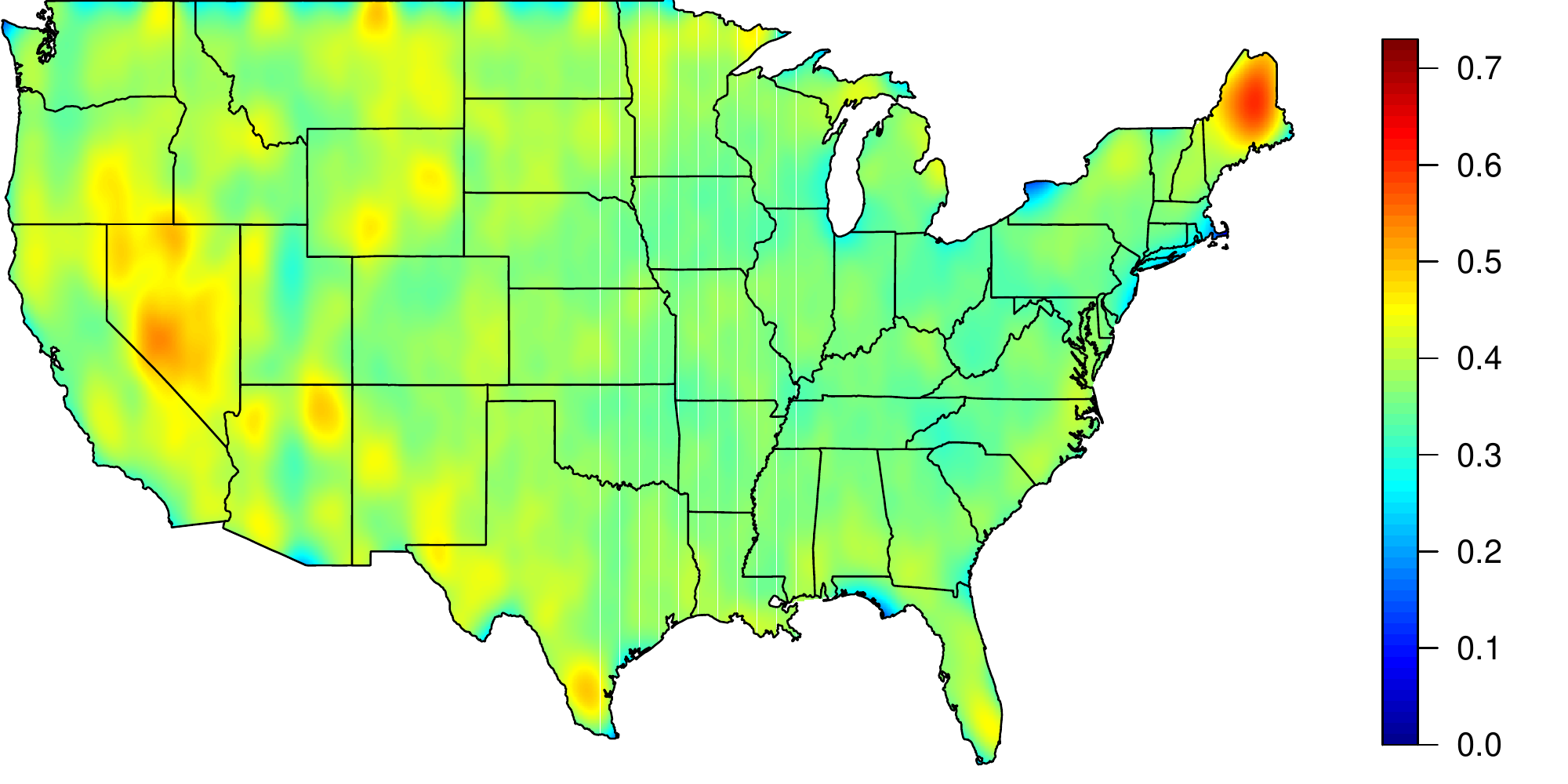}\label{USsigk}}}
	\caption{Mean temperatures in the US in April 1990; (a)
          displays the 5030 observations using 501 knots (black
          triangles) at one resolution, (b) depicts the
          reduced basis kriging estimates only, and (c) shows the sum
          of the spline regression predicted values for elevation, 
          latitude, and longitude with the kriging estimates. 
          (d) shows the kriging standard errors.} 
	\label{USresults}
\end{figure*}

The knots ($m=501$) were selected on a regular triangular grid within
the U.S. and tested on three levels of resolution. Basis functions with a 
single resolution and a bandwidth constant of $b=1.6$ fit the best based 
on minimizing the mean kriging standard error. When run on a
standard laptop with 6MB of RAM and a 64-bit Intel dual-core i5 processor, 
estimation was complete in 4.17 seconds and the entire process of cleaning 
the data, defining the matrix of basis functions, detrending, 
estimation, prediction, and plotting results took around 3 minutes. 
Figure \ref{USresults} summarizes the results. 

From the figure, it is easy to see that elevation, latitude, and longitude 
are clearly shown to be useful predictors. The kriging estimates capture 
the additional warmth of states northwest of Texas, despite their high 
altitude, and the temperature difference between western and eastern sides 
of the Appalachian Mountains. In addition, the mild climate of the Pacific 
coast line is  by achieved by negative kriging values in southern California 
and positive values in northern California, Oregon, and Washington. 
The kriging standard errors show relatively uniform variability, with 
higher variability in areas where weather stations are scarce, such as 
Nevada and Maine. There is also less variability in New York along the Erie 
and Ontario lake coast lines and in other areas that are along borders of 
the US. The systematic pattern visible in the kriging errors can be attributed 
to using basis functions on a regular grid. In particular, the alternating 
pattern visible along the US-Canada border is a result of the placement of
regularly spaced knots near the border of the domain.

\section{Discussion}
\label{discussion}
This paper presents a method for performing spatial prediction efficiently and effectively when data 
sets are large. This is accomplished by first approximating the original spatial process using the 
spatial random effects model which separates the spatial covariance into a linear combination of basis 
functions and a reduced rank spatial process. With the approximate spatial process, a QR-decomposition 
is used to reduce the dimensionality of the covariance matrix that is then solved for maximum likelihood 
estimation. Finally, an alternative Sherman-Morrison-Woodbury formula is invoked to obtain the predicted 
values over a spatial grid with increased efficiency.

Using reduced basis kriging as opposed to the E-M approach requires less memory and, thus, is beneficial when implementing kriging on large data sets. Also, in using basis functions to interpolate from the original observations to the knots, much of the spatial dependence between locations becomes absorbed in the basis functions when the bandwidth is relative large. In these situations and when the primary goal is prediction, a simplistic covariance between knots ($K$) can be sufficient to achieve high accuracy.
Consequently, it is essential that the basis functions are properly constructed, that is to say, the bandwidth and number of resolutions are carefully selected. Given the substantial computational gains achieved by a simplified covariance structure and the ability to utilize parallel processing, optimal bandwidth and number of resolutions can be obtained through model selection characterized by minimizing mean square prediction error through cross-validation. 
Investigation into the proper form of the basis functions is ongoing.

Although reduced basis kriging should be applicable to any Gaussian random field with intrinsic stationarity, the previous conclusions pertain to when the basis functions are local bisquare functions and only applies to situations where a Mat\'{e}rn covariance structure could be assumed on the original spatial process.
Reduced basis kriging provides an algorithm for simplifying the process down to the aspects essential for accurate interpolation while maintaining the advantage of quantifying uncertainty gained through kriging. However, the added flexibility of the E-M approach is preferred when estimation of the covariance between knots is of importance.
Often, when data is collected over space, it is also defined by a temporal component. This work focuses on predictions for spatial data at a fixed point in time, so a spatio-temporal model would be required for more complex data.

There are several topics for future research. As mentioned, the bandwidth constant and the number of resolutions is currently being investigated. Optimal knot placement is another aspect of basis function identification that is of interest. Simulations were attempted using both a regular grid and a stratified sampling technique, without an obvious difference in performance, but these results apply only to randomly located observations and thus a more rigorous analysis would be required to come to any definite conclusions. Alternative basis functions could possibly be used to adjust for directional dependence. Recent work by \citet{katzfuss2016} which allows for flexible basis functions shows promise in this area.
Also, considering that the accuracy of these methods were quantified by means of mean square prediction error, another area for future work would be in investigating the effects of this criteria on kriging predictions errors. Thus, we see that while we have made contributions, there is additional work to be done.

\section*{Acknowledgment}
 This research was supported, in part, by National Science Foundation
(NSF) grant DMS-0707069, and by NSF grant DMS-CAREER-0437555. The
National Center for Atmospheric Research is managed by the University
Corporation for Atmospheric Research under the sponsorship of the
NSF. An earlier version of this manuscript won Karl Pazdernik the 2012
Joint Statistical Meetings (JSM) Student Paper competition award
sponsored by the American Statistical Association (ASA) Sections on
Statistical Computing and Statistical Graphics.

\section{Supplementary Materials}

\begin{description}
	\item[Additional Investigations:] Figures S-1 through S-16 are
          in the supplementary materials archive of the journal
          website.
 	\item[R-code:] R-code used to perform the estimation and prediction methods described in the article is included.
\end{description}

\bibliographystyle{apalike}
\bibliography{References}

\end{document}


\title{Supplement to ``Reduced Basis Kriging for Massive Spatial Fields''} 
\author[1]{Karl Pazdernik}
\author[1]{Ranjan Maitra}
\author[2]{Douglas Nychka}
\author[2]{Stephan Sain}
\affil[1]{Department of Statistics and Statistical Laboratory, Iowa State University}
\affil[2]{Institute for Mathematics Applied to Geosciences, National Center for Atmospheric Research}
\date{}
\maketitle

\section{Additional Experimental Evaluations}

A simulation study was conducted to assess the value of reduced basis kriging 
in comparison to obtaining maximum likelihood estimates of the Spatial Random 
Effects model via the E-M algorithm. One hundred 2-dimensional Gaussian random 
fields were simulated using the \texttt{R} package {\tt fields} 
\citep{fields2015} on either a $50 \times 50$ or $200 \times 200$ grid. The 
spatial process at the locations of the observations were assumed to follow a 
Gaussian distribution with a Mat\'ern covariance structure. The Mat\'ern 
parameter values $\nu$ and $\theta$ were varied in unison to be such that $(\nu,\theta) \in
\{(0.5,0.205), (1,0.137), (1.5,0.110), (2,0.095)\}$, holding $\rho = 1$ constant. 
For each field, simulated measurement errors ($\sigma^2 = \{0, 0.1, 0.25, 0.4\}$) 
were added to each response at 300 randomly selected locations, resulting in 
values that represented data. 

ML estimates and the corresponding predicted values for each simulated field 
were then computed using three different methods: reduced basis kriging using an 
identity covariance for $\bK=\delta \bI$ (blue), by the E-M approach approximating 
$\bK$ with the identity matrix, $\delta \bI$ (gold), and by the E-M approach using 
the full form of $\bK$ with constant $b=1.5$ (white). Basis functions with a single resolution and either 
23, 77, or 175 knots were used during estimation and prediction, with the bandwidth 
constant allowed to vary from 0.5 to 2.5 by 0.1 for both methods that use the 
identity covariance. As mentioned in the full paper, these computations were 
done using sparse matrix methods \citep{davis2006} implemented as per the 
\texttt{R} package ``Matrix'' \citep{matrix2015}.

\begin{figure}[H]
	\begin{center}
		\subfigure[$50 \times 50$ grid with $m=23$]{\includegraphics[scale=.23]{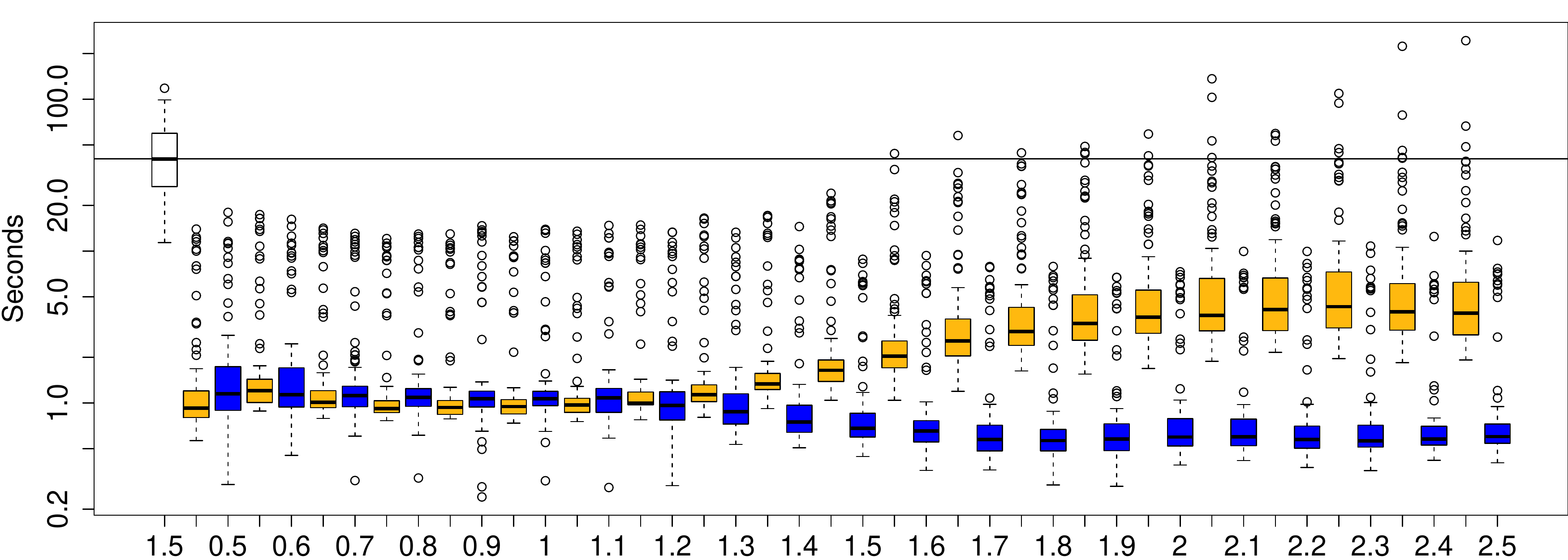}}
		\subfigure[$50 \times 50$ grid with $m=23$]{\includegraphics[scale=.23]{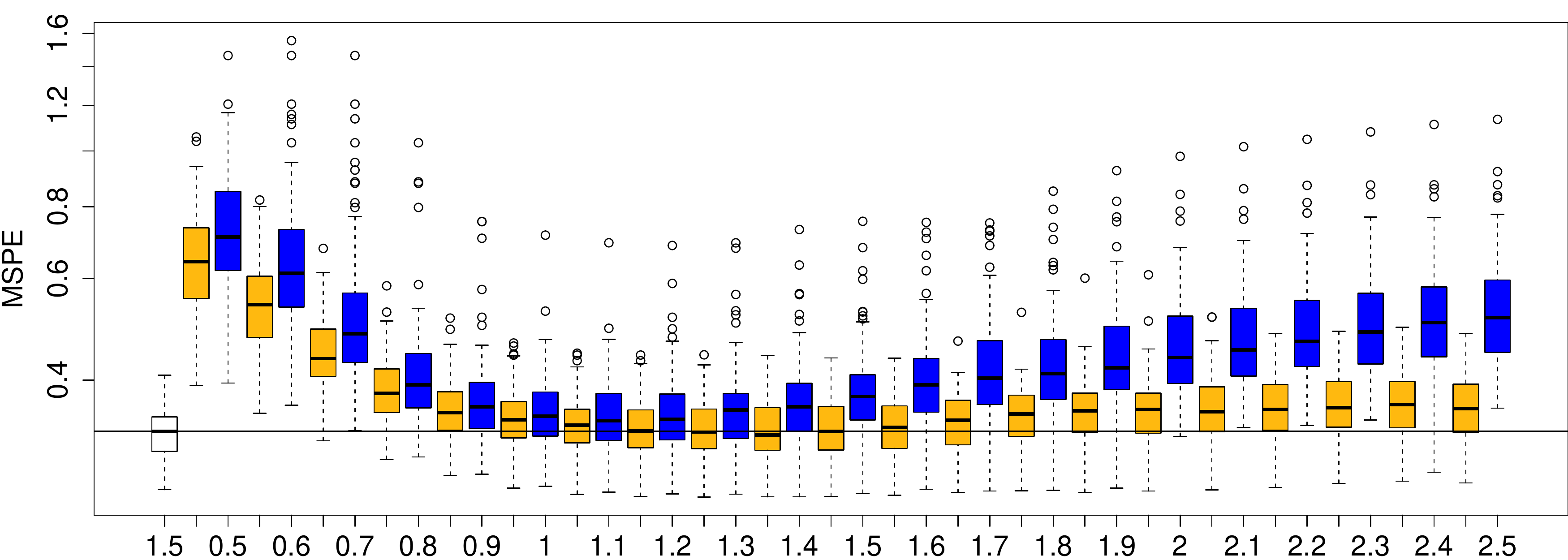}}
		\subfigure[$50 \times 50$ grid with $m=77$]{\label{time77} \includegraphics[scale=.23]{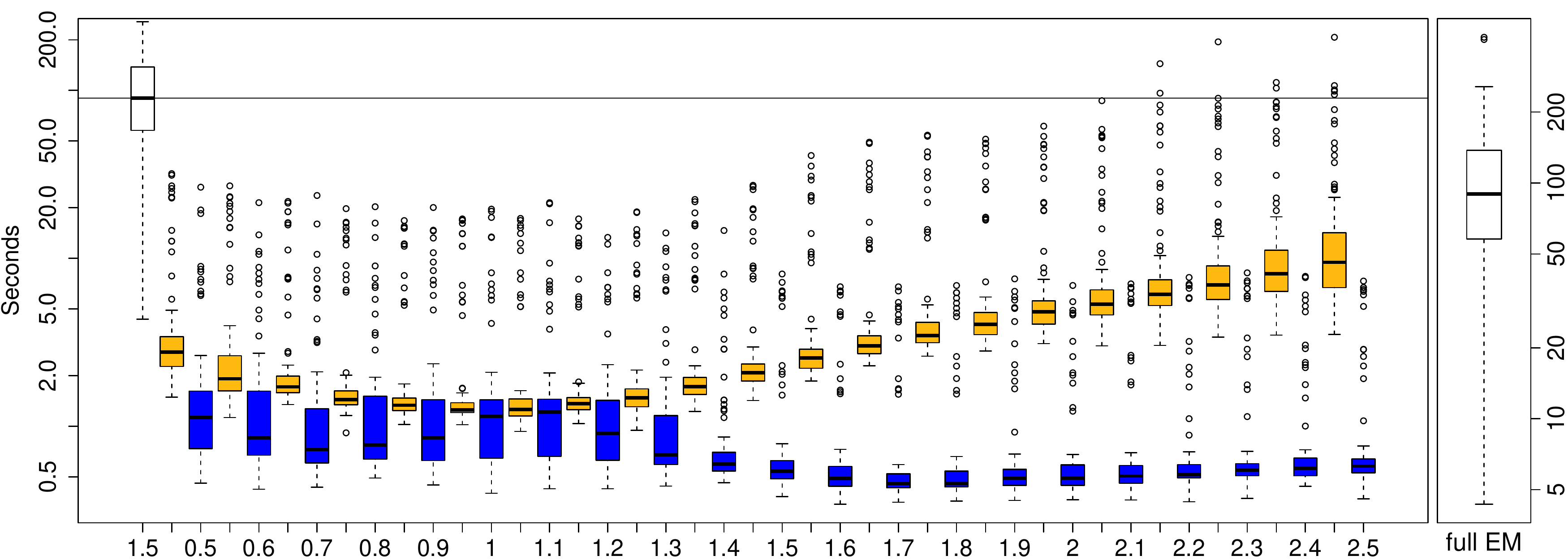}}
		\subfigure[$50 \times 50$ grid with $m=77$]{\includegraphics[scale=.23]{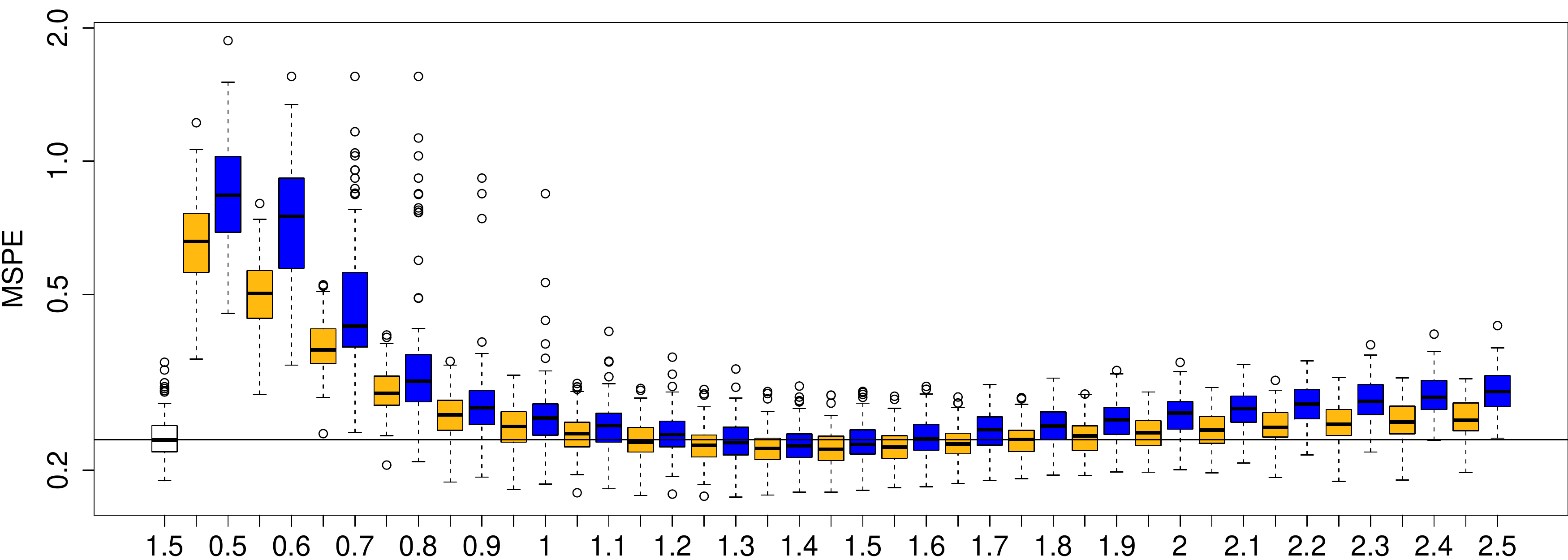}}
		\subfigure[$50 \times 50$ grid with $m=175$]{\includegraphics[scale=.23]{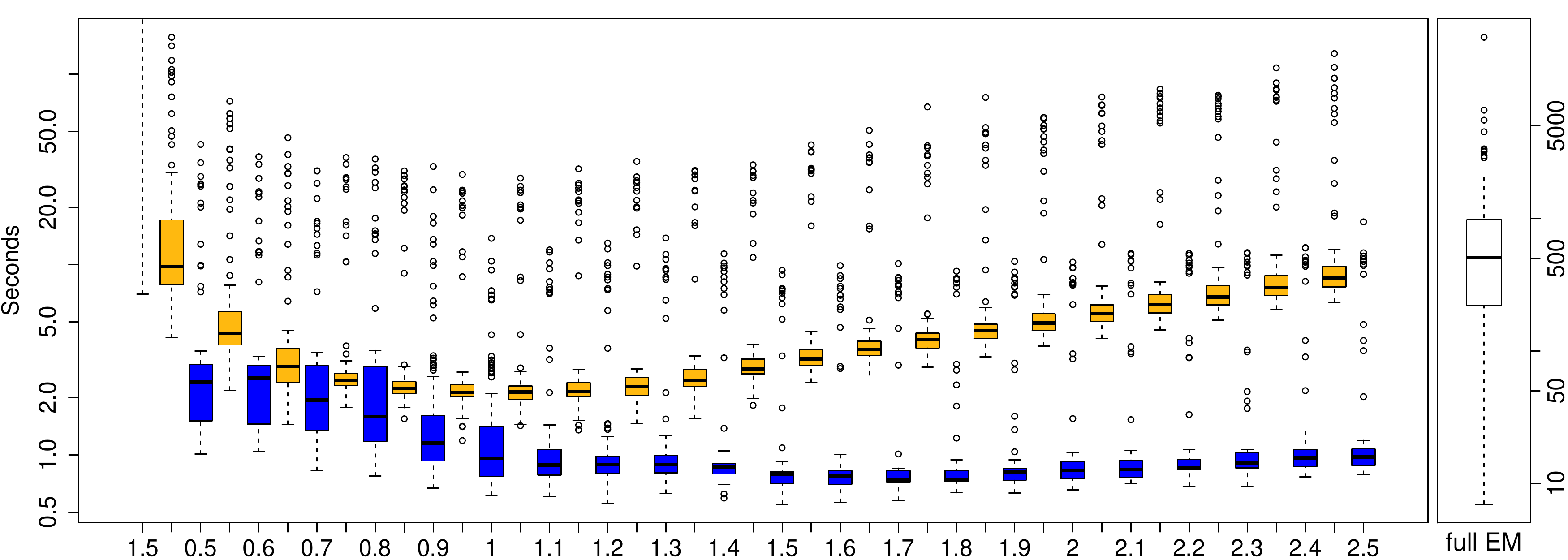}}		
		\subfigure[$50 \times 50$ grid with $m=175$]{\includegraphics[scale=.23]{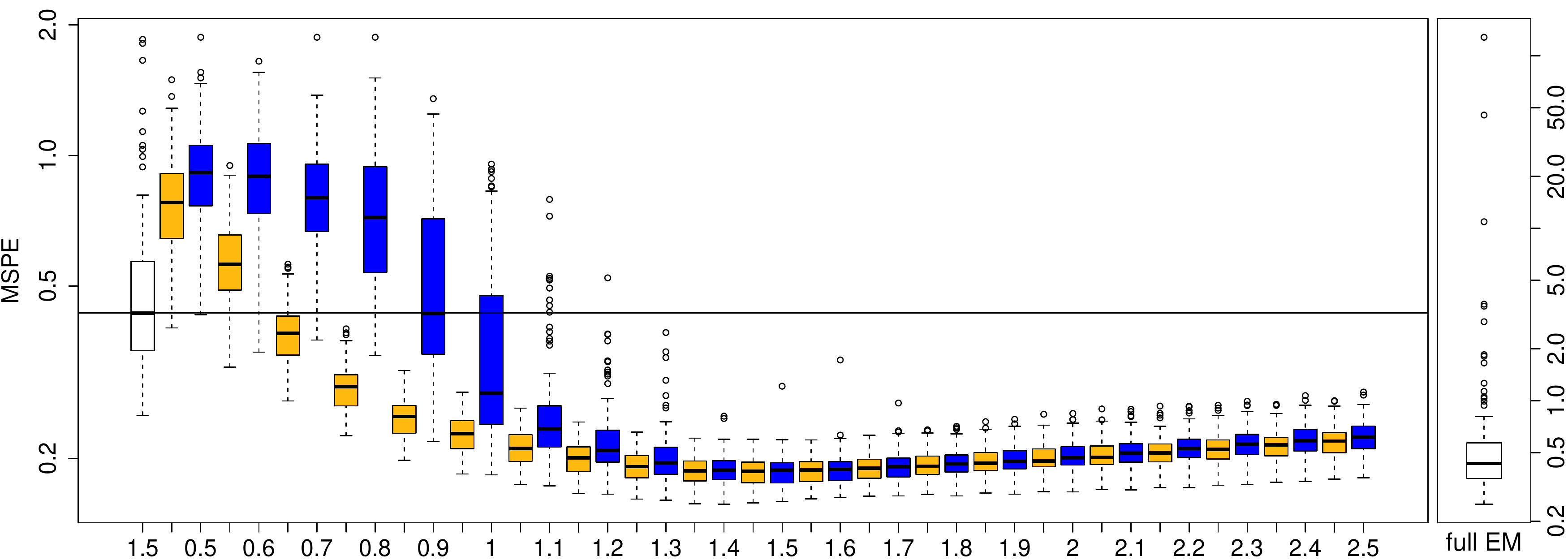}}
		\subfigure[$200 \times 200$ grid with $m=23$]{\includegraphics[scale=.23]{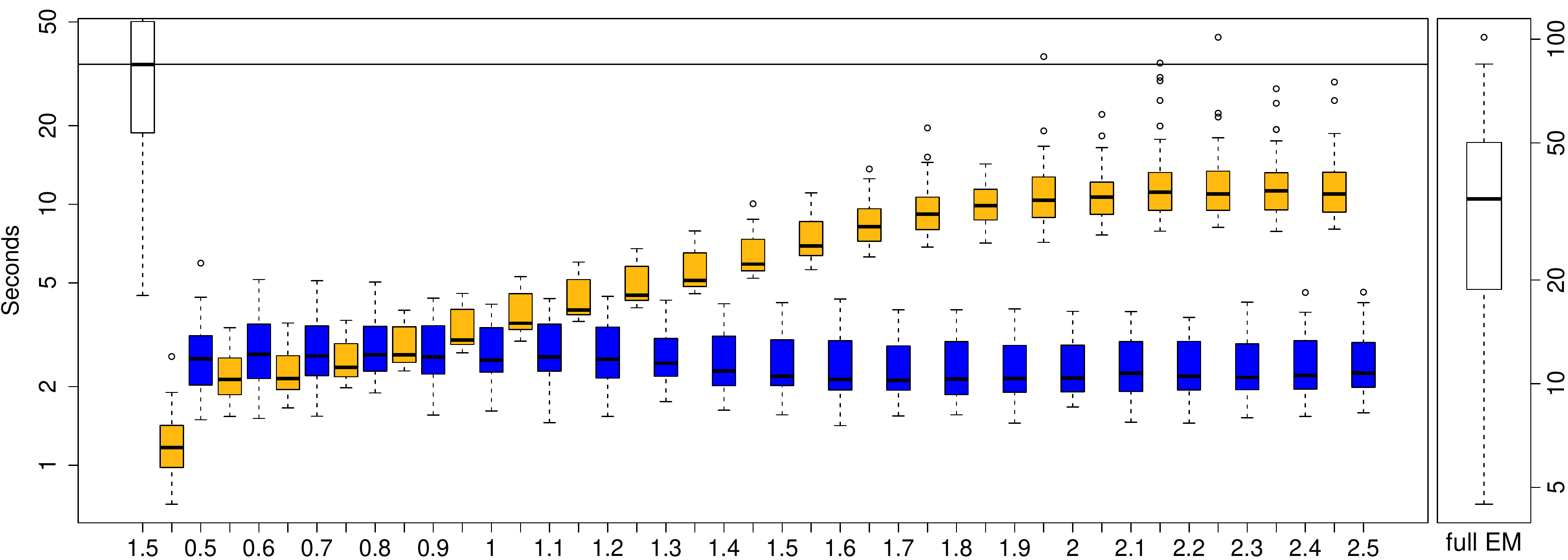}}
		\subfigure[$200 \times 200$ grid with $m=23$]{\includegraphics[scale=.23]{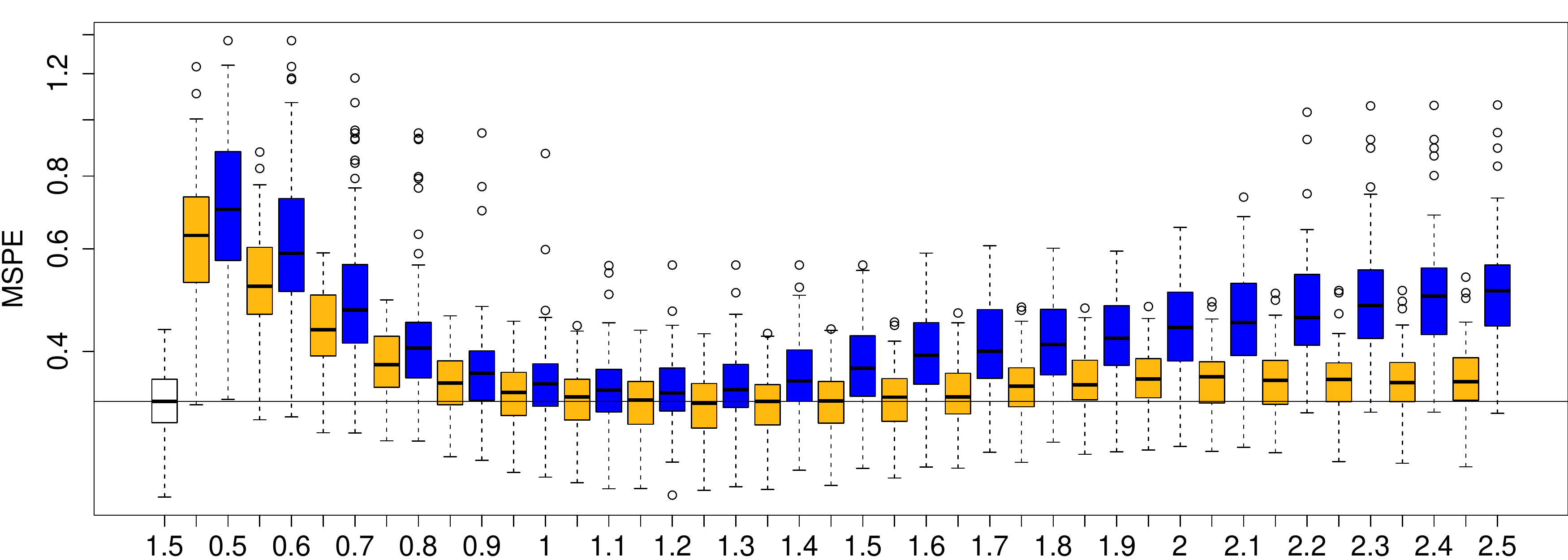}}
		\subfigure[$200 \times 200$ grid with $m=77$]{\includegraphics[scale=.23]{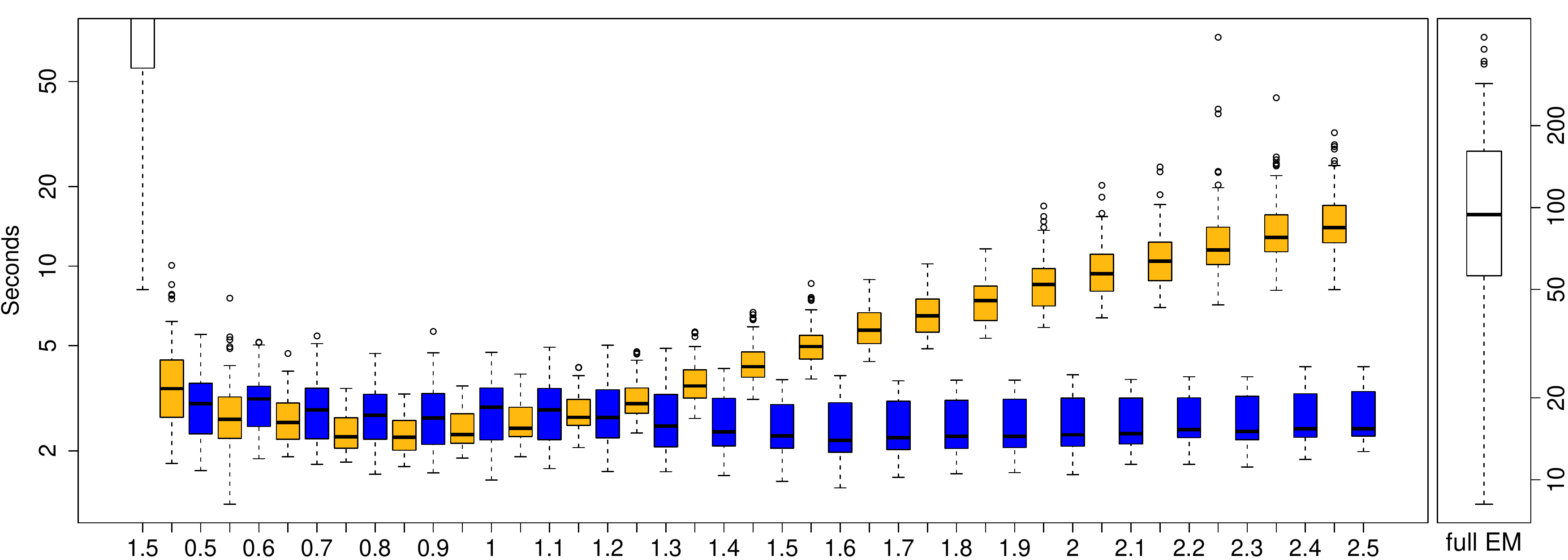}}
		\subfigure[$200 \times 200$ grid with $m=77$]{\includegraphics[scale=.23]{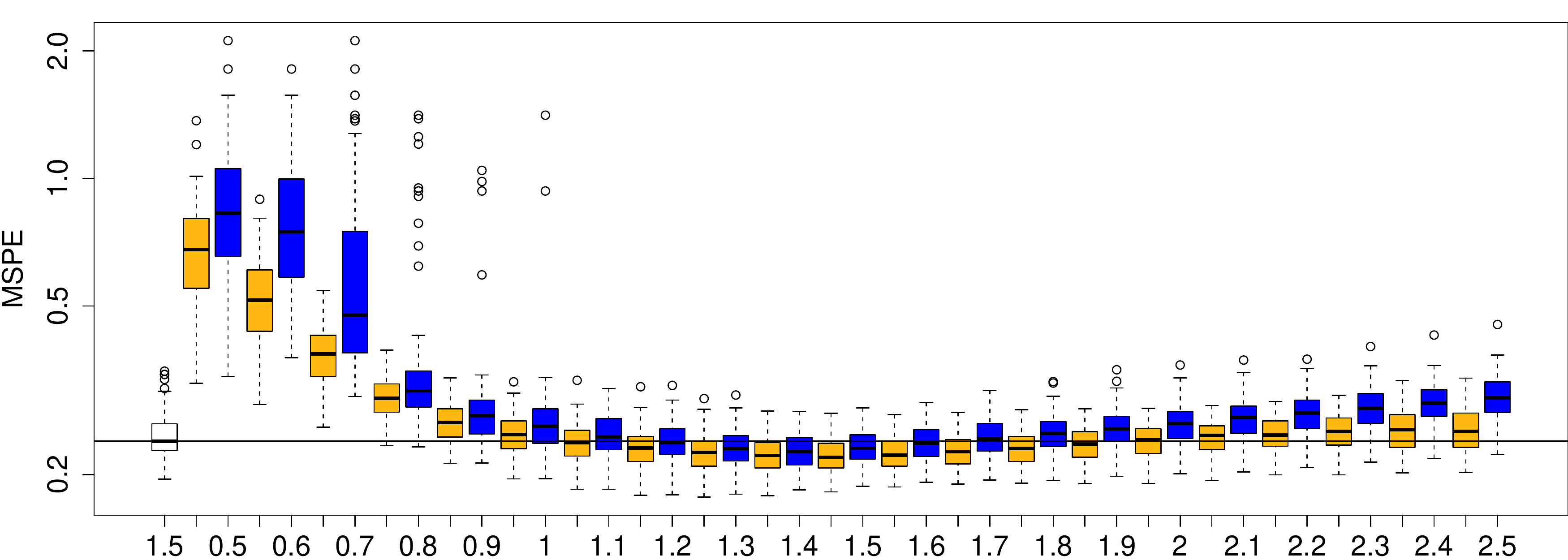}}
		\subfigure[$200 \times 200$ grid with $m=175$]{\includegraphics[scale=.23]{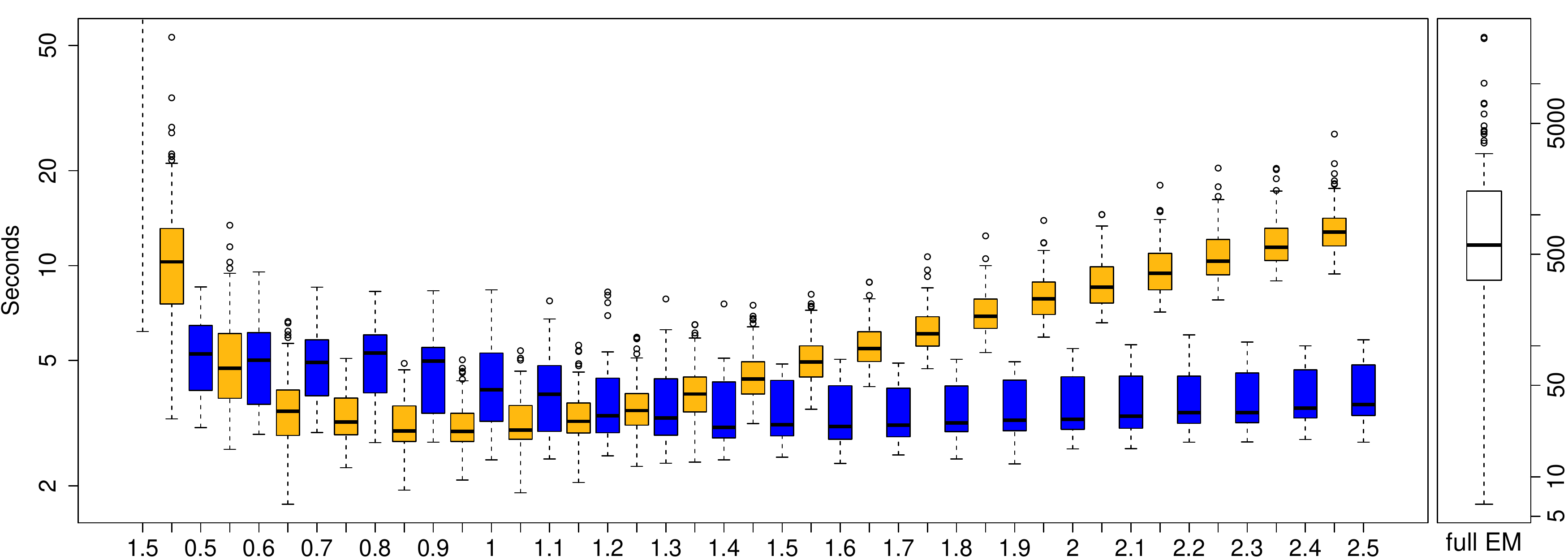}}		
		\subfigure[$200 \times 200$ grid with $m=175$]{ \includegraphics[scale=.23]{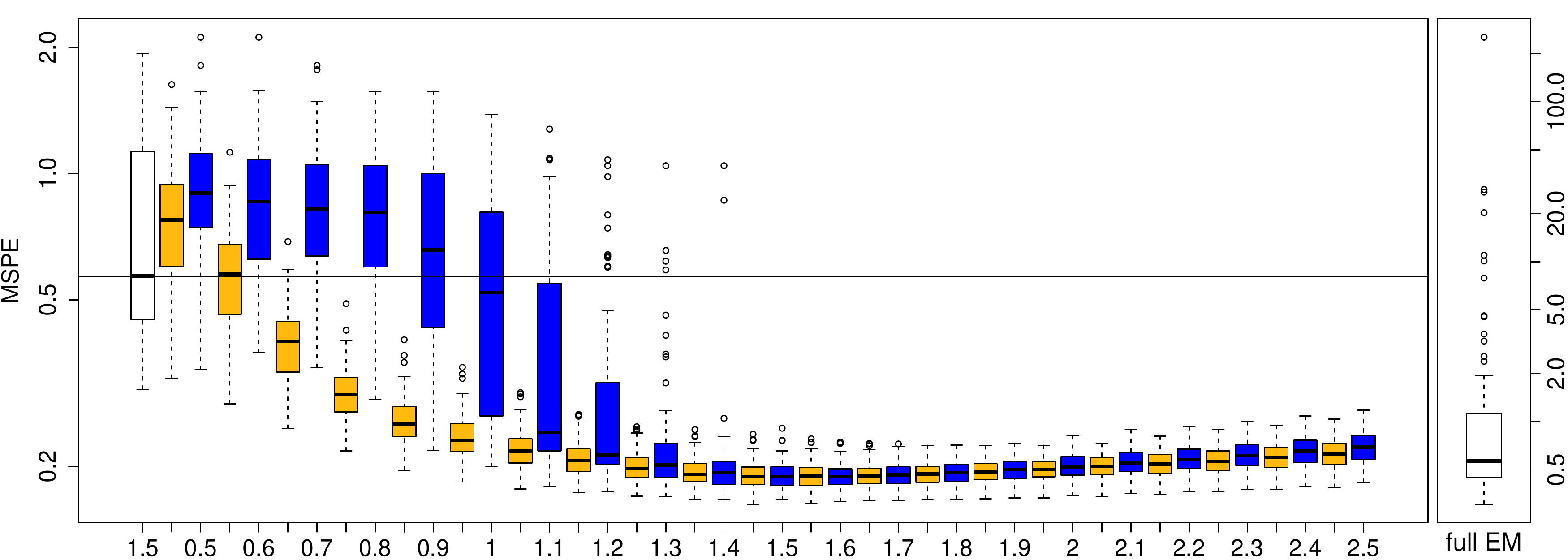}}
	\end{center}
	\caption{Seconds (left panel) and Mean Squared Prediction
          Errors (MSPEs, right panel) for varying bandwidths constants
          (x-axis) and number of knots ($m$) for all three estimation
          methods on both grids when $\nu=0.5$ and $\sigma^2=0$.}
	\label{bp1}
\end{figure}

\begin{figure}[H]
	\begin{center}
		\subfigure[$50 \times 50$ grid with $m=23$]{\includegraphics[scale=.23]{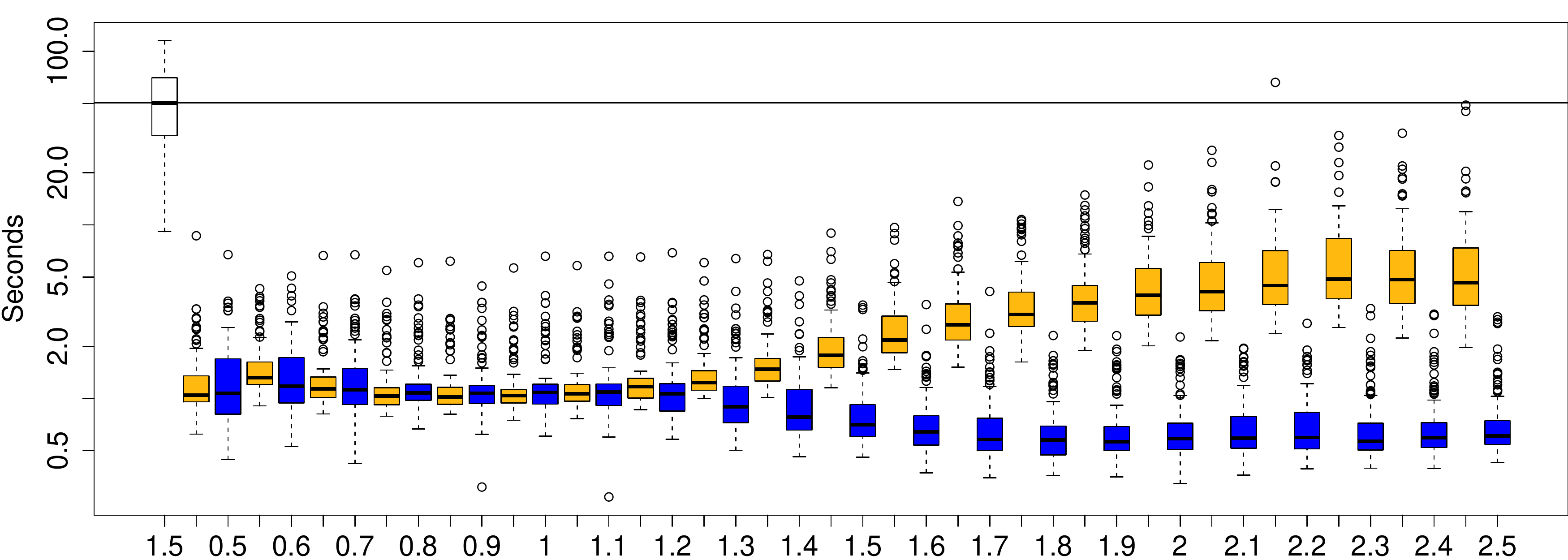}}
		\subfigure[$50 \times 50$ grid with $m=23$]{\includegraphics[scale=.23]{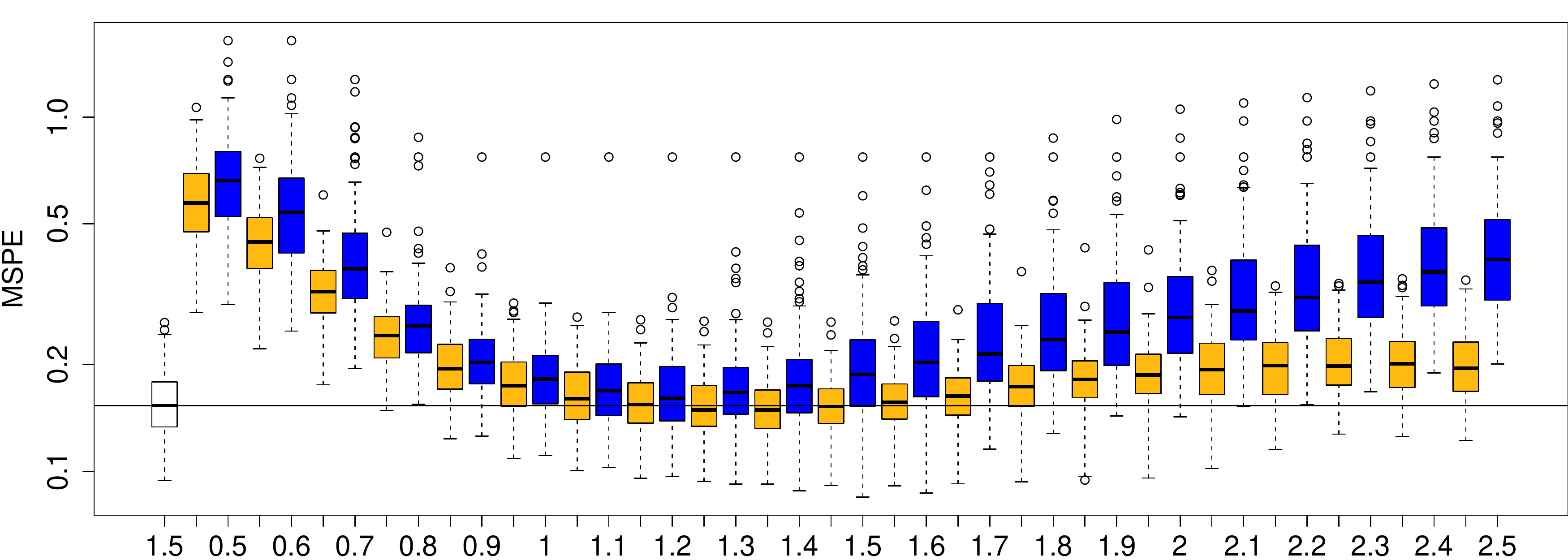}}
		\subfigure[$50 \times 50$ grid with $m=77$]{\includegraphics[scale=.23]{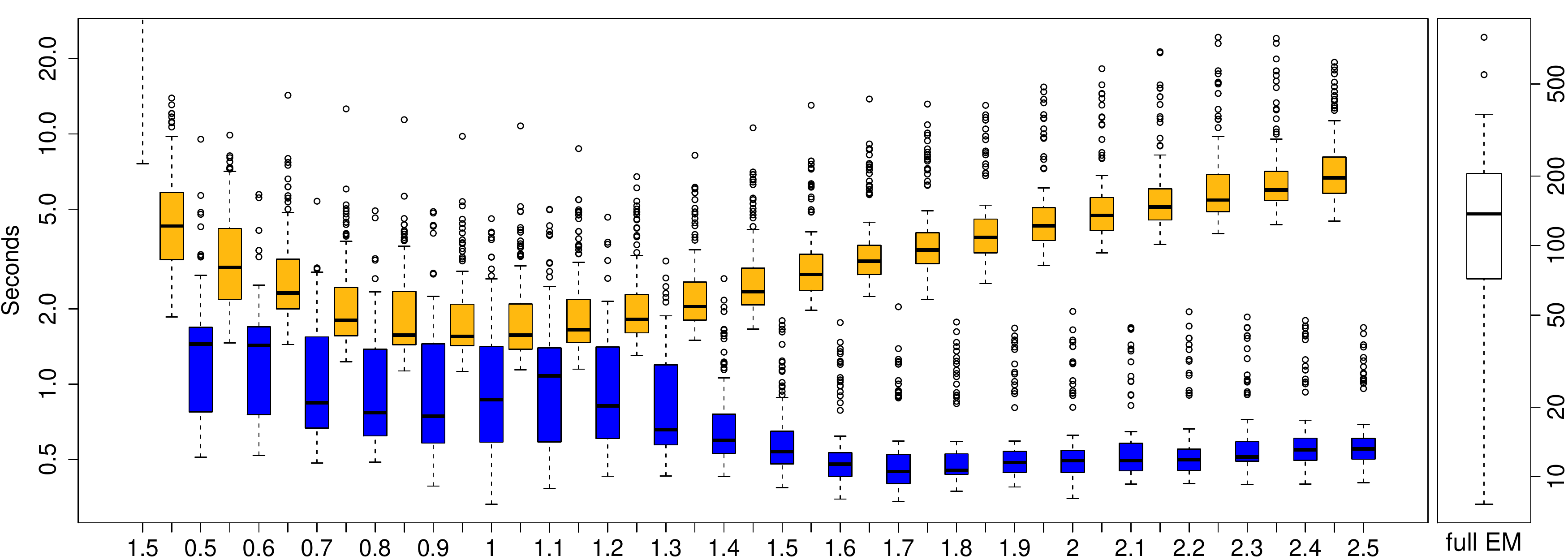}}
		\subfigure[$50 \times 50$ grid with $m=77$]{\includegraphics[scale=.23]{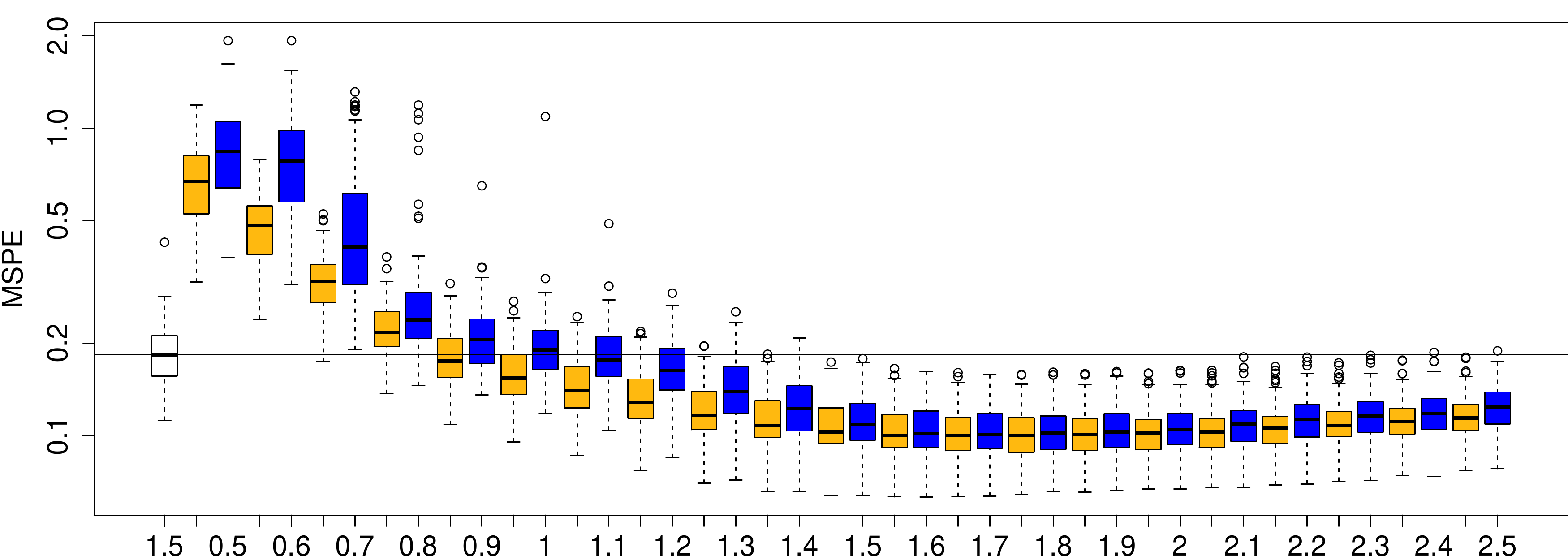}}
		\subfigure[$50 \times 50$ grid with $m=175$]{\includegraphics[scale=.23]{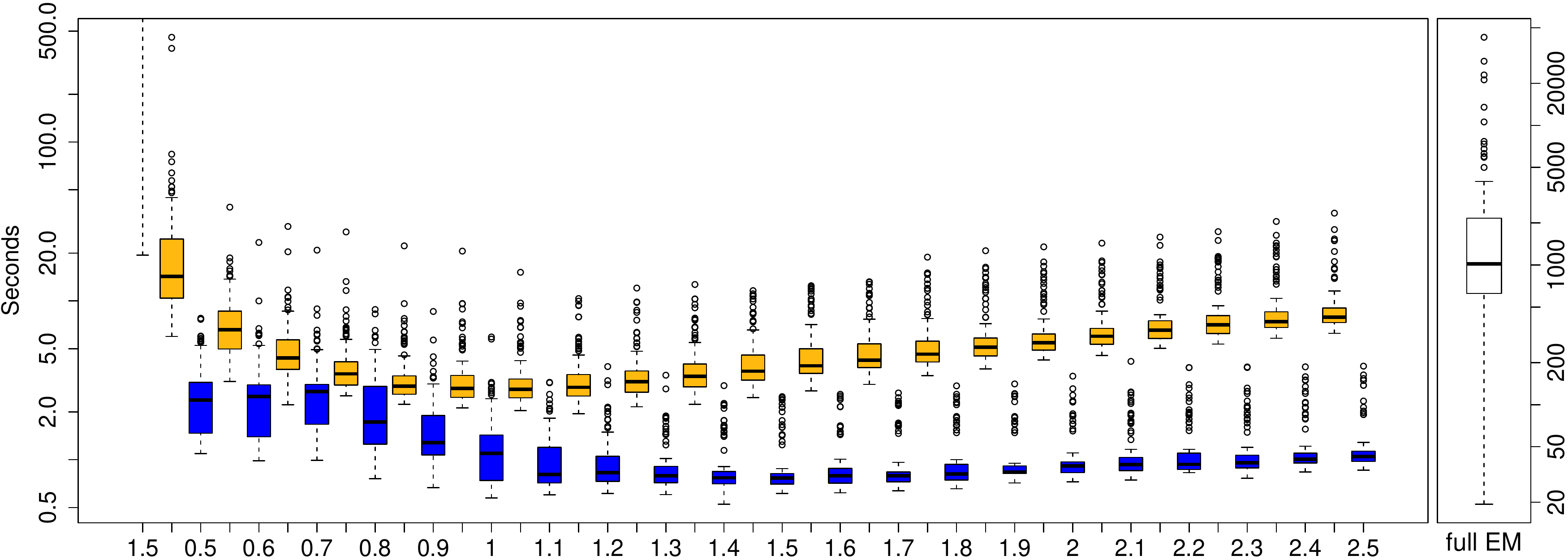}}		
		\subfigure[$50 \times 50$ grid with $m=175$]{\includegraphics[scale=.23]{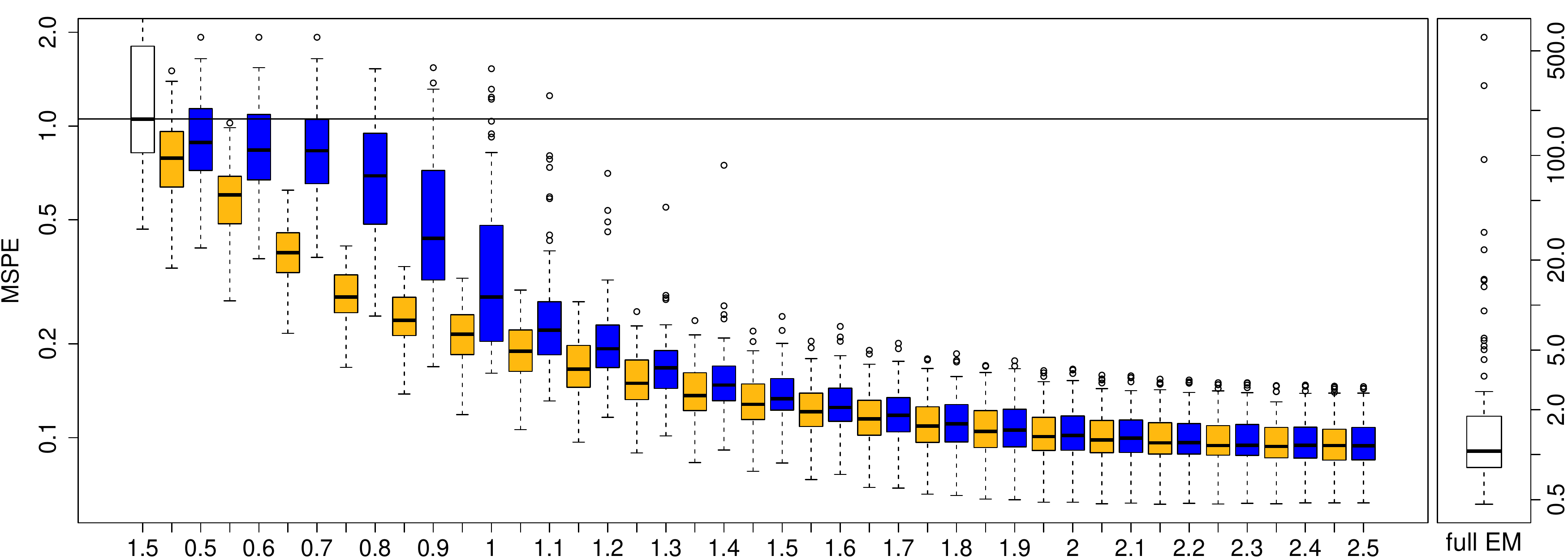}}
		\subfigure[$200 \times 200$ grid with $m=23$]{\includegraphics[scale=.23]{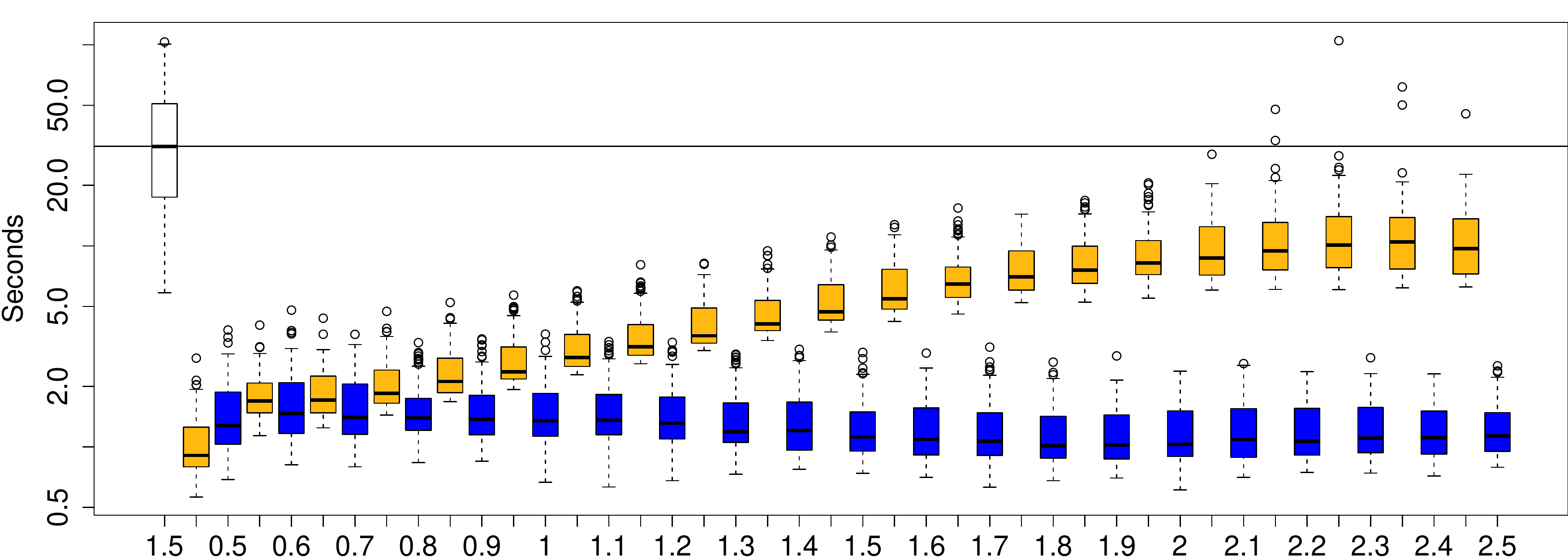}}
		\subfigure[$200 \times 200$ grid with $m=23$]{\includegraphics[scale=.23]{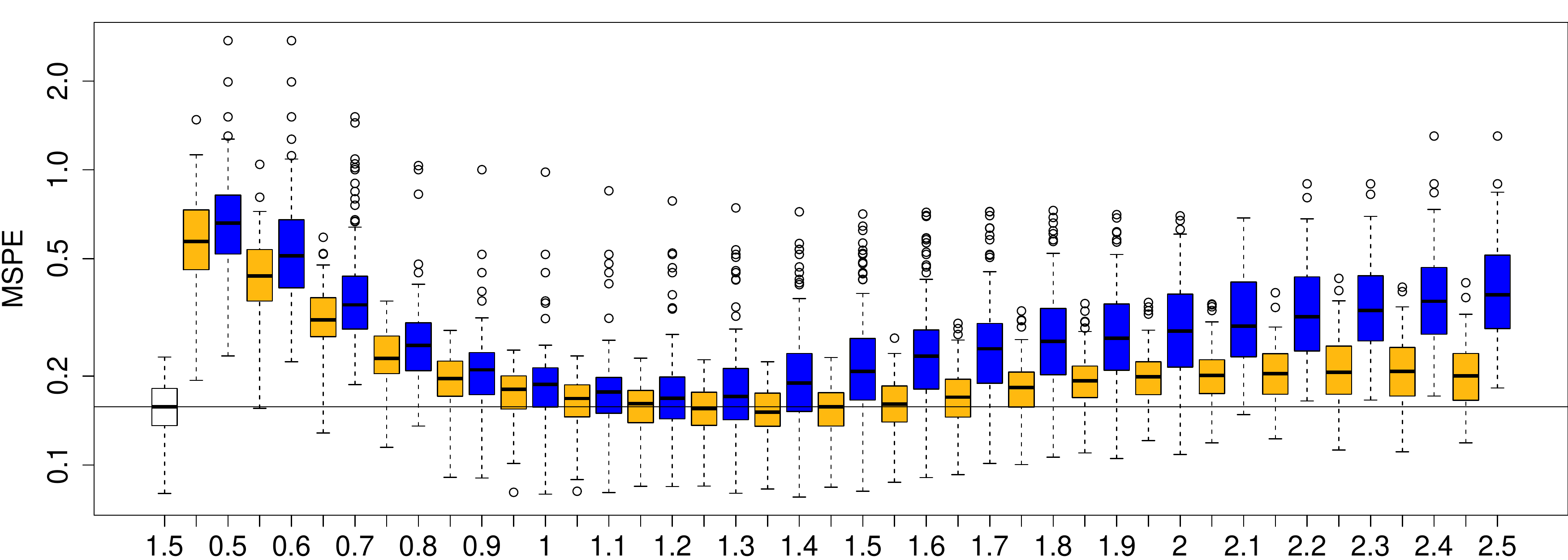}}
		\subfigure[$200 \times 200$ grid with $m=77$]{\includegraphics[scale=.23]{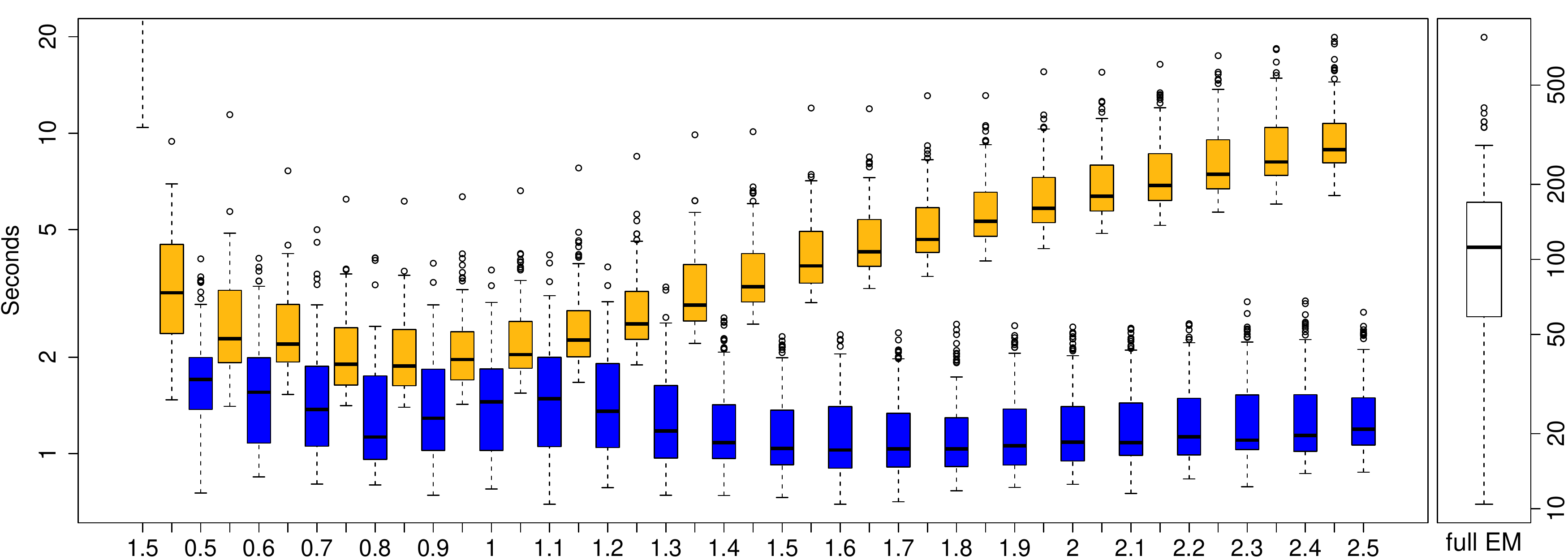}}
		\subfigure[$200 \times 200$ grid with $m=77$]{\includegraphics[scale=.23]{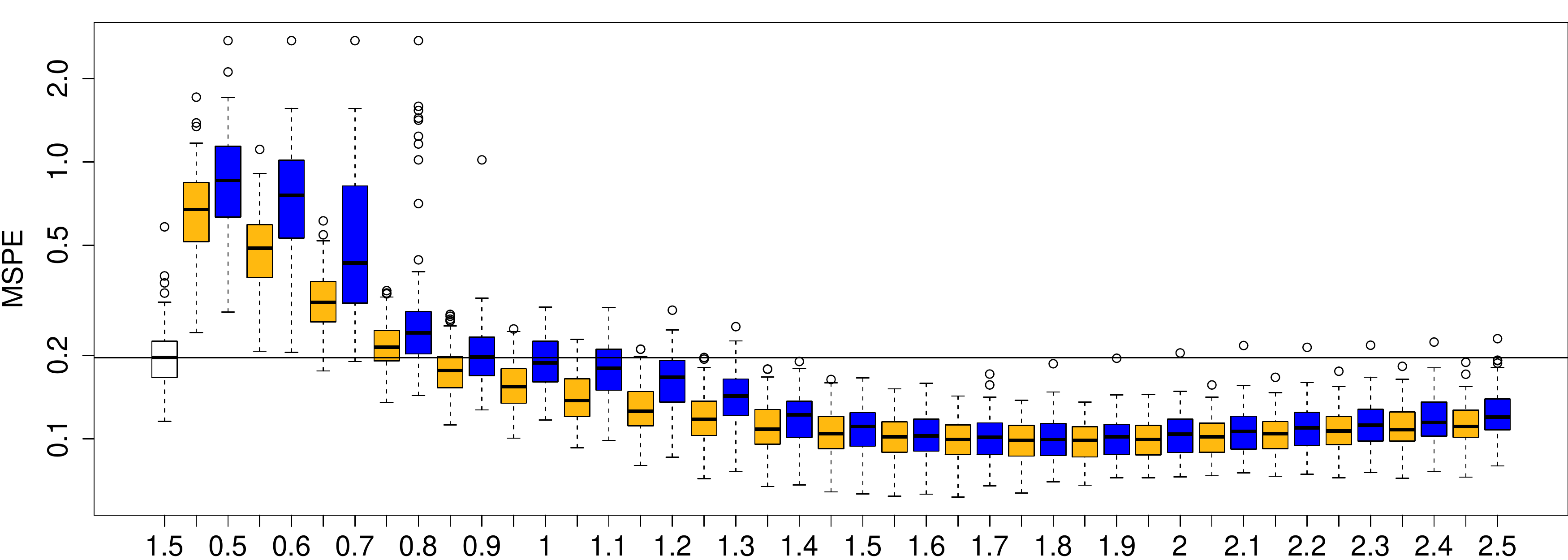}}
		\subfigure[$200 \times 200$ grid with $m=175$]{\includegraphics[scale=.23]{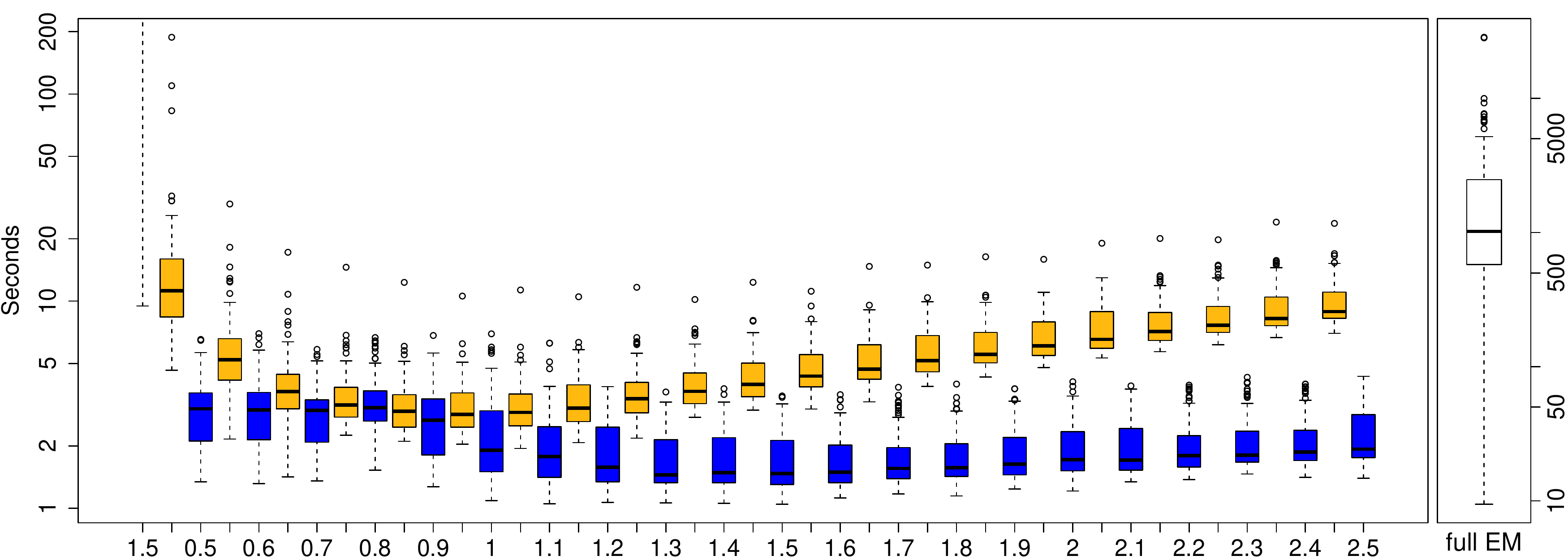}}		
		\subfigure[$200 \times 200$ grid with $m=175$]{\includegraphics[scale=.23]{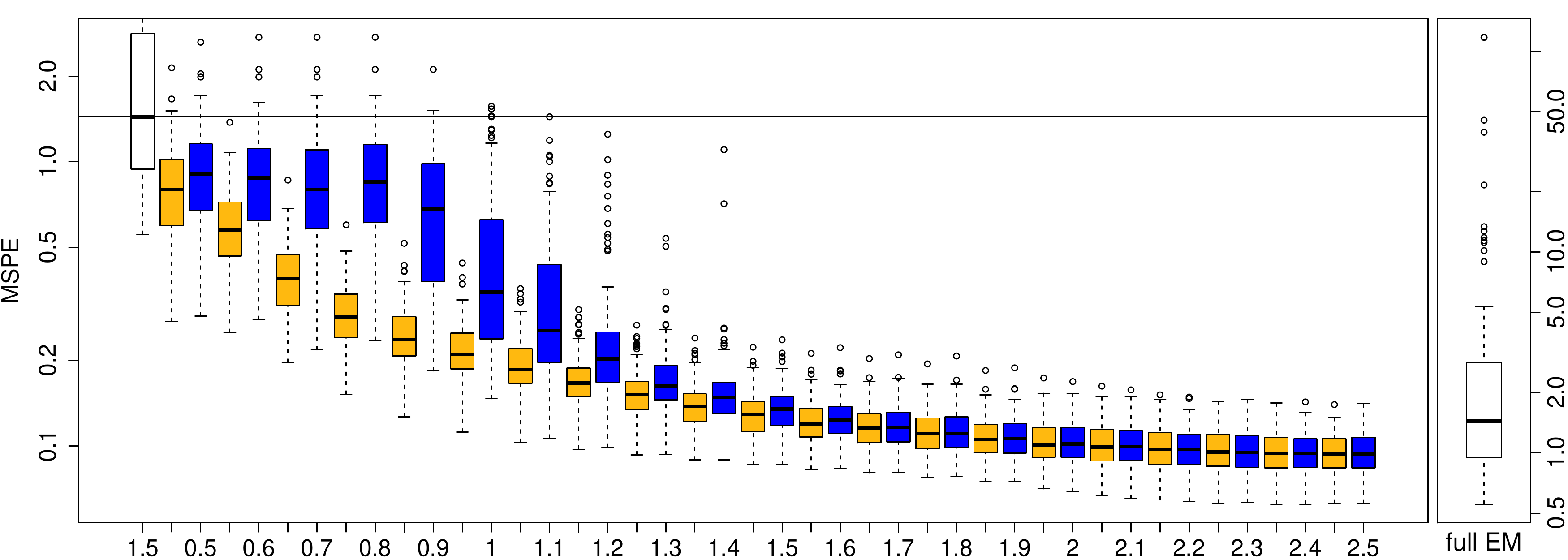}}
	\end{center}
	\caption{Seconds (left panel) and Mean Squared Prediction
          Errors (MSPEs, right panel) for varying bandwidths constants
          (x-axis) and number of knots ($m$) for all three estimation
          methods on both grids when $\nu=2$ and $\sigma^2=0.4$.}
	\label{bp2} 
\end{figure}

The value of reduced basis kriging was quantified by two summaries: the time 
required to iterate to convergence followed by computing kriging estimates 
and the mean square prediction error of the predicted field. 
The distributions of the one hundred simulated fields are 
represented with boxplots on the log scale in Figures \ref{bp1} through \ref{bp2}. 
For clarity of presentation, the distribution of seconds or MSPE for the 
E-M approach using the full form of $\bK$ are truncated on the main figure and 
presented in full separately when the maximum is significantly beyond the range 
of the corresponding results for the identity form of $\bK$, as in Figure \sref{time77}.

Two expected gains in computational efficiency were achieved by either using 
the identity covariance for $\bK$ or by using fewer knots. The value of the 
bandwidth constant, $b$, was also critical to minimizing CPU time because it 
controlled the level of sparsity in $\bS$. The effect of $b$ on the E-M approach 
was predictable as lower values resulted in decreased CPU time, with an optimal 
level of sparsity occurring near $b=1$ for the $50 \times 50$ grid. An 
interesting difference between the E-M approach and reduced basis kriging is the 
robustness of reduced basis kriging to increasing $m$ or a poor choice of bandwidth 
constant. Reduced basis kriging is efficient regardless of the specifics of the 
basis functions used and is at least as efficient as the E-M approach with the 
identity covariance for any reasonable choice of $b$. The E-M approach is consistently more 
computationally expensive for larger values of $b$. When the $200 \times 200$ grid 
size was used, the E-M approach would occasionally reach convergence sooner than 
reduced basis kriging for low values of the bandwidth constant ($b < 1$), with a 
dramatic reduction occurring when $b = 0.5$ and $m=23$. However, the resulting MSPE 
of these estimations were particularly poor.

The distributions of MSPE for both $50 \times 50$ and $200 \times 200$ 
grid sizes show that for a reasonable $b$, the identity covariance 
structure is at least as accurate in terms of prediction as the full 
covariance structure. In fact, as $m$ approaches $n$, the identity 
covariance performs substantially better than the full covariance 
with regards to both median and maximum MSPE. This suggests that added 
complexity to $\bK$ is unnecessary when our focus is on prediction. 
Using the identity covariance for $\bK$, reduced basis kriging and the
E-M approach are comparable for choices of bandwidth constant where
MSPE is minimized, however an unreasonable $b$ favors the E-M approach
in terms of MSPE. This is opposite of computational efficiency, in
which reduced basis kriging was more robust to poor choices of
$b$.

\begin{figure}[H]
	\begin{center}
		\subfigure[$50 \times 50$ grid with $m=23$]{\includegraphics[scale=.23]{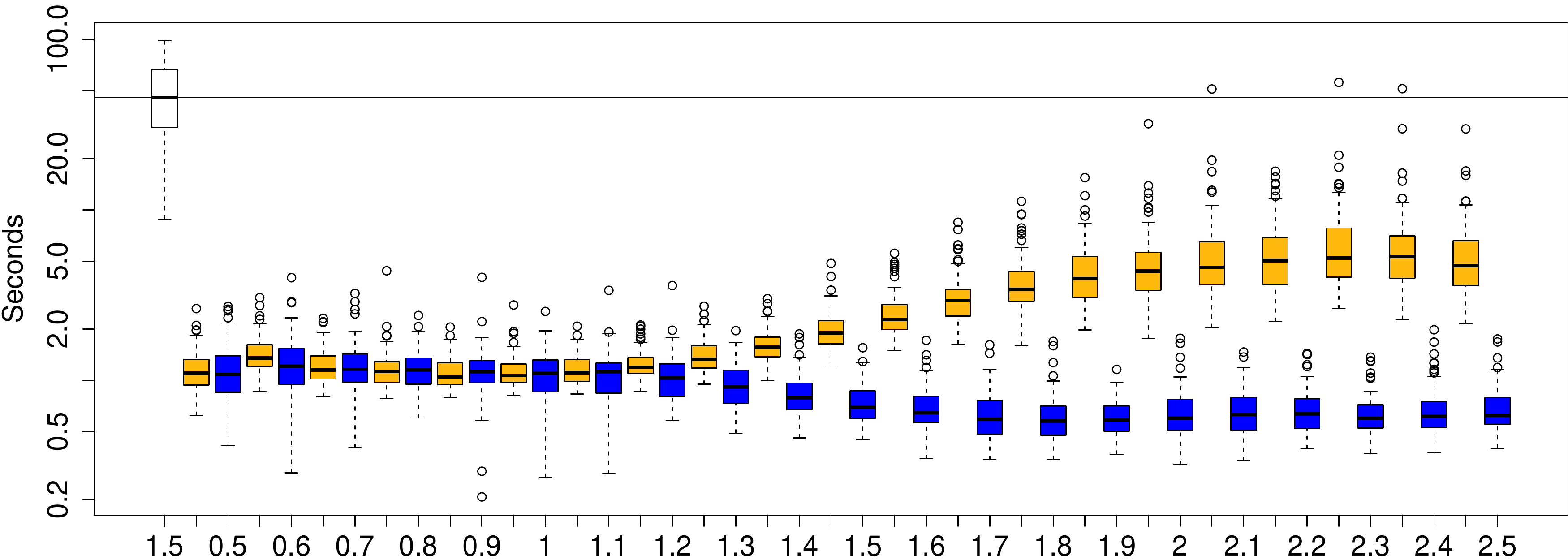}}
		\subfigure[$50 \times 50$ grid with $m=23$]{\includegraphics[scale=.23]{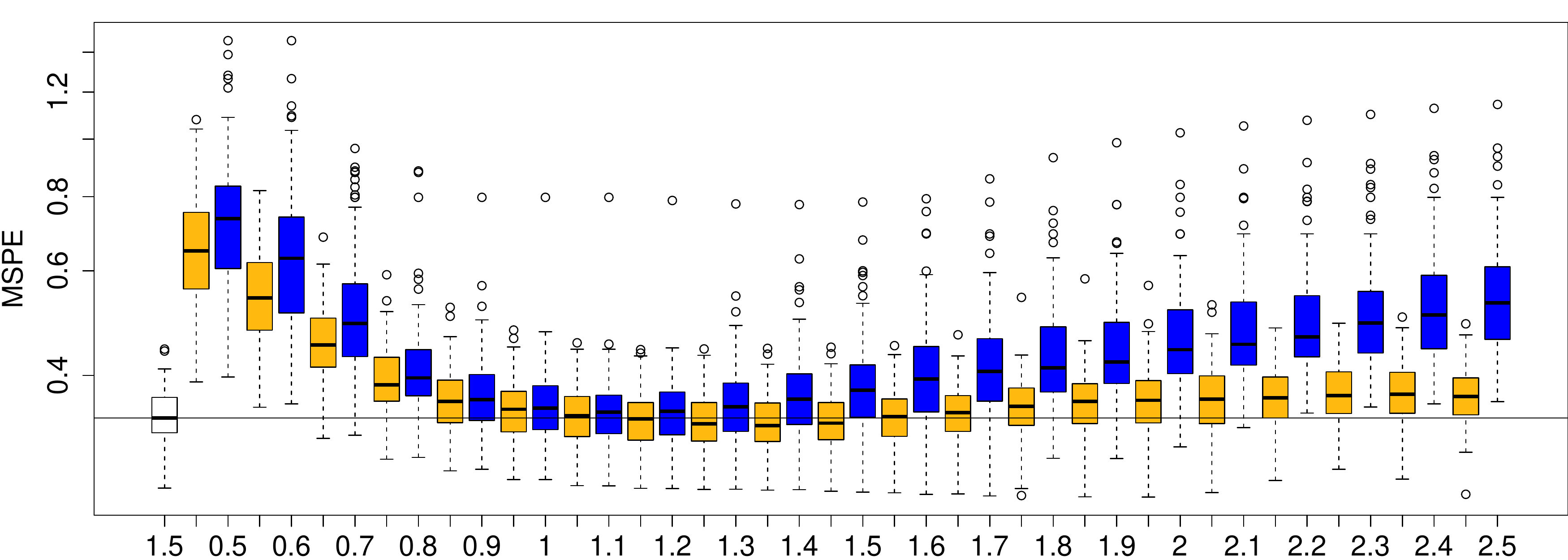}}
		\subfigure[$50 \times 50$ grid with $m=77$]{\includegraphics[scale=.23]{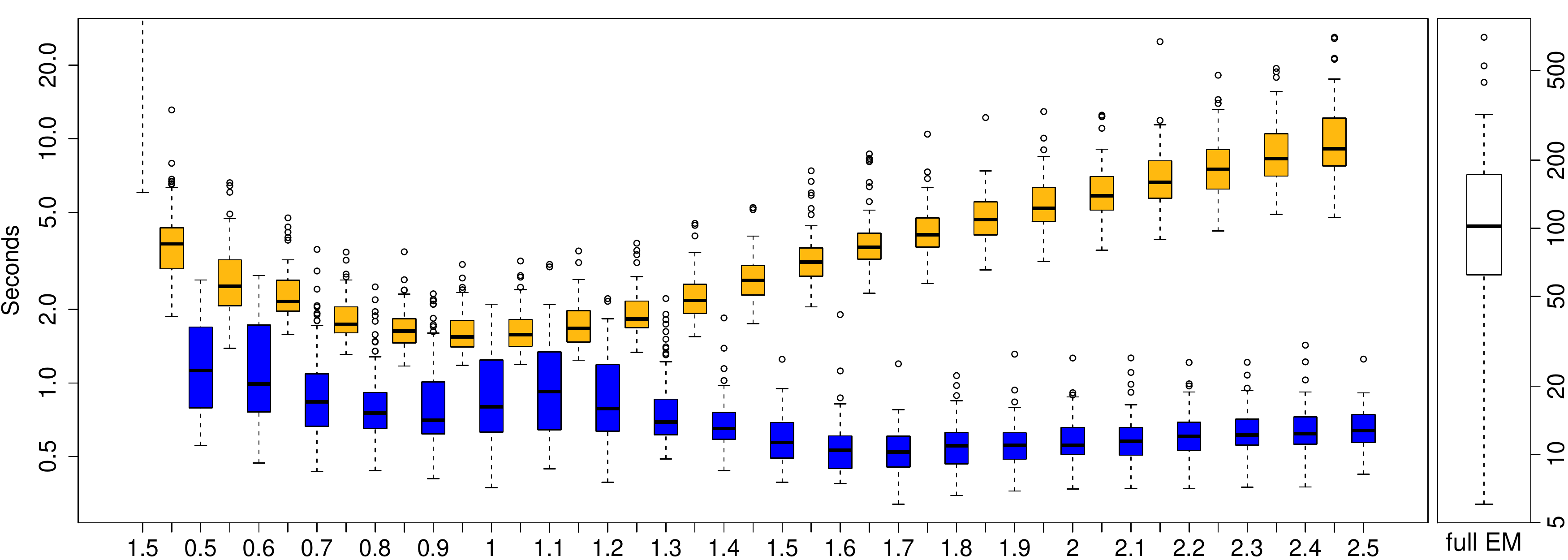}}
		\subfigure[$50 \times 50$ grid with $m=77$]{\includegraphics[scale=.23]{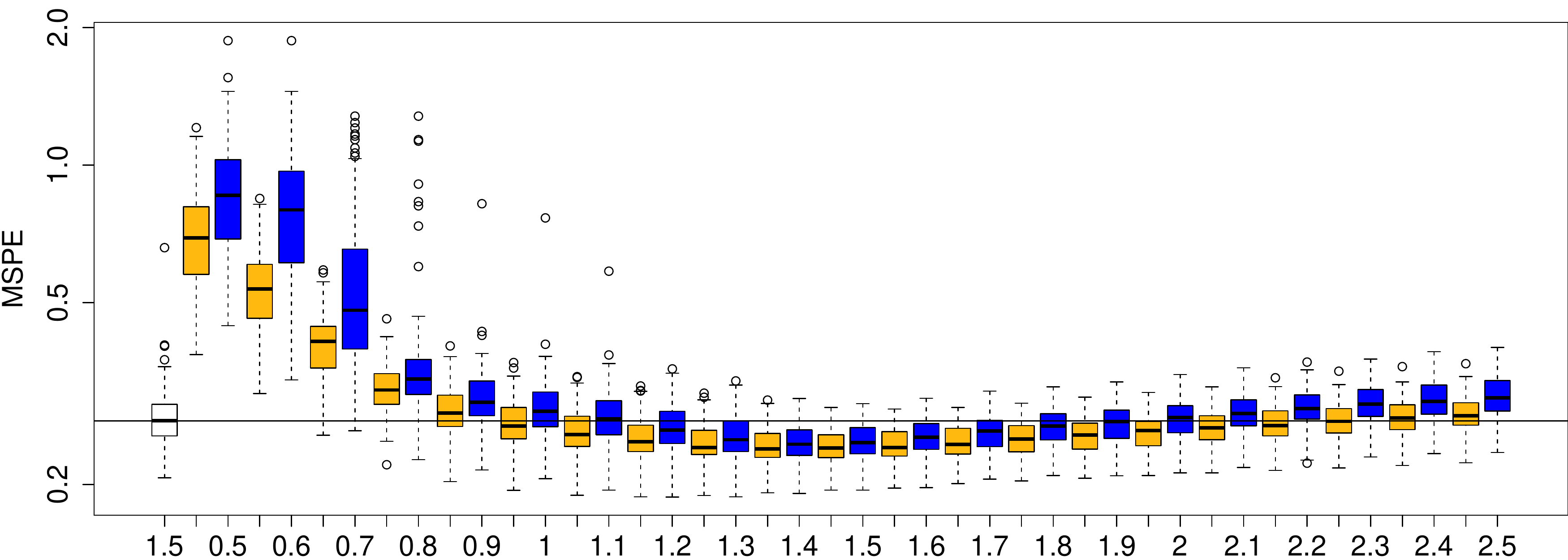}}
		\subfigure[$50 \times 50$ grid with $m=175$]{\includegraphics[scale=.23]{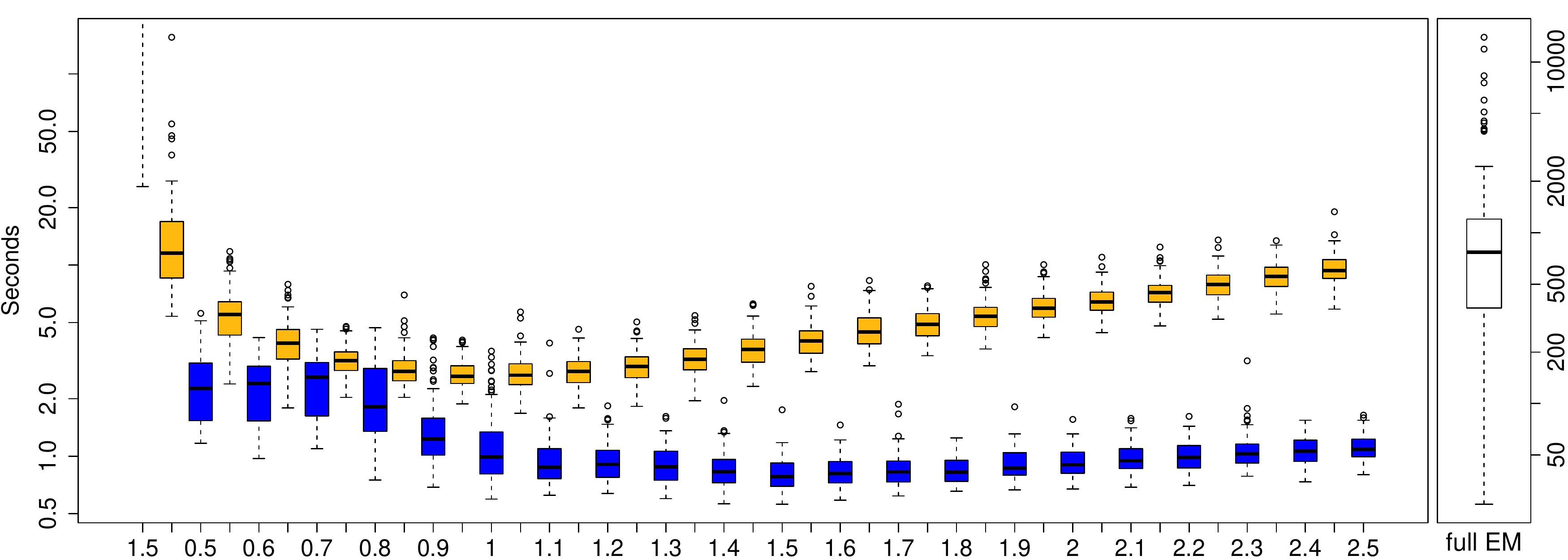}}		
		\subfigure[$50 \times 50$ grid with $m=175$]{\includegraphics[scale=.23]{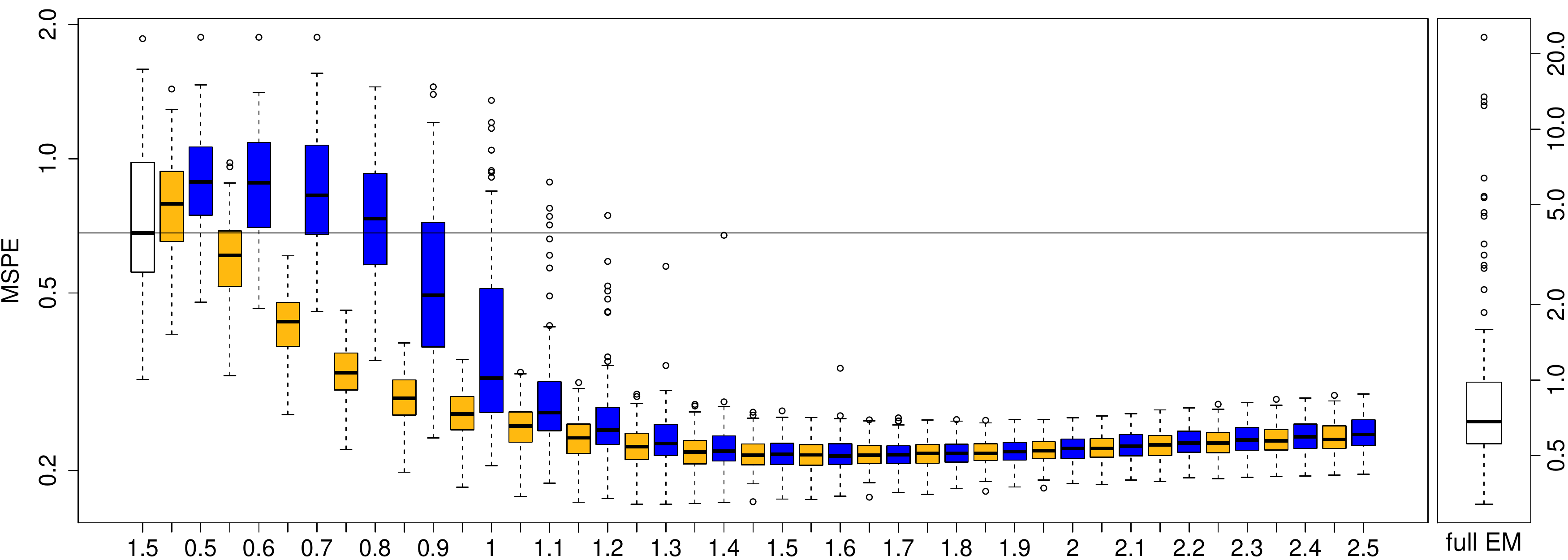}}
		\subfigure[$200 \times 200$ grid with $m=23$]{\includegraphics[scale=.23]{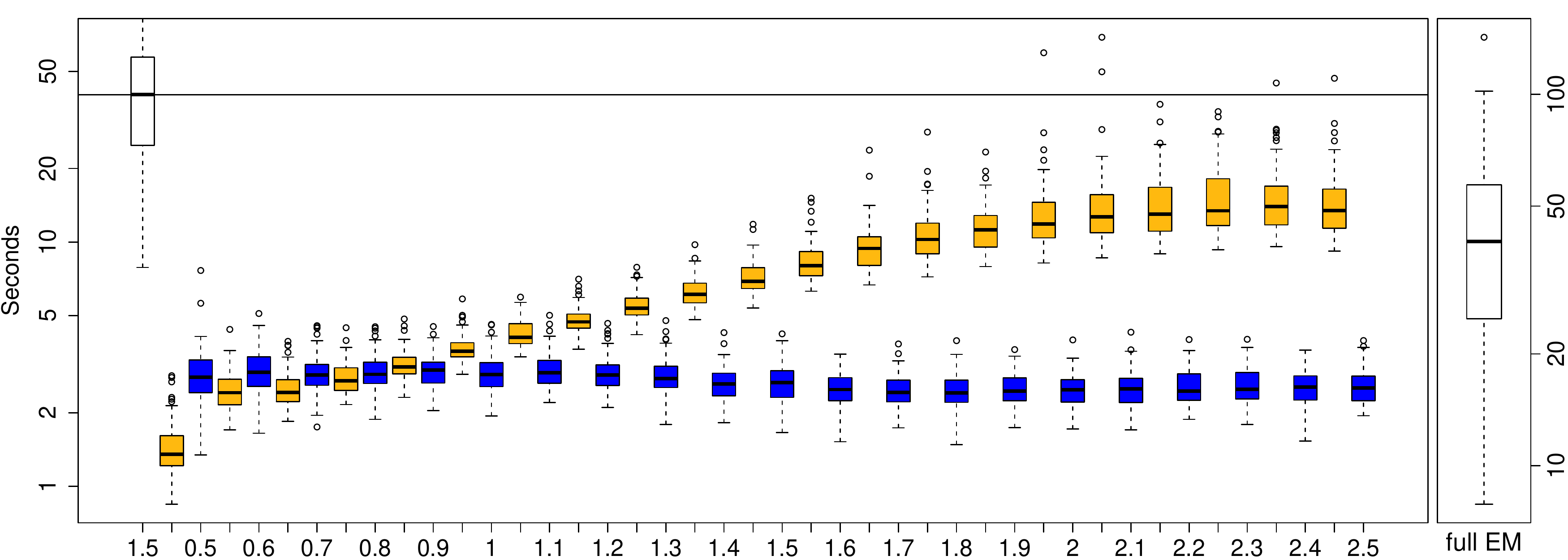}}
		\subfigure[$200 \times 200$ grid with $m=23$]{\includegraphics[scale=.23]{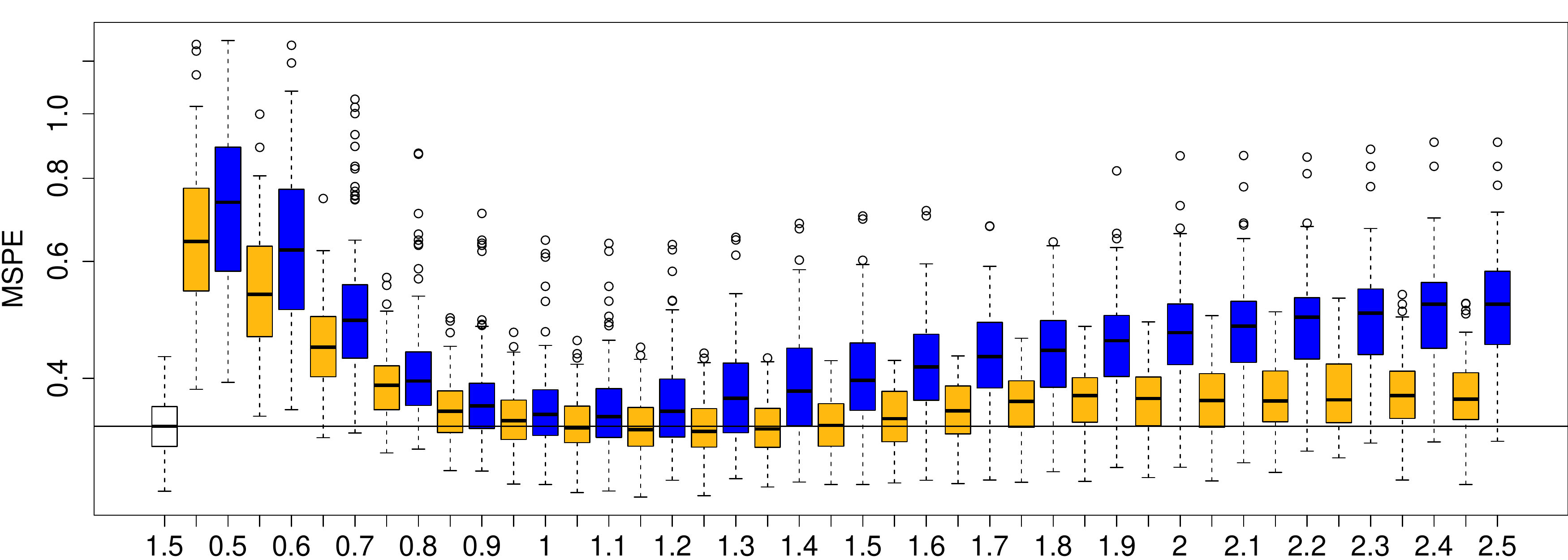}}
		\subfigure[$200 \times 200$ grid with $m=77$]{\includegraphics[scale=.23]{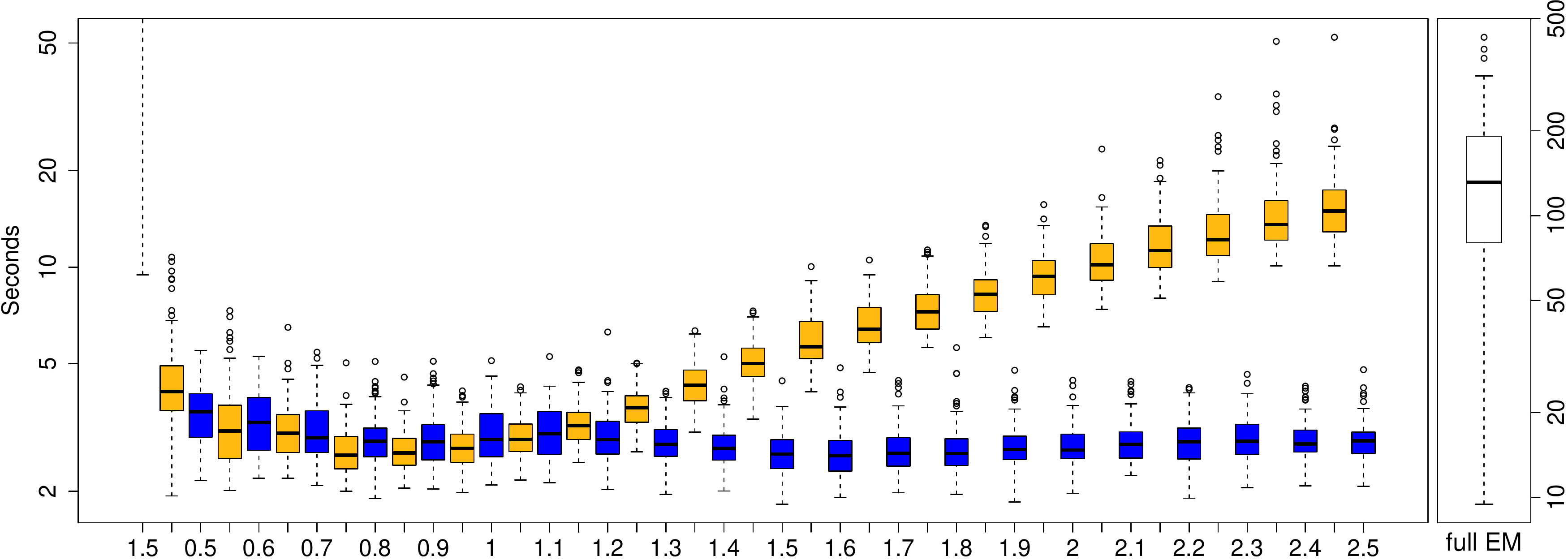}}
		\subfigure[$200 \times 200$ grid with $m=77$]{\includegraphics[scale=.23]{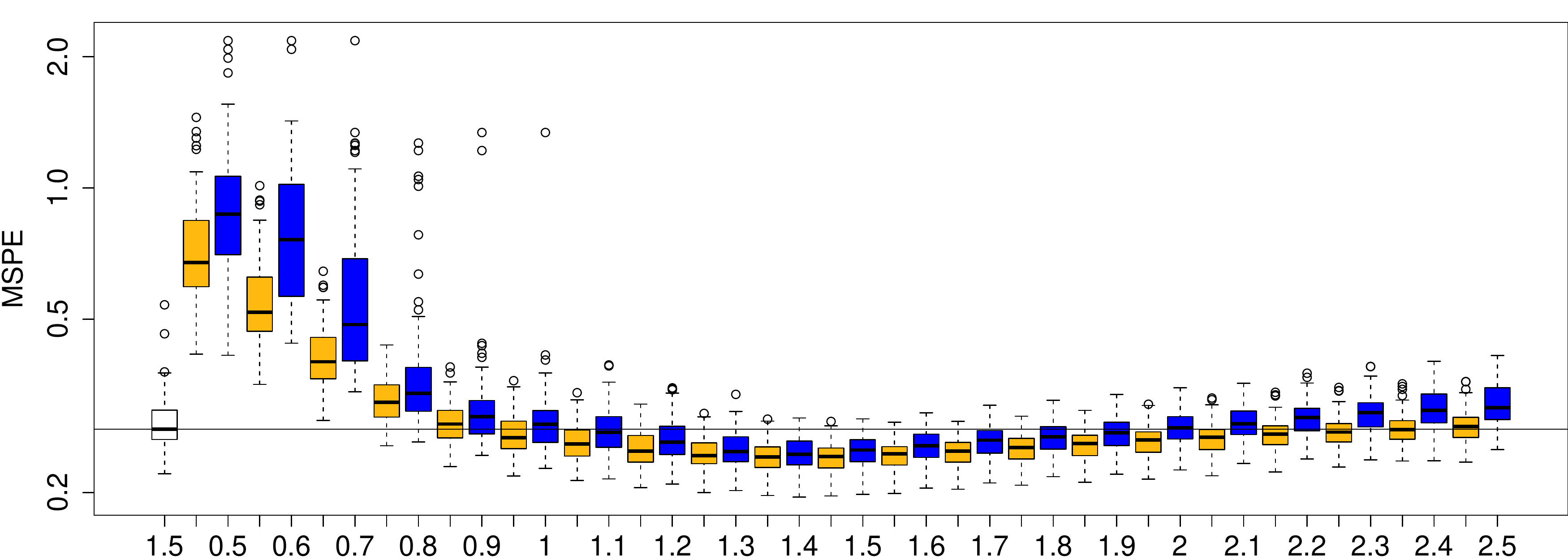}}
		\subfigure[$200 \times 200$ grid with $m=175$]{\includegraphics[scale=.23]{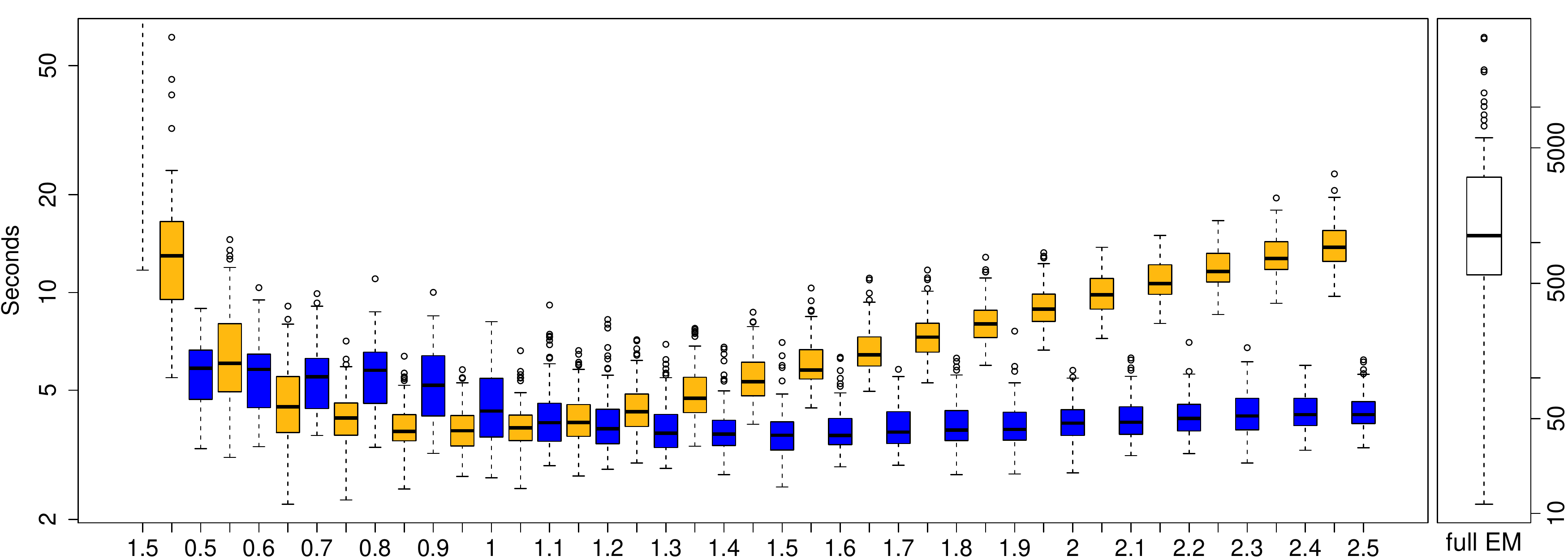}}		
		\subfigure[$200 \times 200$ grid with $m=175$]{\includegraphics[scale=.23]{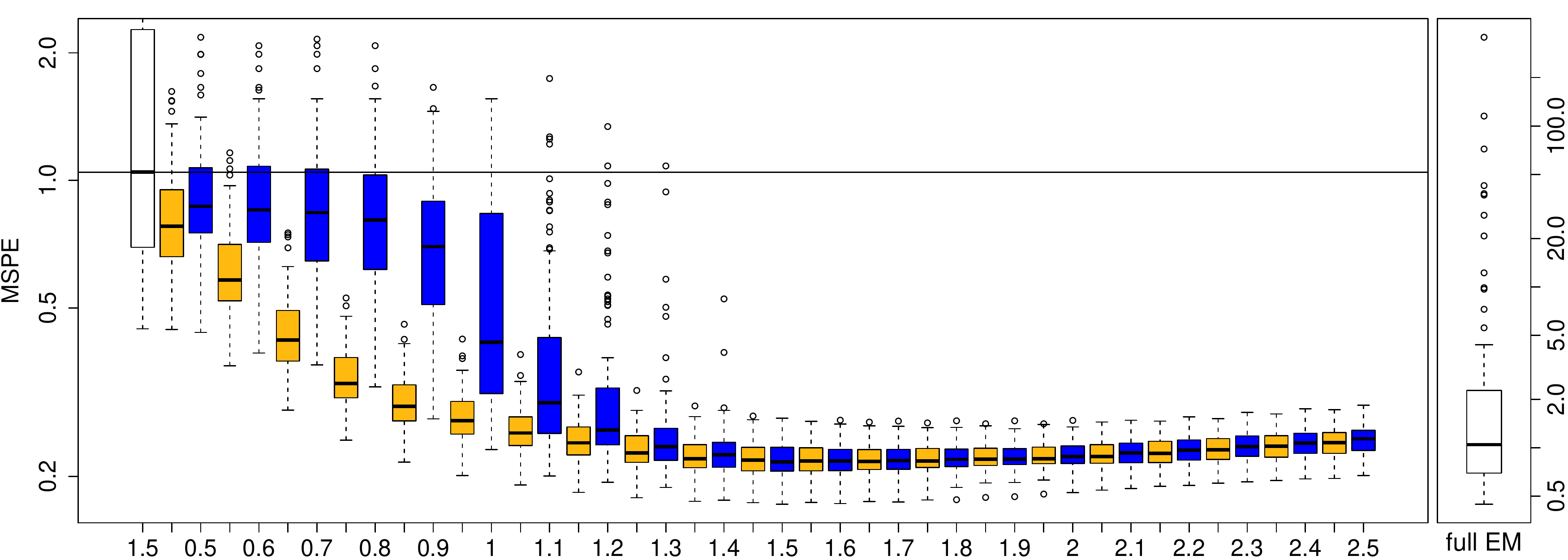}}
	\end{center}
	\caption{Seconds (left panel) and Mean Squared Prediction
          Errors (MSPEs, right panel) for varying bandwidths constants
          (x-axis) and number of knots ($m$) for all three estimation
          methods on both grids when $\nu=0.5$ and $\sigma^2=0.1$.} 
\end{figure}

\begin{figure}[H]
	\begin{center}
		\subfigure[$50 \times 50$ grid with $m=23$]{\includegraphics[scale=.23]{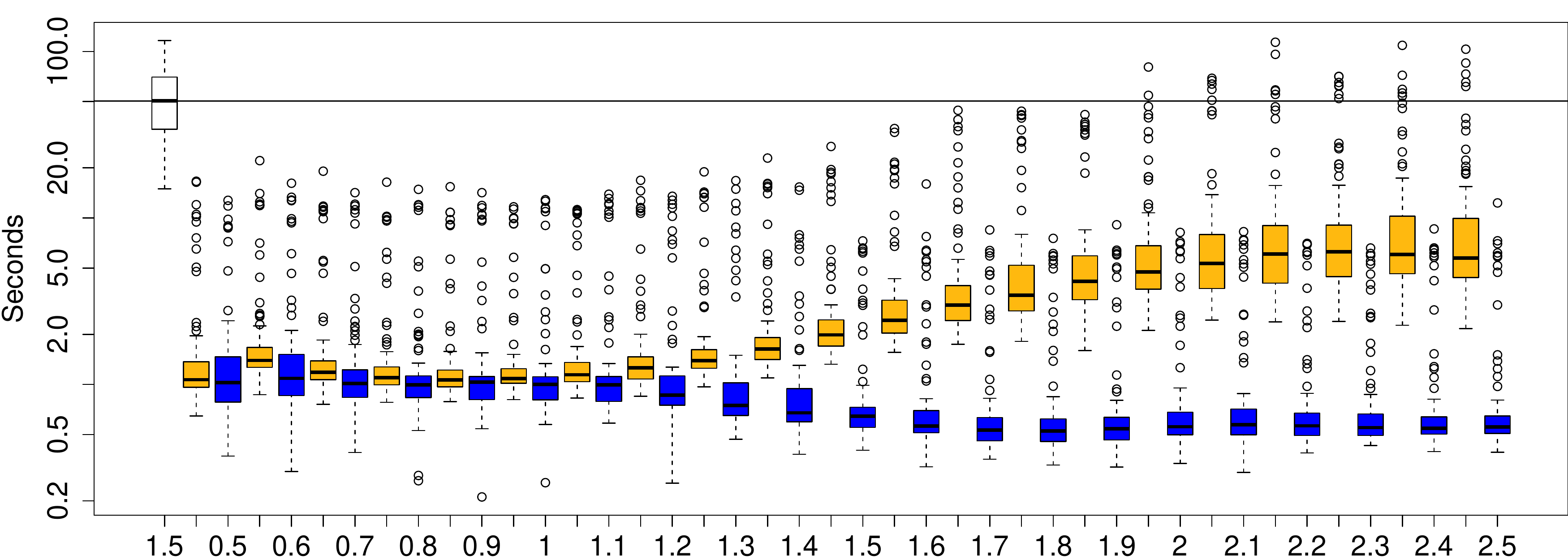}}
		\subfigure[$50 \times 50$ grid with $m=23$]{\includegraphics[scale=.23]{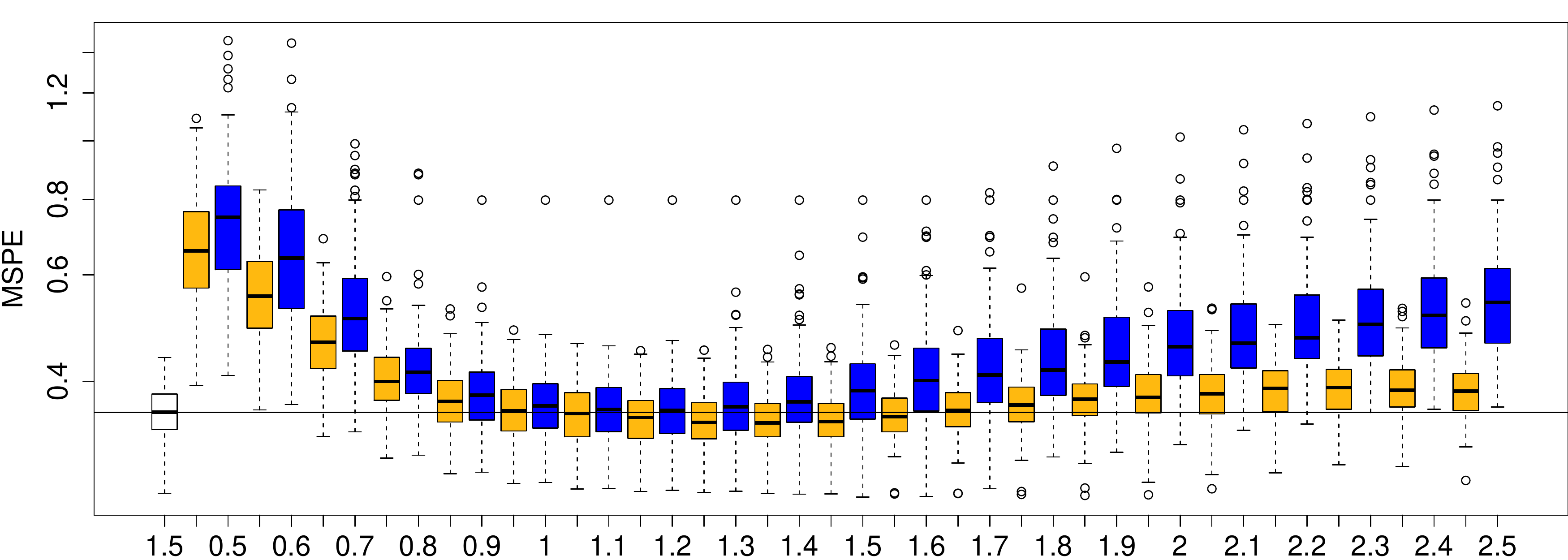}}
		\subfigure[$50 \times 50$ grid with $m=77$]{\includegraphics[scale=.23]{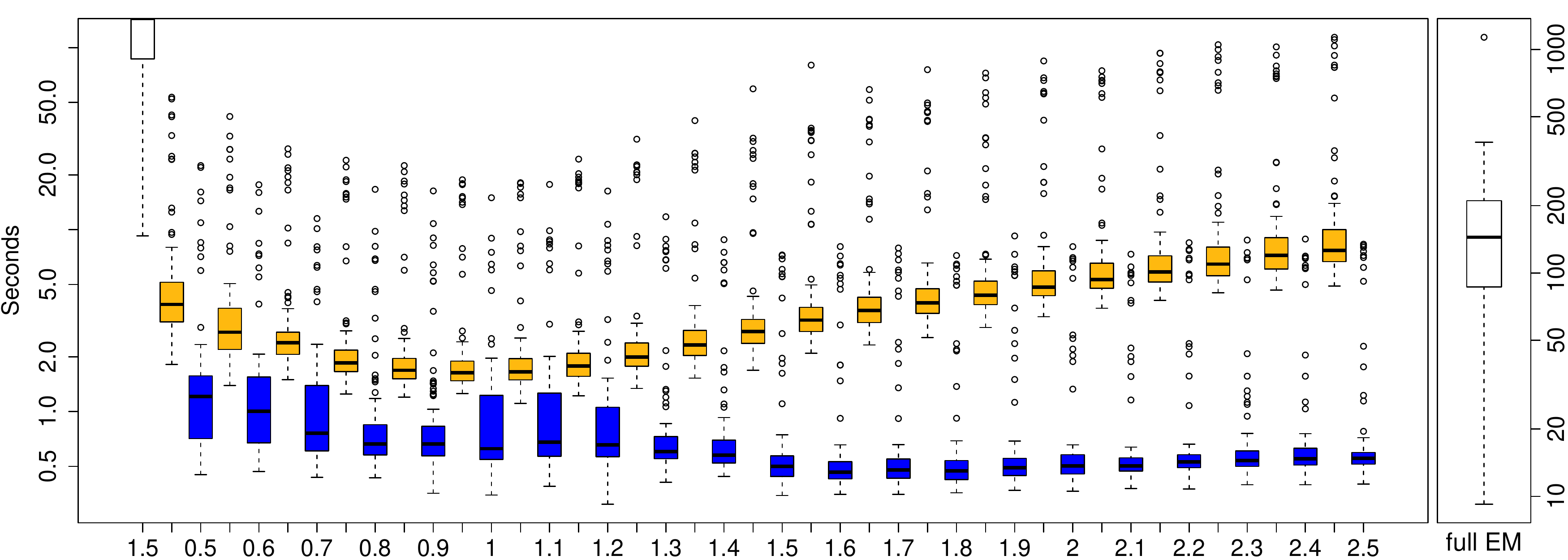}}
		\subfigure[$50 \times 50$ grid with $m=77$]{\includegraphics[scale=.23]{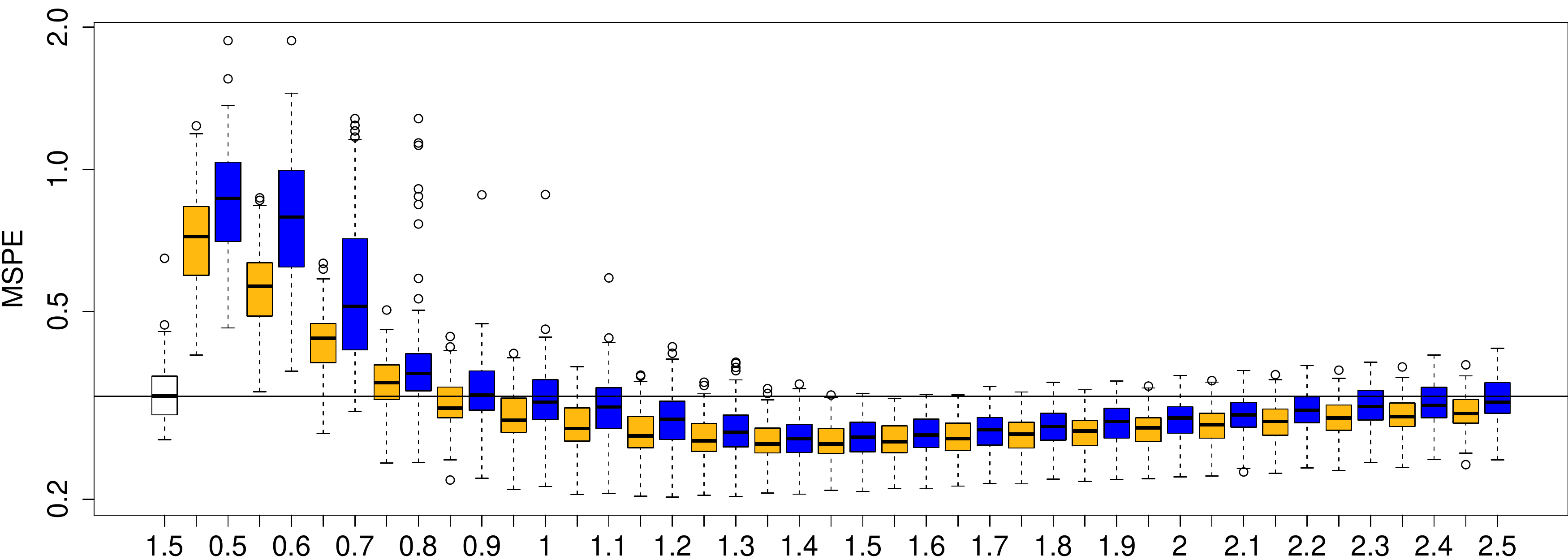}}
		\subfigure[$50 \times 50$ grid with $m=175$]{\includegraphics[scale=.23]{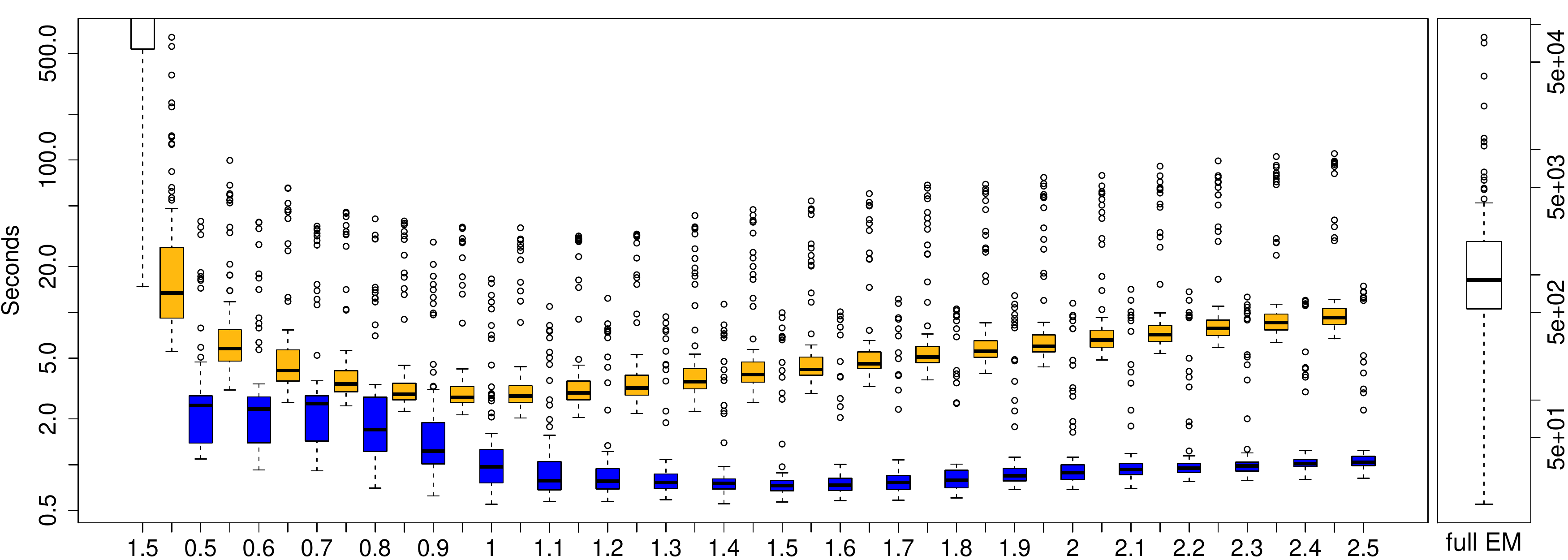}}		
		\subfigure[$50 \times 50$ grid with $m=175$]{\includegraphics[scale=.23]{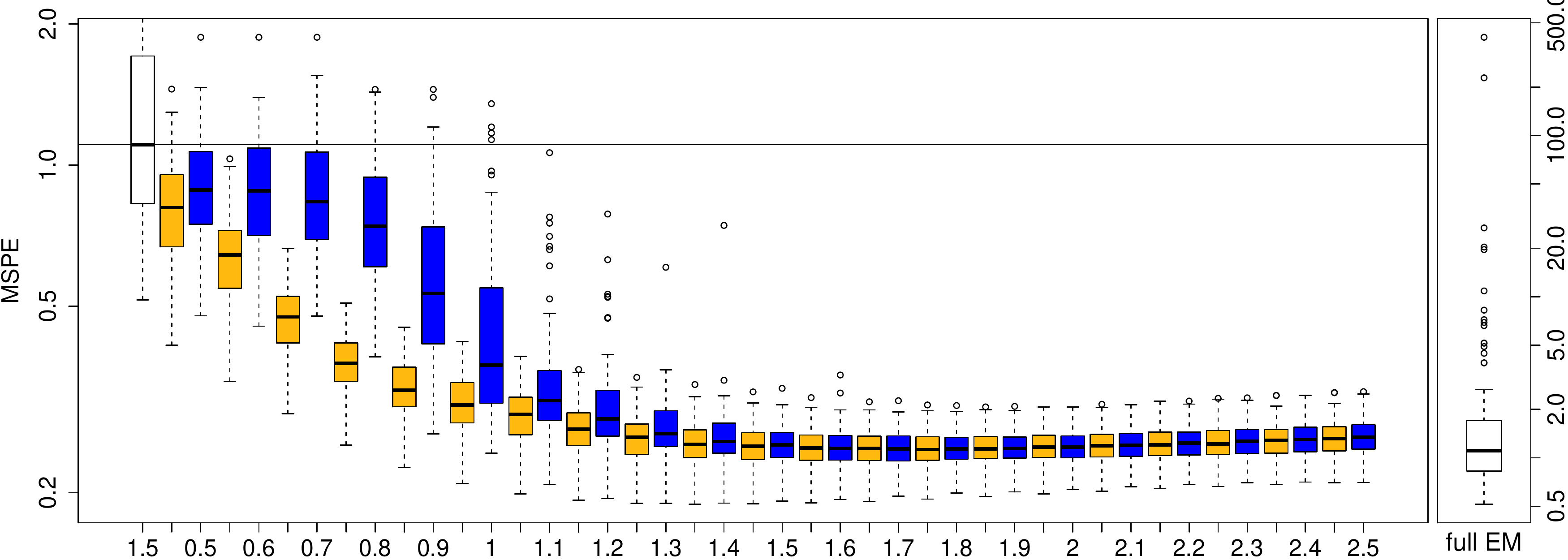}}
		\subfigure[$200 \times 200$ grid with $m=23$]{\includegraphics[scale=.23]{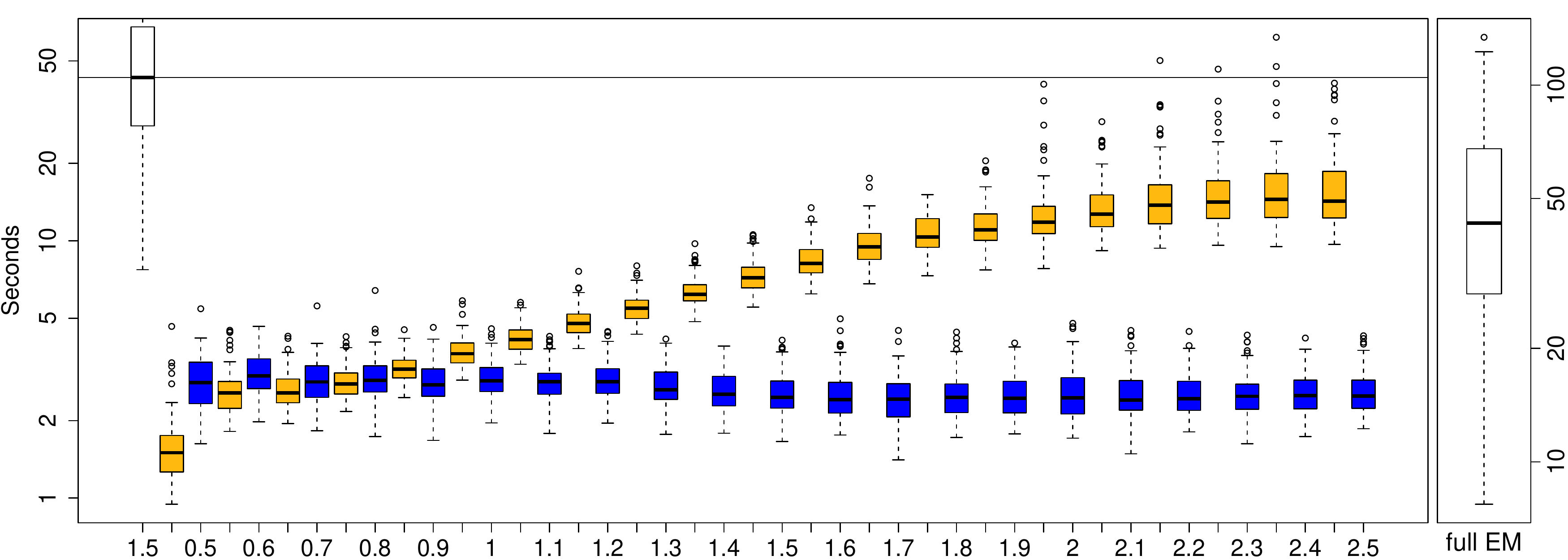}}
		\subfigure[$200 \times 200$ grid with $m=23$]{\includegraphics[scale=.23]{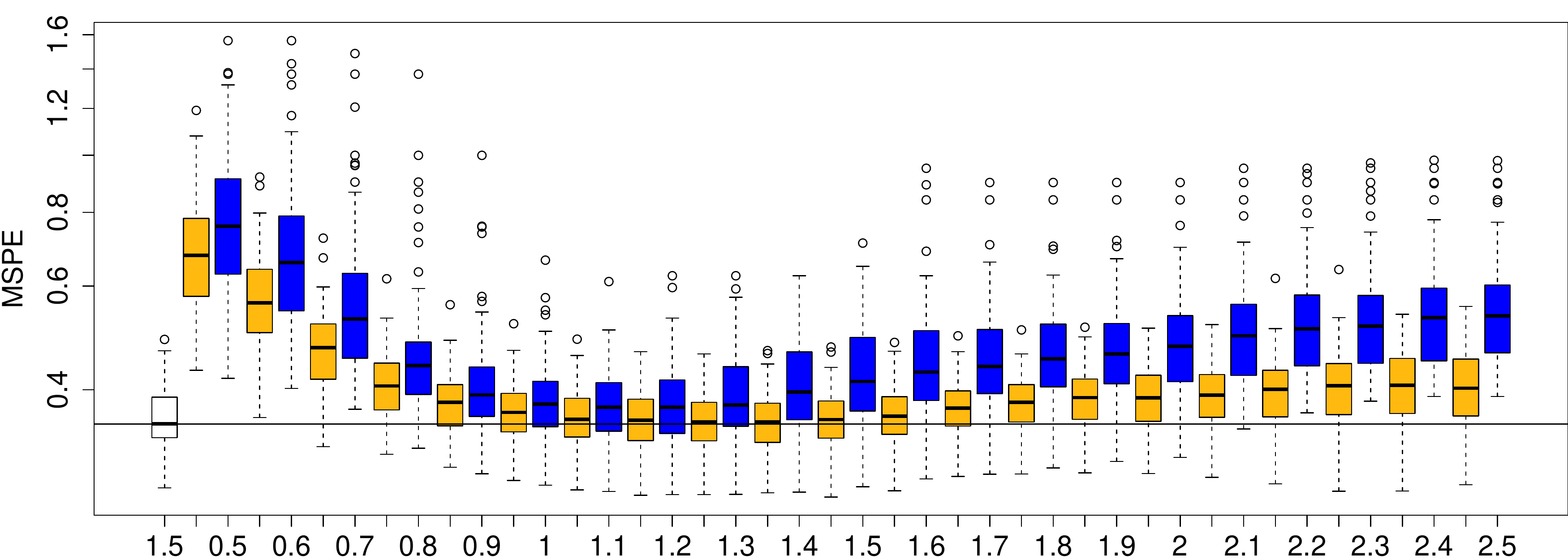}}
		\subfigure[$200 \times 200$ grid with $m=77$]{\includegraphics[scale=.23]{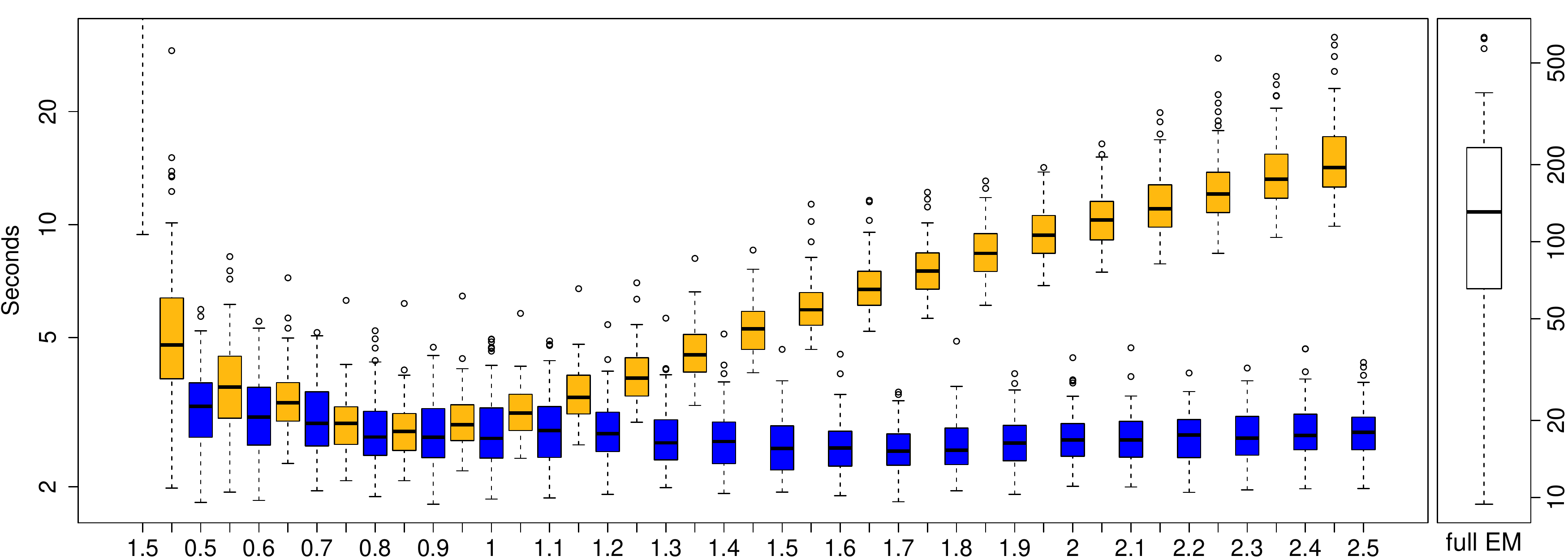}}
		\subfigure[$200 \times 200$ grid with $m=77$]{\includegraphics[scale=.23]{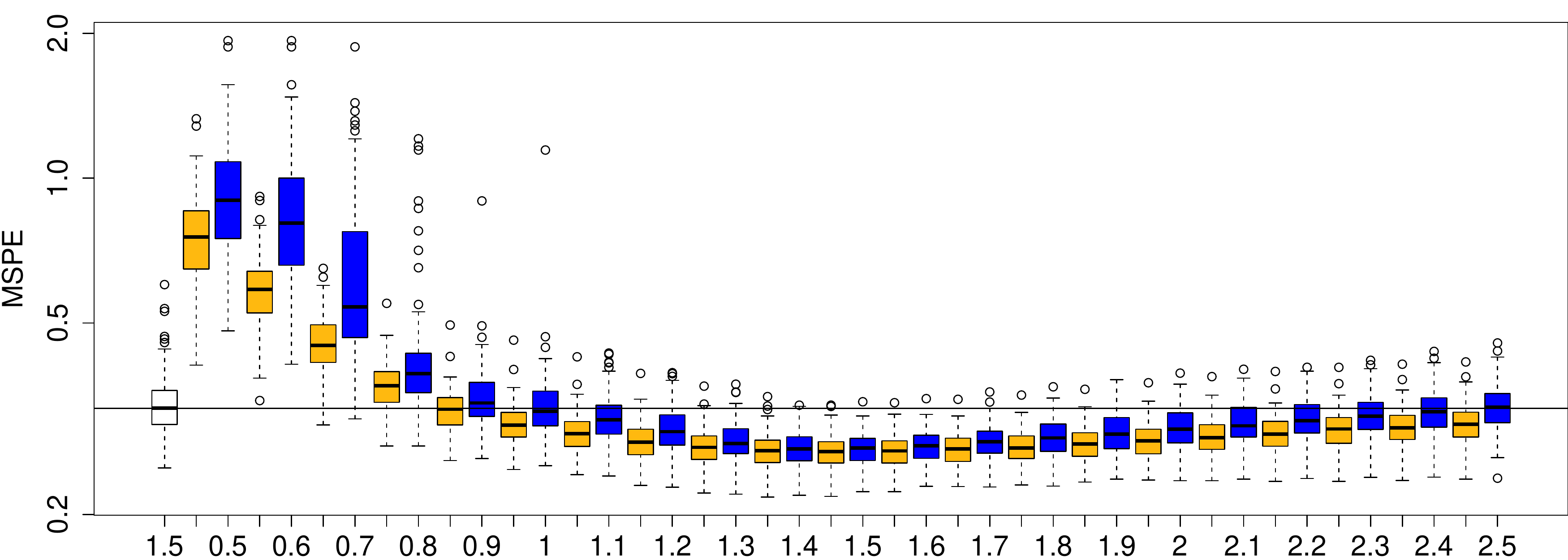}}
		\subfigure[$200 \times 200$ grid with $m=175$]{\includegraphics[scale=.23]{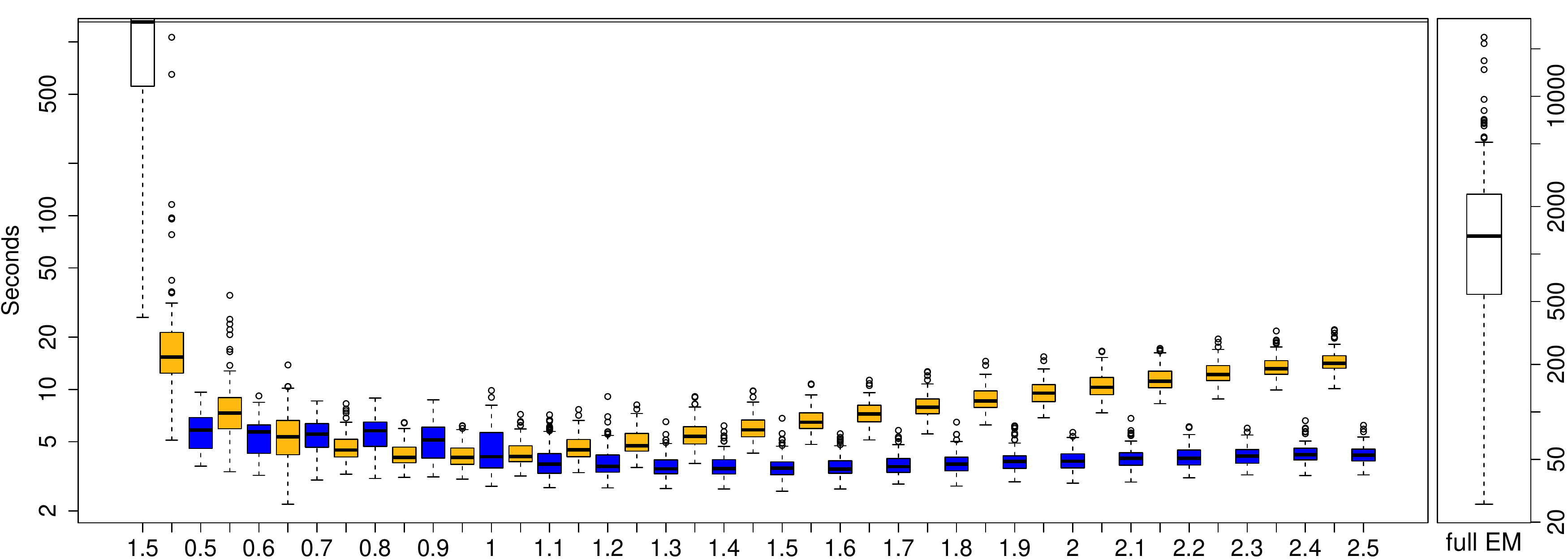}}		
		\subfigure[$200 \times 200$ grid with $m=175$]{\includegraphics[scale=.23]{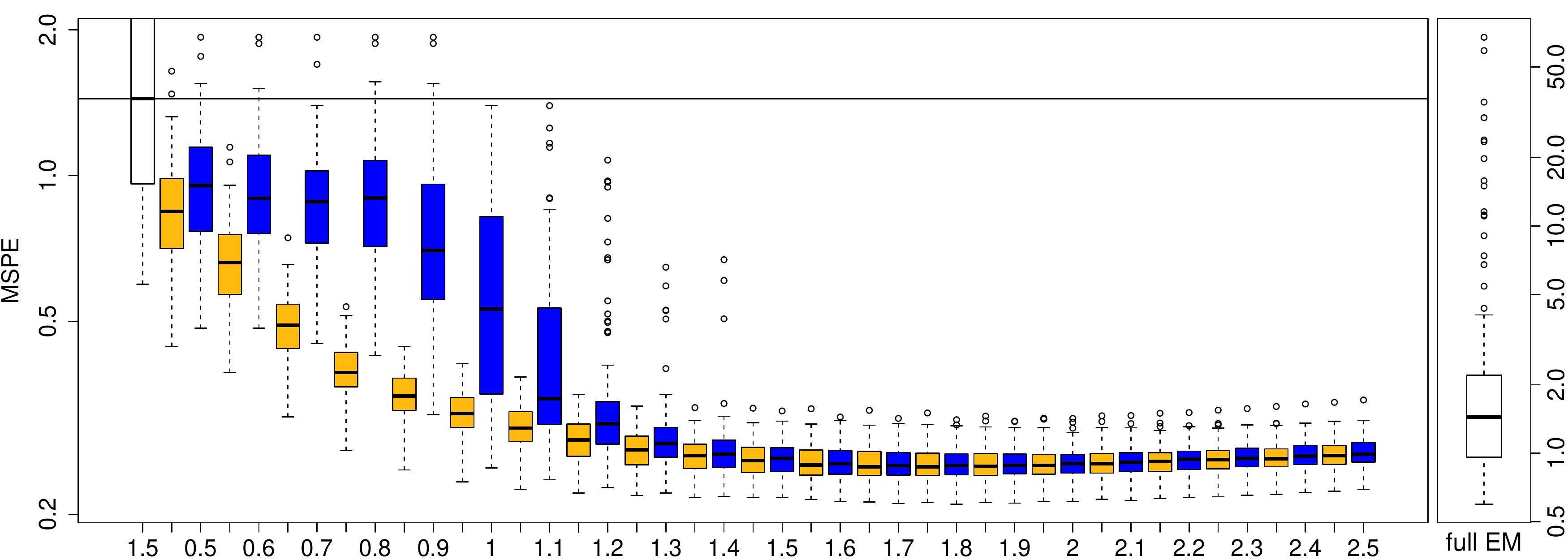}}
	\end{center}
	\caption{Seconds (left panel) and Mean Squared Prediction
          Errors (MSPEs, right panel) for varying bandwidths constants
          (x-axis) and number of knots ($m$) for all three estimation
          methods on both grids when $\nu=0.5$ and $\sigma^2=0.25$.} 
\end{figure}

\begin{figure}[H]
	\begin{center}
		\subfigure[$50 \times 50$ grid with $m=23$]{\includegraphics[scale=.23]{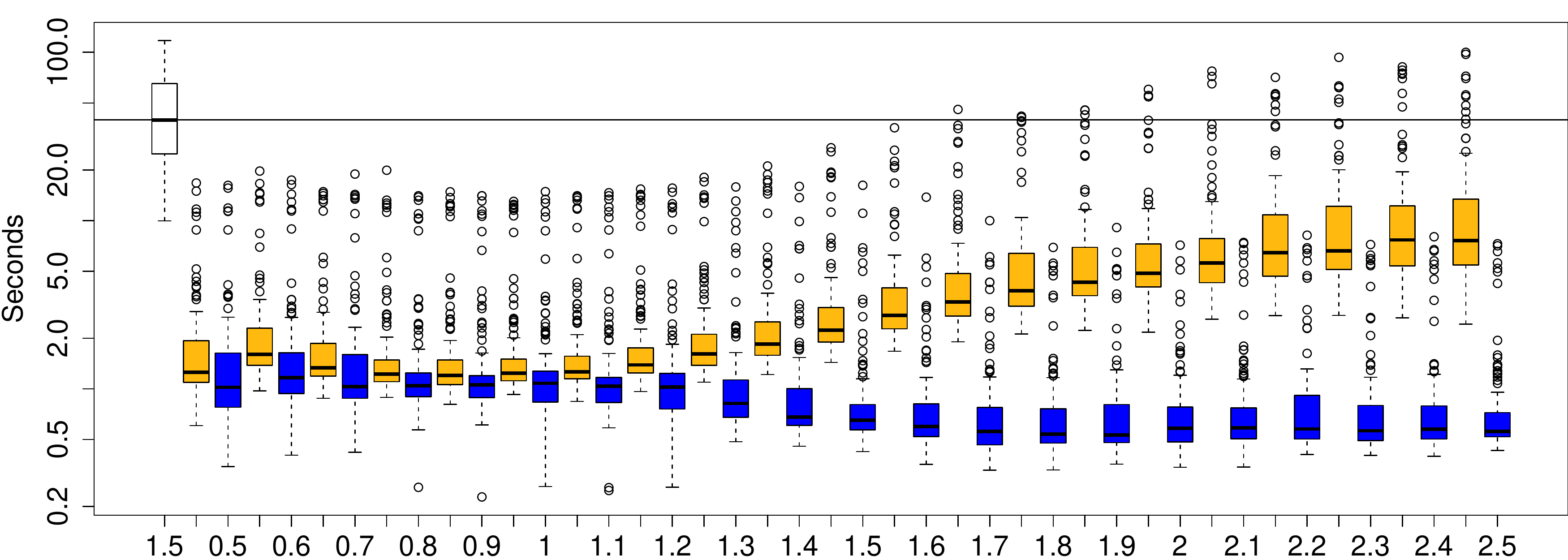}}
		\subfigure[$50 \times 50$ grid with $m=23$]{\includegraphics[scale=.23]{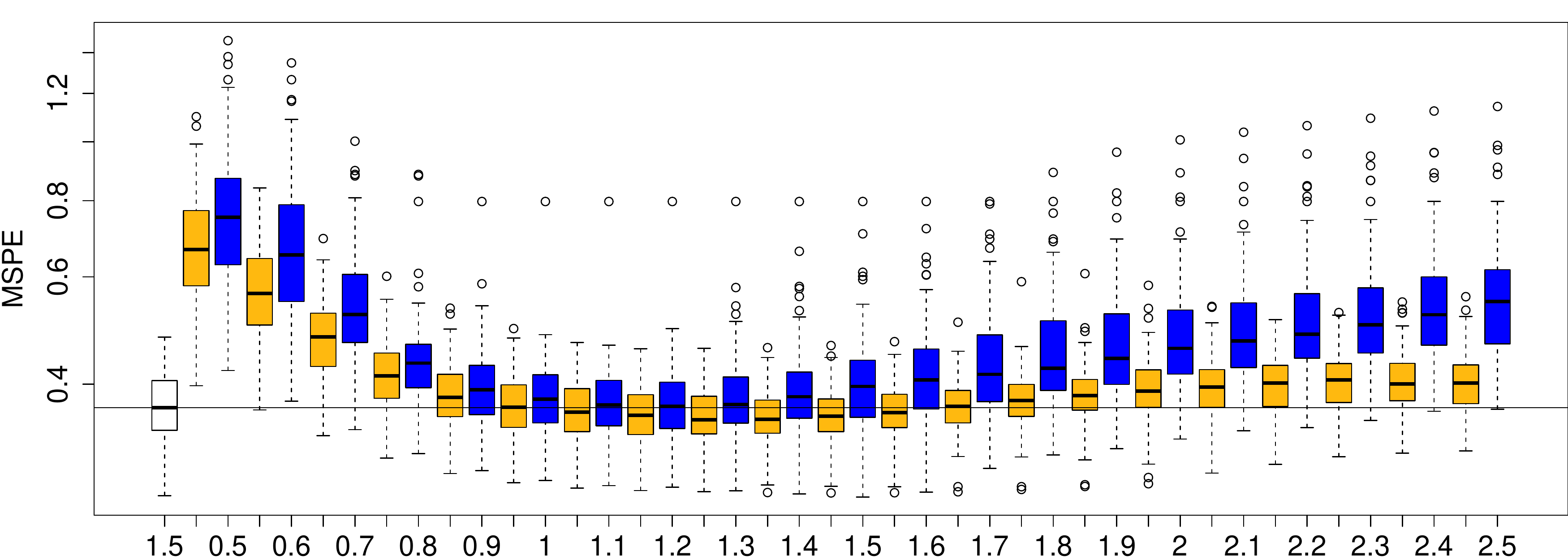}}
		\subfigure[$50 \times 50$ grid with $m=77$]{\includegraphics[scale=.23]{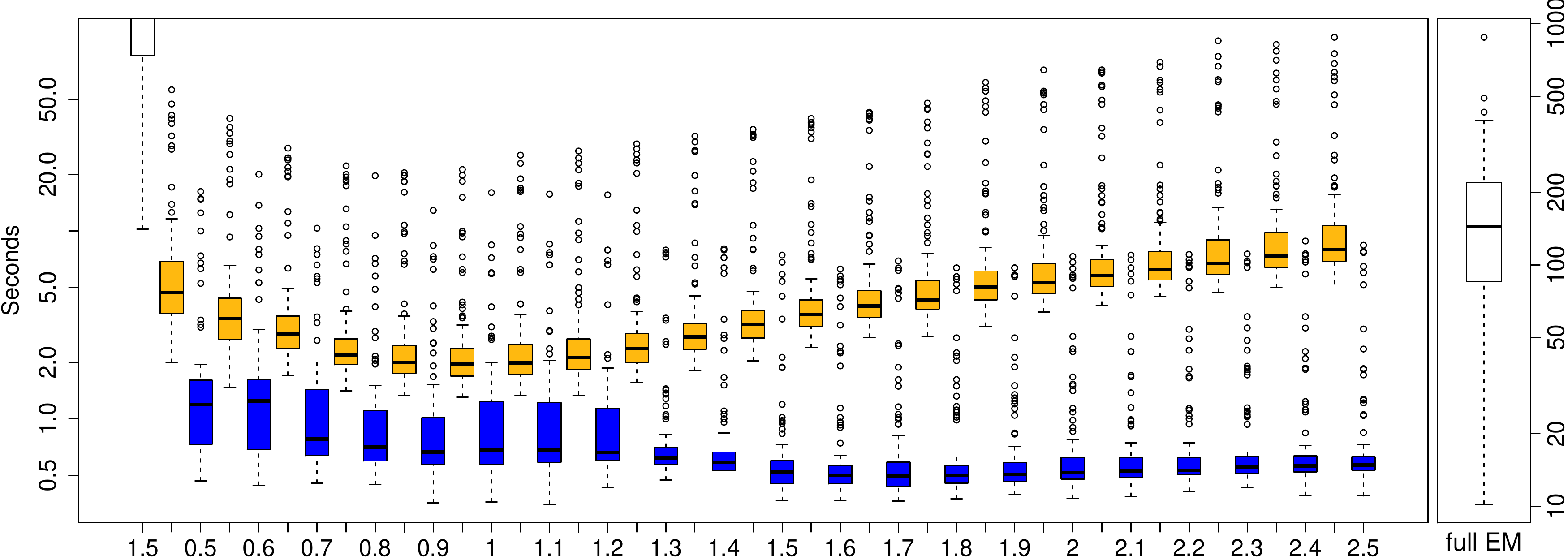}}
		\subfigure[$50 \times 50$ grid with $m=77$]{\includegraphics[scale=.23]{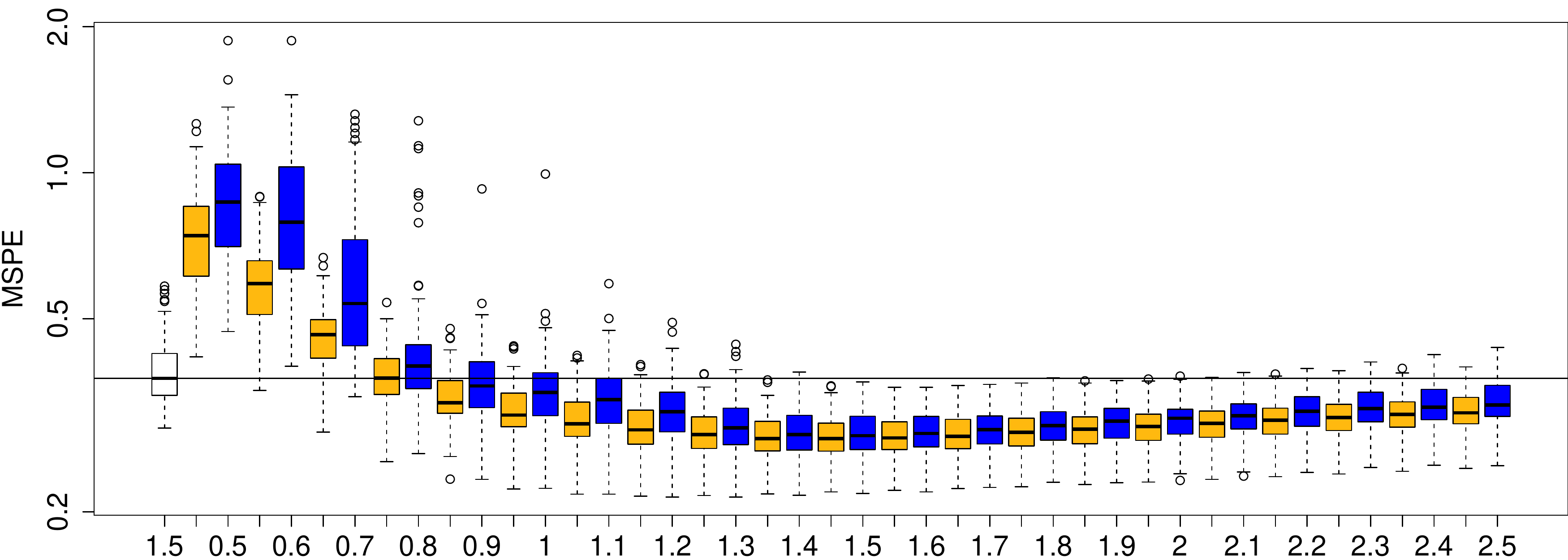}}
		\subfigure[$50 \times 50$ grid with $m=175$]{\includegraphics[scale=.23]{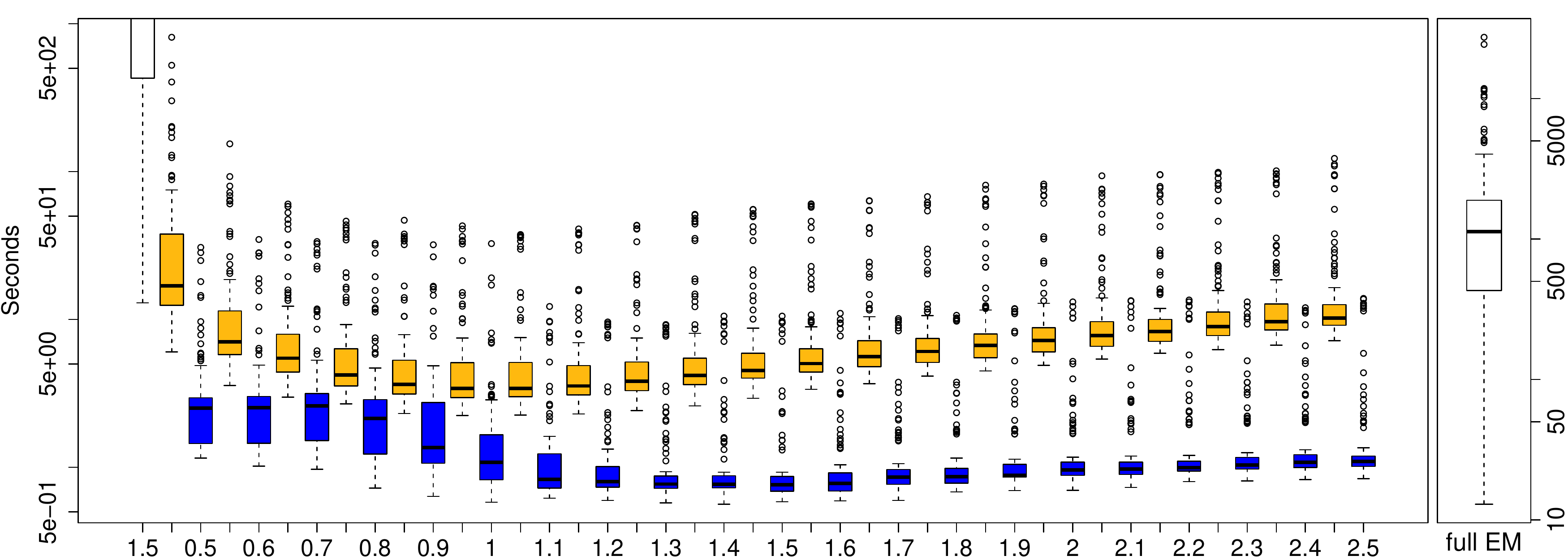}}		
		\subfigure[$50 \times 50$ grid with $m=175$]{\includegraphics[scale=.23]{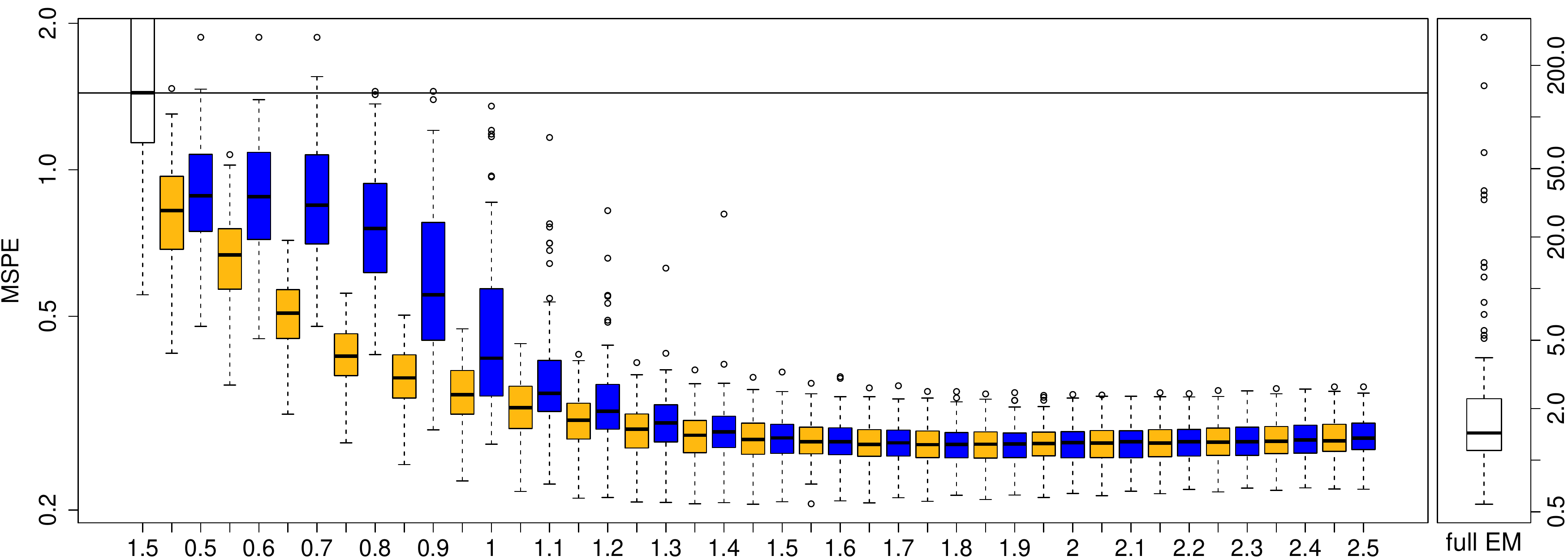}}
		\subfigure[$200 \times 200$ grid with $m=23$]{\includegraphics[scale=.23]{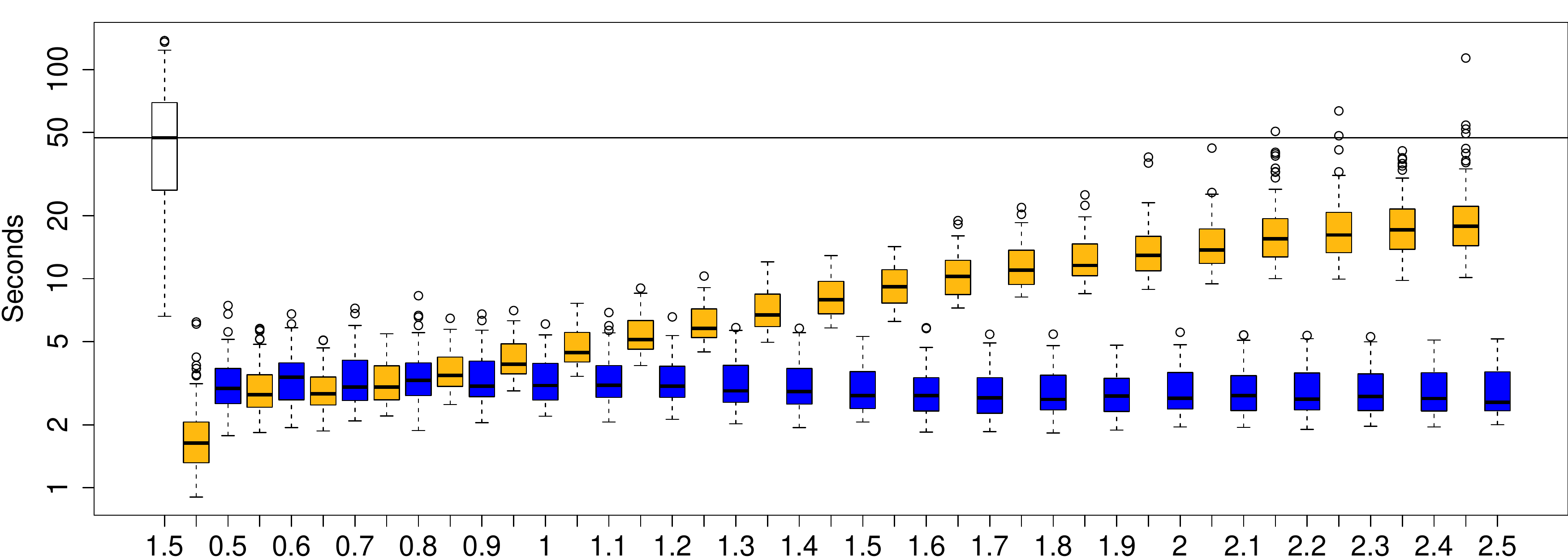}}
		\subfigure[$200 \times 200$ grid with $m=23$]{\includegraphics[scale=.23]{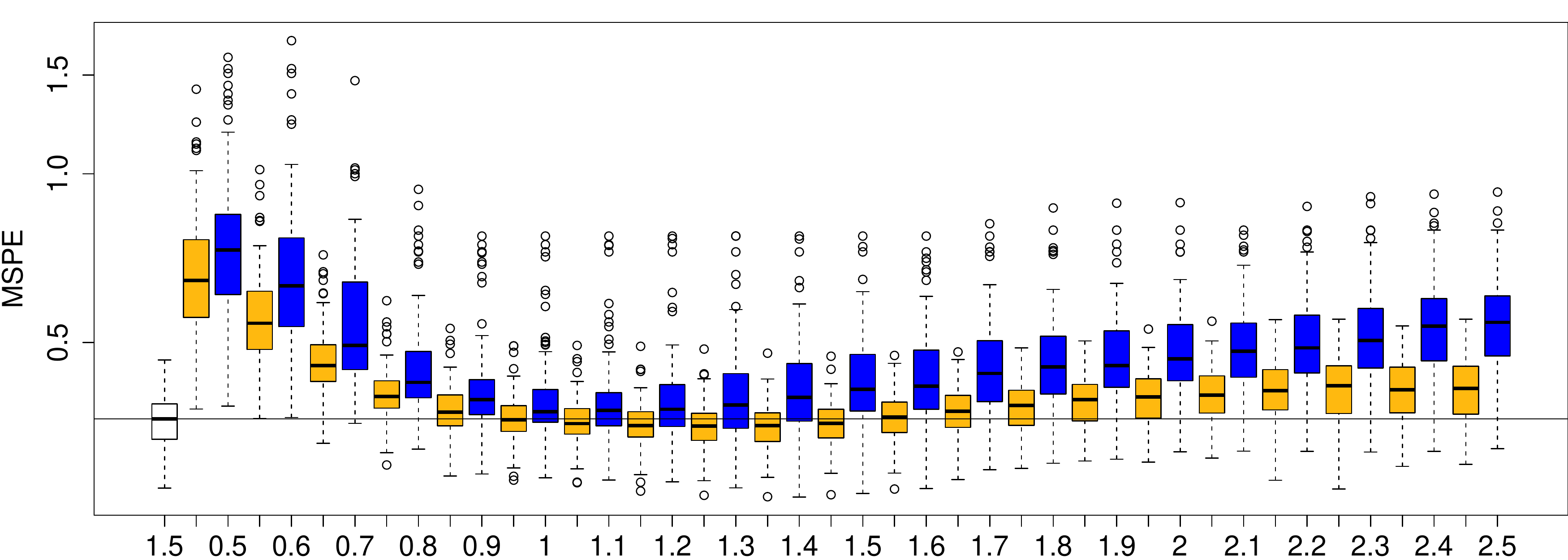}}
		\subfigure[$200 \times 200$ grid with $m=77$]{\includegraphics[scale=.23]{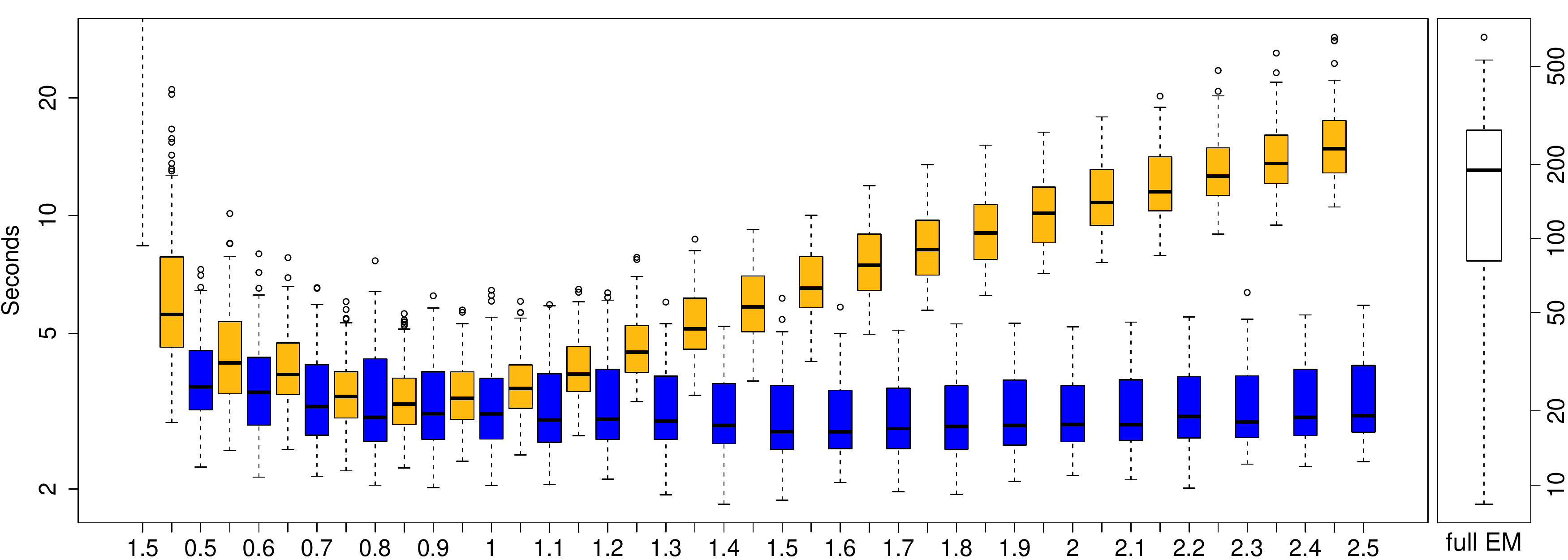}}
		\subfigure[$200 \times 200$ grid with $m=77$]{\includegraphics[scale=.23]{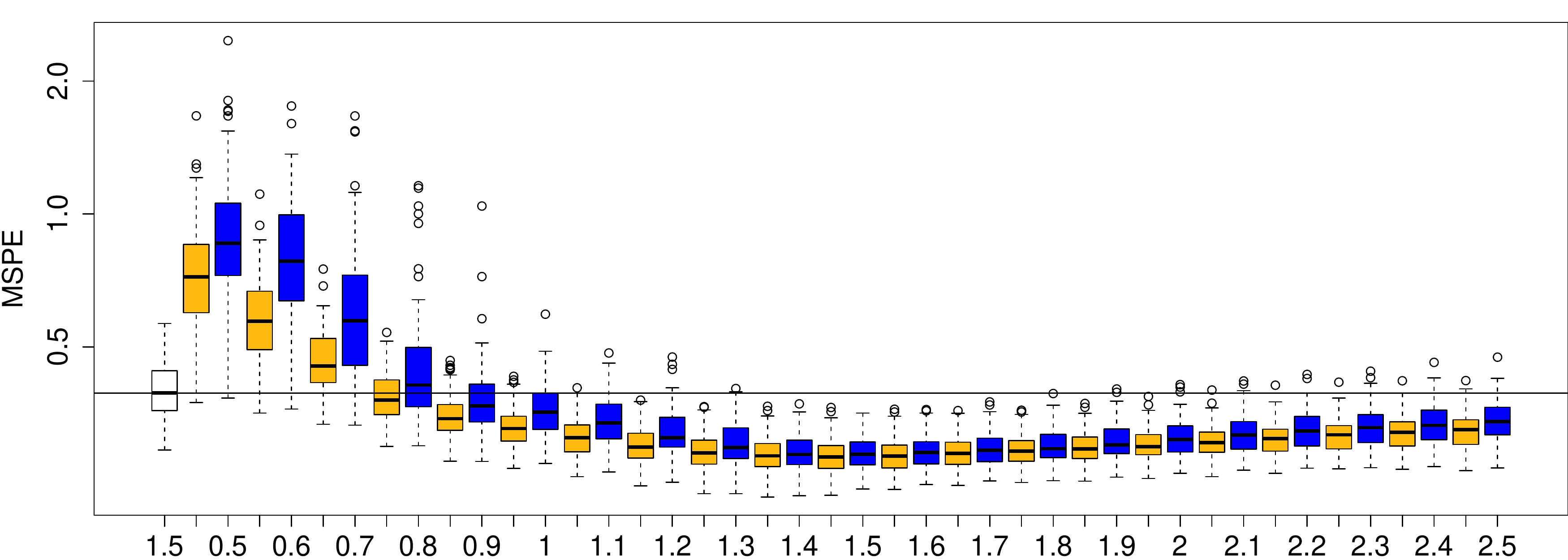}}
		\subfigure[$200 \times 200$ grid with $m=175$]{\includegraphics[scale=.23]{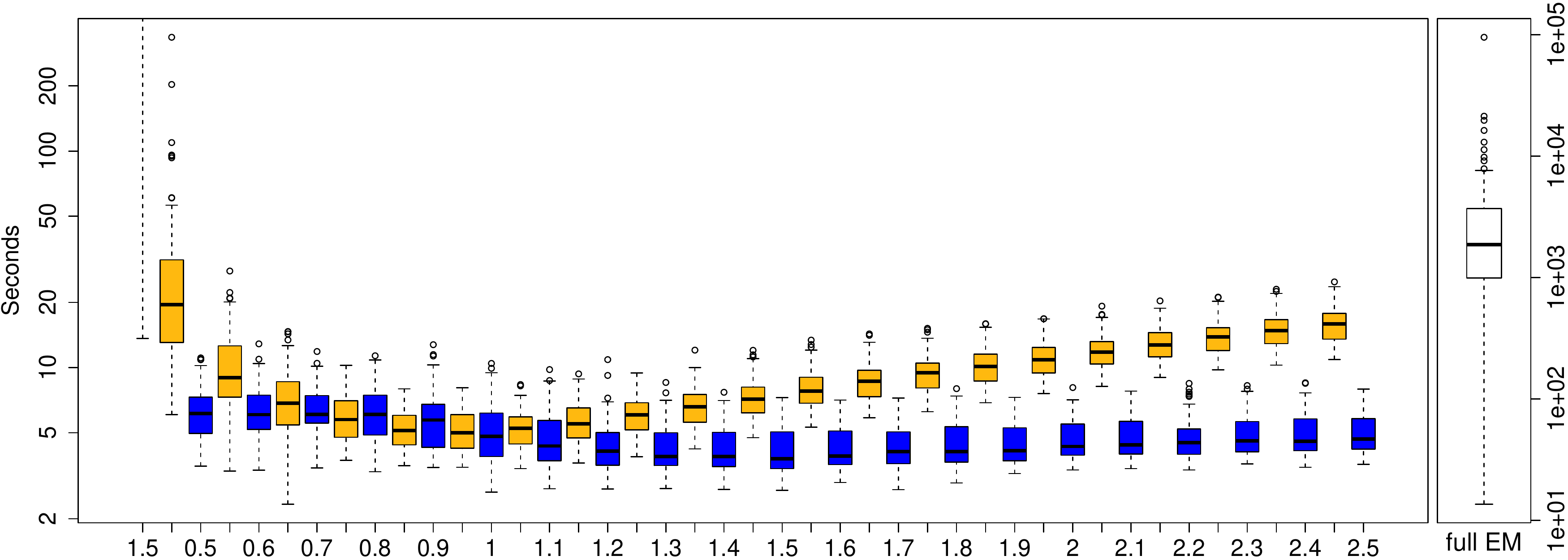}}		
		\subfigure[$200 \times 200$ grid with $m=175$]{\includegraphics[scale=.23]{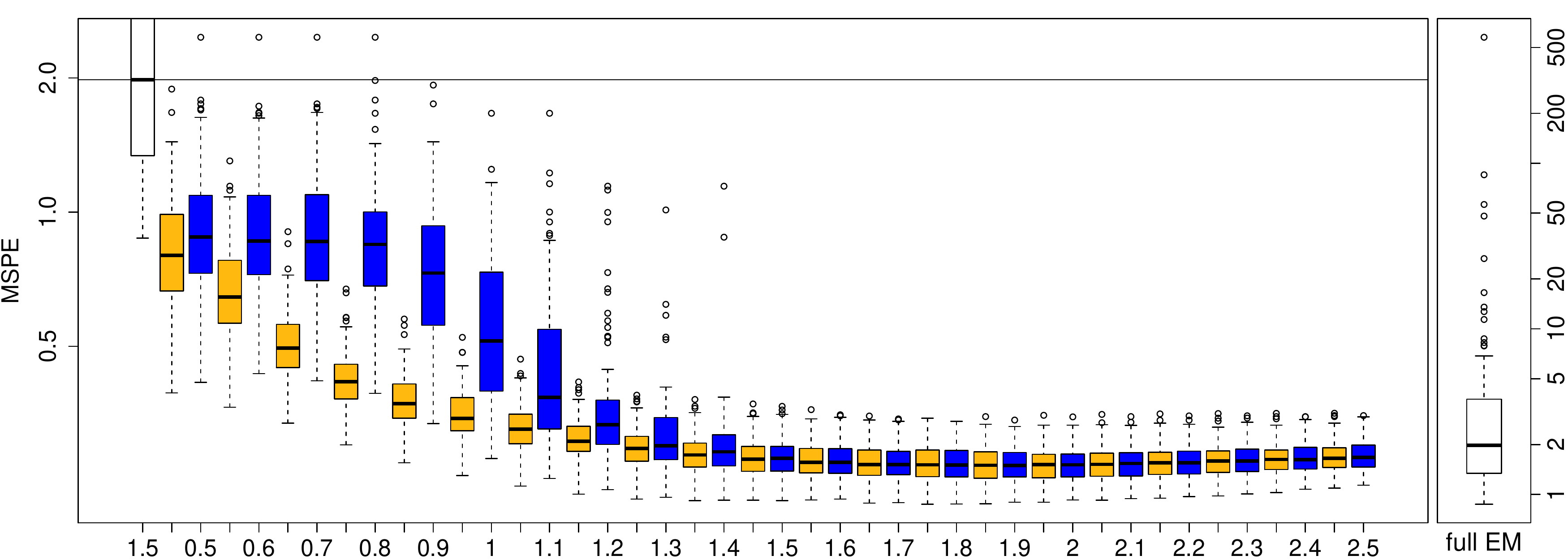}}
	\end{center}
	\caption{Seconds (left panel) and Mean Squared Prediction
          Errors (MSPEs, right panel) for varying bandwidths constants
          (x-axis) and number of knots ($m$) for all three estimation
          methods on both grids when $\nu=0.5$ and $\sigma^2=0.4$.} 
\end{figure}

\begin{figure}[H]
	\begin{center}
		\subfigure[$50 \times 50$ grid with $m=23$]{\includegraphics[scale=.23]{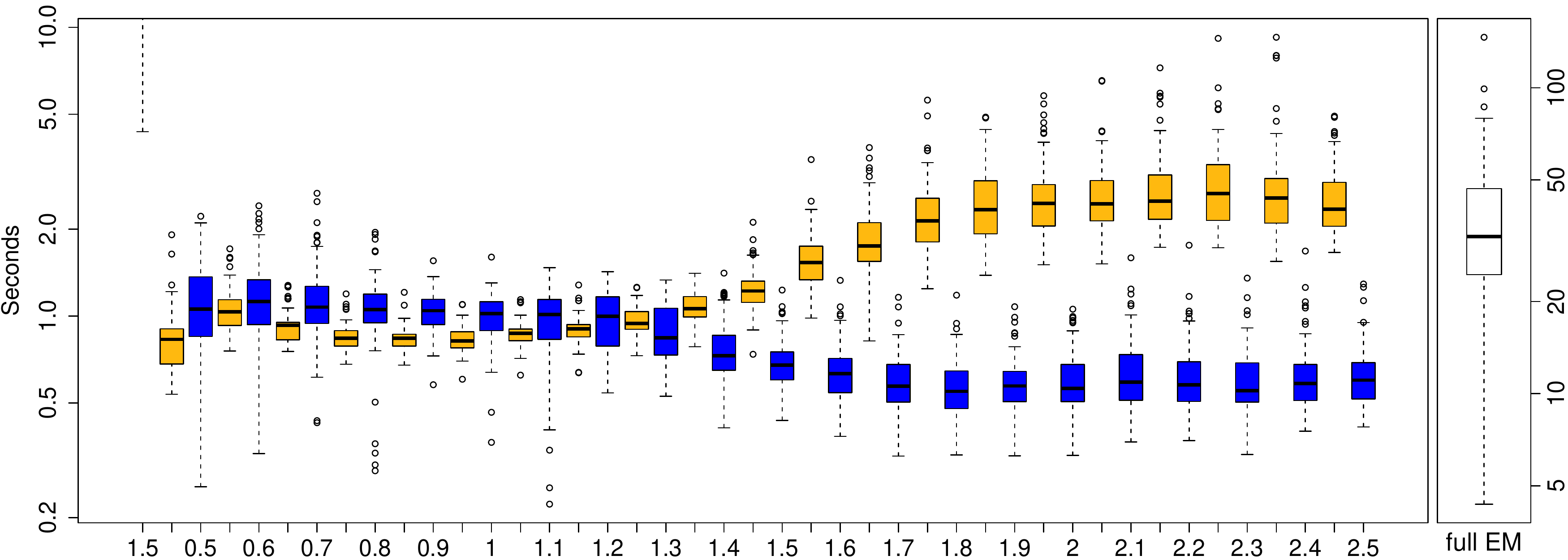}}
		\subfigure[$50 \times 50$ grid with $m=23$]{\includegraphics[scale=.23]{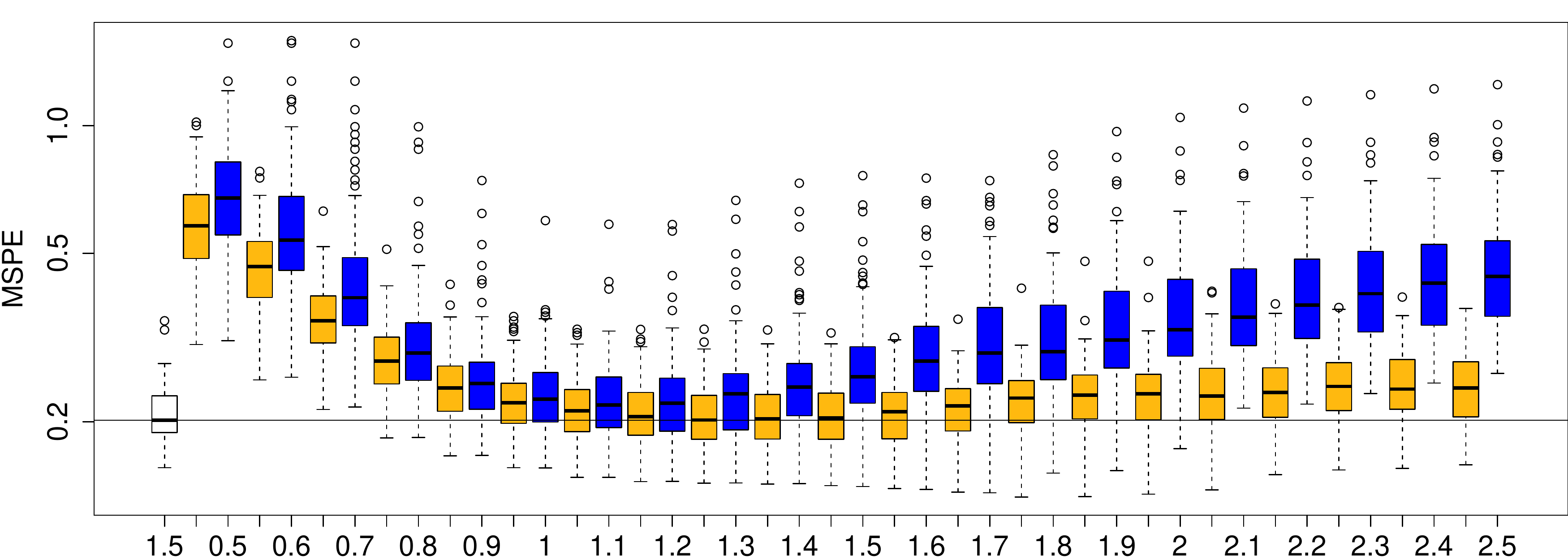}}
		\subfigure[$50 \times 50$ grid with $m=77$]{\includegraphics[scale=.23]{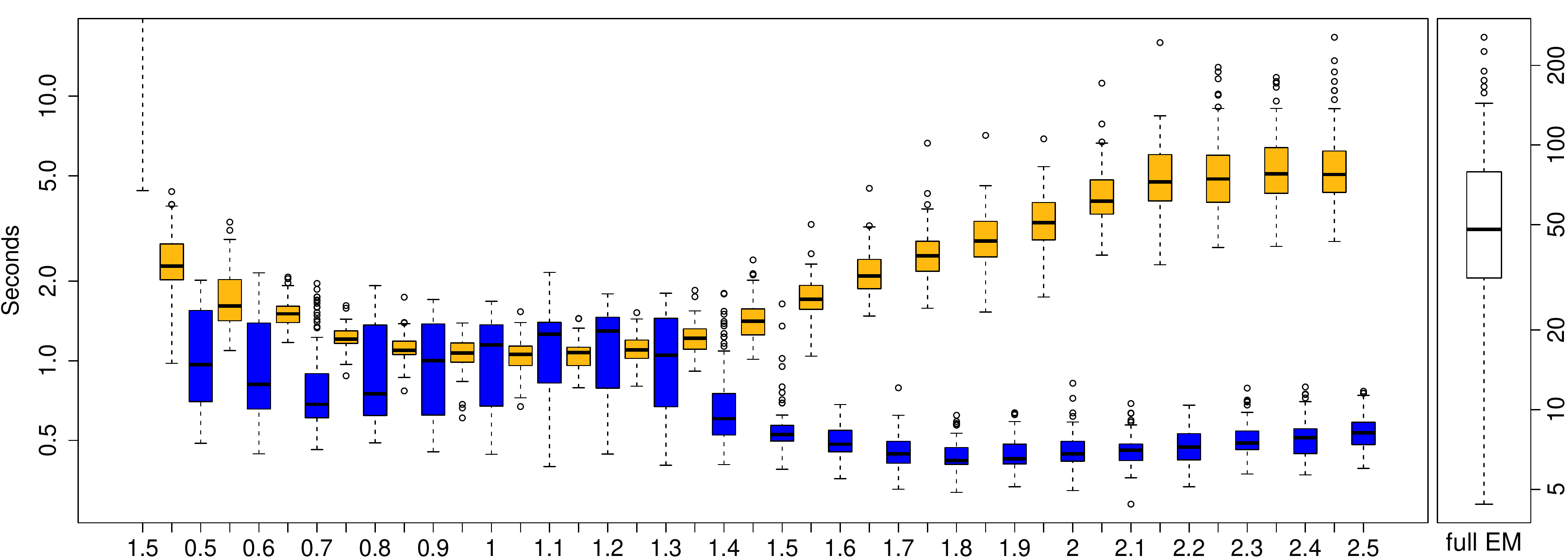}}
		\subfigure[$50 \times 50$ grid with $m=77$]{\includegraphics[scale=.23]{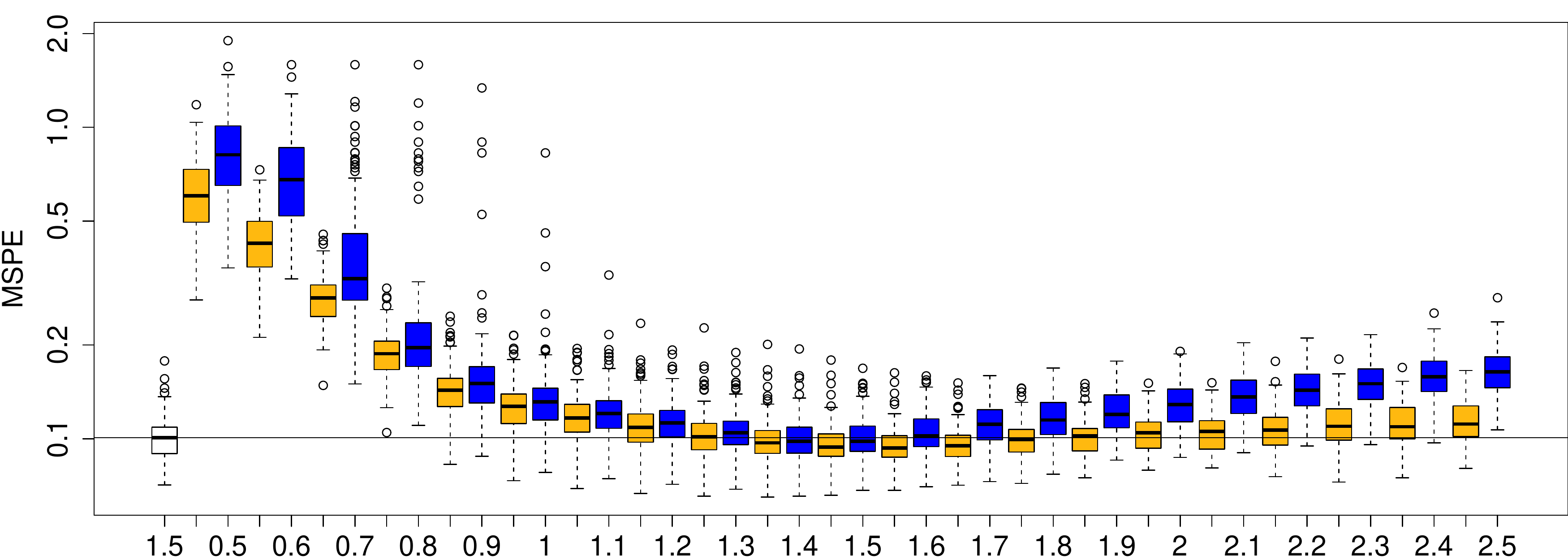}}
		\subfigure[$50 \times 50$ grid with $m=175$]{\includegraphics[scale=.23]{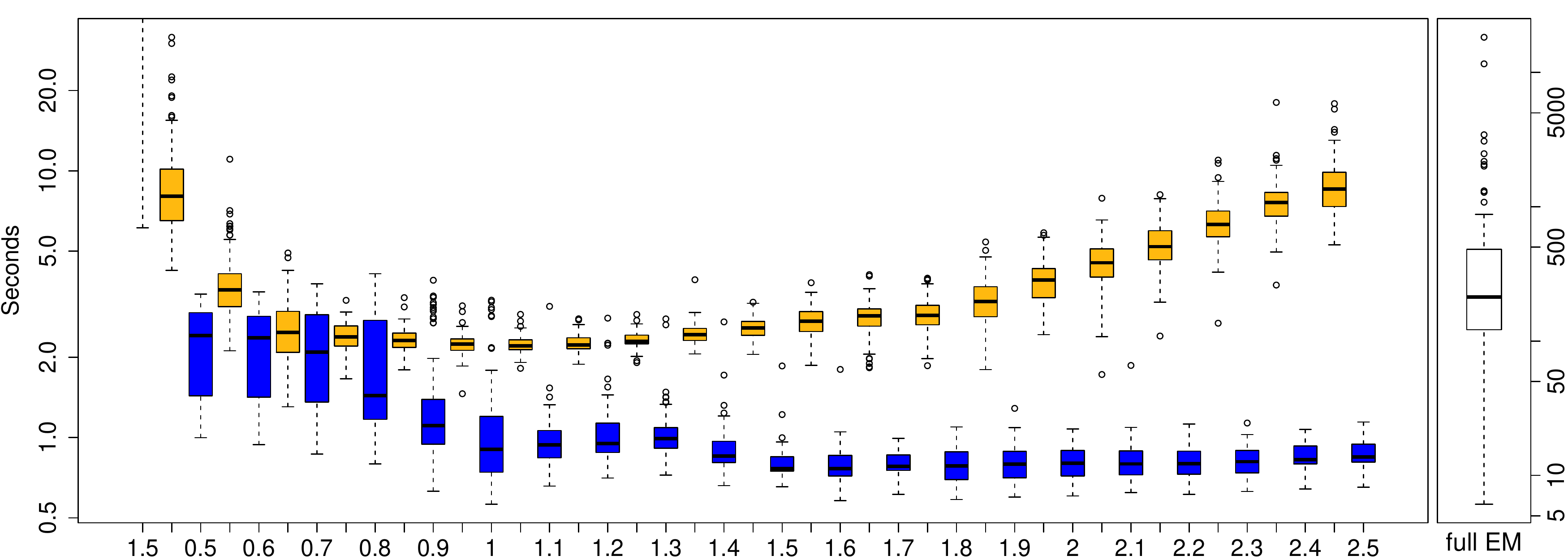}}		
		\subfigure[$50 \times 50$ grid with $m=175$]{\includegraphics[scale=.23]{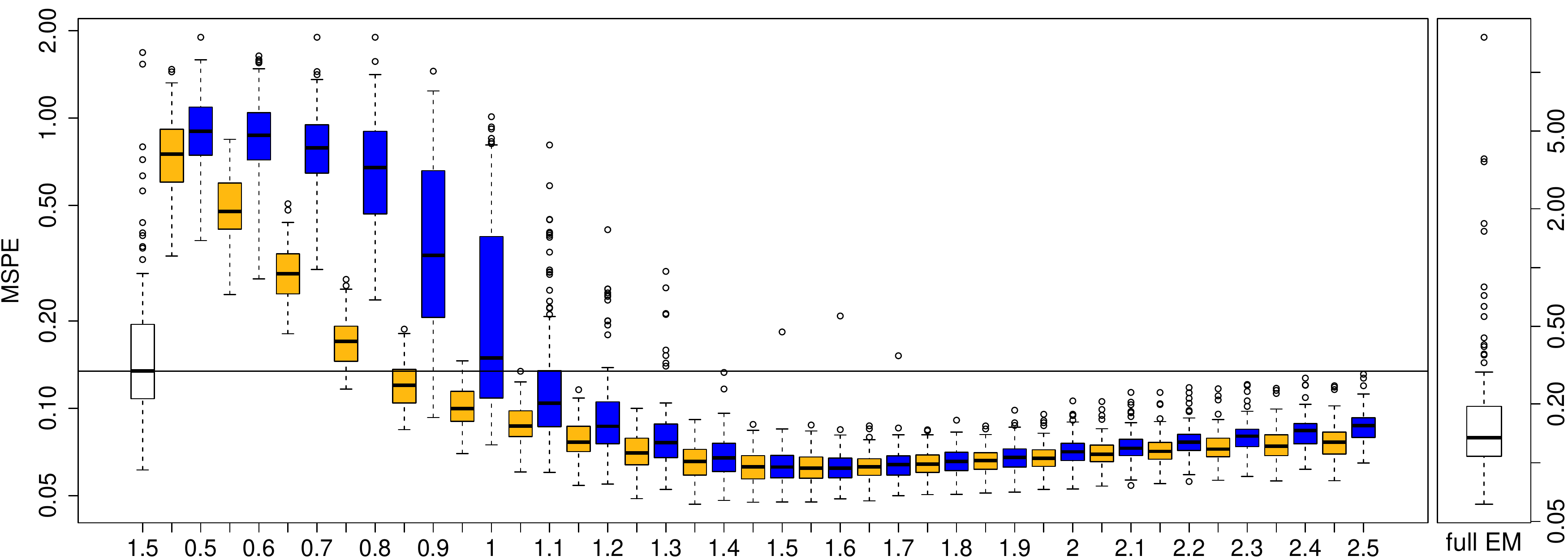}}
		\subfigure[$200 \times 200$ grid with $m=23$]{\includegraphics[scale=.23]{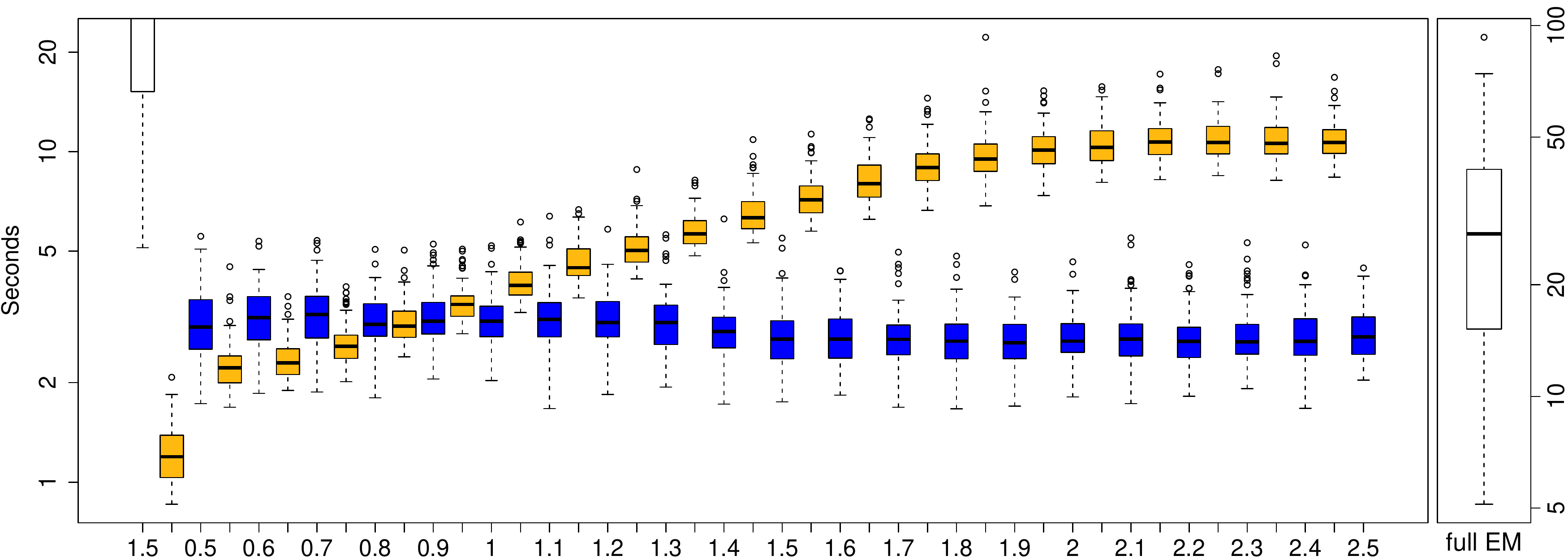}}
		\subfigure[$200 \times 200$ grid with $m=23$]{\includegraphics[scale=.23]{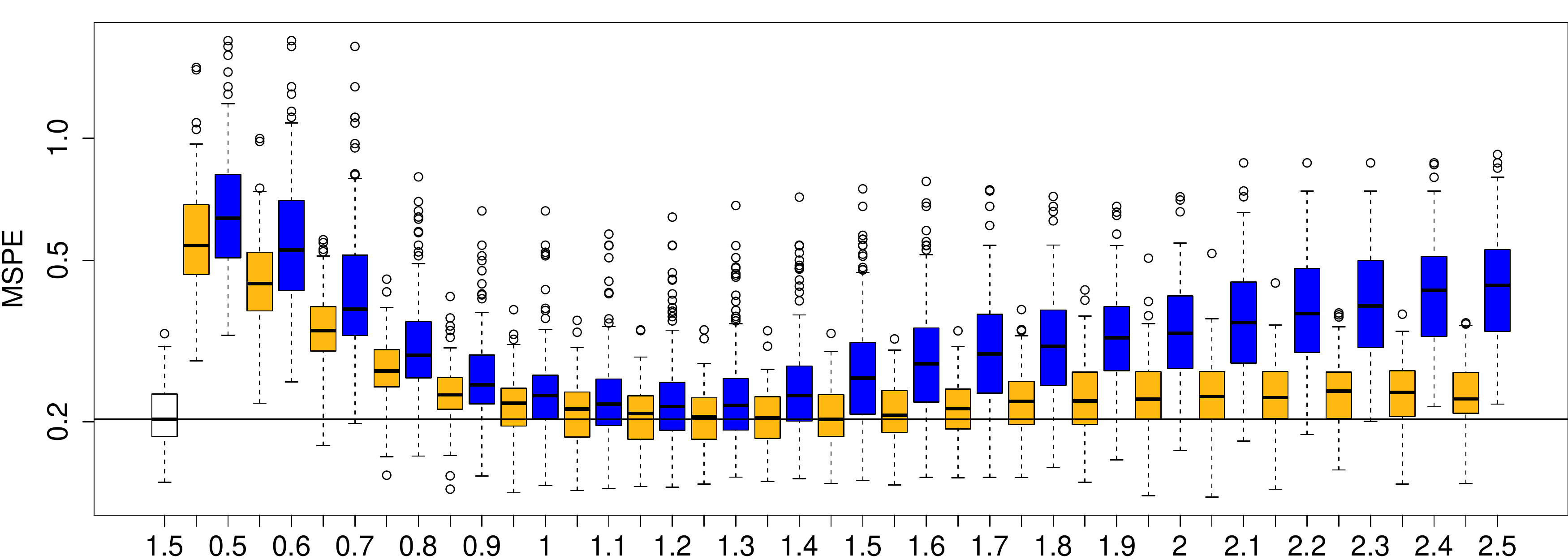}}
		\subfigure[$200 \times 200$ grid with $m=77$]{\includegraphics[scale=.23]{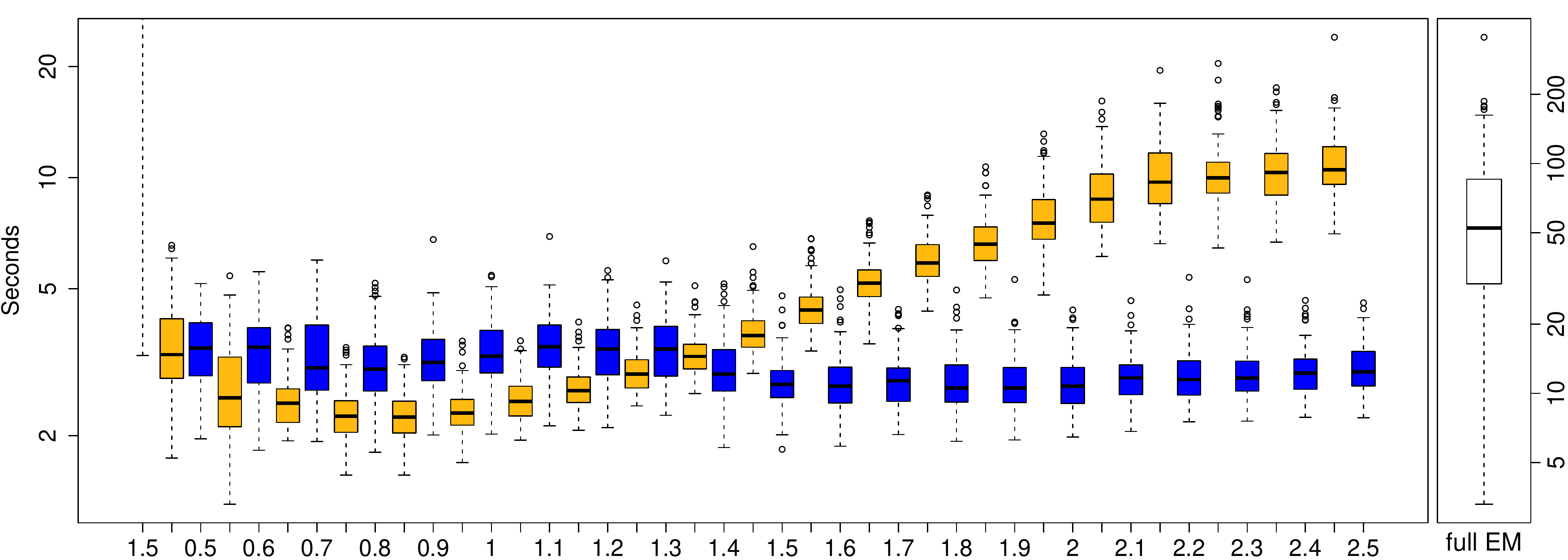}}
		\subfigure[$200 \times 200$ grid with $m=77$]{\includegraphics[scale=.23]{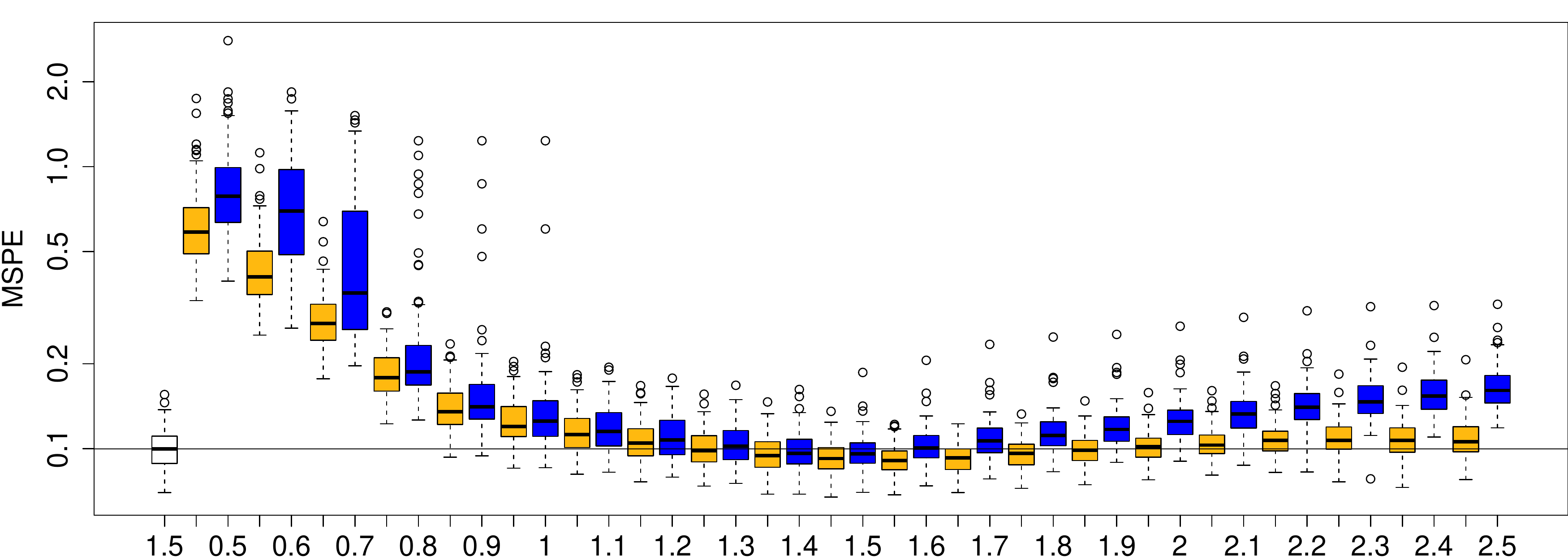}}
		\subfigure[$200 \times 200$ grid with $m=175$]{\includegraphics[scale=.23]{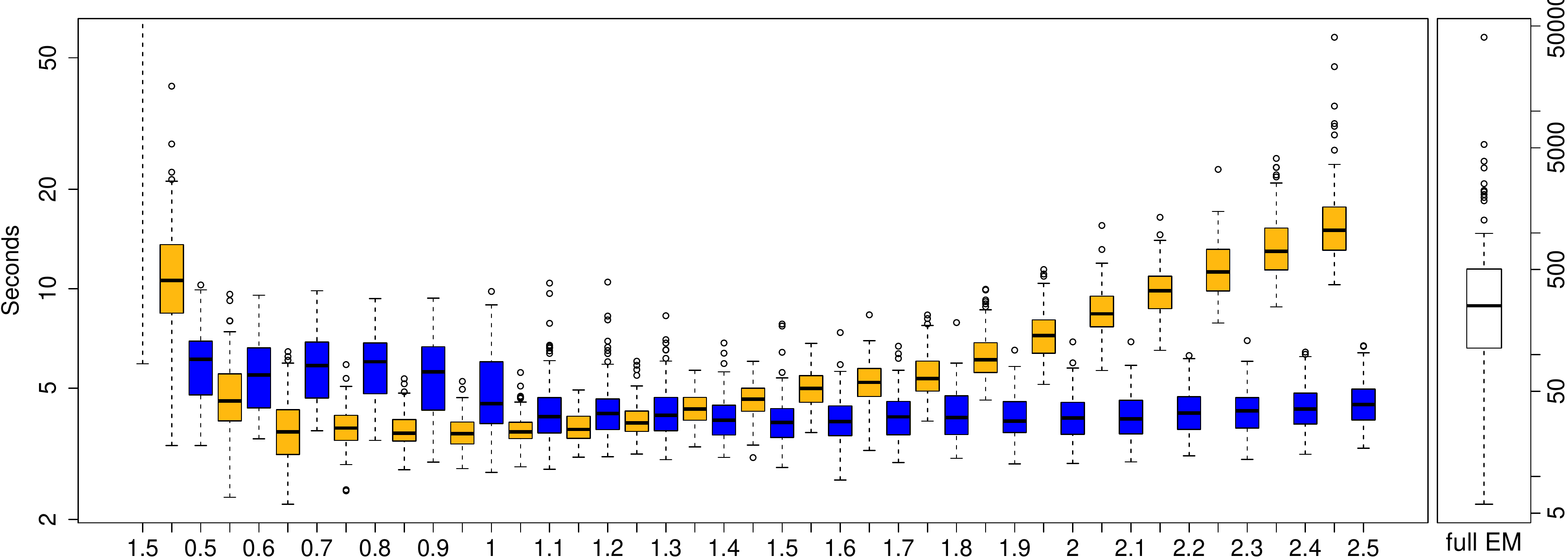}}		
		\subfigure[$200 \times 200$ grid with $m=175$]{ \includegraphics[scale=.23]{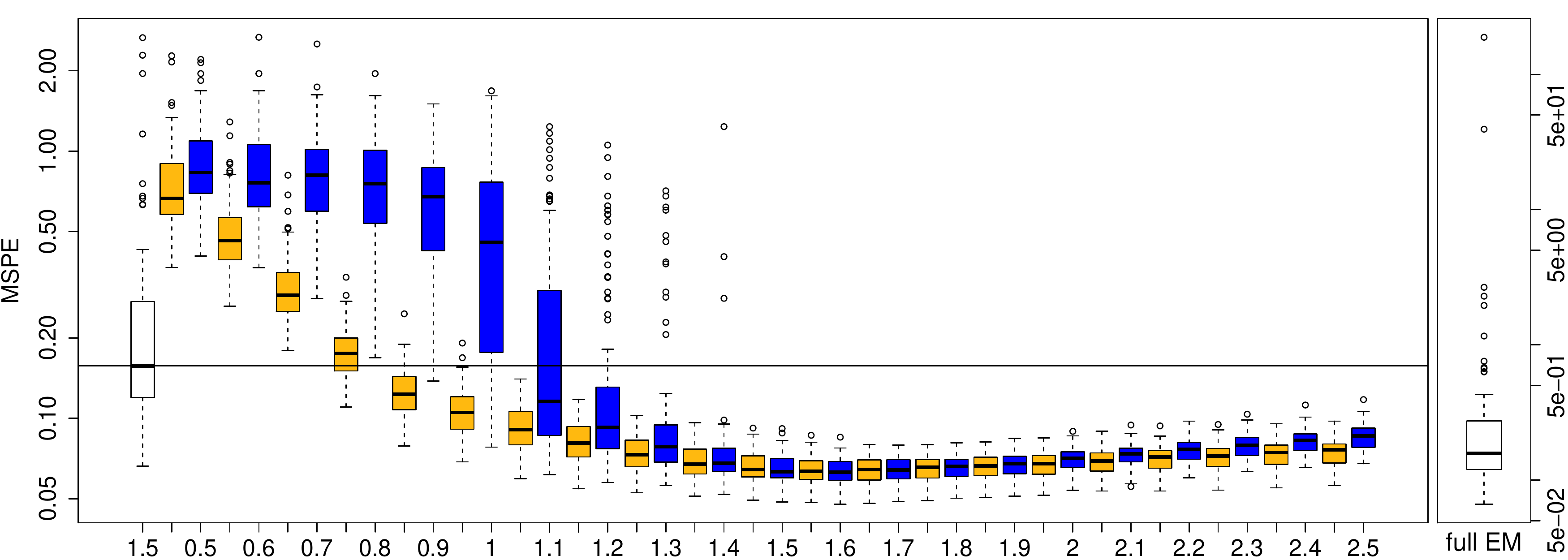}}
	\end{center}
	\caption{Seconds (left panel) and Mean Squared Prediction
          Errors (MSPEs, right panel) for varying bandwidths constants
          (x-axis) and number of knots ($m$) for all three estimation
          methods on both grids when $\nu=1$ and $\sigma^2=0$.}
\end{figure}

\begin{figure}[H]
	\begin{center}
		\subfigure[$50 \times 50$ grid with $m=23$]{\includegraphics[scale=.23]{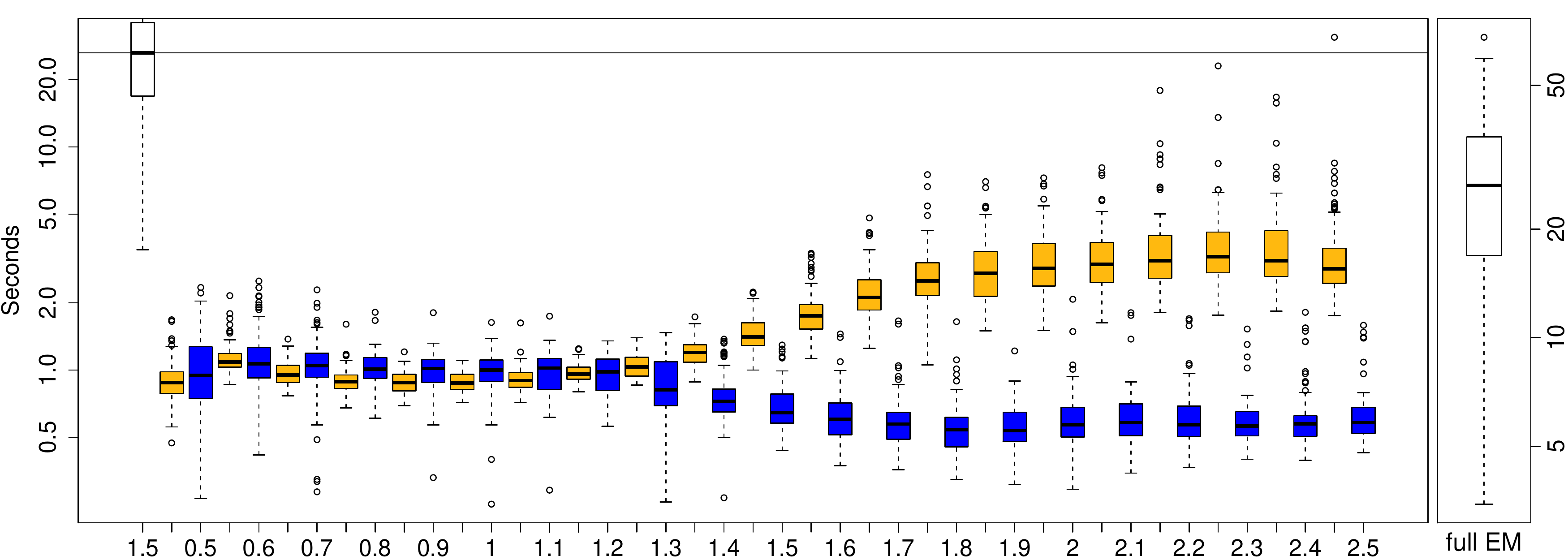}}
		\subfigure[$50 \times 50$ grid with $m=23$]{\includegraphics[scale=.23]{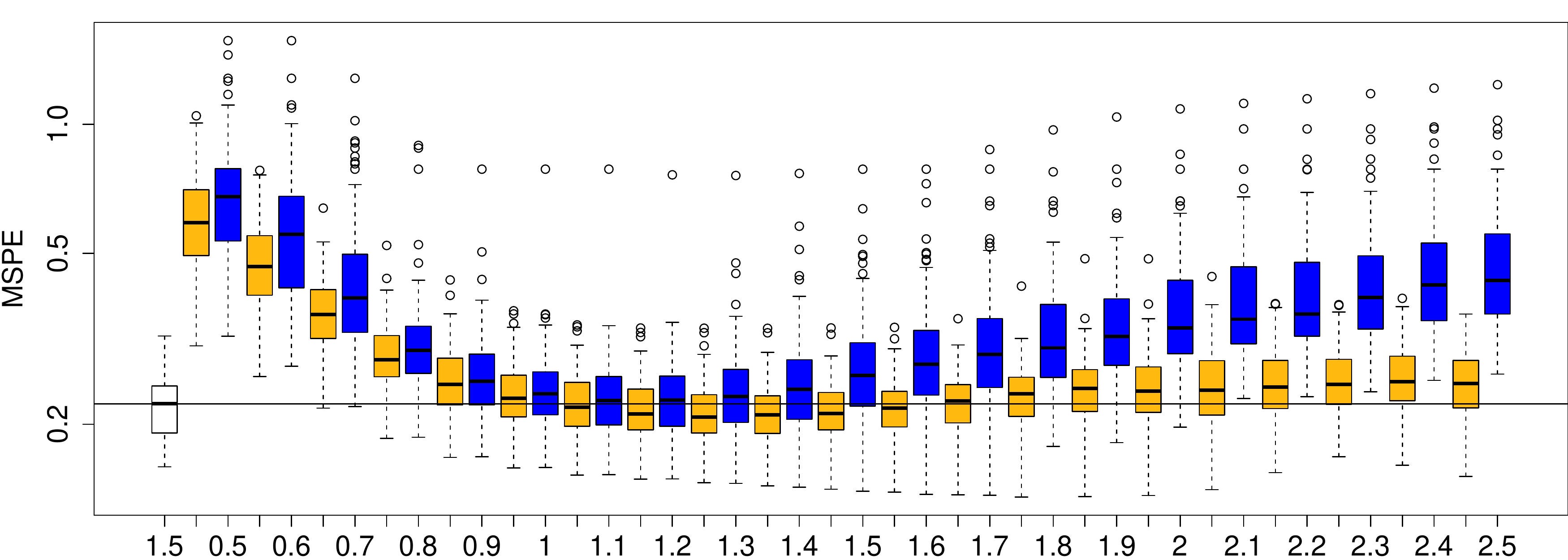}}
		\subfigure[$50 \times 50$ grid with $m=77$]{\includegraphics[scale=.23]{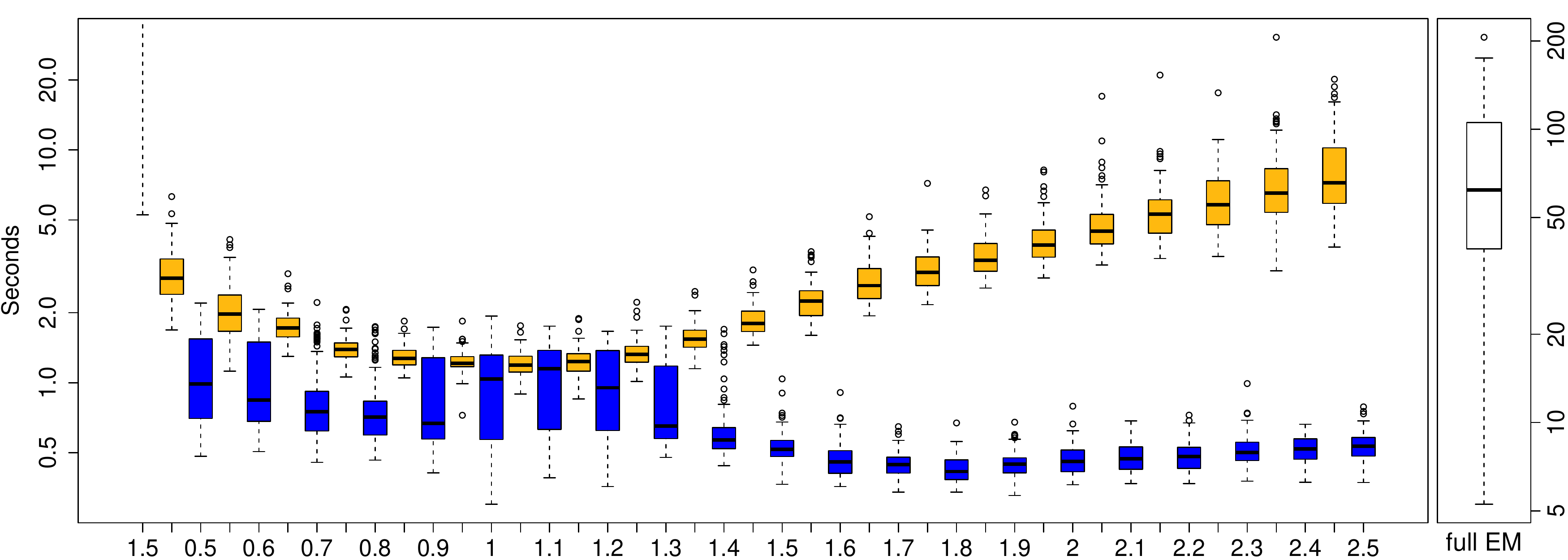}}
		\subfigure[$50 \times 50$ grid with $m=77$]{\includegraphics[scale=.23]{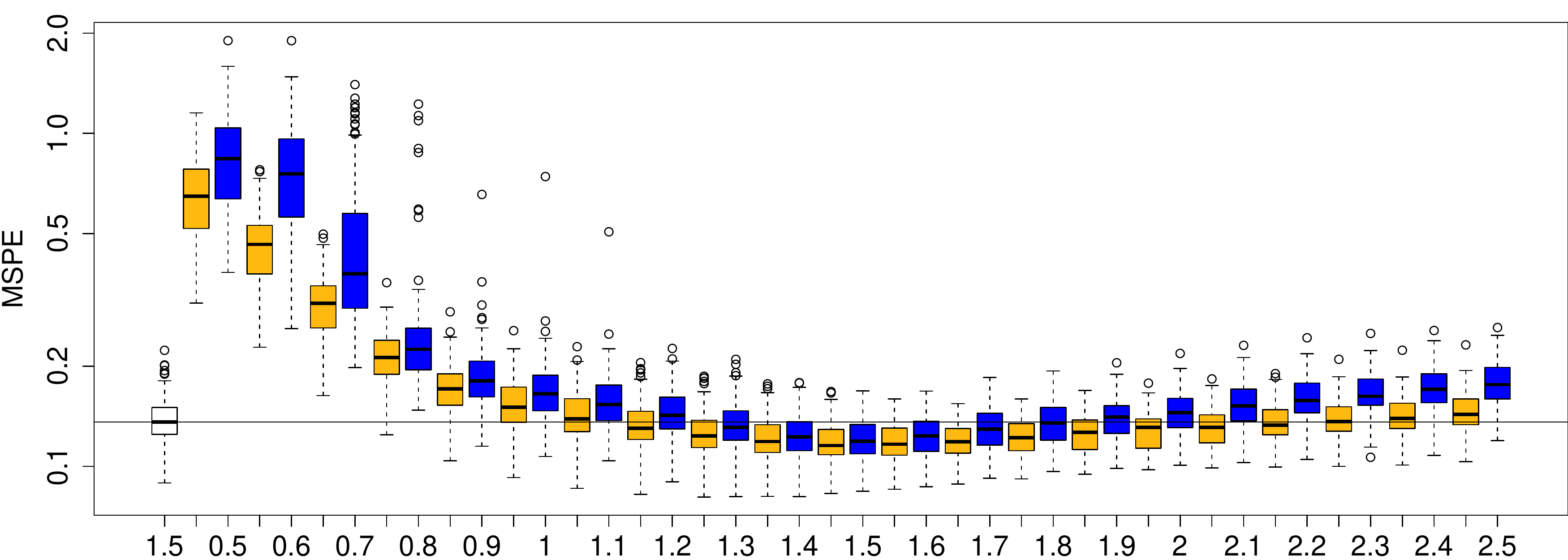}}
		\subfigure[$50 \times 50$ grid with $m=175$]{\includegraphics[scale=.23]{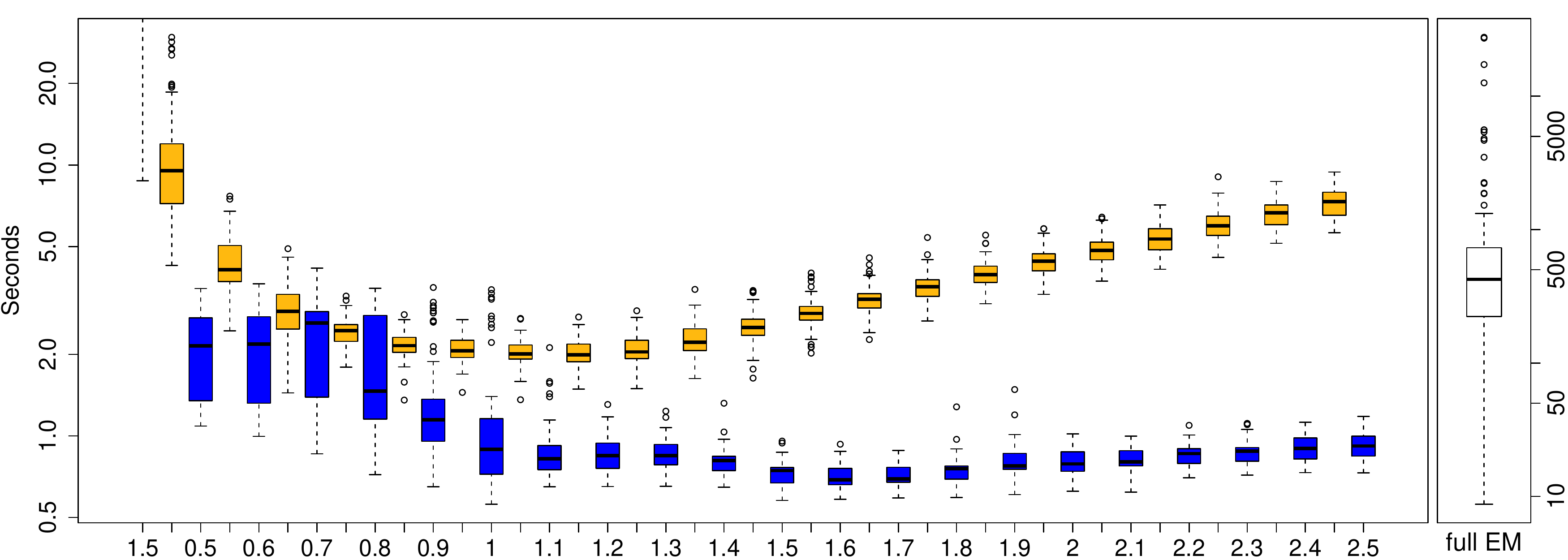}}		
		\subfigure[$50 \times 50$ grid with $m=175$]{\includegraphics[scale=.23]{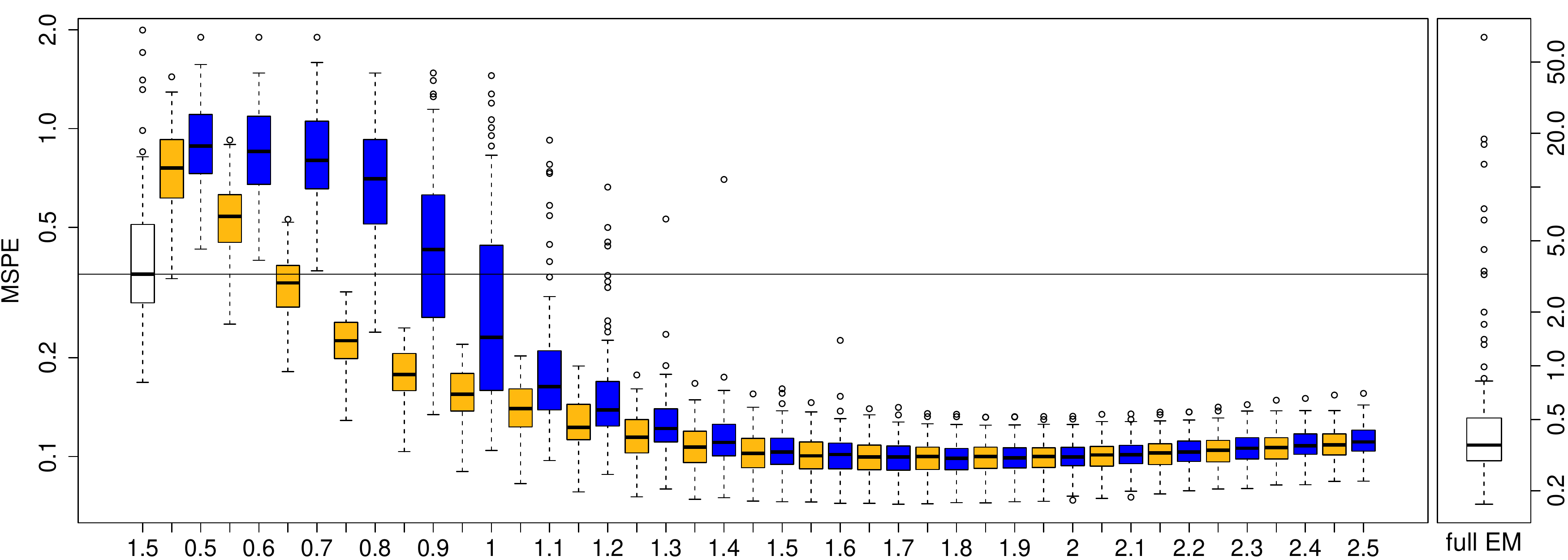}}
		\subfigure[$200 \times 200$ grid with $m=23$]{\includegraphics[scale=.23]{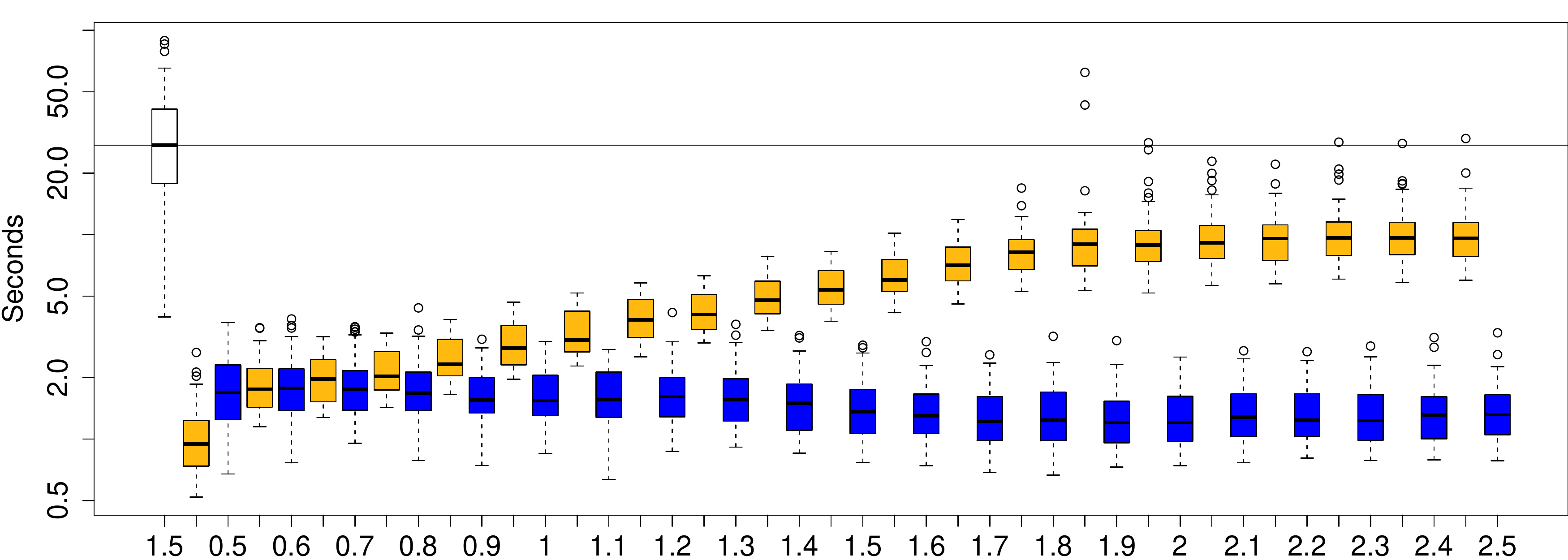}}
		\subfigure[$200 \times 200$ grid with $m=23$]{\includegraphics[scale=.23]{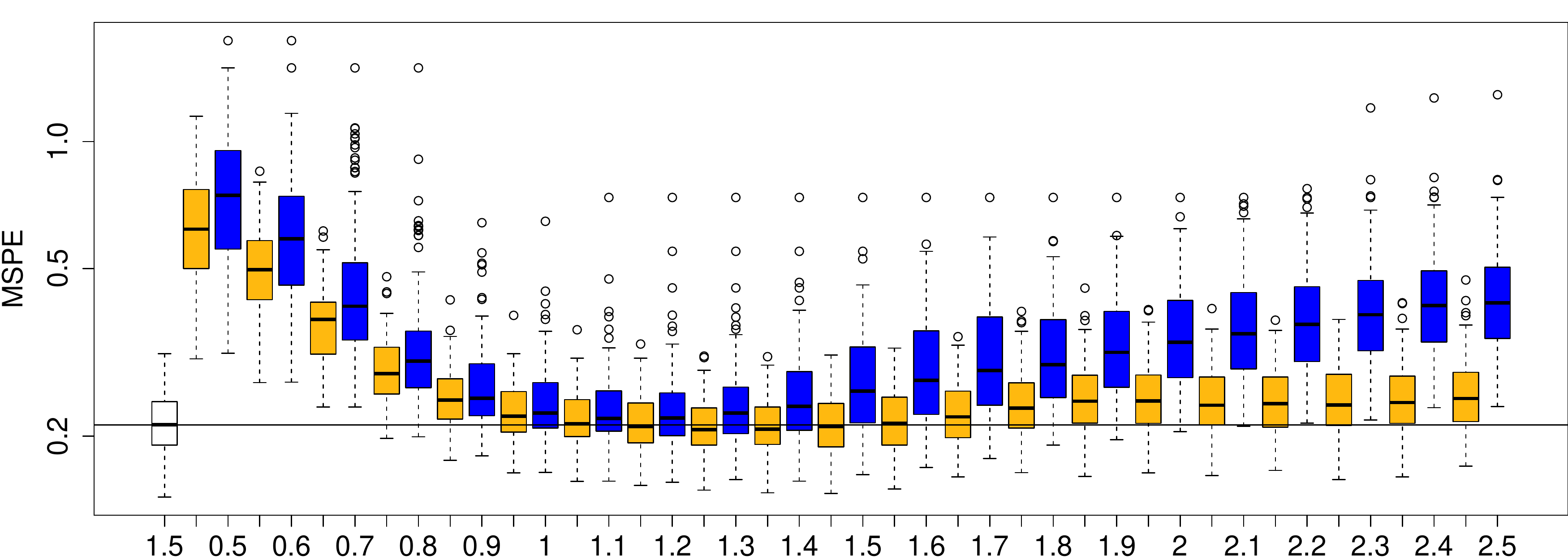}}
		\subfigure[$200 \times 200$ grid with $m=77$]{\includegraphics[scale=.23]{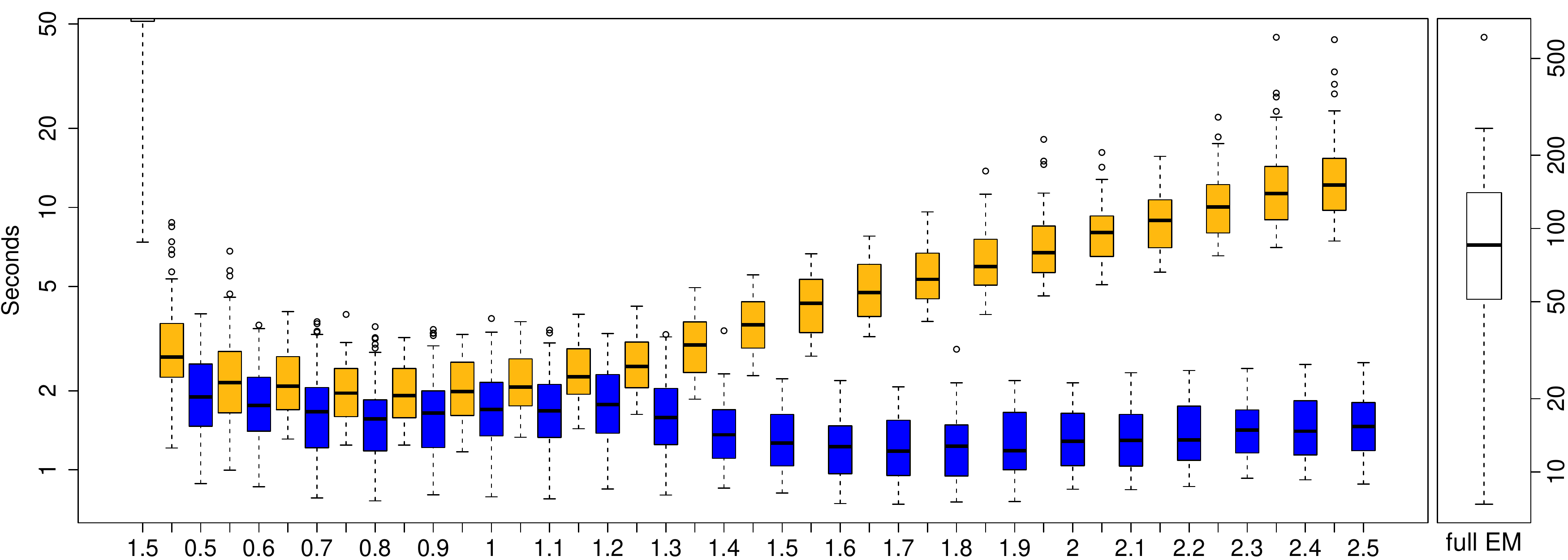}}
		\subfigure[$200 \times 200$ grid with $m=77$]{\includegraphics[scale=.23]{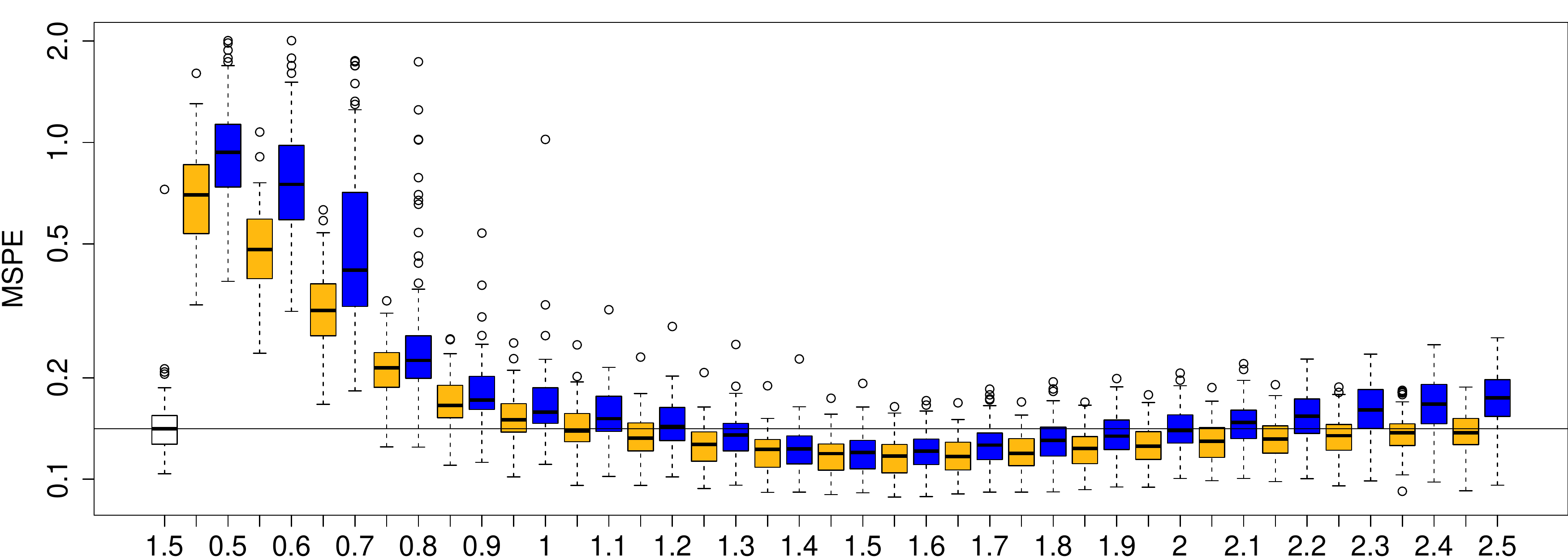}}
		\subfigure[$200 \times 200$ grid with $m=175$]{\includegraphics[scale=.23]{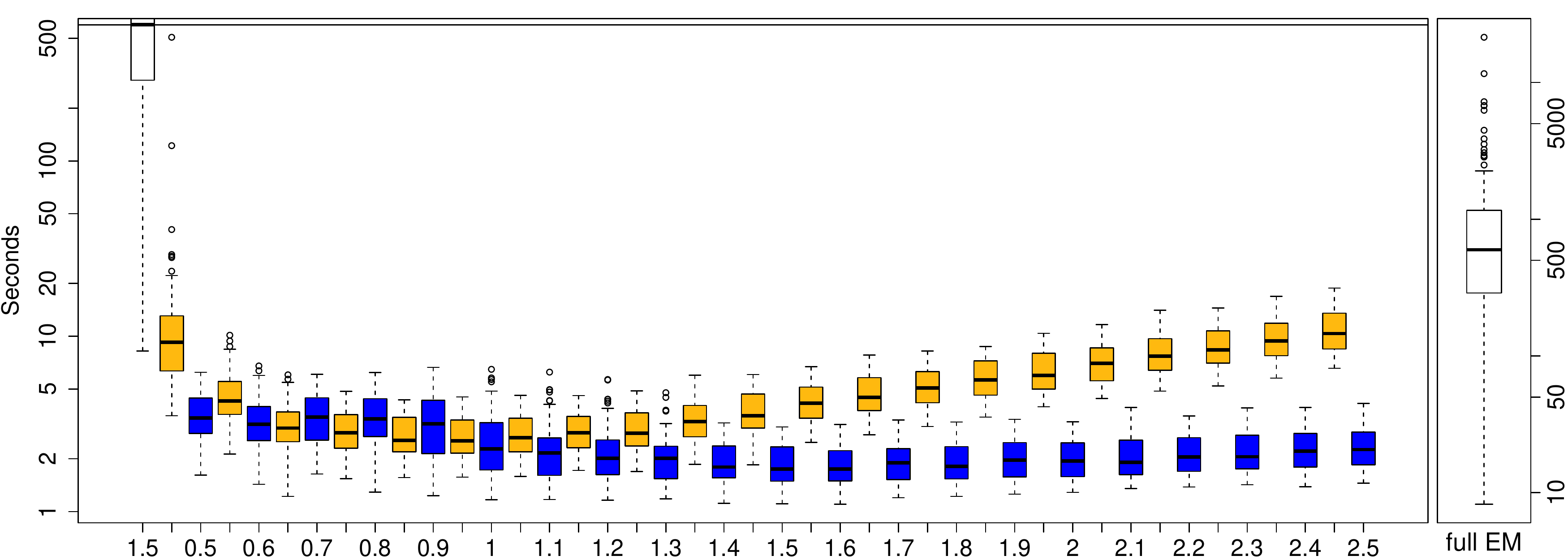}}		
		\subfigure[$200 \times 200$ grid with $m=175$]{\includegraphics[scale=.23]{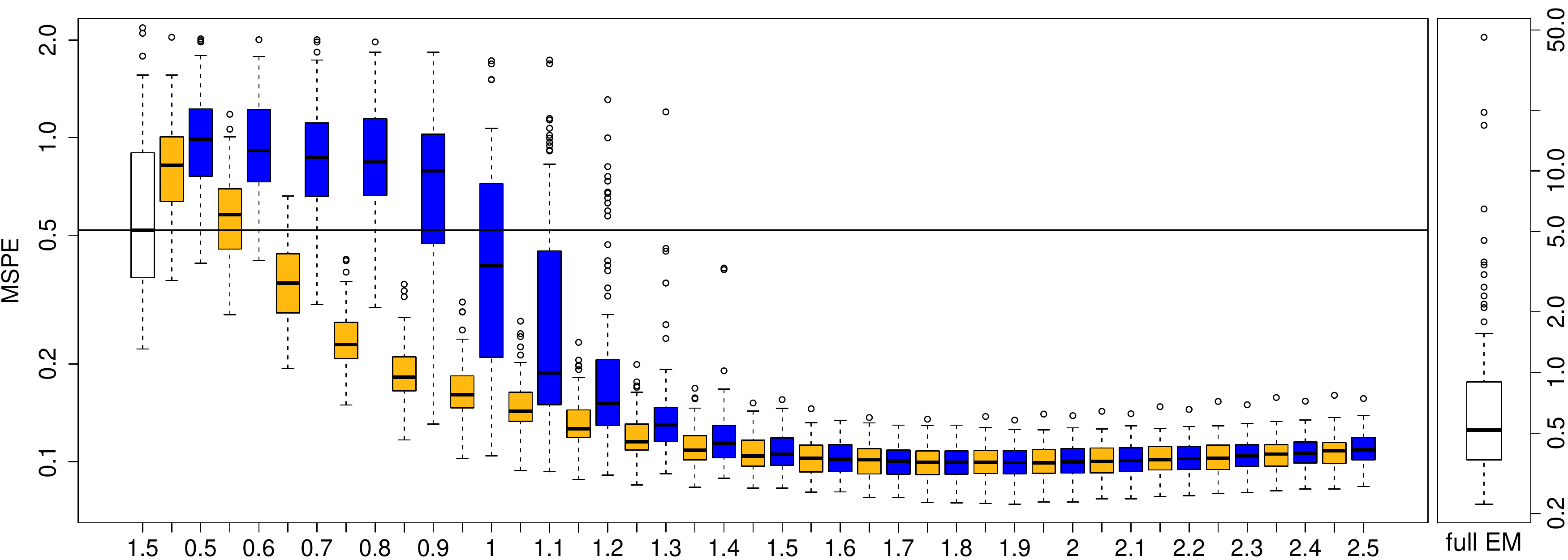}}
	\end{center}
	\caption{Seconds (left panel) and Mean Squared Prediction
          Errors (MSPEs, right panel) for varying bandwidths constants
          (x-axis) and number of knots ($m$) for all three estimation
          methods on both grids when $\nu=1$ and $\sigma^2=0.1$.} 
\end{figure}

\begin{figure}[H]
	\begin{center}
		\subfigure[$50 \times 50$ grid with $m=23$]{\includegraphics[scale=.23]{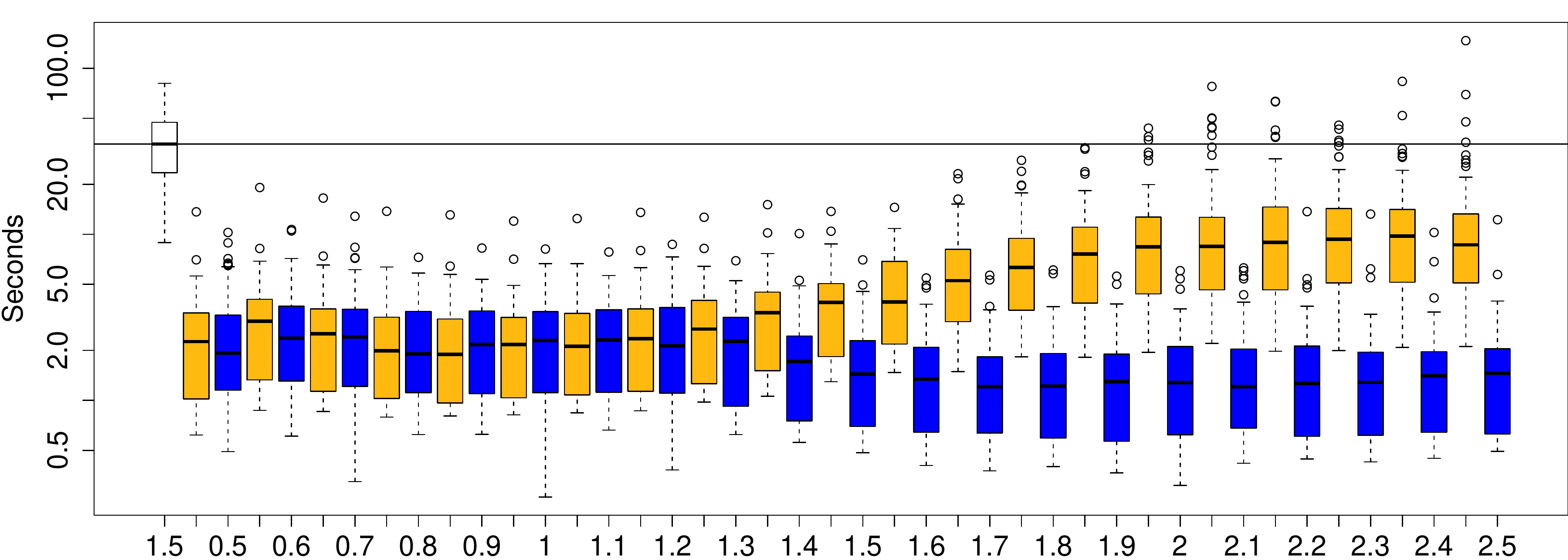}}
		\subfigure[$50 \times 50$ grid with $m=23$]{\includegraphics[scale=.23]{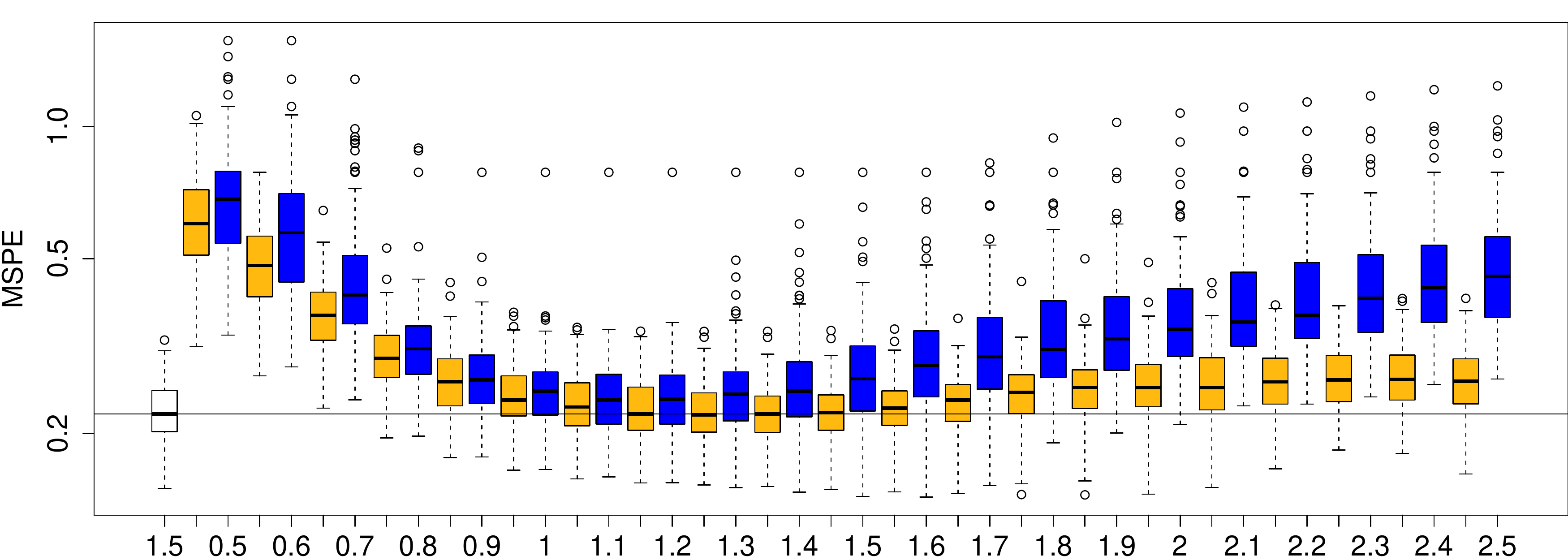}}
		\subfigure[$50 \times 50$ grid with $m=77$]{\includegraphics[scale=.23]{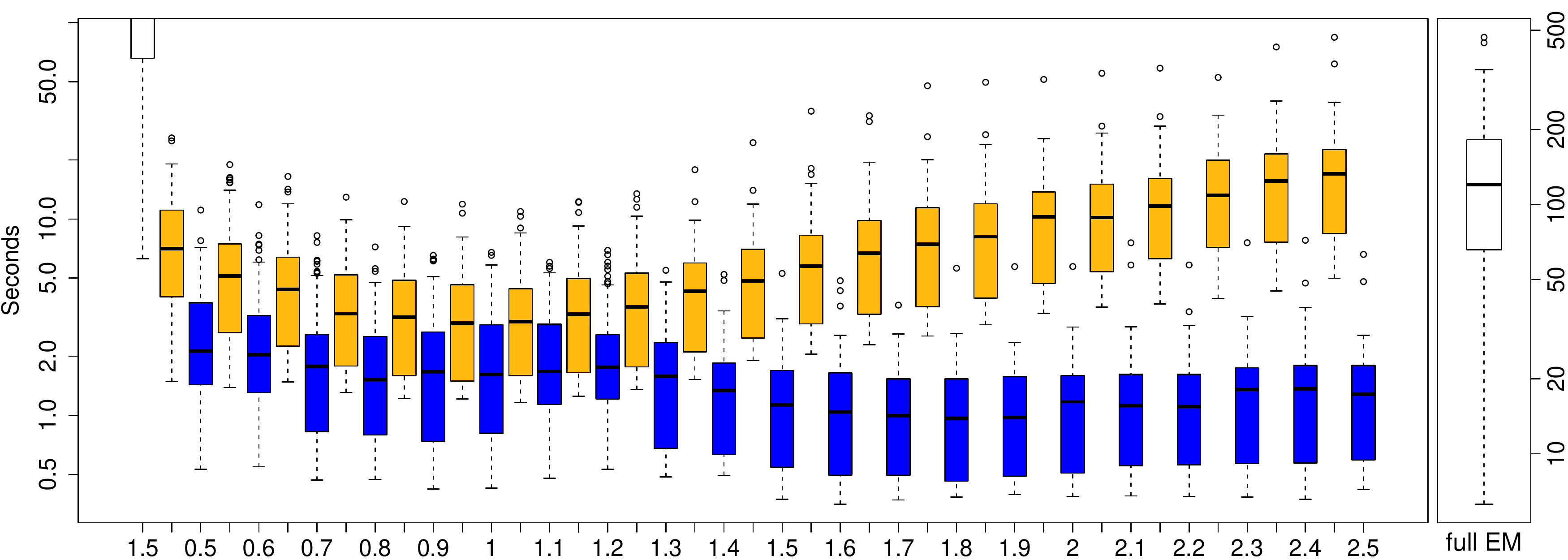}}
		\subfigure[$50 \times 50$ grid with $m=77$]{\includegraphics[scale=.23]{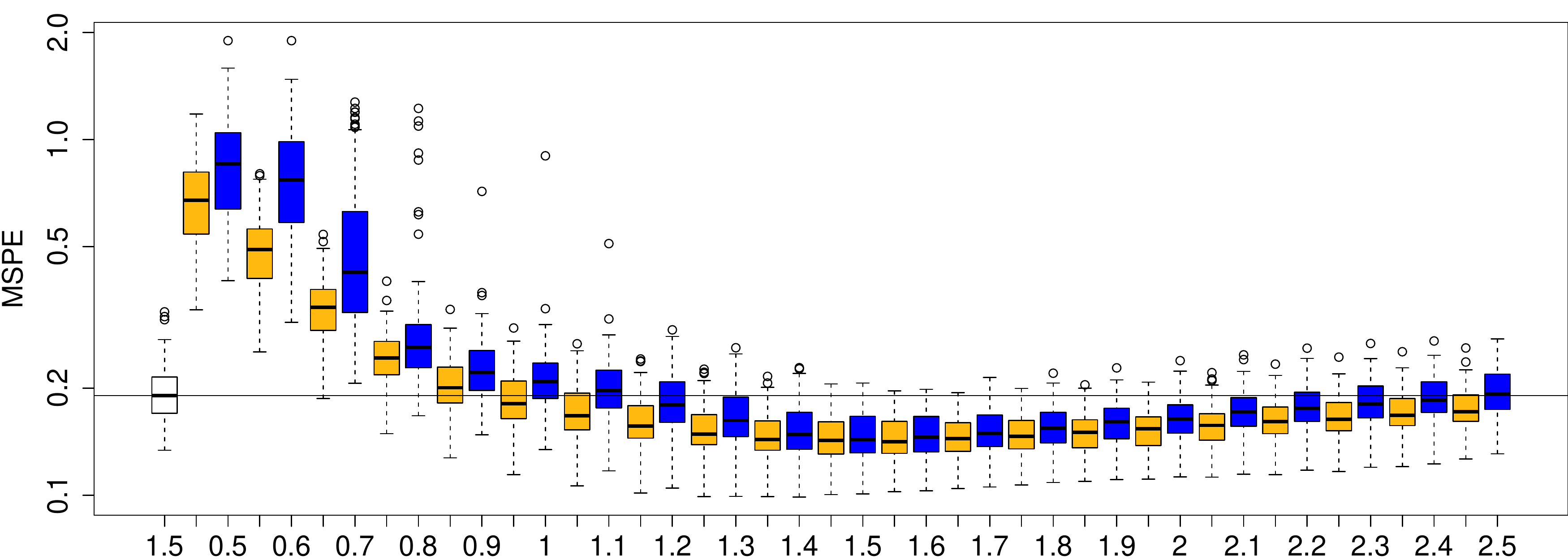}}
		\subfigure[$50 \times 50$ grid with $m=175$]{\includegraphics[scale=.23]{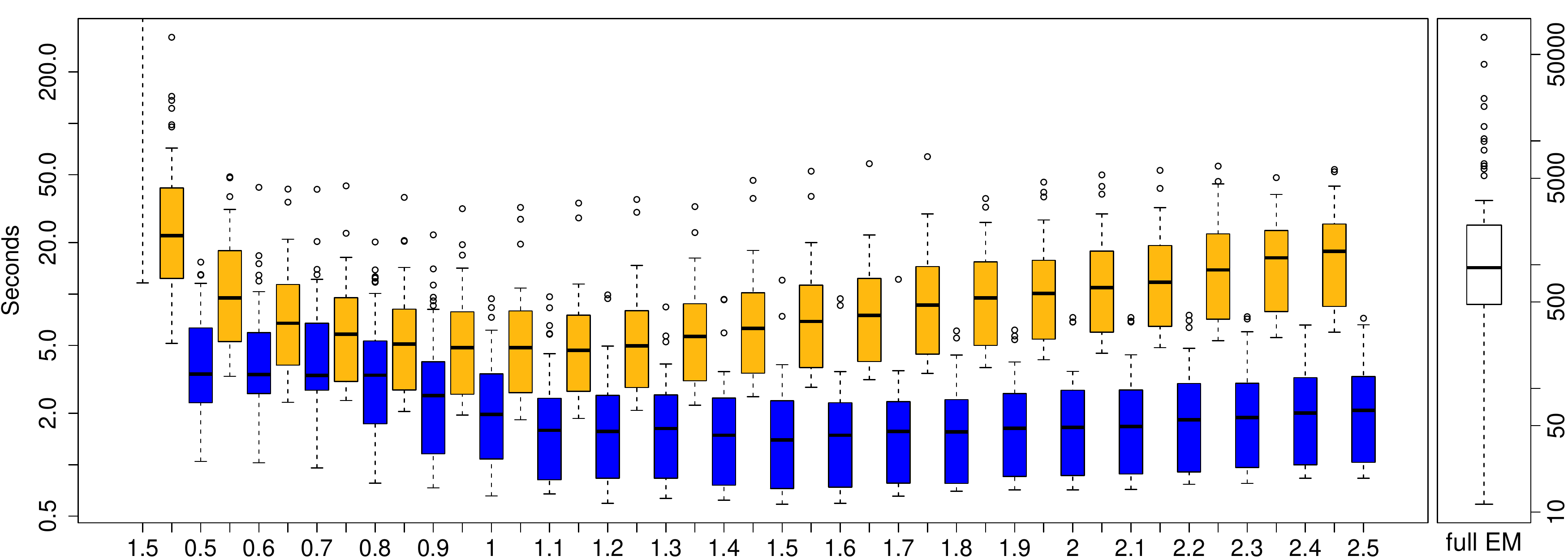}}		
		\subfigure[$50 \times 50$ grid with $m=175$]{\includegraphics[scale=.23]{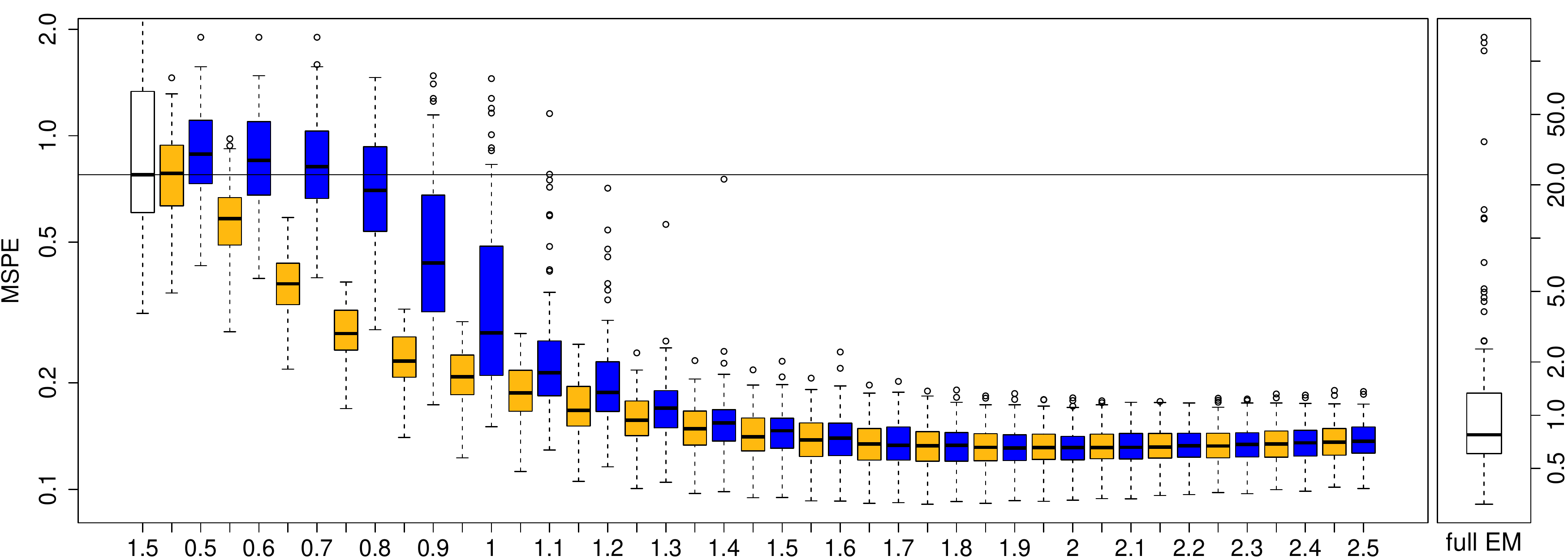}}
		\subfigure[$200 \times 200$ grid with $m=23$]{\includegraphics[scale=.23]{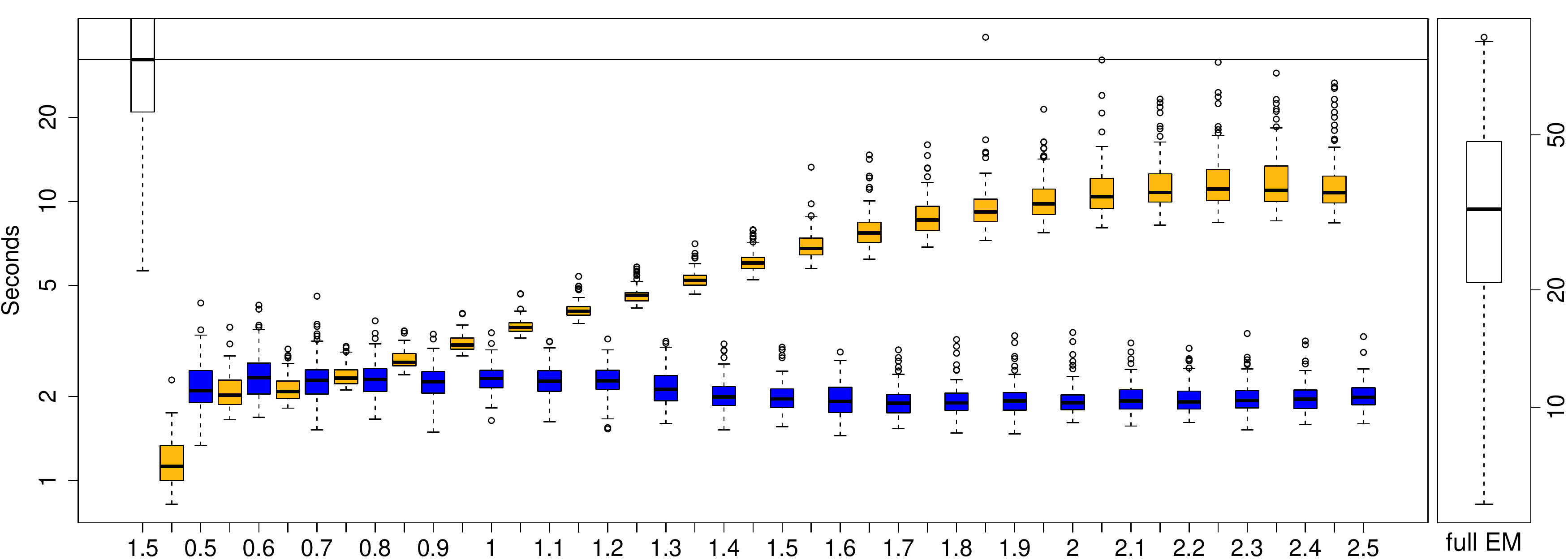}}
		\subfigure[$200 \times 200$ grid with $m=23$]{\includegraphics[scale=.23]{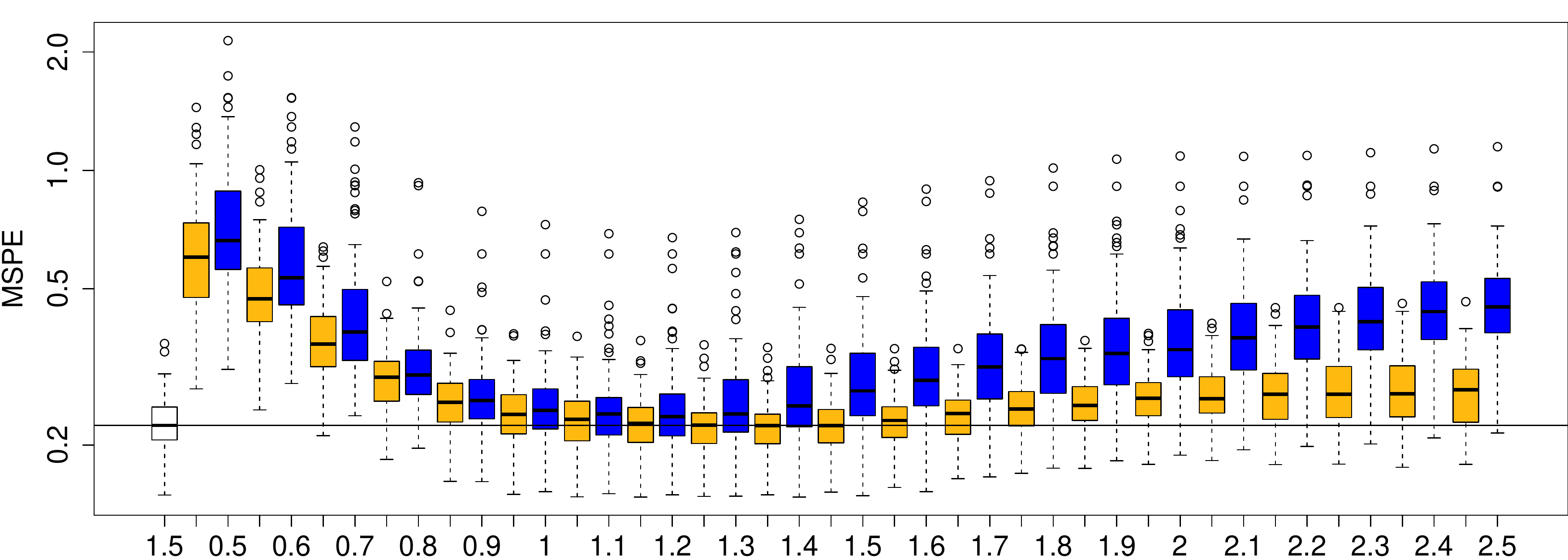}}
		\subfigure[$200 \times 200$ grid with $m=77$]{\includegraphics[scale=.23]{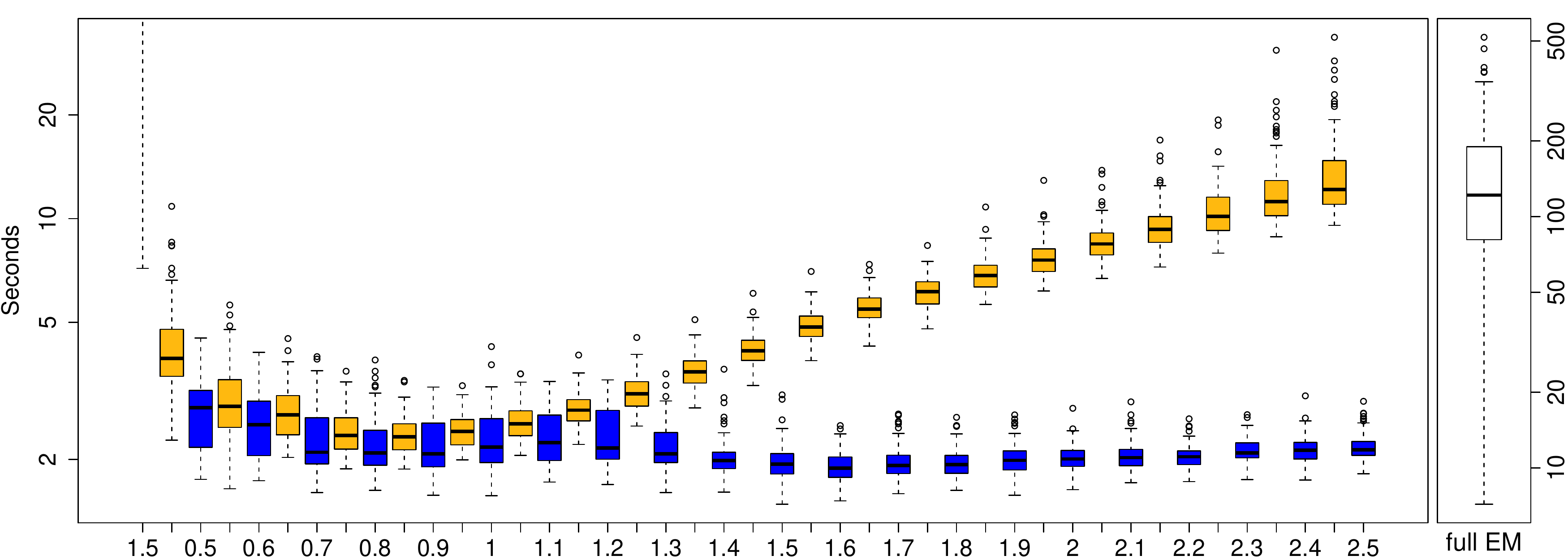}}
		\subfigure[$200 \times 200$ grid with $m=77$]{\includegraphics[scale=.23]{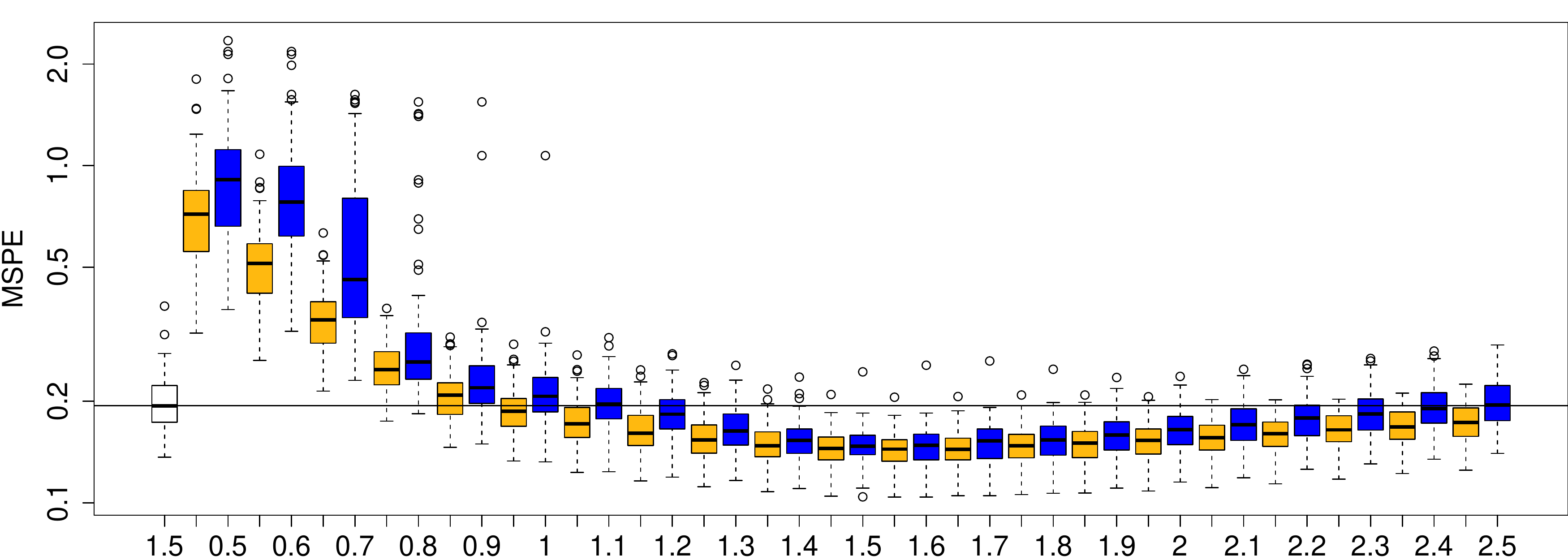}}
		\subfigure[$200 \times 200$ grid with $m=175$]{\includegraphics[scale=.23]{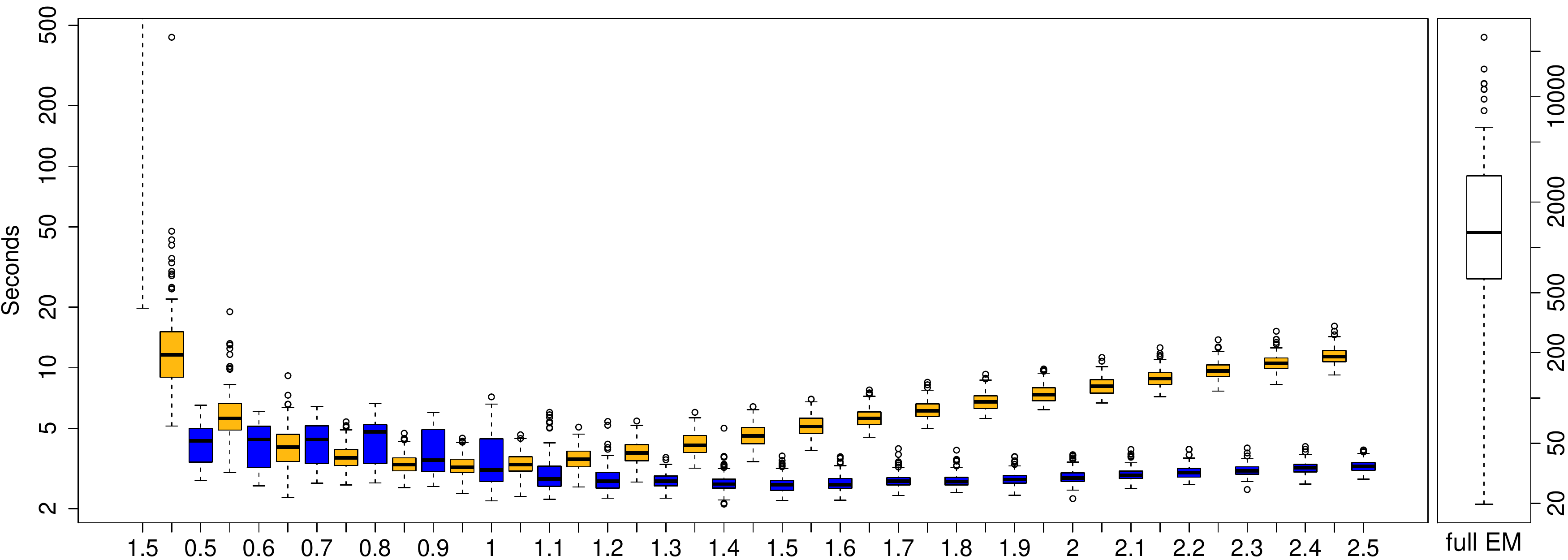}}		
		\subfigure[$200 \times 200$ grid with $m=175$]{\includegraphics[scale=.23]{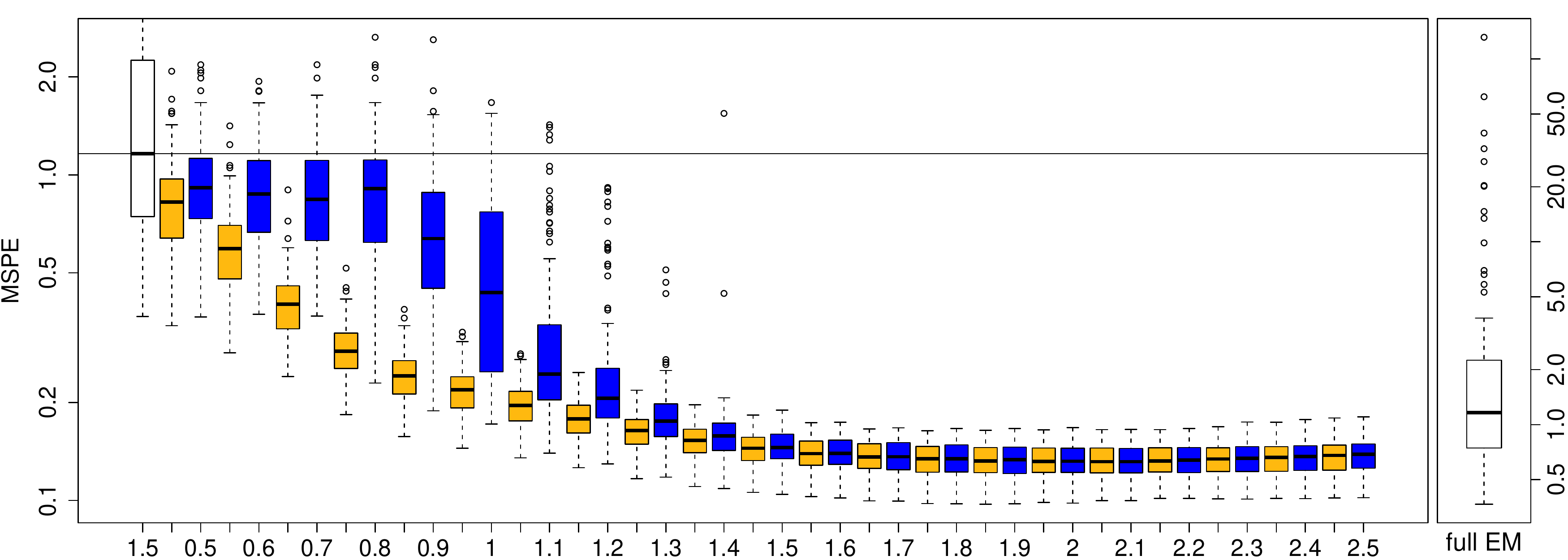}}
	\end{center}
	\caption{Seconds (left panel) and Mean Squared Prediction
          Errors (MSPEs, right panel) for varying bandwidths constants
          (x-axis) and number of knots ($m$) for all three estimation
          methods on both grids when $\nu=1$ and $\sigma^2=0.25$.} 
\end{figure}

\begin{figure}[H]
	\begin{center}
		\subfigure[$50 \times 50$ grid with $m=23$]{\includegraphics[scale=.23]{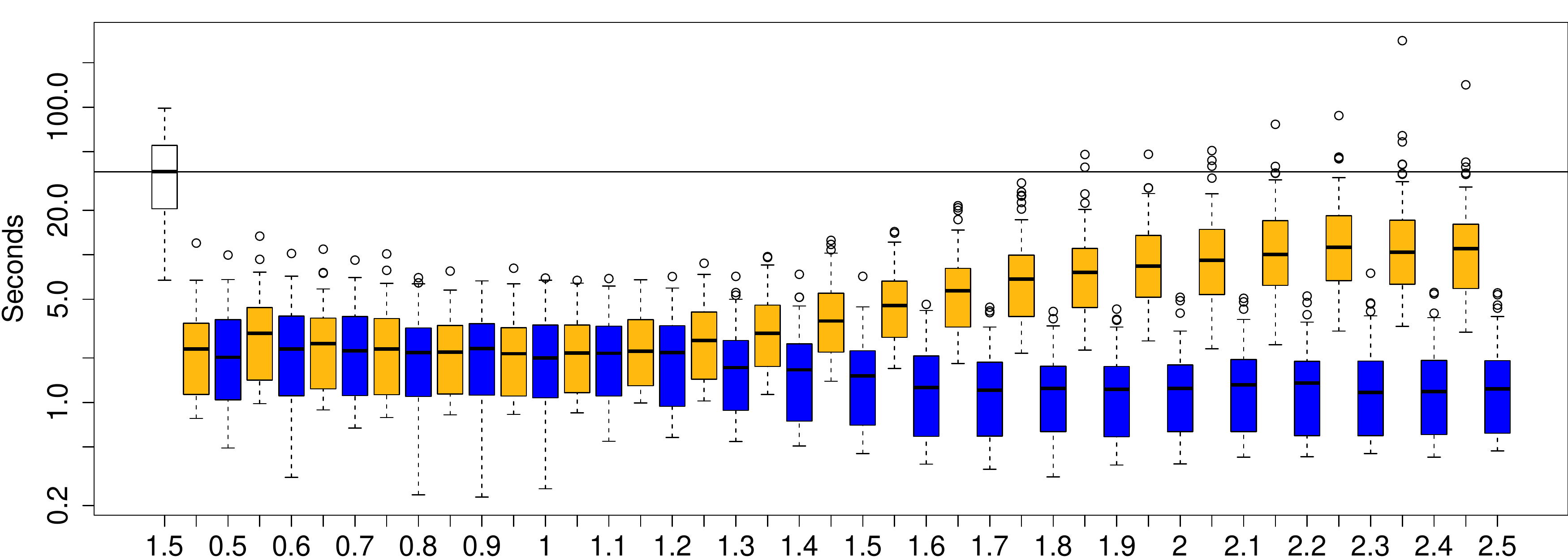}}
		\subfigure[$50 \times 50$ grid with $m=23$]{\includegraphics[scale=.23]{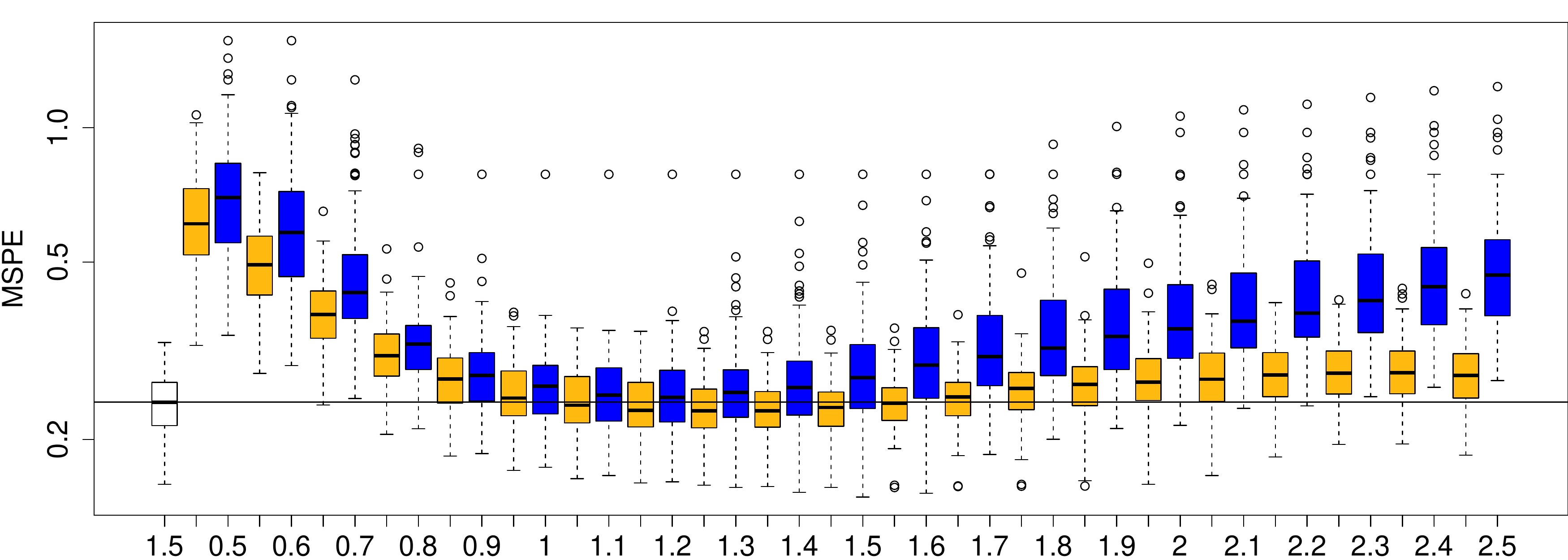}}
		\subfigure[$50 \times 50$ grid with $m=77$]{\includegraphics[scale=.23]{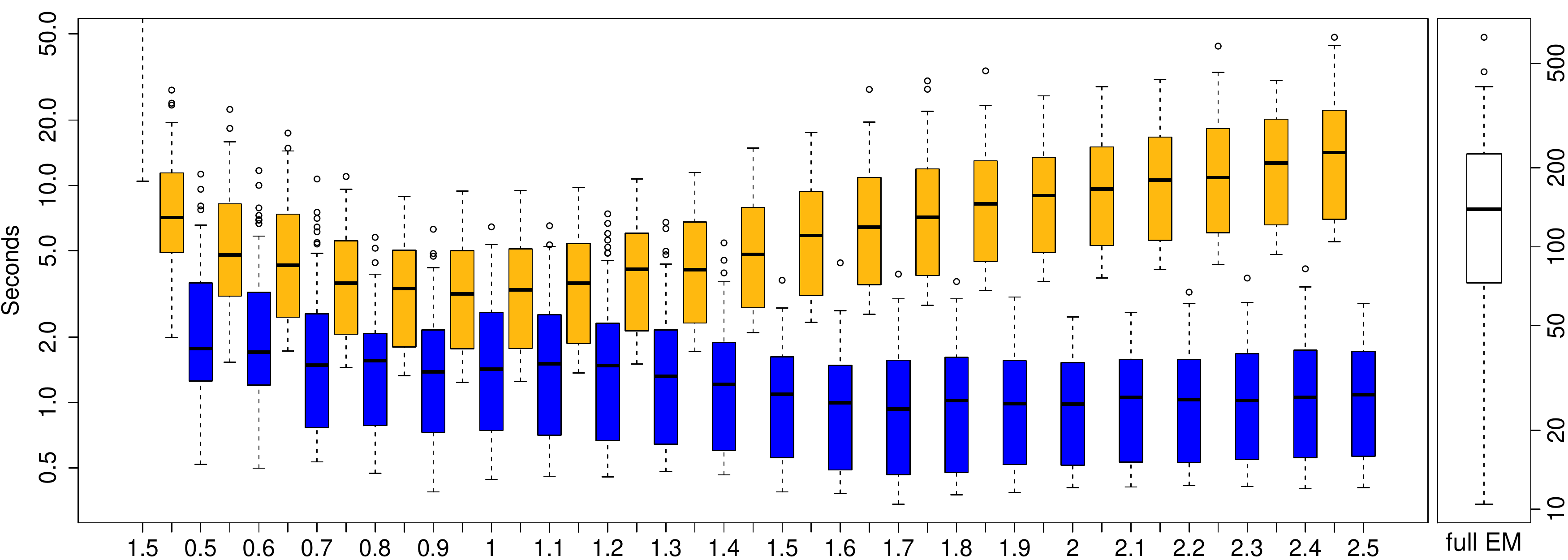}}
		\subfigure[$50 \times 50$ grid with $m=77$]{\includegraphics[scale=.23]{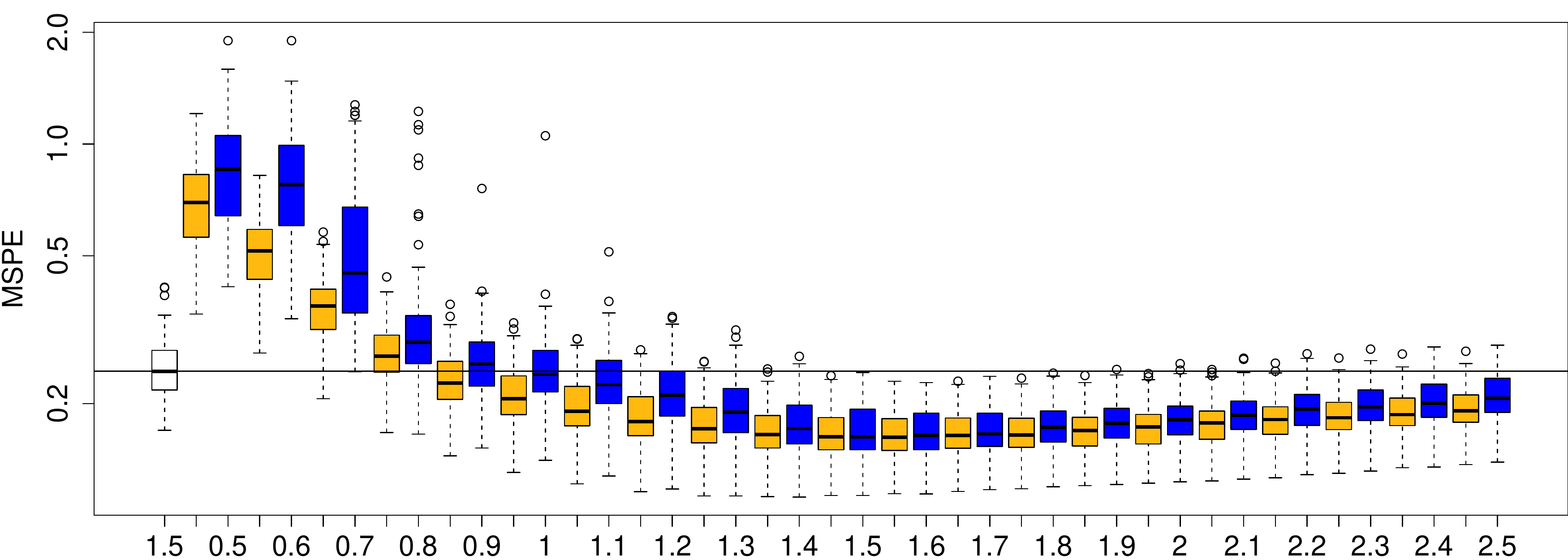}}
		\subfigure[$50 \times 50$ grid with $m=175$]{\includegraphics[scale=.23]{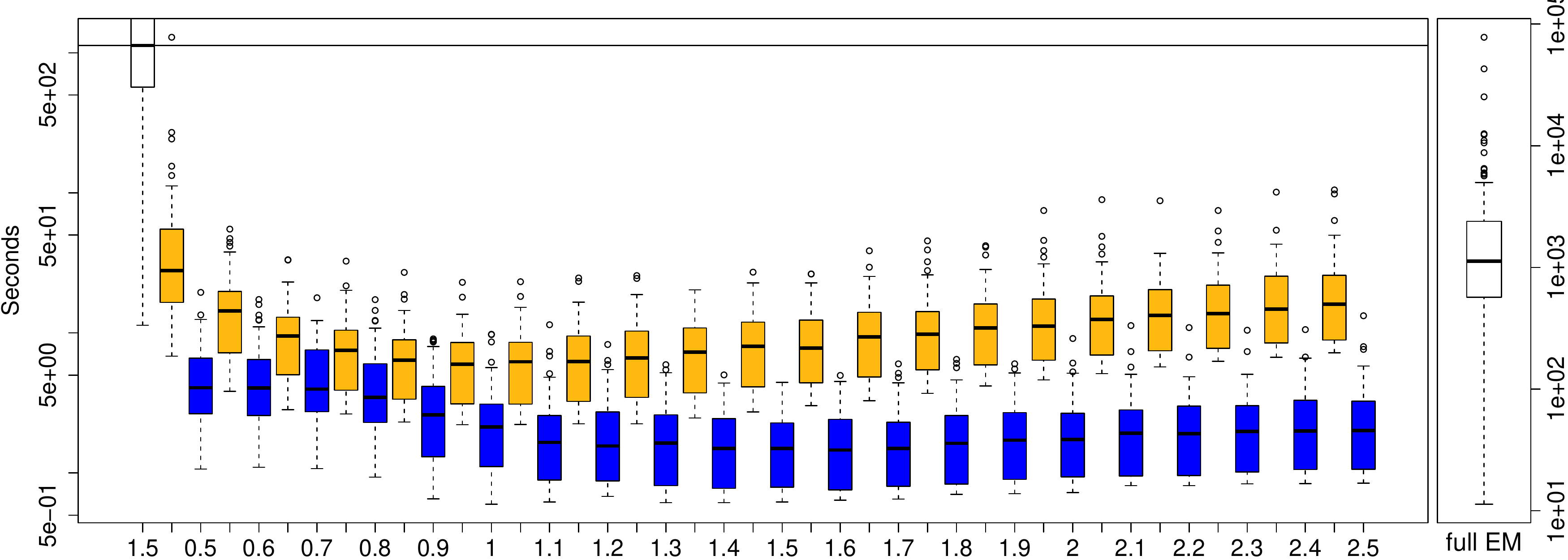}}		
		\subfigure[$50 \times 50$ grid with $m=175$]{\includegraphics[scale=.23]{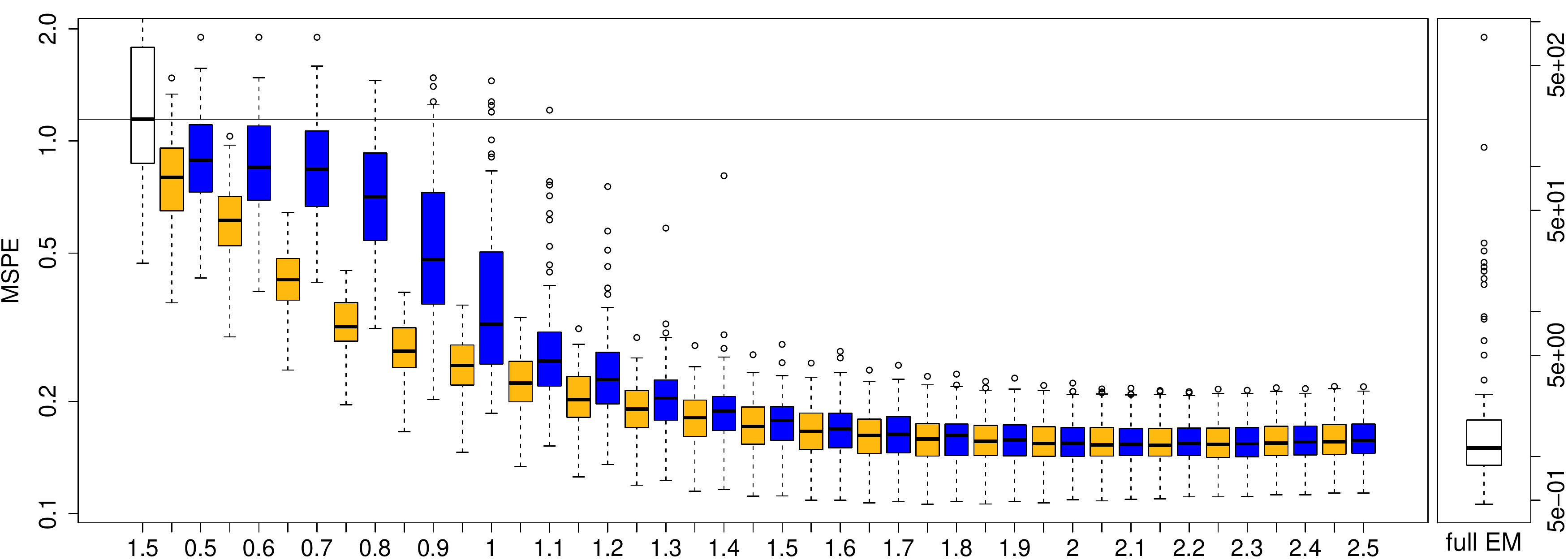}}
		\subfigure[$200 \times 200$ grid with $m=23$]{\includegraphics[scale=.23]{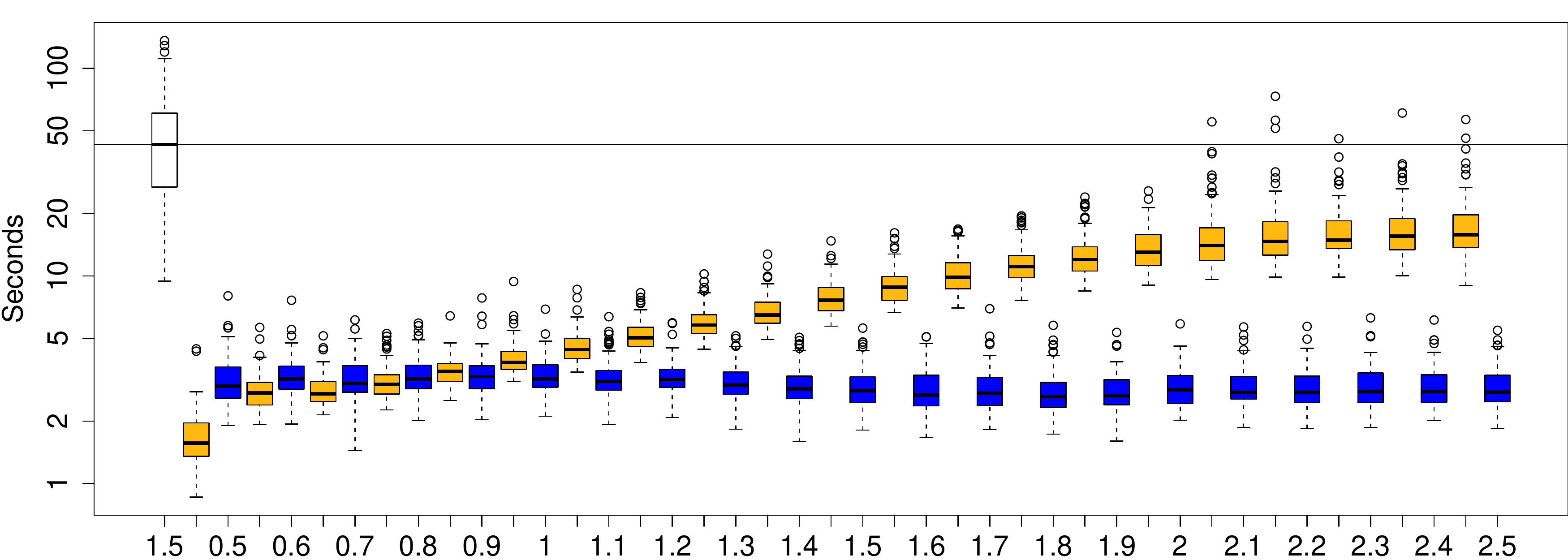}}
		\subfigure[$200 \times 200$ grid with $m=23$]{\includegraphics[scale=.23]{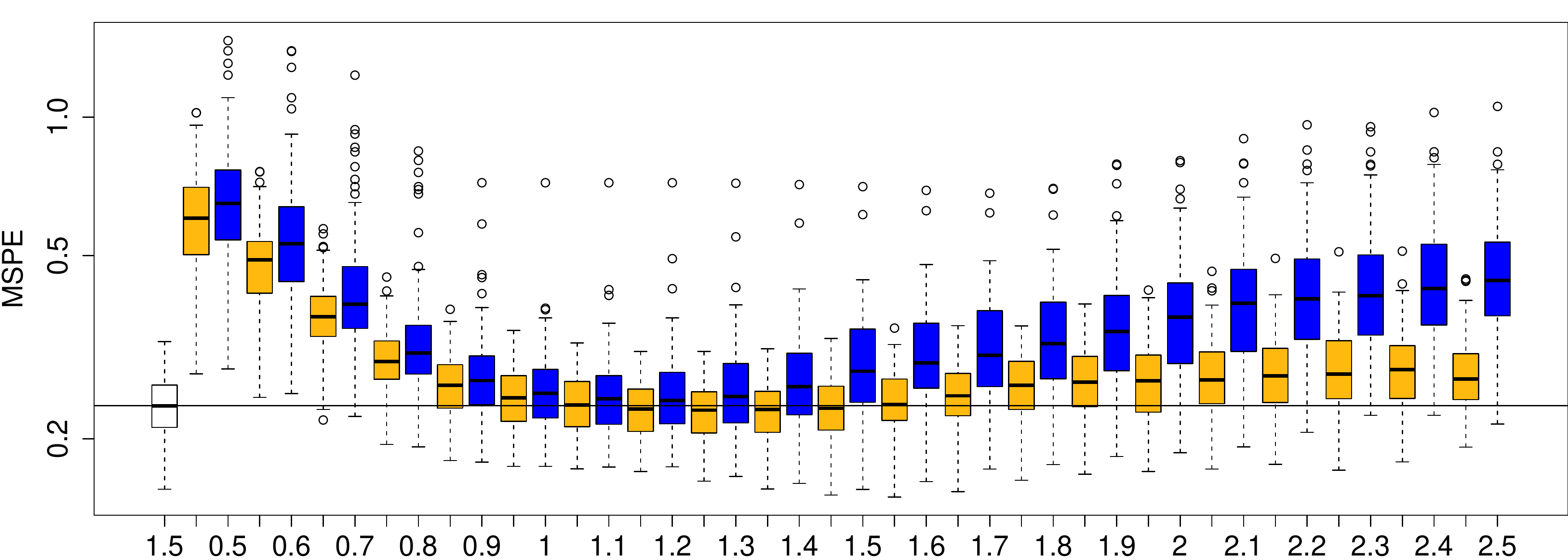}}
		\subfigure[$200 \times 200$ grid with $m=77$]{\includegraphics[scale=.23]{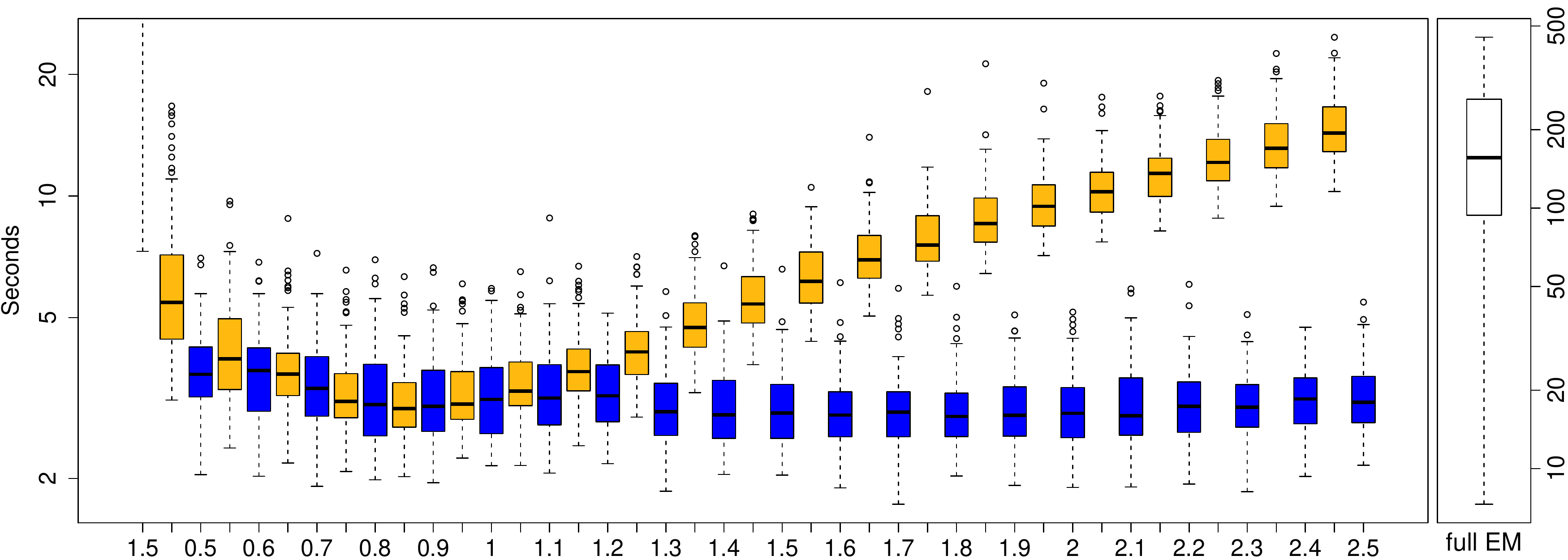}}
		\subfigure[$200 \times 200$ grid with $m=77$]{\includegraphics[scale=.23]{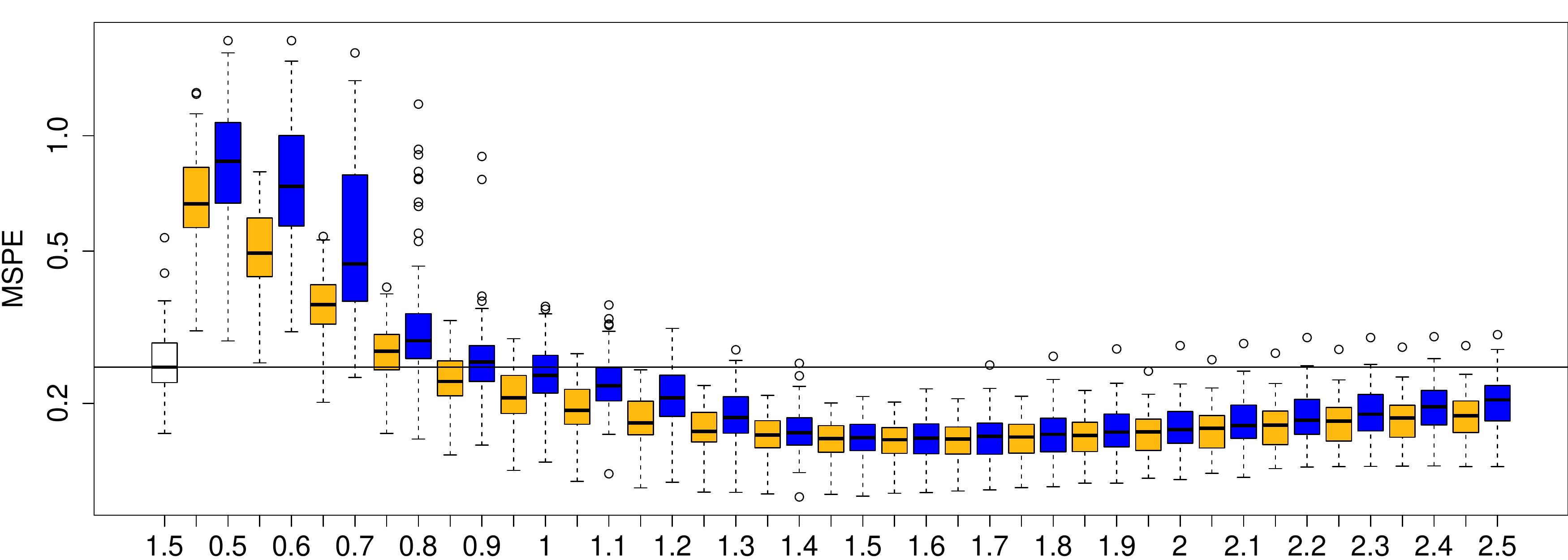}}
		\subfigure[$200 \times 200$ grid with $m=175$]{\includegraphics[scale=.23]{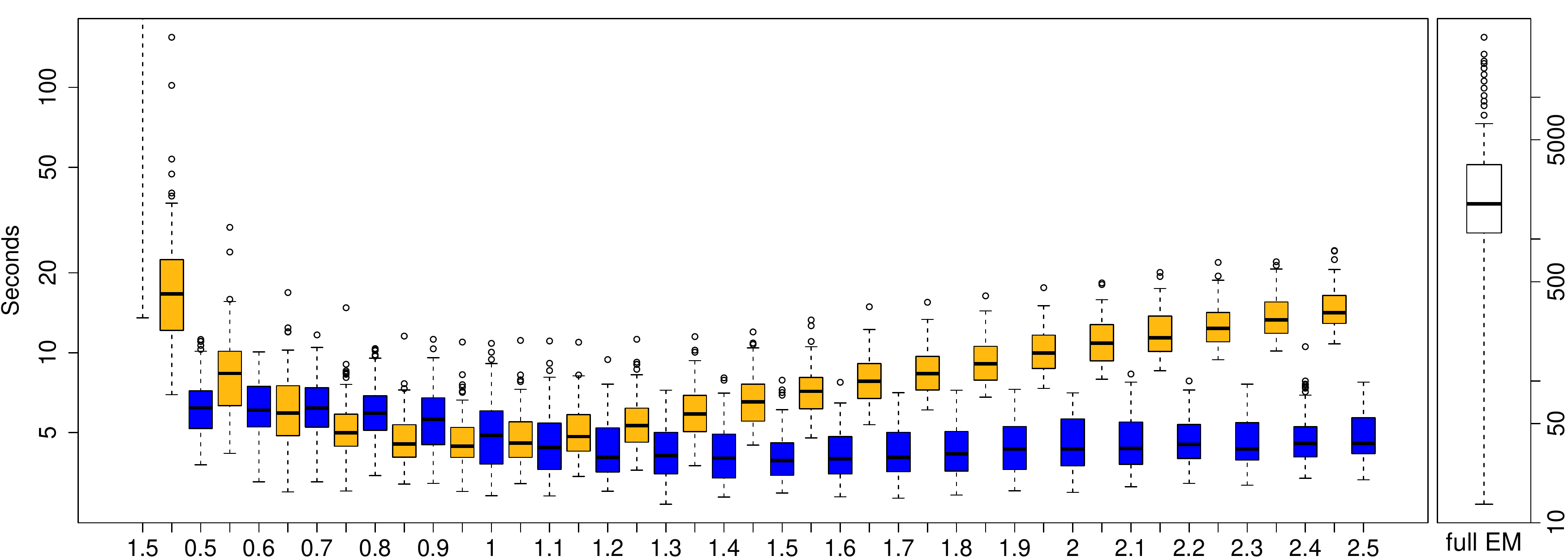}}		
		\subfigure[$200 \times 200$ grid with $m=175$]{\includegraphics[scale=.23]{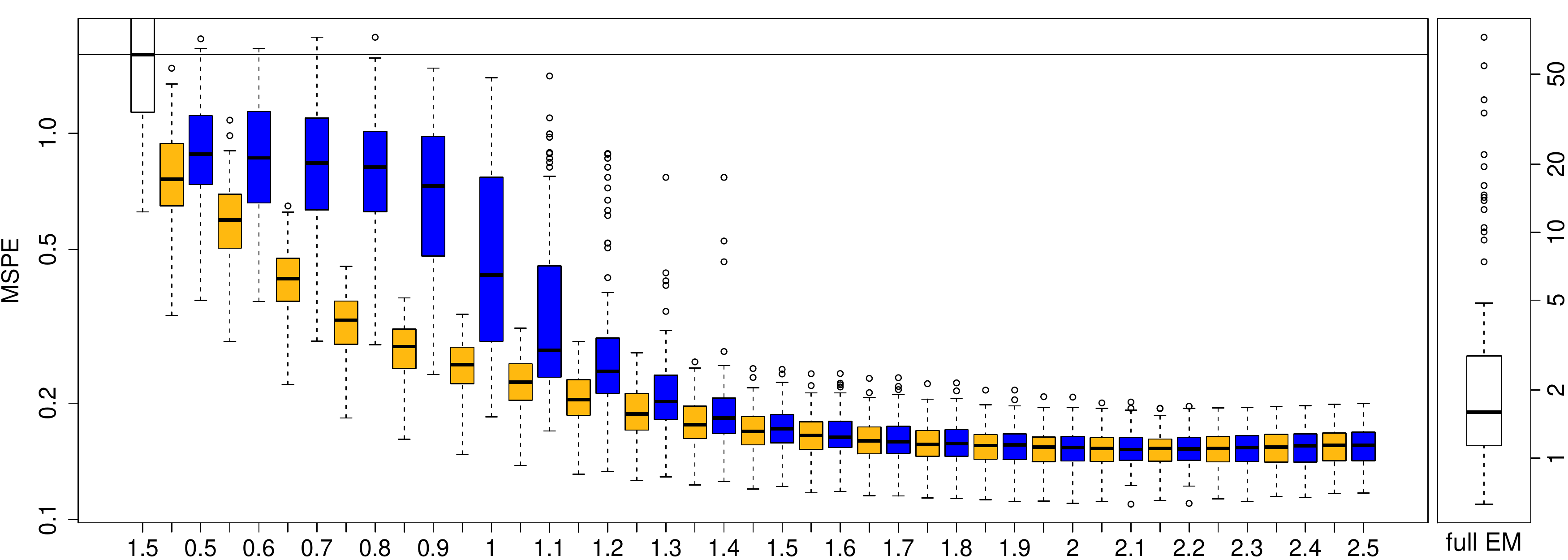}}
	\end{center}
	\caption{Seconds (left panel) and Mean Squared Prediction
          Errors (MSPEs, right panel) for varying bandwidths constants
          (x-axis) and number of knots ($m$) for all three estimation
          methods on both grids when $\nu=1$ and $\sigma^2=0.4$.} 
\end{figure}

\begin{figure}[H]
	\begin{center}
		\subfigure[$50 \times 50$ grid with $m=23$]{\includegraphics[scale=.23]{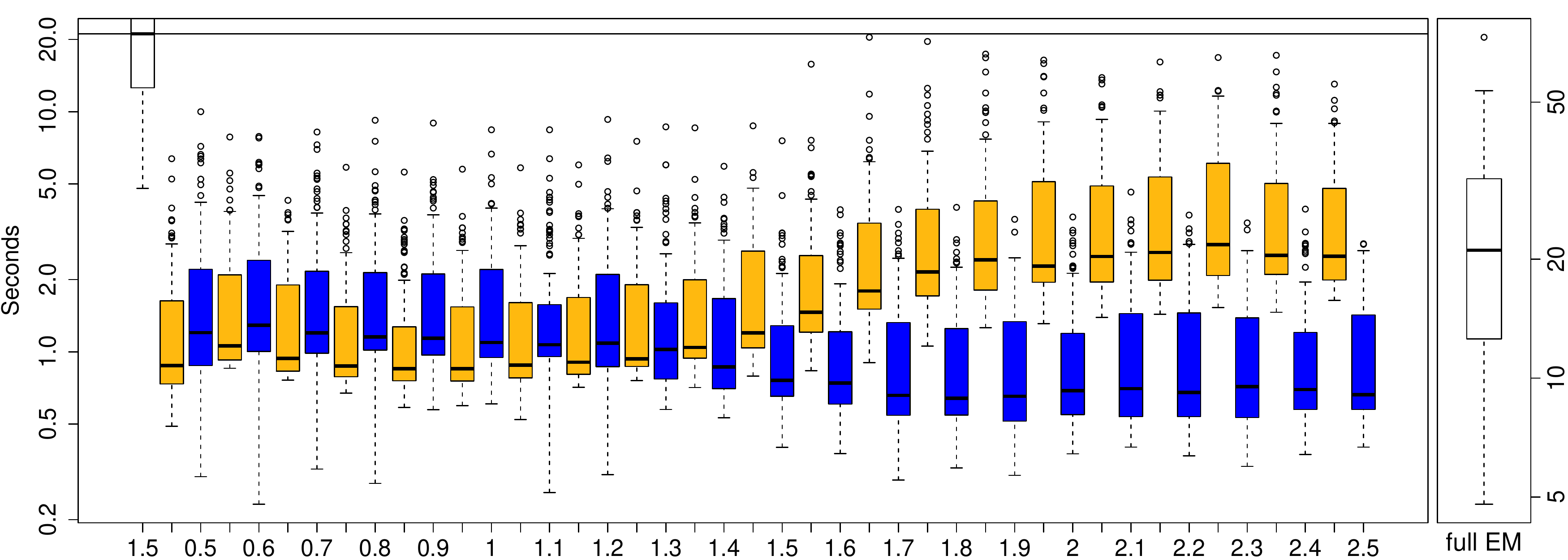}}
		\subfigure[$50 \times 50$ grid with $m=23$]{\includegraphics[scale=.23]{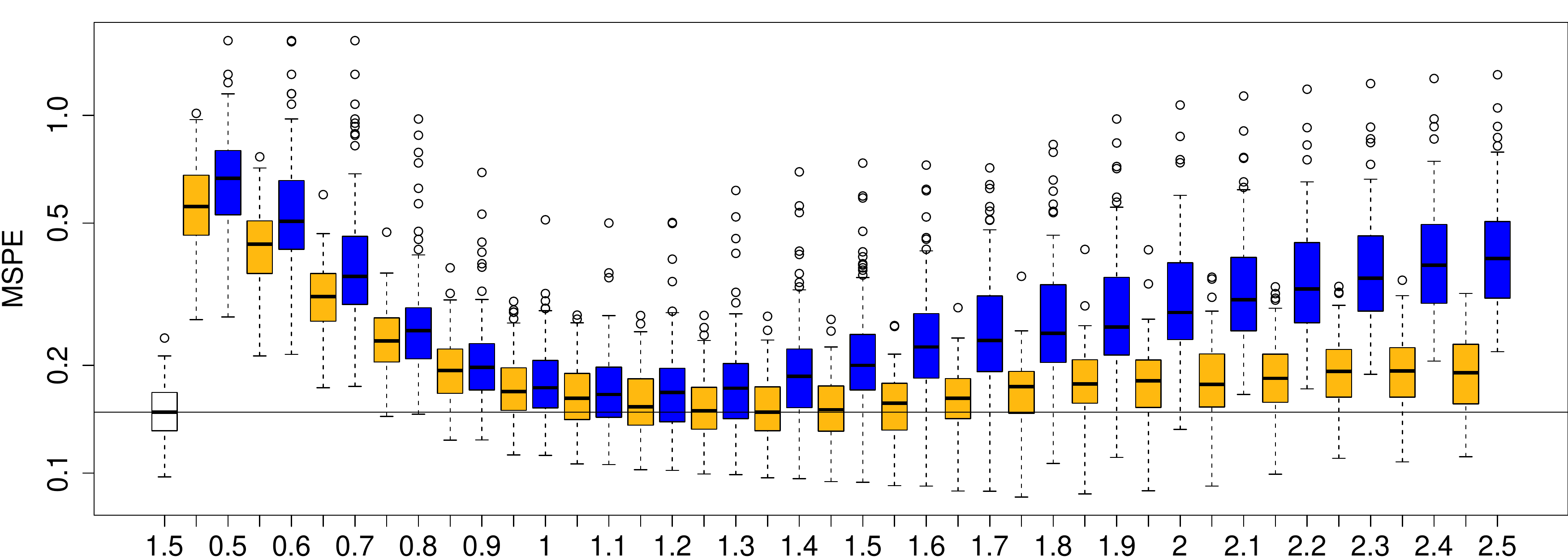}}
		\subfigure[$50 \times 50$ grid with $m=77$]{\includegraphics[scale=.23]{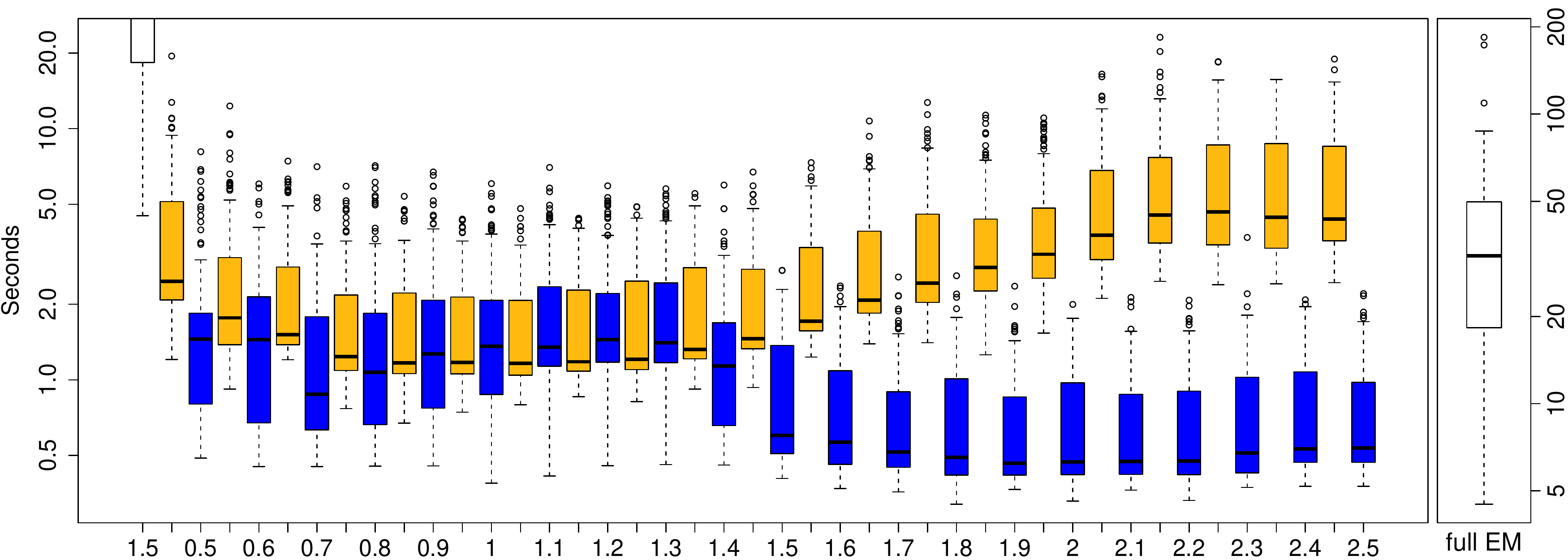}}
		\subfigure[$50 \times 50$ grid with $m=77$]{\includegraphics[scale=.23]{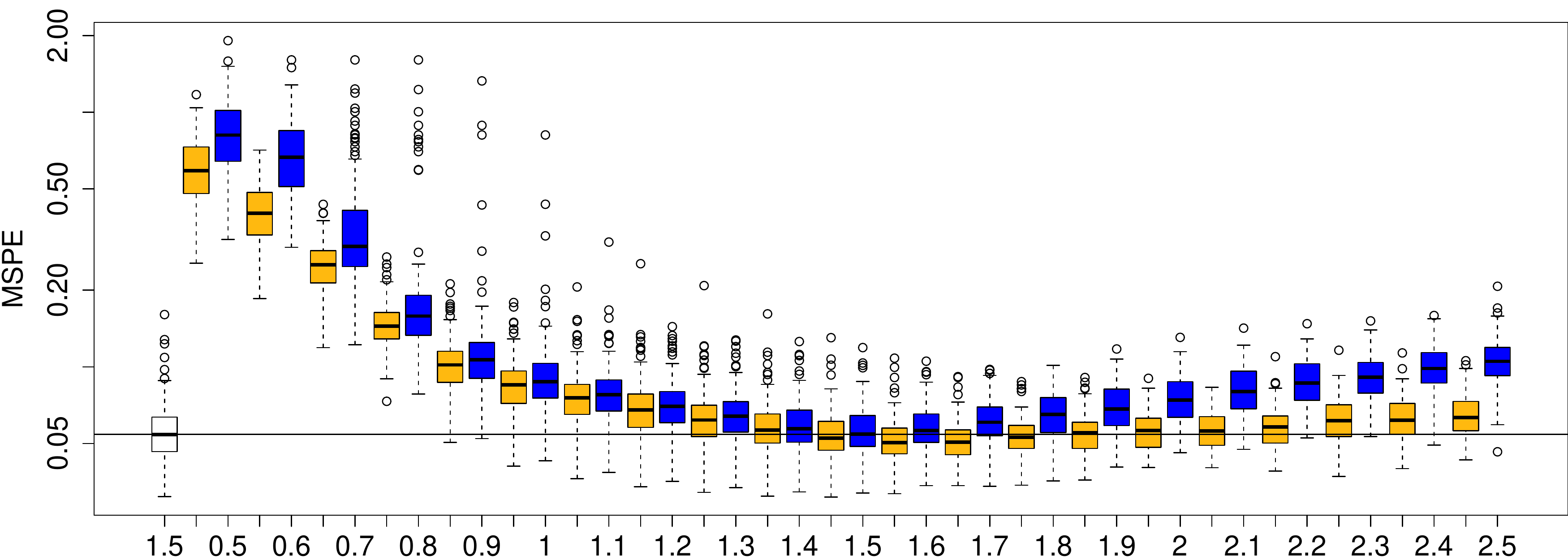}}
		\subfigure[$50 \times 50$ grid with $m=175$]{\includegraphics[scale=.23]{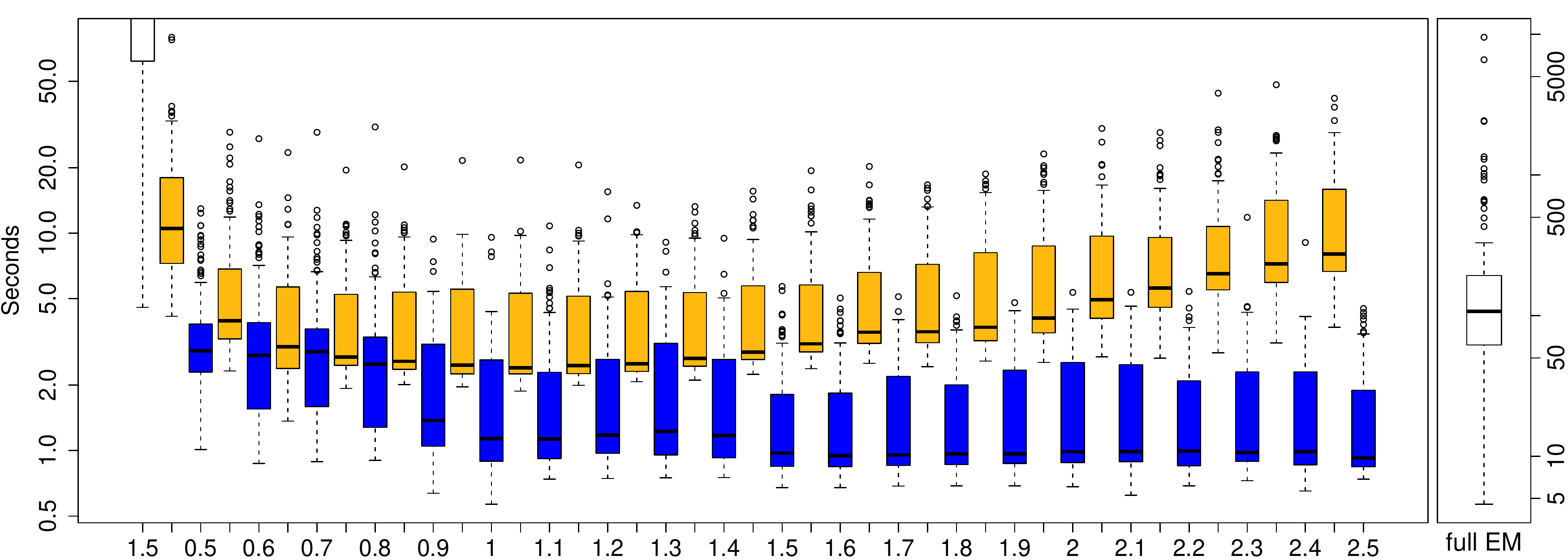}}		
		\subfigure[$50 \times 50$ grid with $m=175$]{\includegraphics[scale=.23]{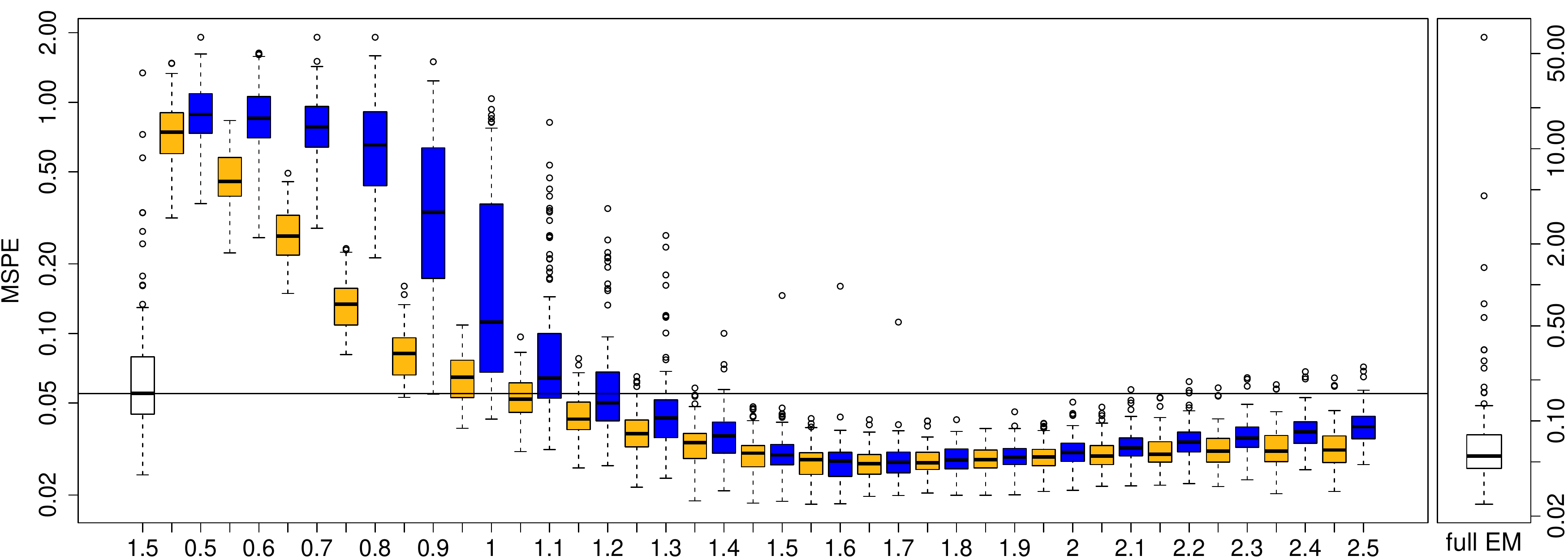}}
		\subfigure[$200 \times 200$ grid with $m=23$]{\includegraphics[scale=.23]{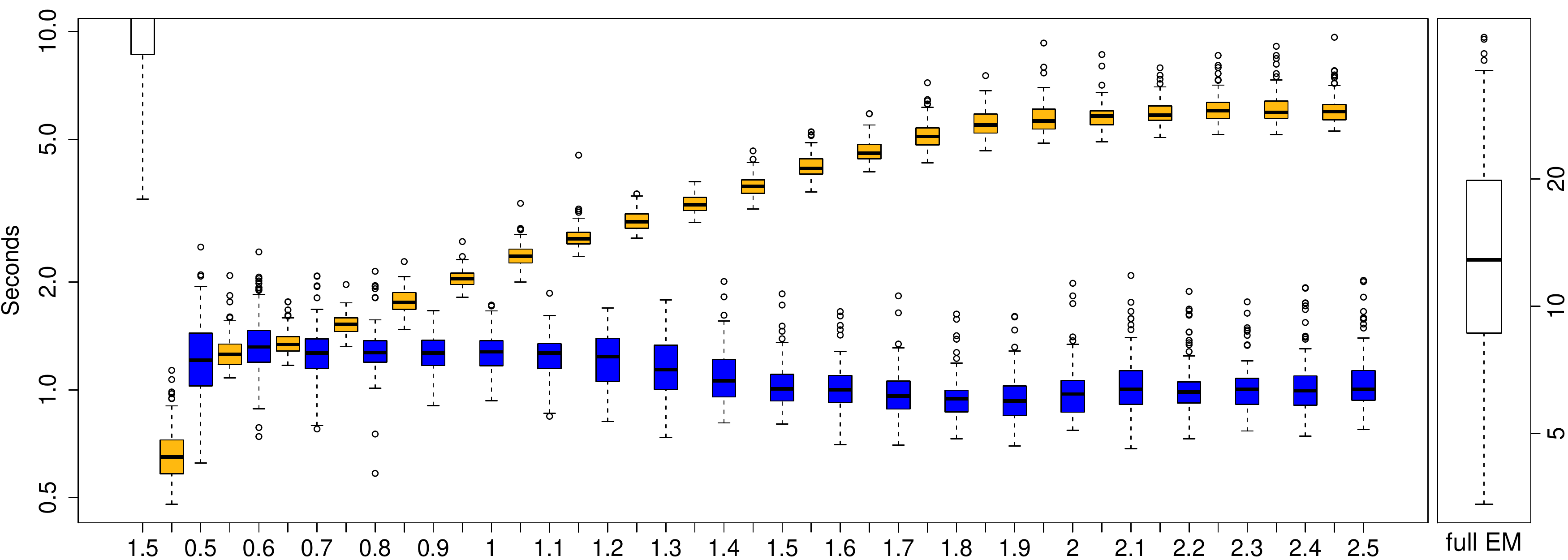}}
		\subfigure[$200 \times 200$ grid with $m=23$]{\includegraphics[scale=.23]{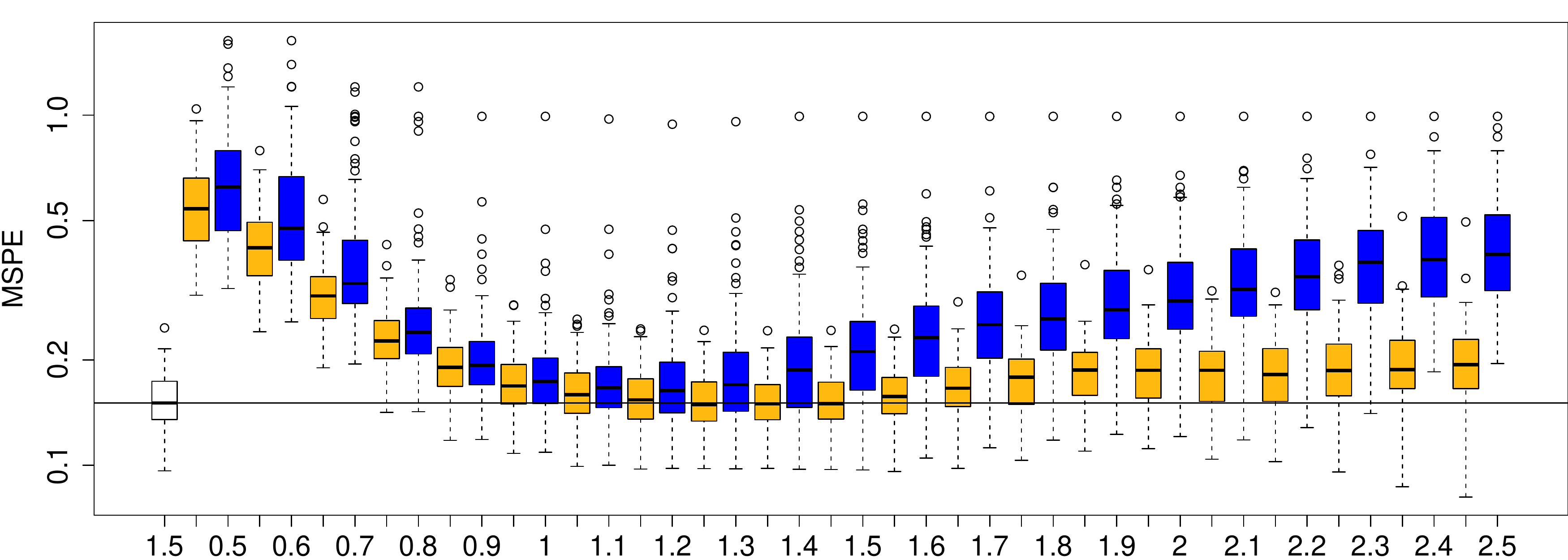}}
		\subfigure[$200 \times 200$ grid with $m=77$]{\includegraphics[scale=.23]{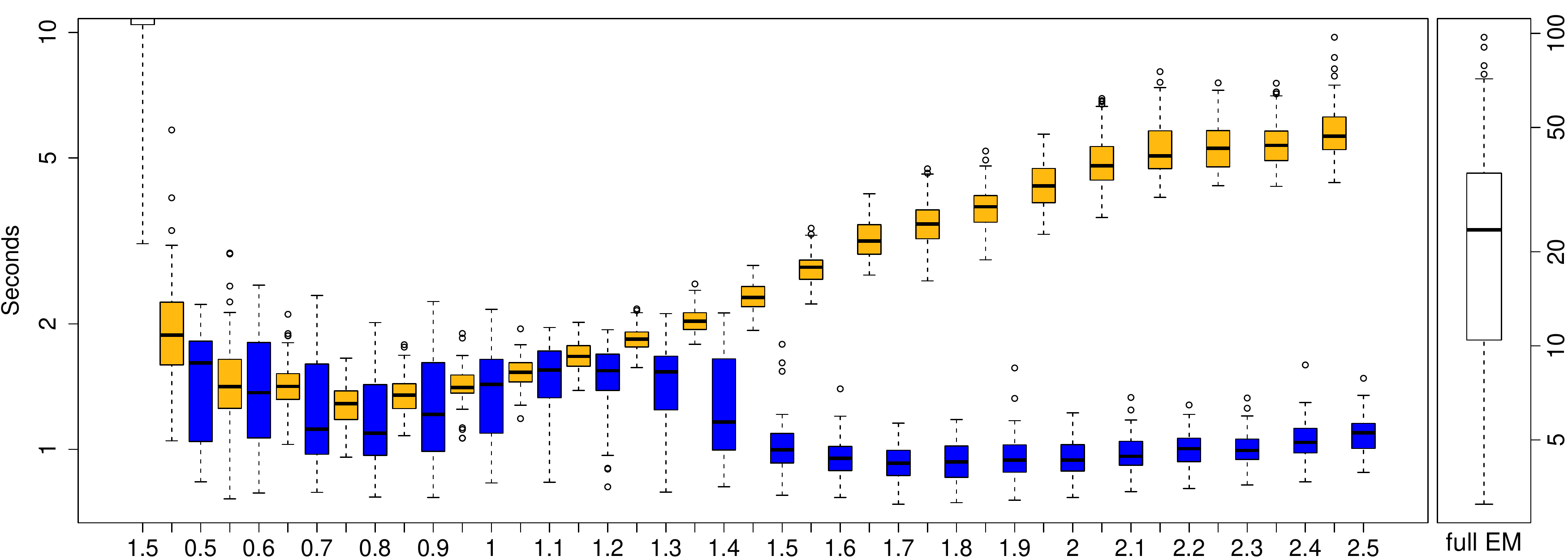}}
		\subfigure[$200 \times 200$ grid with $m=77$]{\includegraphics[scale=.23]{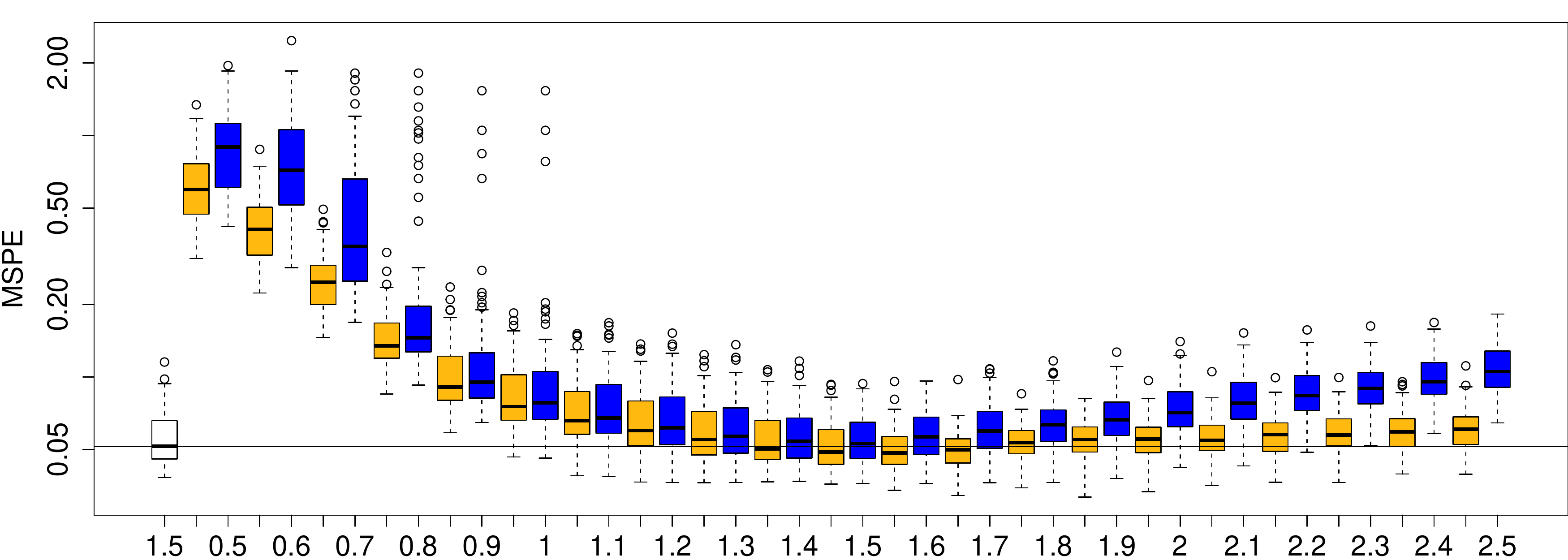}}
		\subfigure[$200 \times 200$ grid with $m=175$]{\includegraphics[scale=.23]{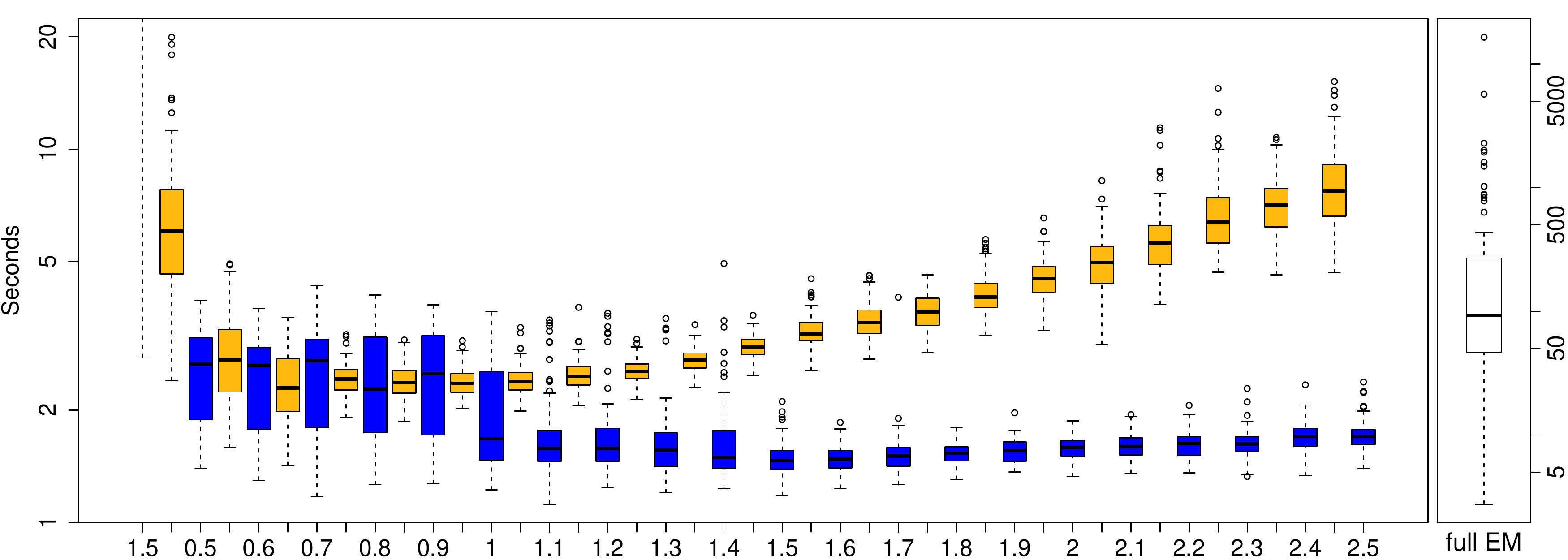}}		
		\subfigure[$200 \times 200$ grid with $m=175$]{ \includegraphics[scale=.23]{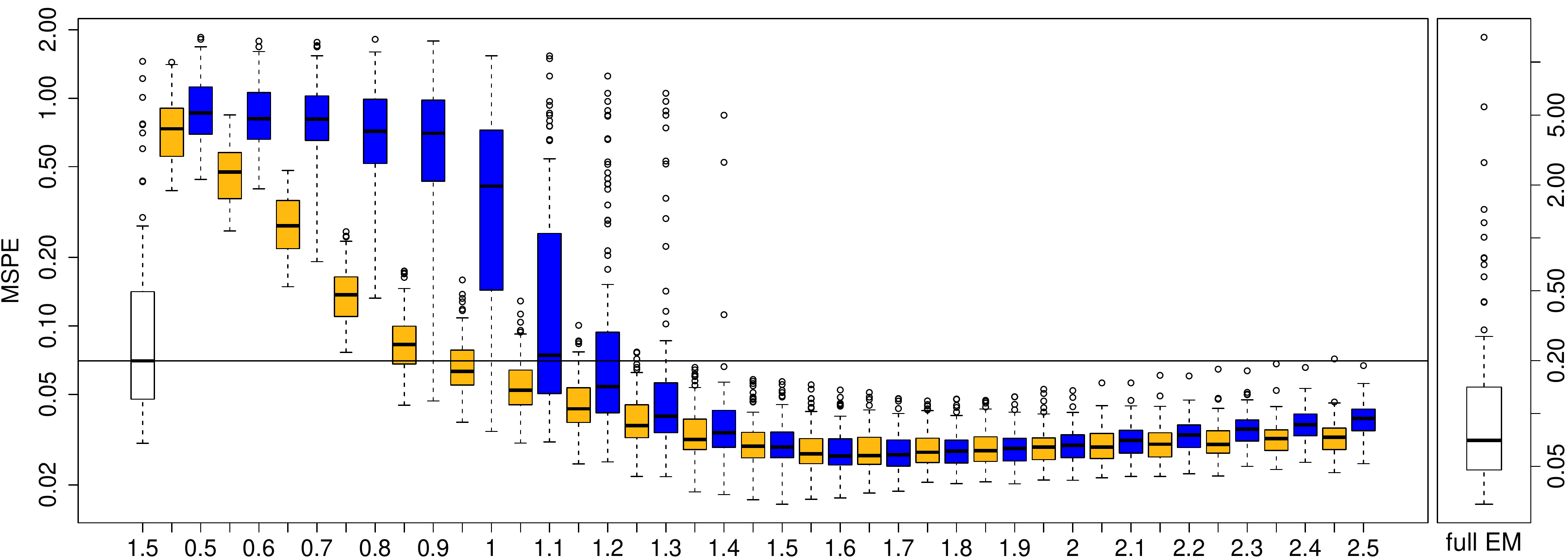}}
	\end{center}
	\caption{Seconds (left panel) and Mean Squared Prediction
          Errors (MSPEs, right panel) for varying bandwidths constants
          (x-axis) and number of knots ($m$) for all three estimation
          methods on both grids when $\nu=1.5$ and $\sigma^2=0$.}
\end{figure}

\begin{figure}[H]
	\begin{center}
		\subfigure[$50 \times 50$ grid with $m=23$]{\includegraphics[scale=.23]{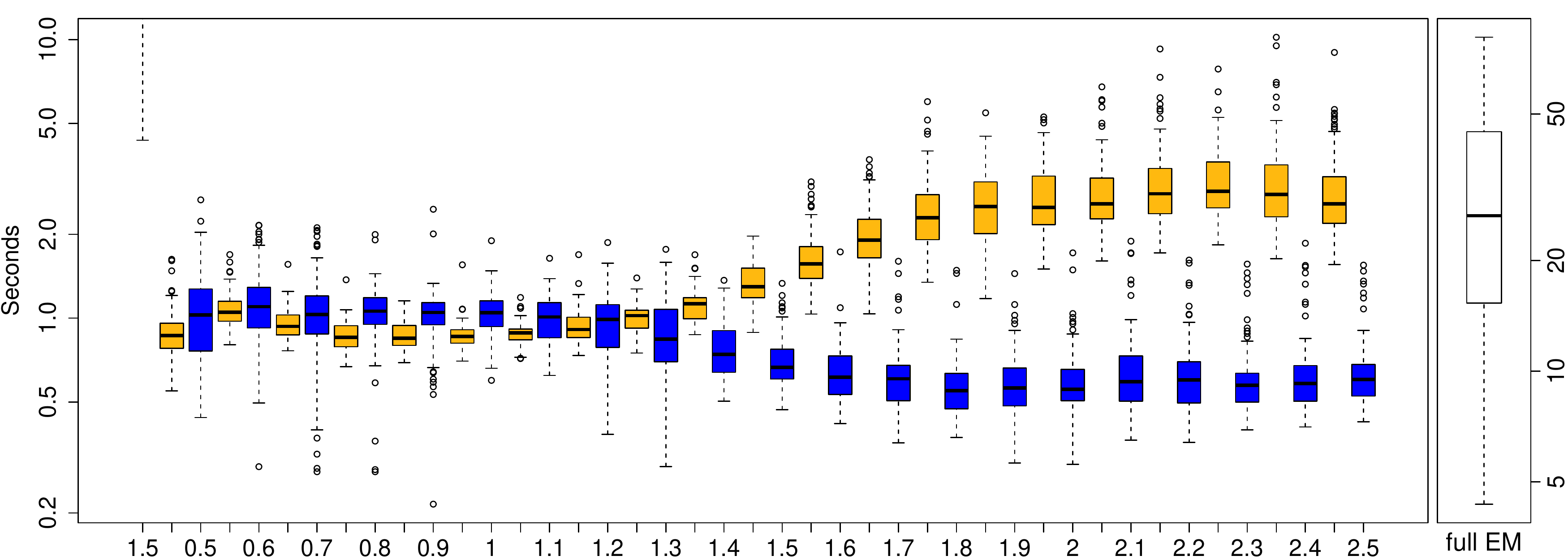}}
		\subfigure[$50 \times 50$ grid with $m=23$]{\includegraphics[scale=.23]{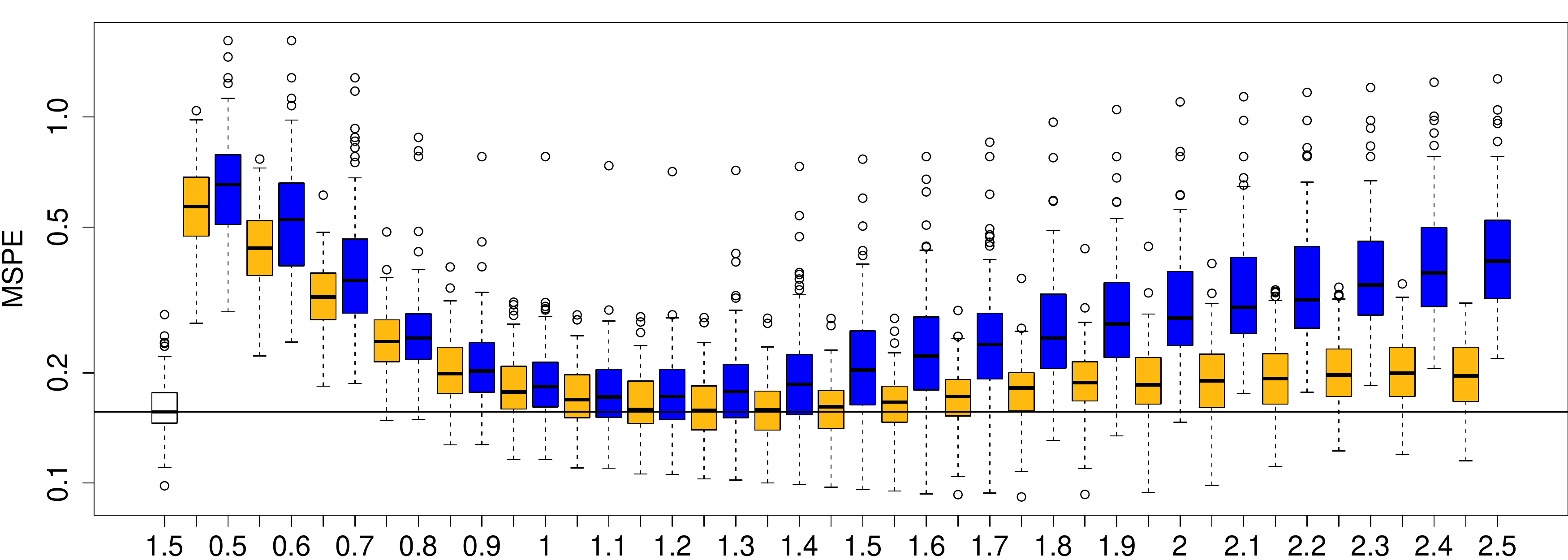}}
		\subfigure[$50 \times 50$ grid with $m=77$]{\includegraphics[scale=.23]{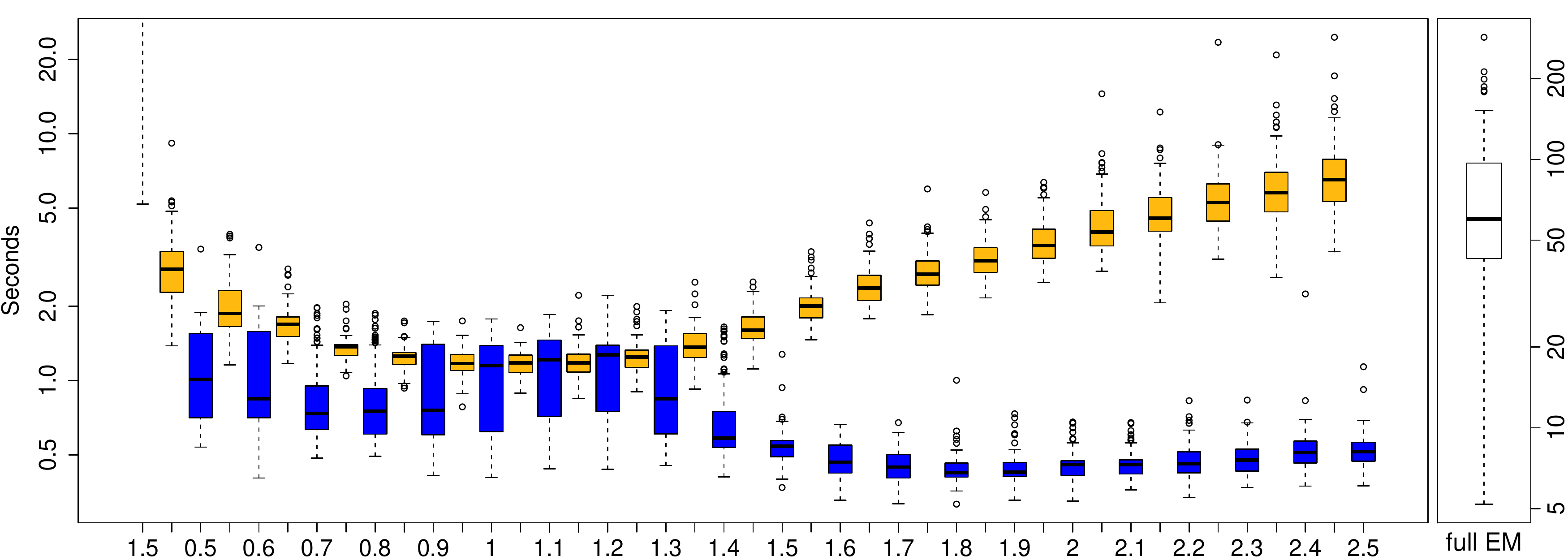}}
		\subfigure[$50 \times 50$ grid with $m=77$]{\includegraphics[scale=.23]{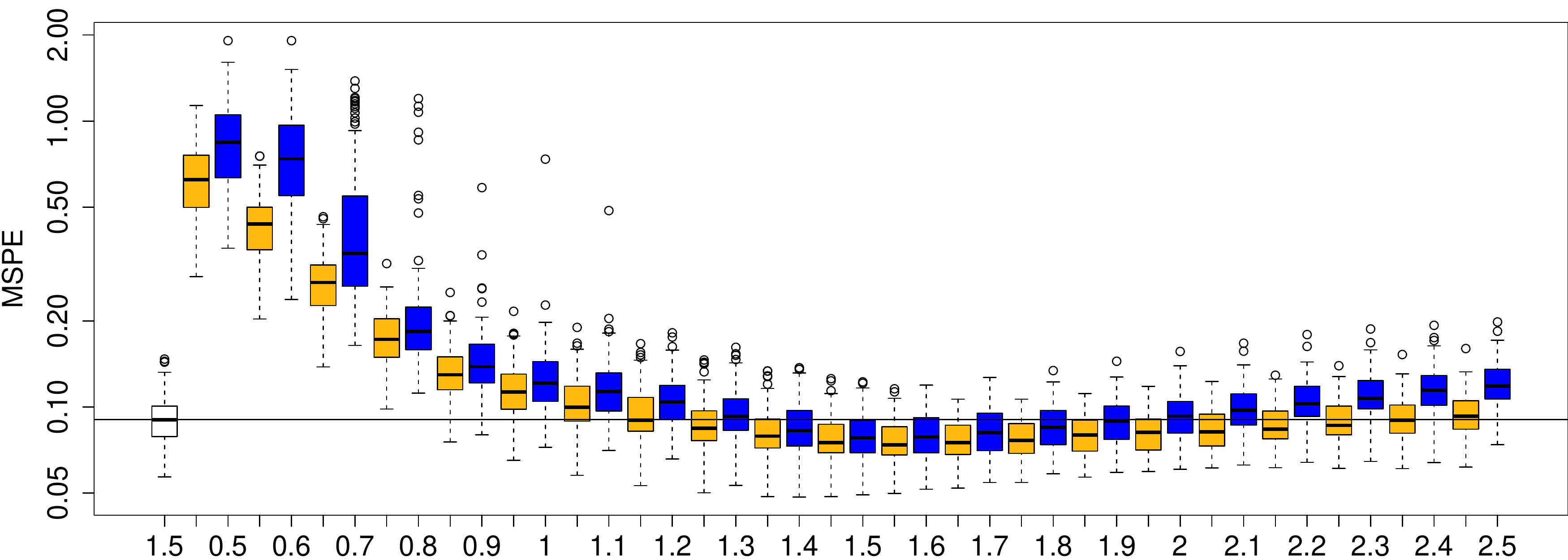}}
		\subfigure[$50 \times 50$ grid with $m=175$]{\includegraphics[scale=.23]{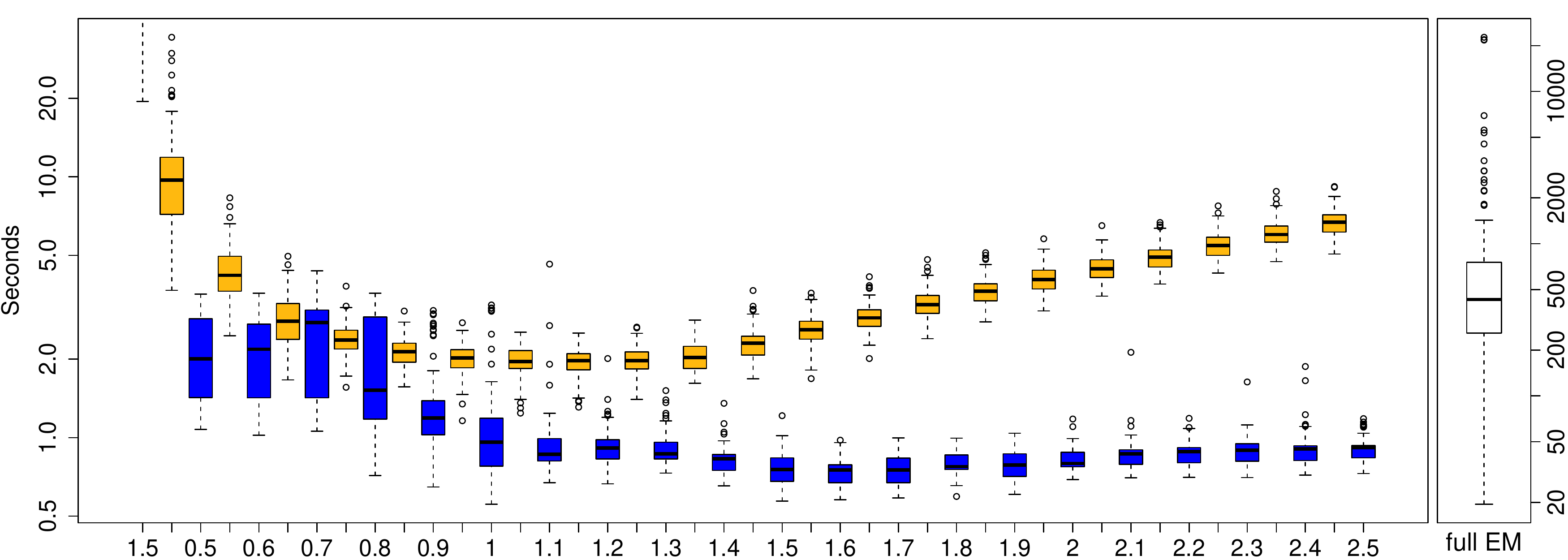}}		
		\subfigure[$50 \times 50$ grid with $m=175$]{\includegraphics[scale=.23]{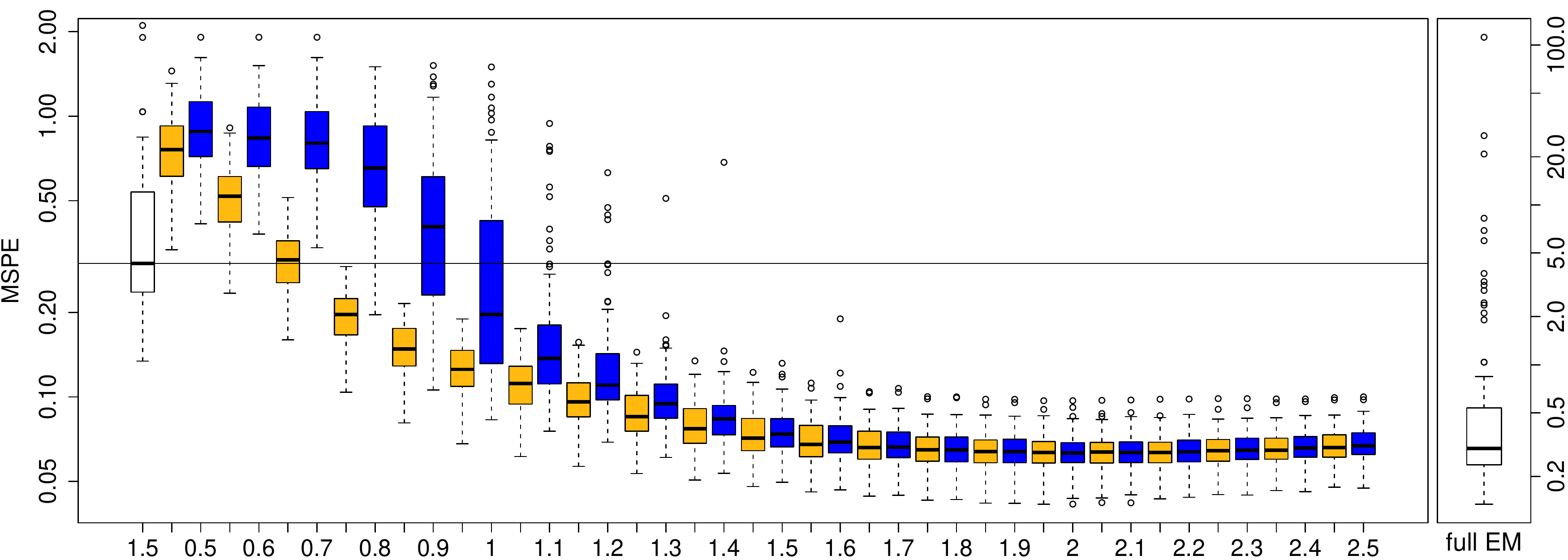}}
		\subfigure[$200 \times 200$ grid with $m=23$]{\includegraphics[scale=.23]{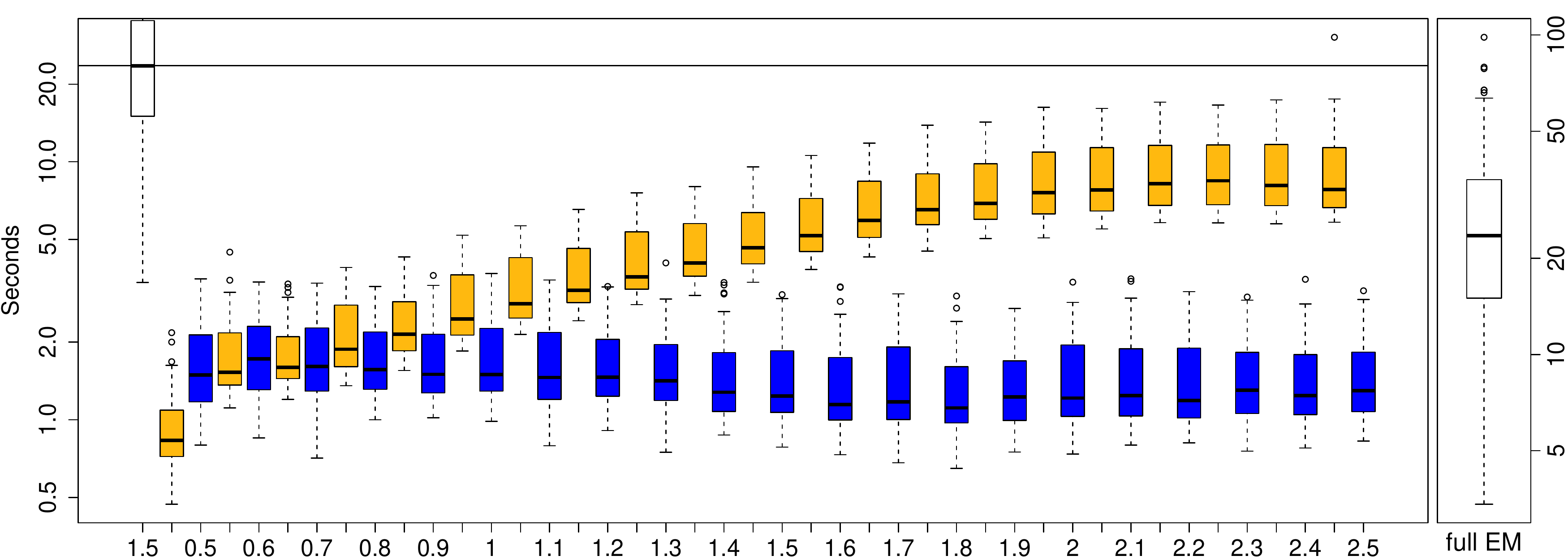}}
		\subfigure[$200 \times 200$ grid with $m=23$]{\includegraphics[scale=.23]{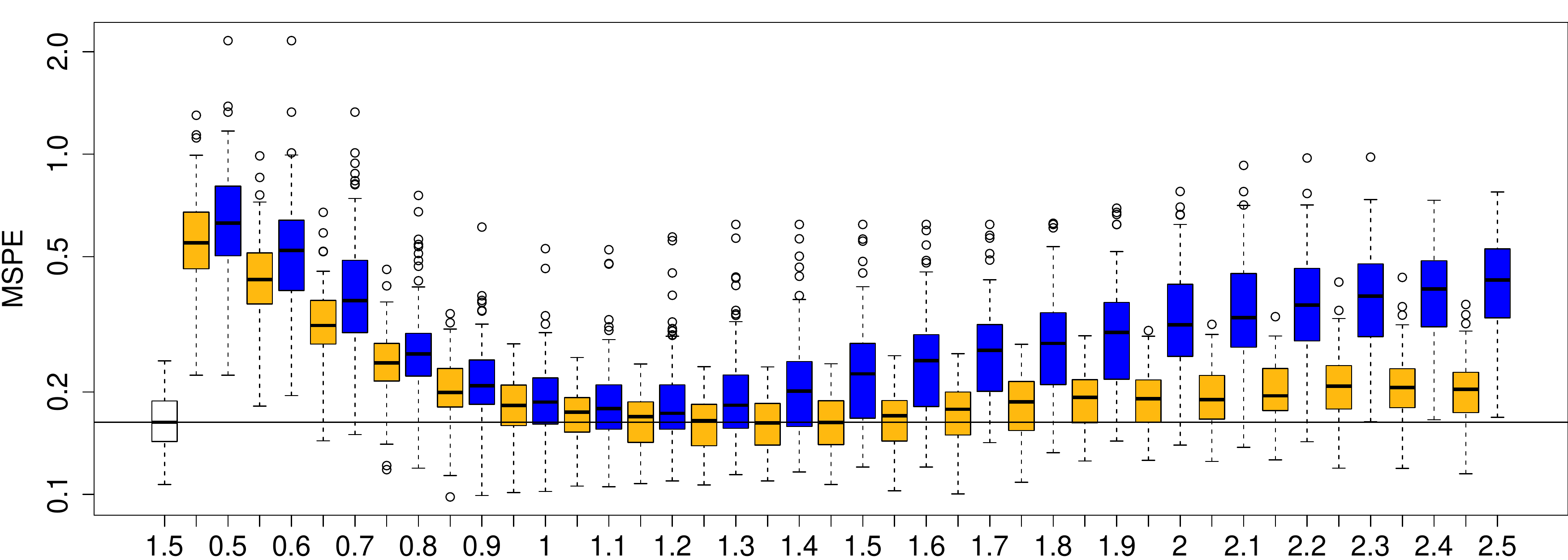}}
		\subfigure[$200 \times 200$ grid with $m=77$]{\includegraphics[scale=.23]{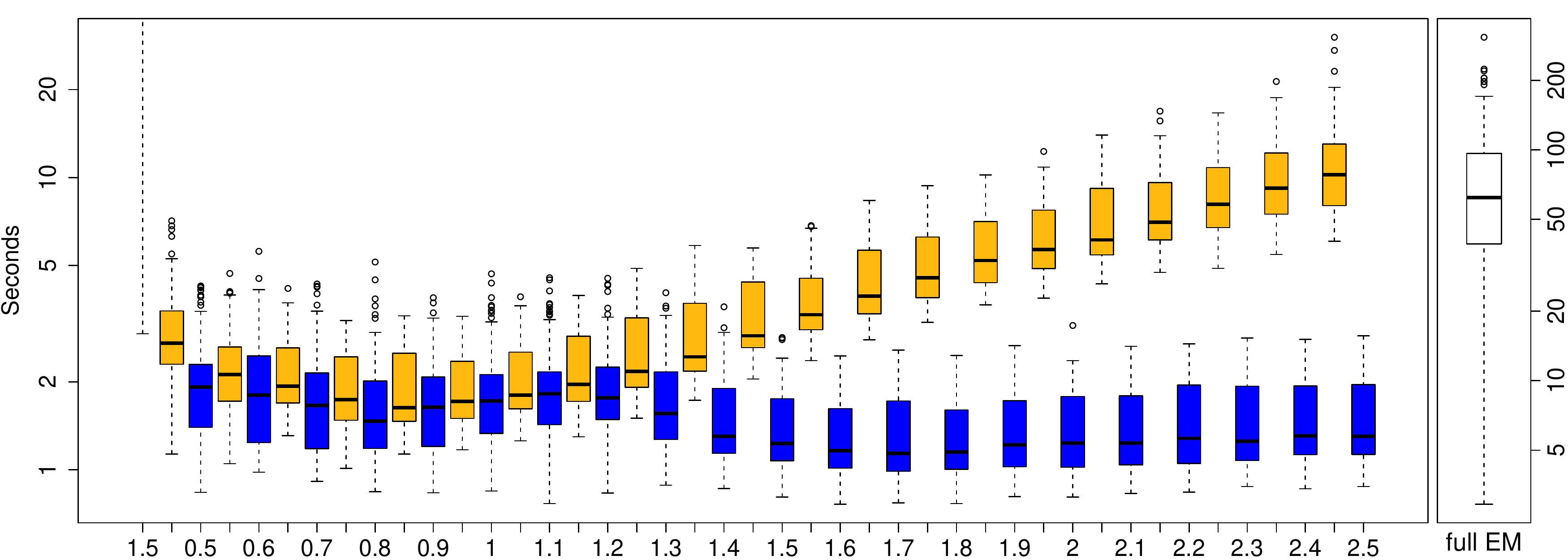}}
		\subfigure[$200 \times 200$ grid with $m=77$]{\includegraphics[scale=.23]{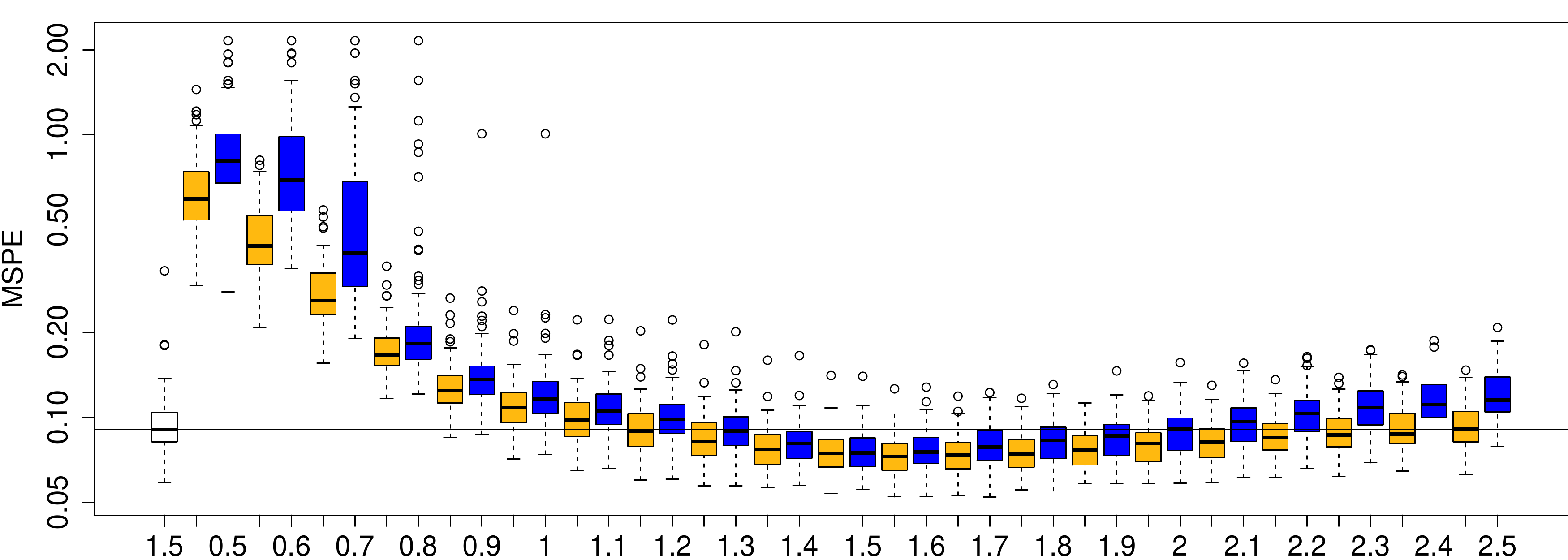}}
		\subfigure[$200 \times 200$ grid with $m=175$]{\includegraphics[scale=.23]{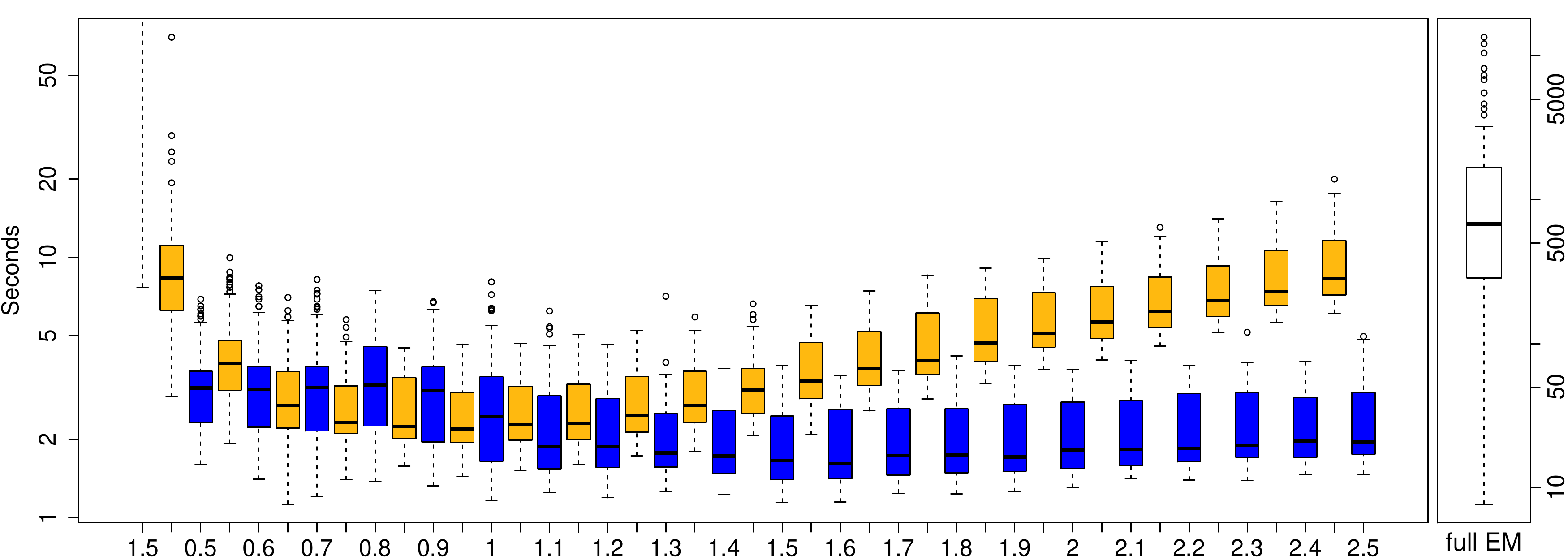}}		
		\subfigure[$200 \times 200$ grid with $m=175$]{\includegraphics[scale=.23]{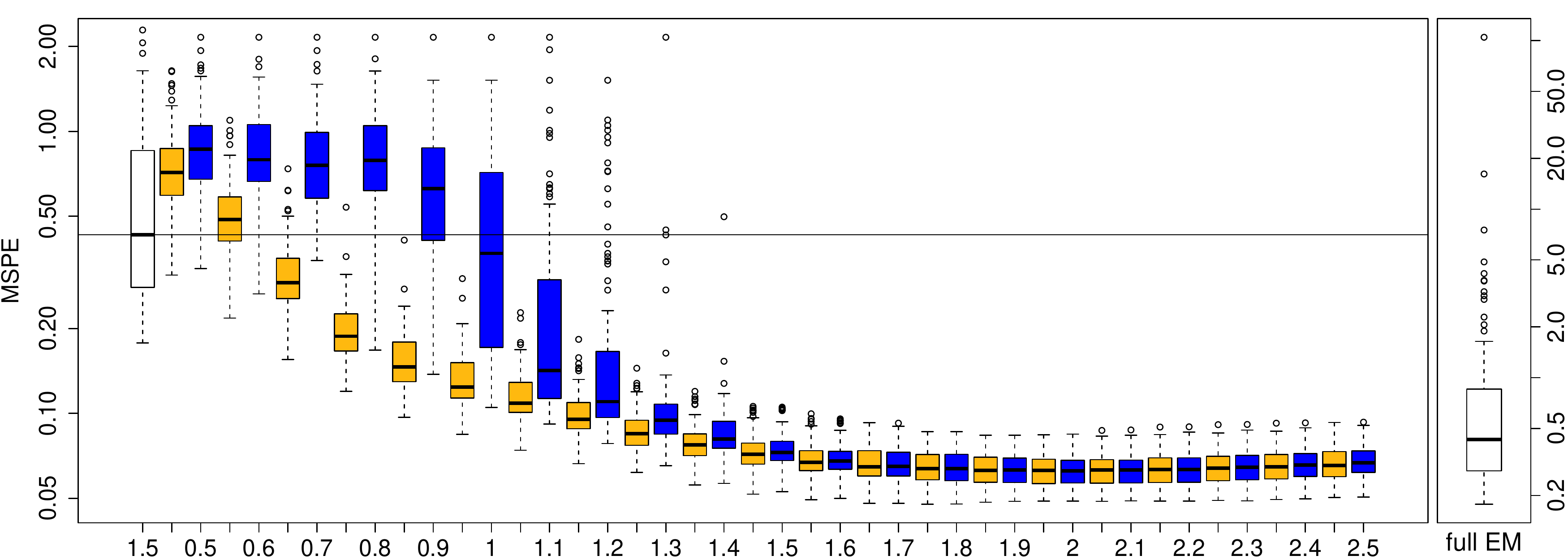}}
	\end{center}
	\caption{Seconds (left panel) and Mean Squared Prediction
          Errors (MSPEs, right panel) for varying bandwidths constants
          (x-axis) and number of knots ($m$) for all three estimation
          methods on both grids when $\nu=1.5$ and $\sigma^2=0.1$.} 
\end{figure}

\begin{figure}[H]
	\begin{center}
		\subfigure[$50 \times 50$ grid with $m=23$]{\includegraphics[scale=.23]{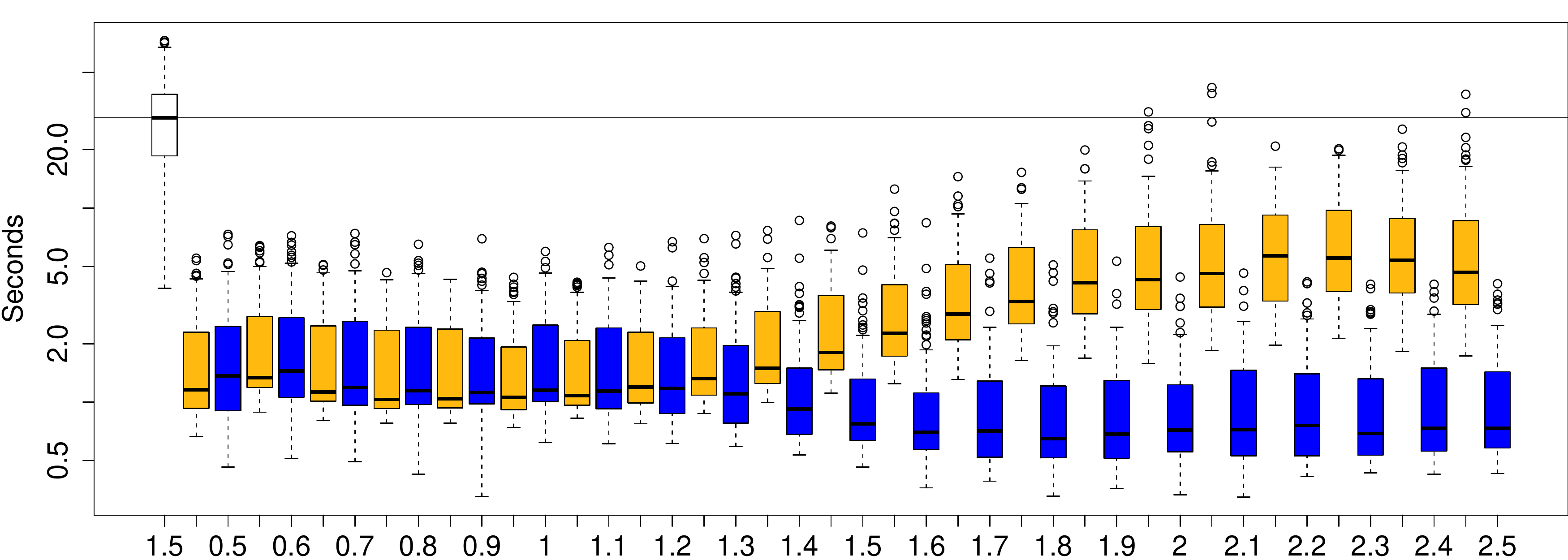}}
		\subfigure[$50 \times 50$ grid with $m=23$]{\includegraphics[scale=.23]{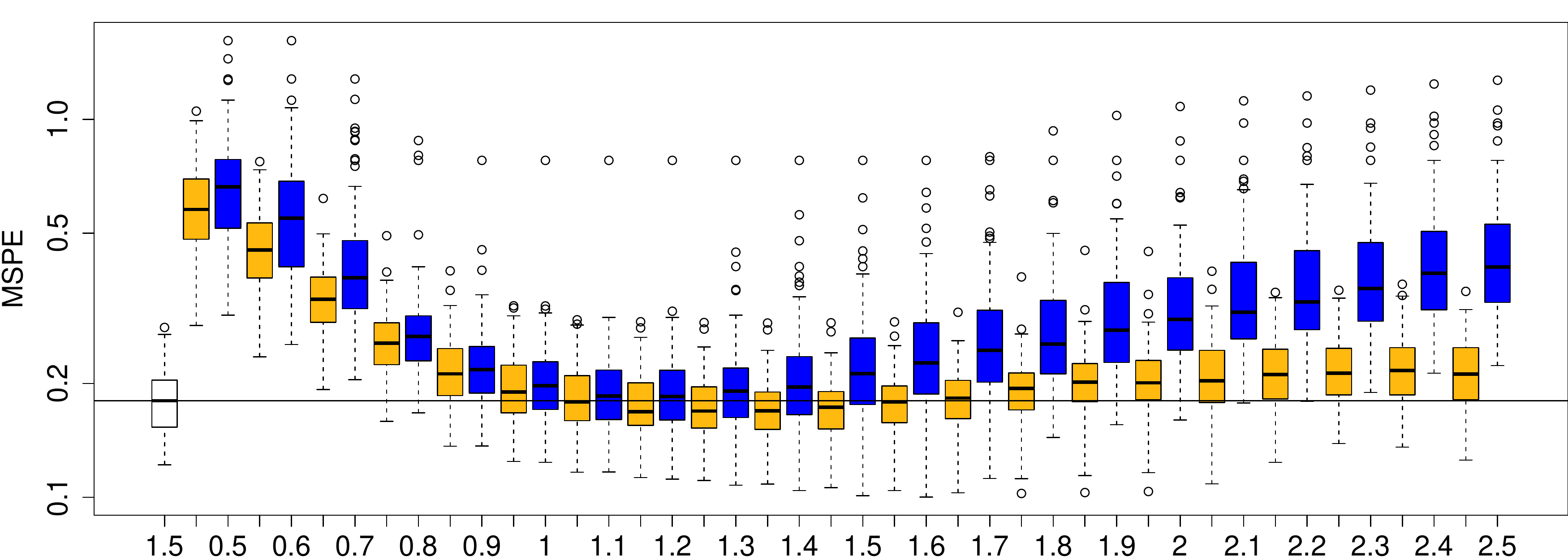}}
		\subfigure[$50 \times 50$ grid with $m=77$]{\includegraphics[scale=.23]{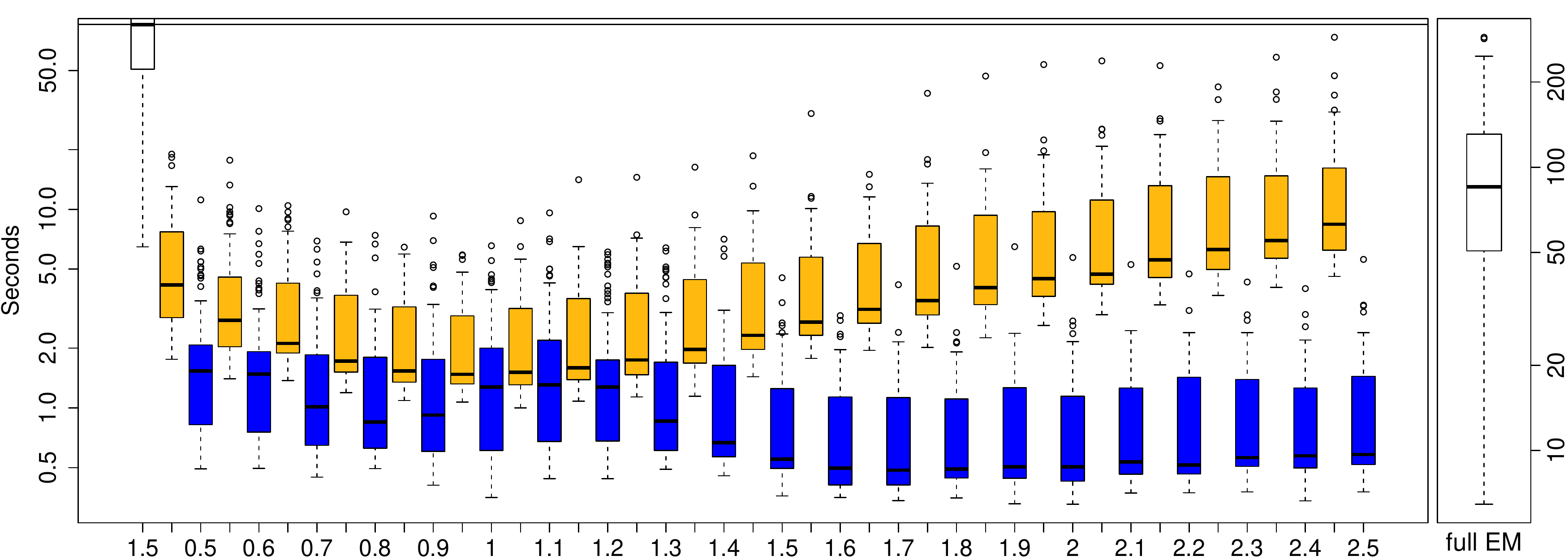}}
		\subfigure[$50 \times 50$ grid with $m=77$]{\includegraphics[scale=.23]{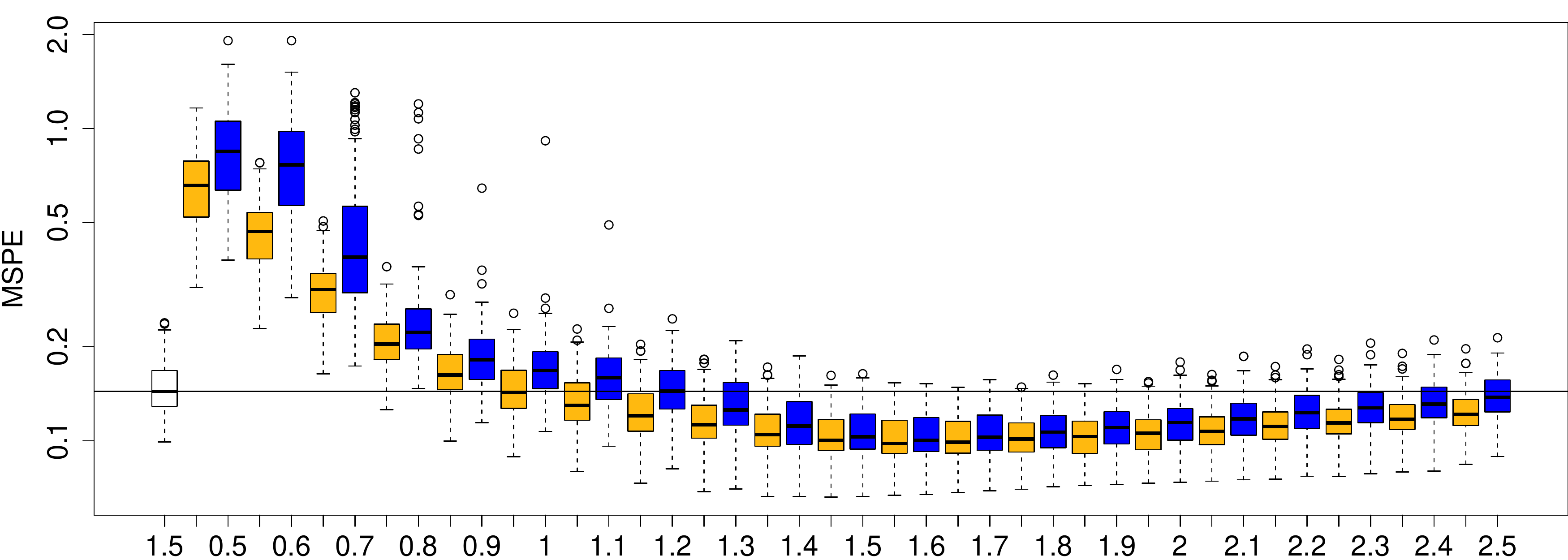}}
		\subfigure[$50 \times 50$ grid with $m=175$]{\includegraphics[scale=.23]{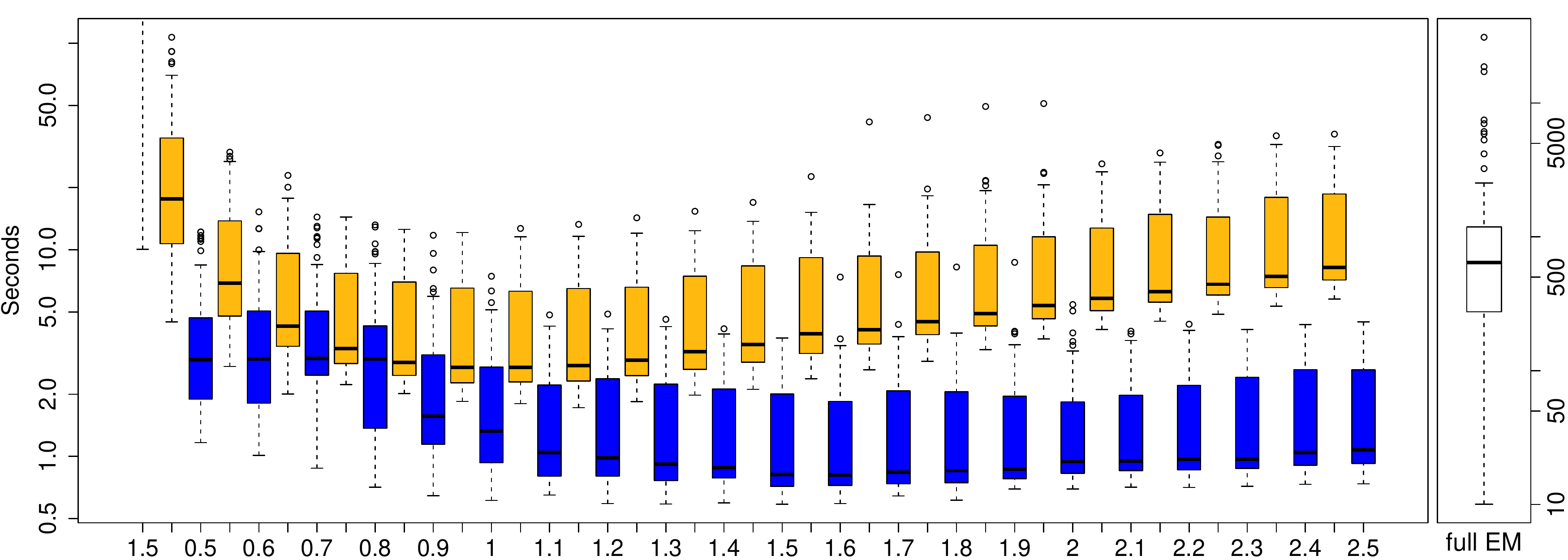}}		
		\subfigure[$50 \times 50$ grid with $m=175$]{\includegraphics[scale=.23]{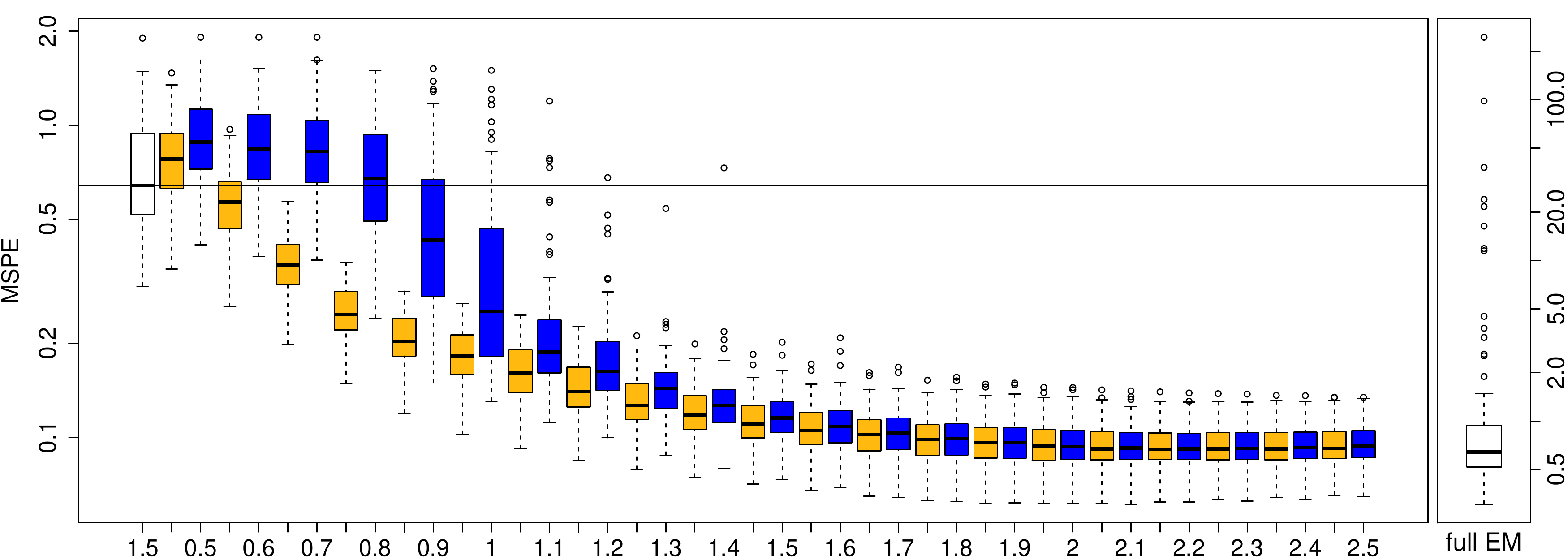}}
		\subfigure[$200 \times 200$ grid with $m=23$]{\includegraphics[scale=.23]{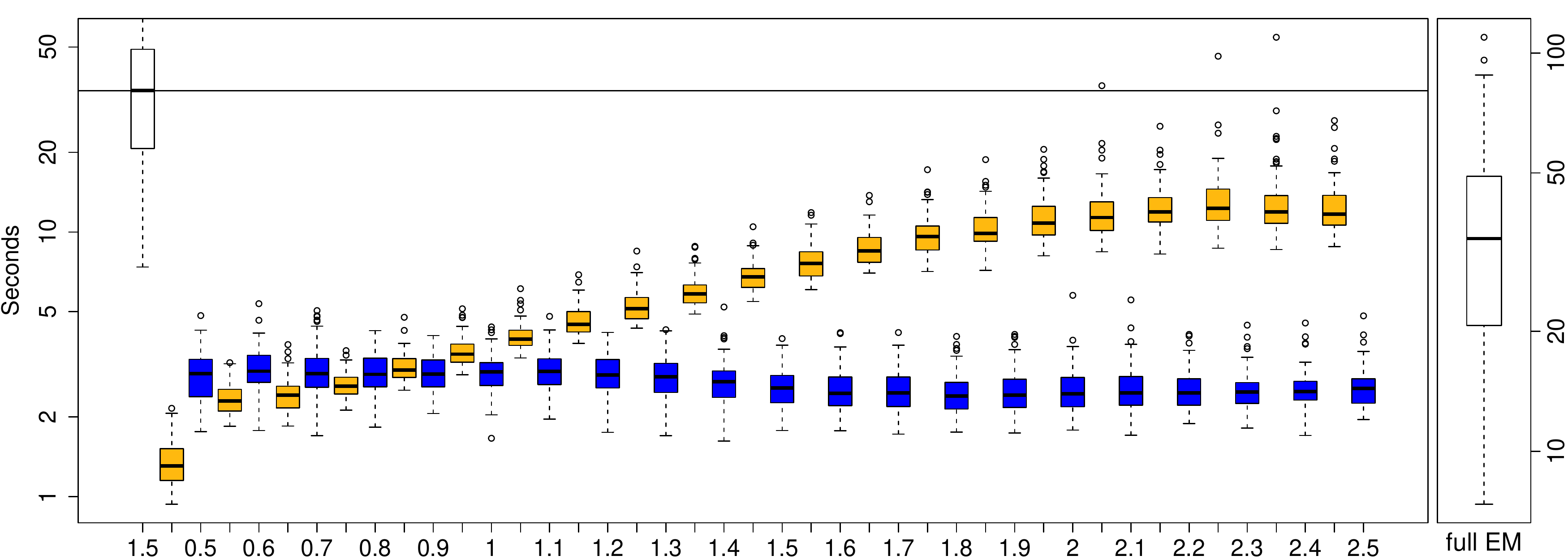}}
		\subfigure[$200 \times 200$ grid with $m=23$]{\includegraphics[scale=.23]{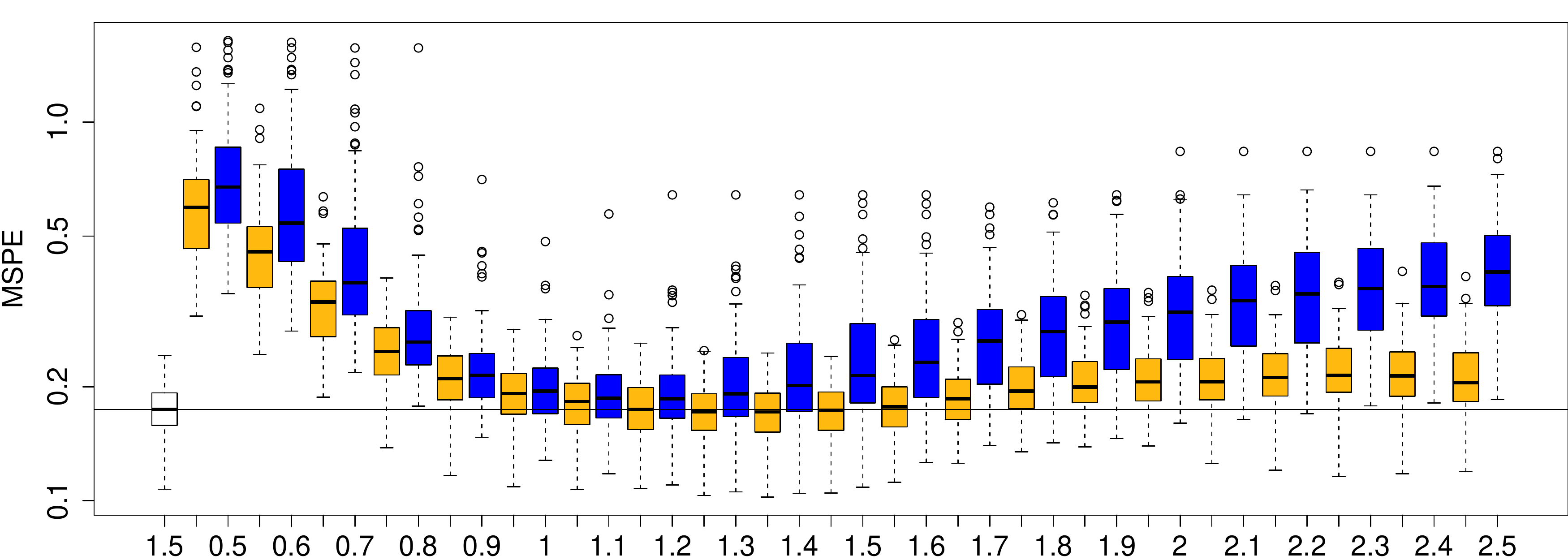}}
		\subfigure[$200 \times 200$ grid with $m=77$]{\includegraphics[scale=.23]{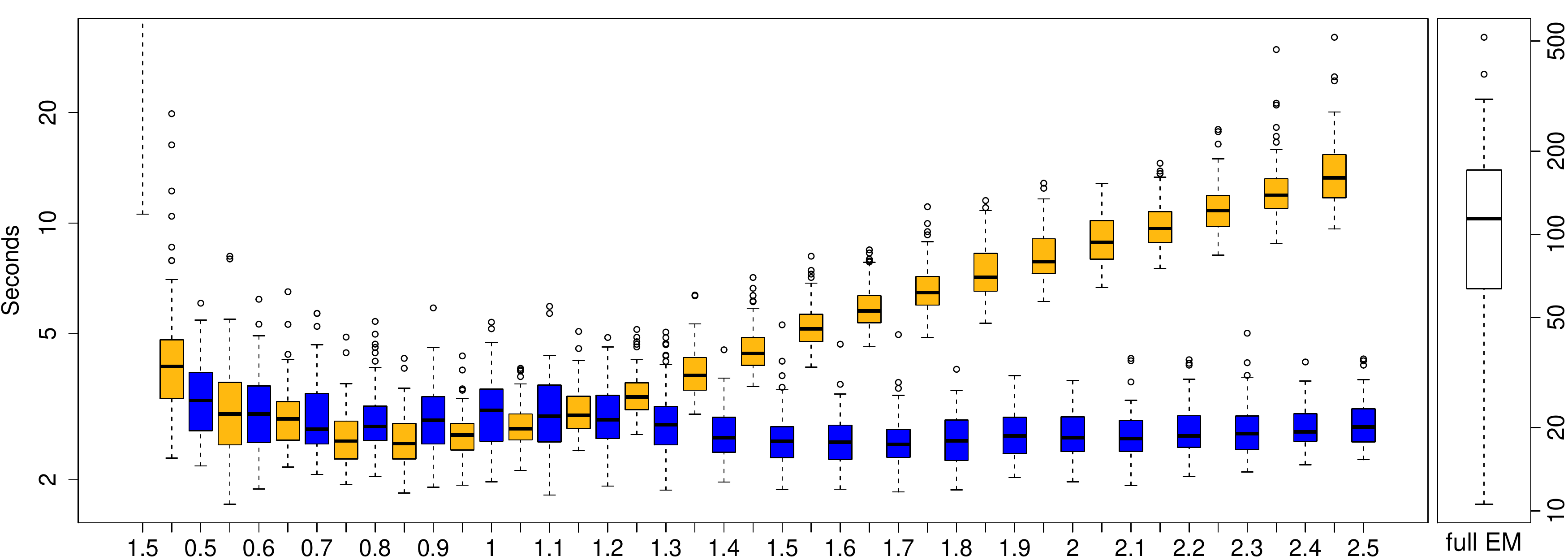}}
		\subfigure[$200 \times 200$ grid with $m=77$]{\includegraphics[scale=.23]{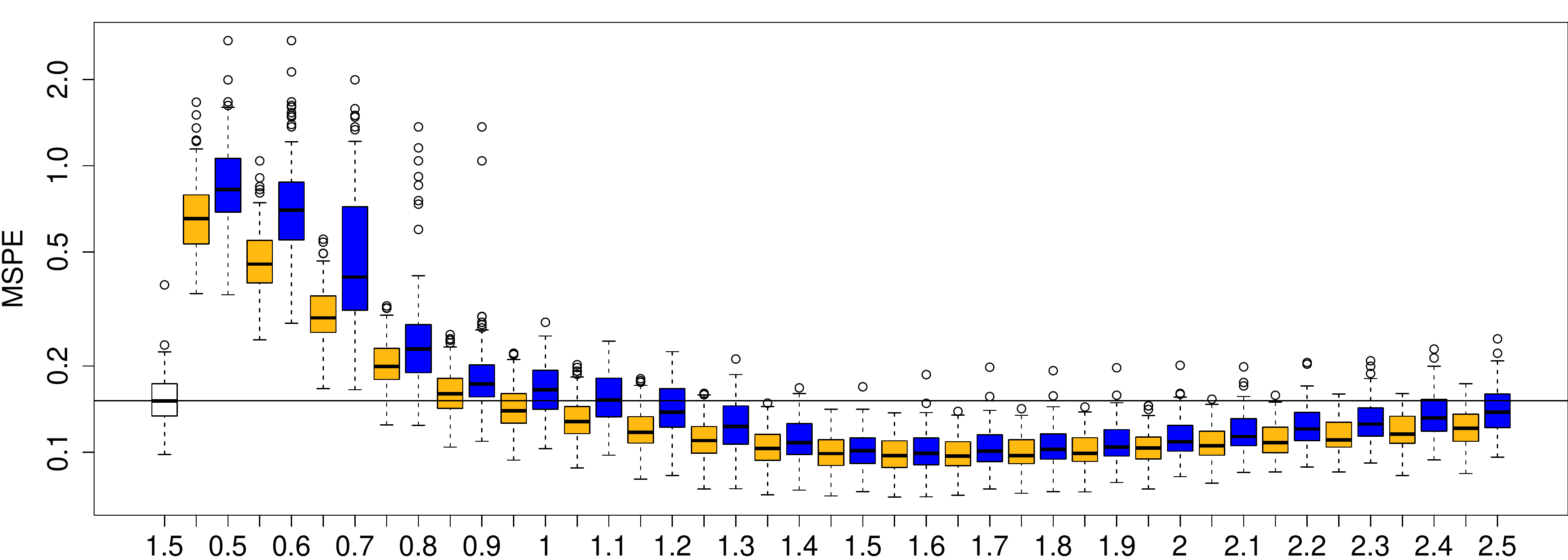}}
		\subfigure[$200 \times 200$ grid with $m=175$]{\includegraphics[scale=.23]{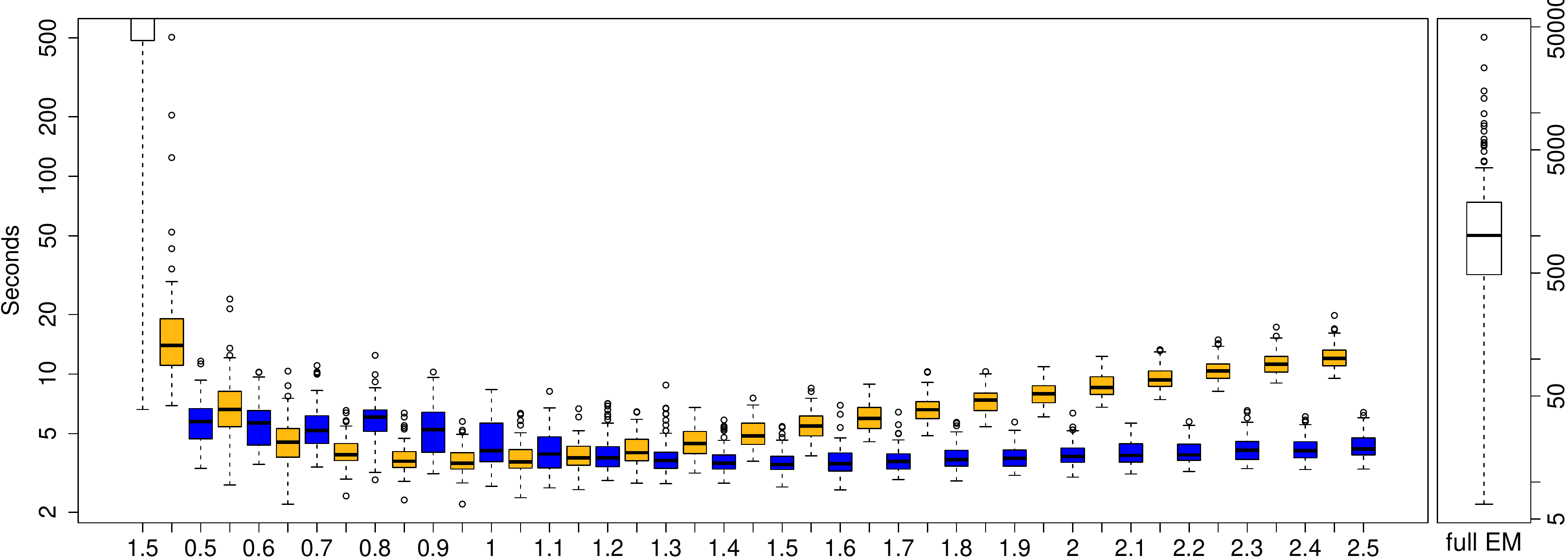}}		
		\subfigure[$200 \times 200$ grid with $m=175$]{\includegraphics[scale=.23]{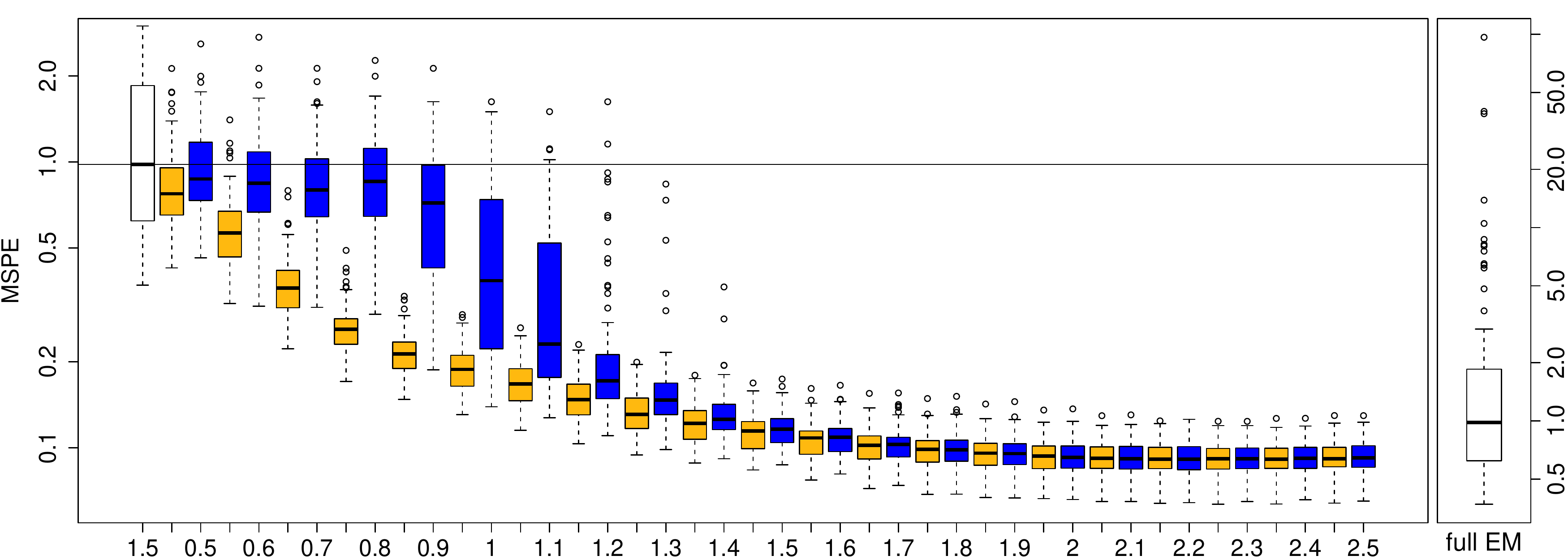}}
	\end{center}
	\caption{Seconds (left panel) and Mean Squared Prediction
          Errors (MSPEs, right panel) for varying bandwidths constants
          (x-axis) and number of knots ($m$) for all three estimation
          methods on both grids when $\nu=1.5$ and $\sigma^2=0.25$.} 
\end{figure}

\begin{figure}[H]
	\begin{center}
		\subfigure[$50 \times 50$ grid with $m=23$]{\includegraphics[scale=.23]{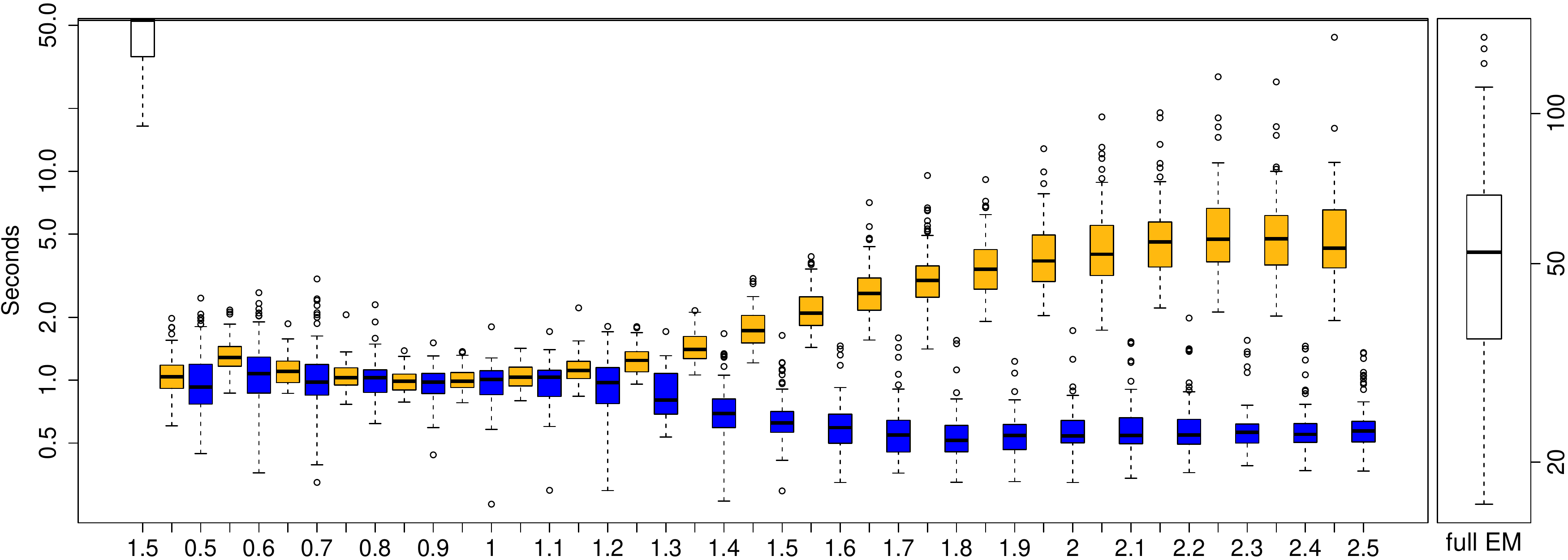}}
		\subfigure[$50 \times 50$ grid with $m=23$]{\includegraphics[scale=.23]{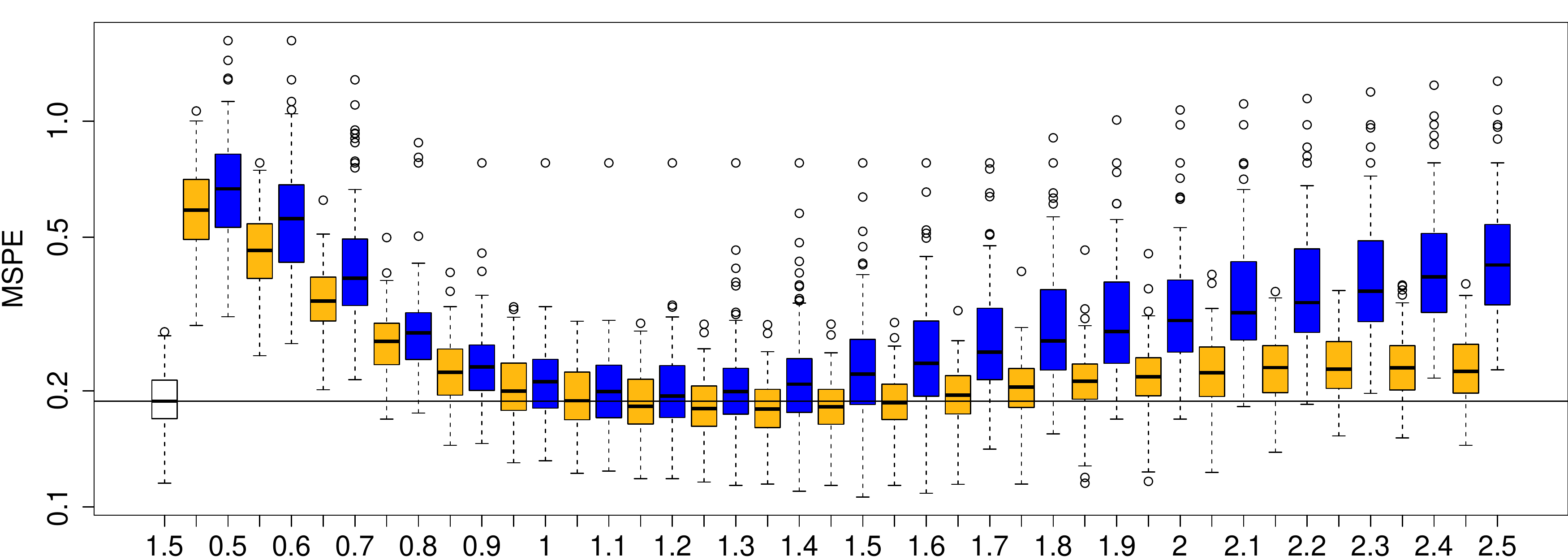}}
		\subfigure[$50 \times 50$ grid with $m=77$]{\includegraphics[scale=.23]{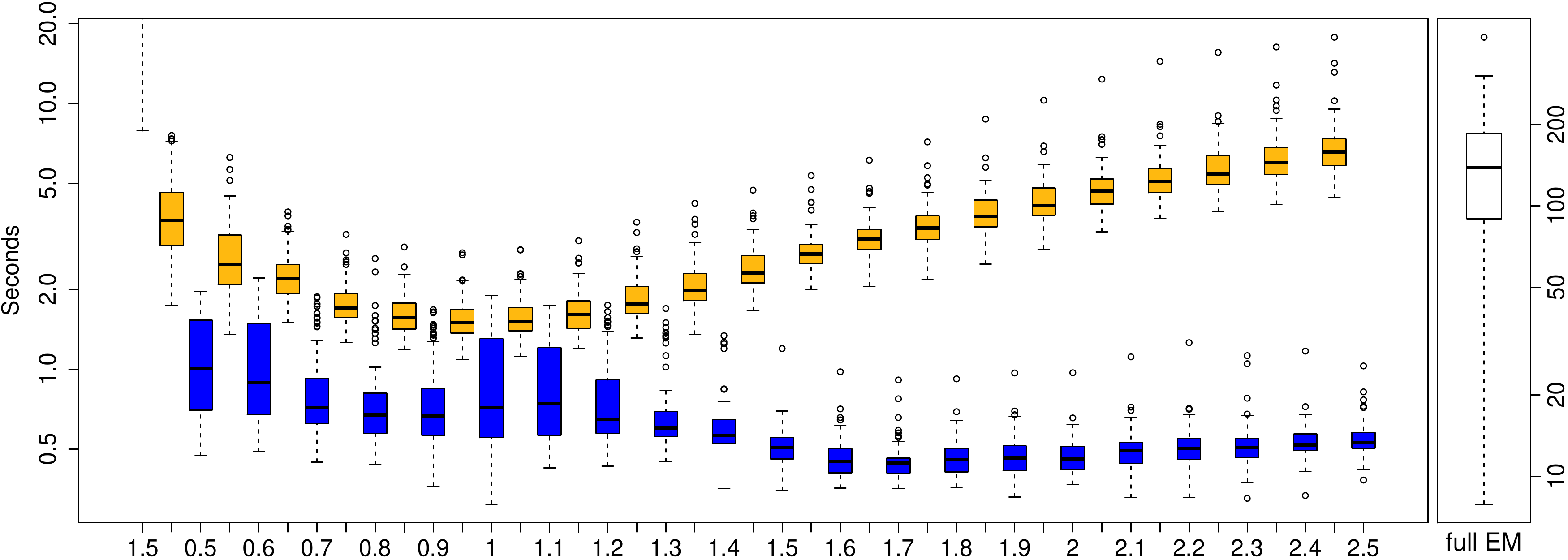}}
		\subfigure[$50 \times 50$ grid with $m=77$]{\includegraphics[scale=.23]{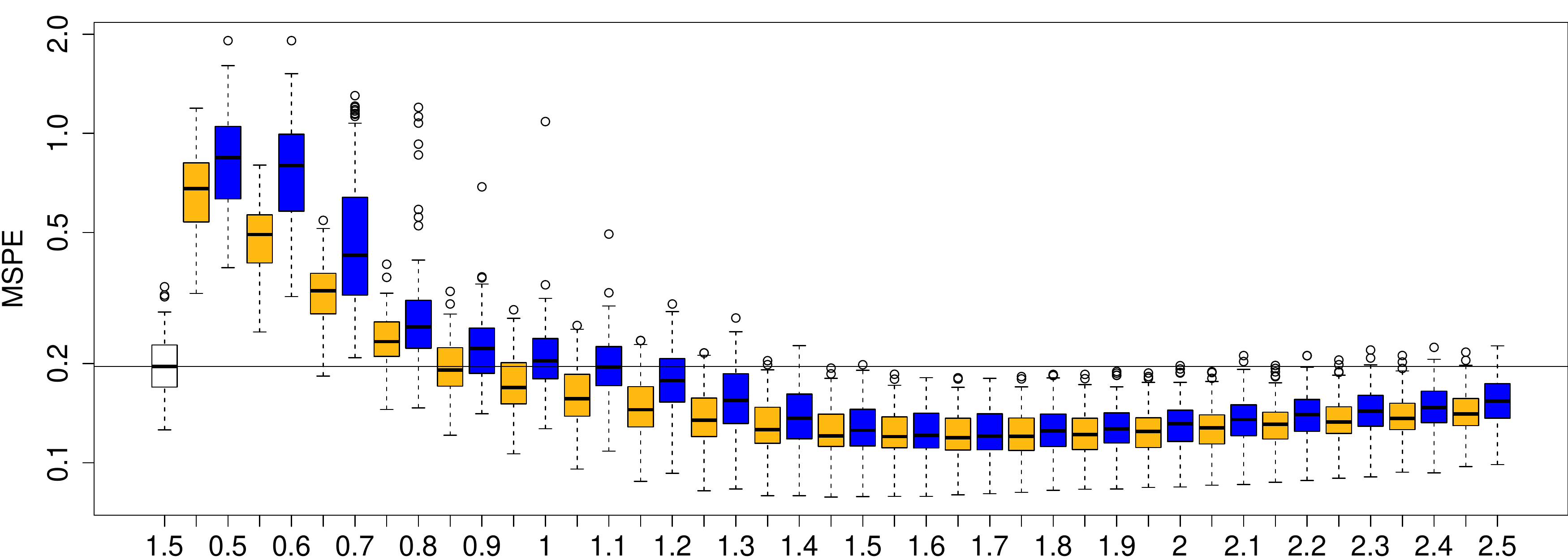}}
		\subfigure[$50 \times 50$ grid with $m=175$]{\includegraphics[scale=.23]{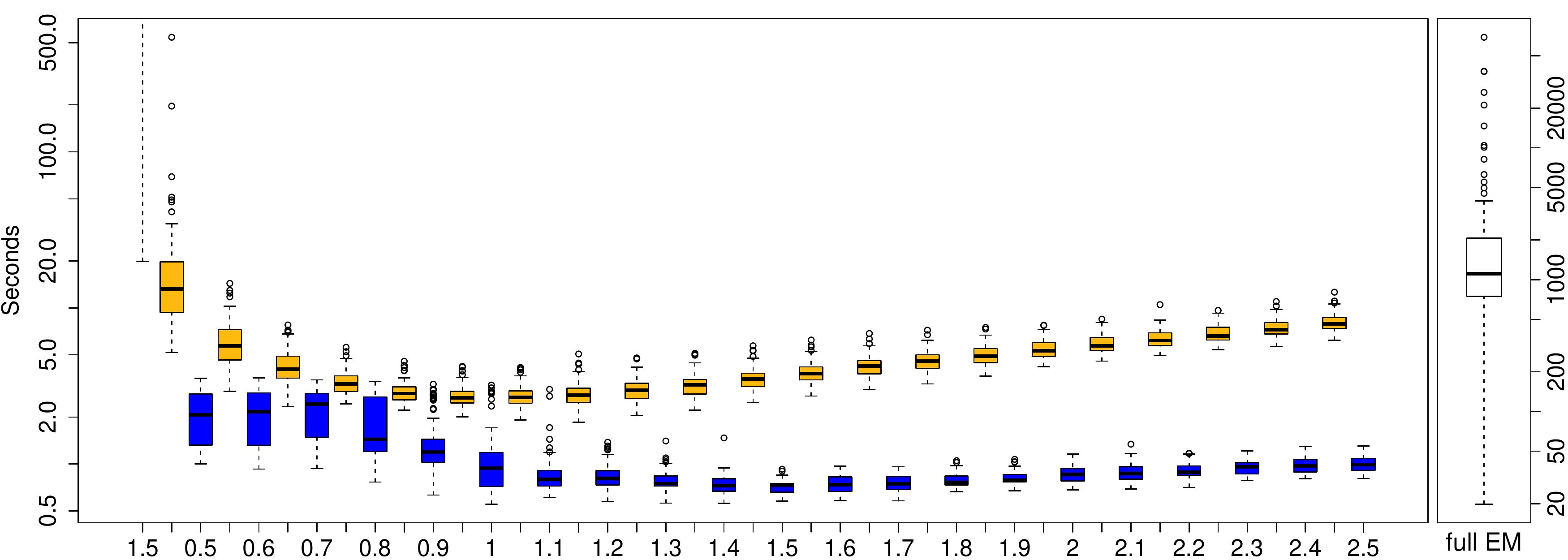}}		
		\subfigure[$50 \times 50$ grid with $m=175$]{\includegraphics[scale=.23]{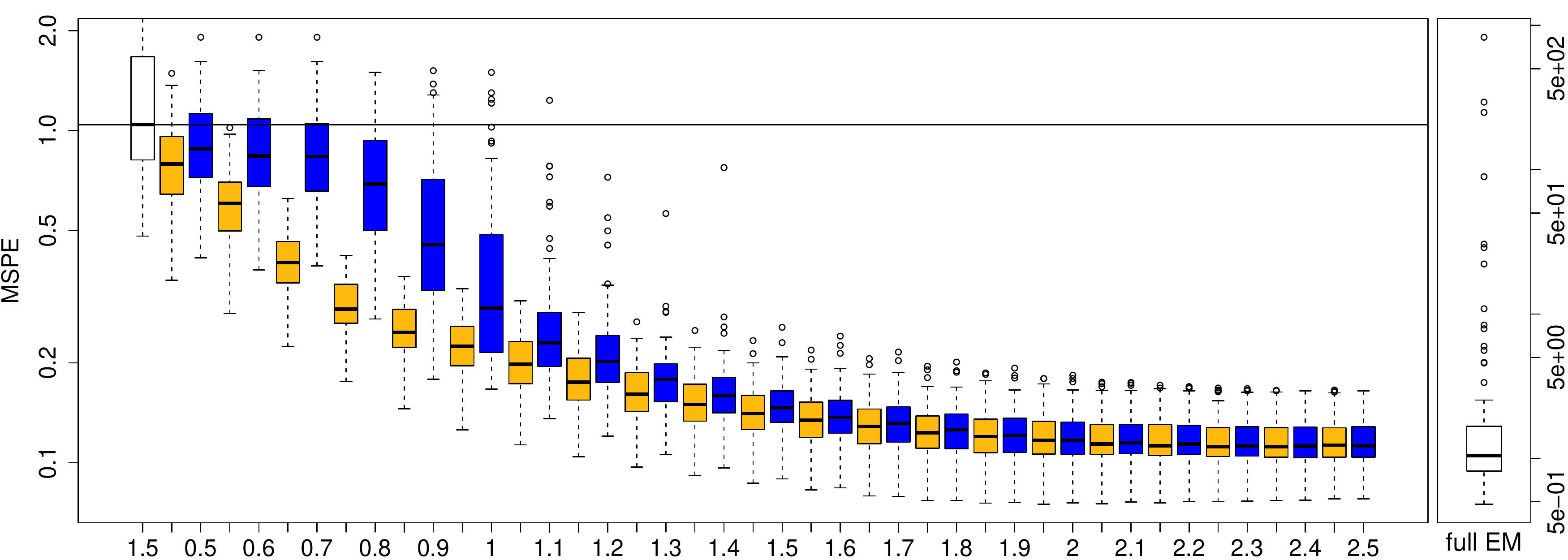}}
		\subfigure[$200 \times 200$ grid with $m=23$]{\includegraphics[scale=.23]{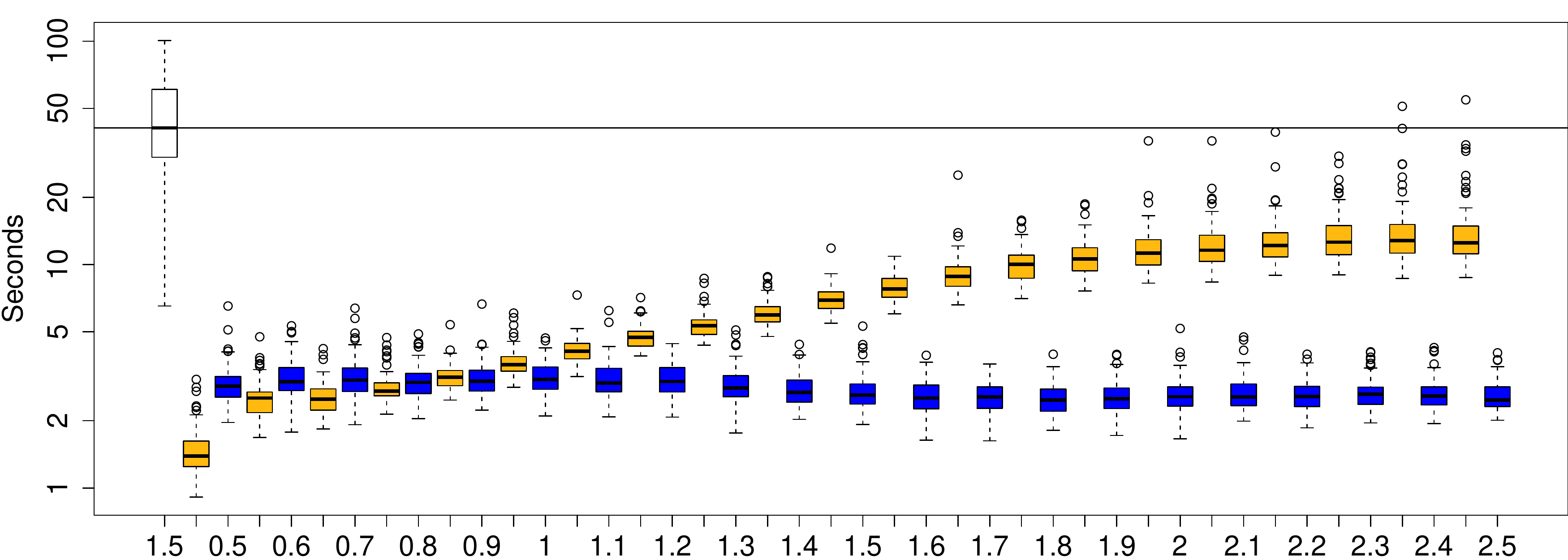}}
		\subfigure[$200 \times 200$ grid with $m=23$]{\includegraphics[scale=.23]{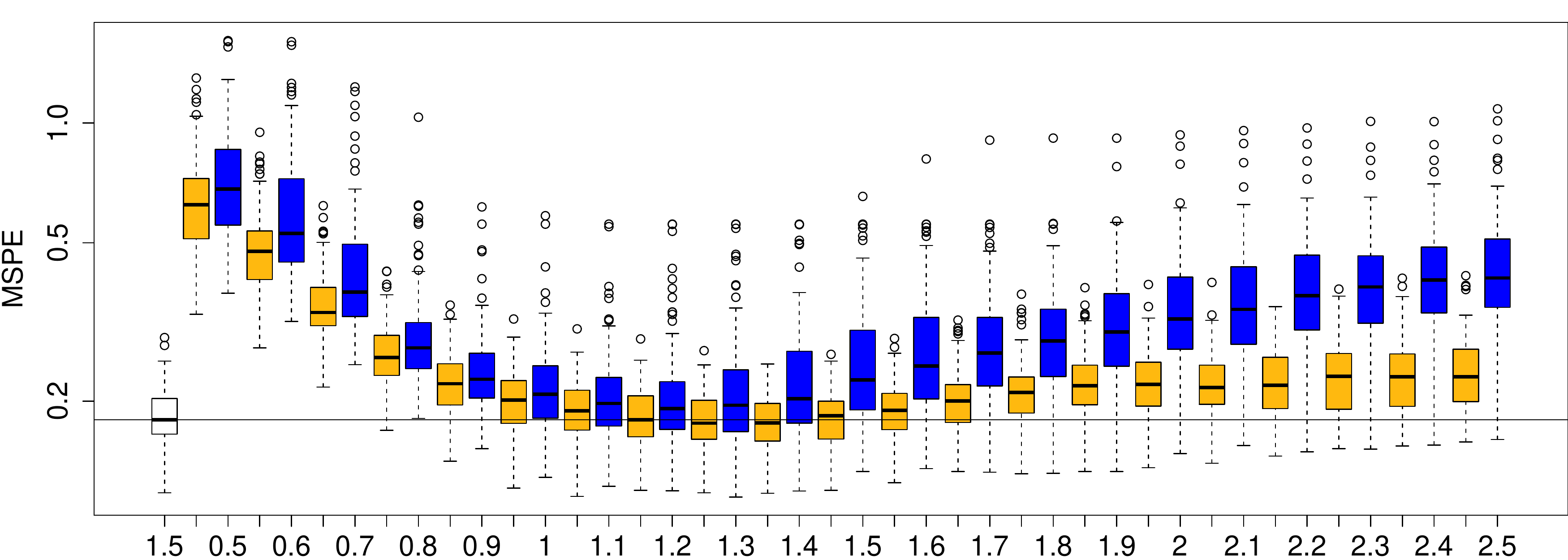}}
		\subfigure[$200 \times 200$ grid with $m=77$]{\includegraphics[scale=.23]{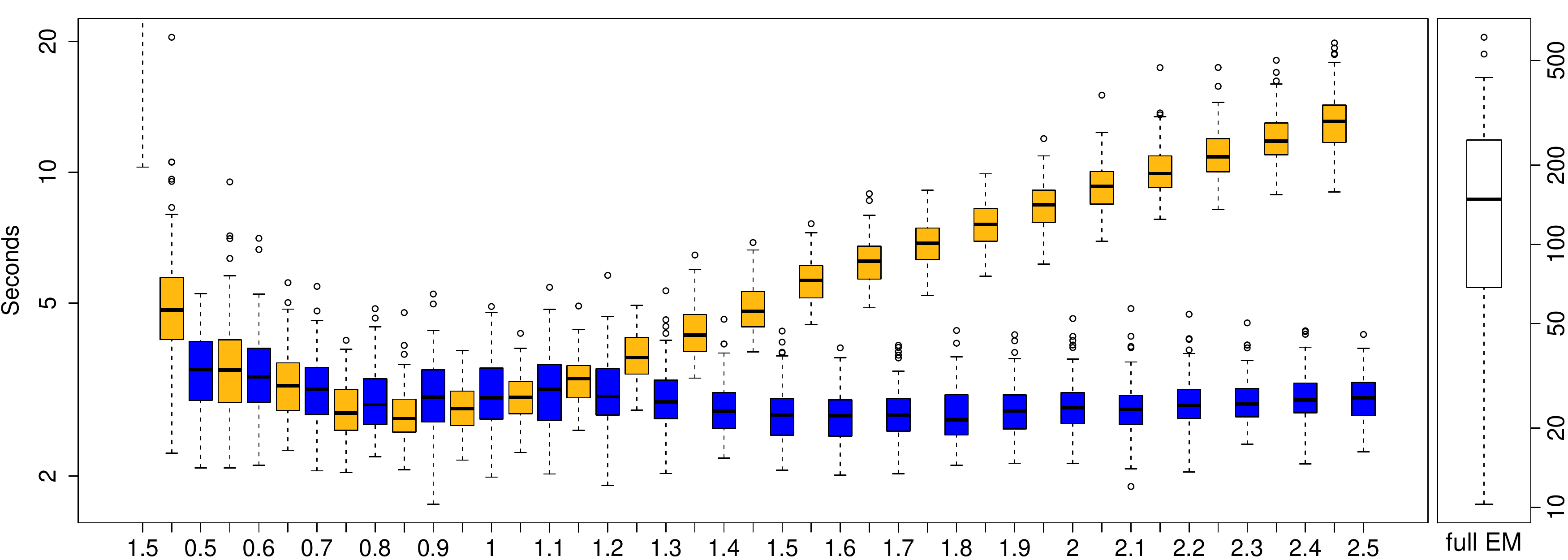}}
		\subfigure[$200 \times 200$ grid with $m=77$]{\includegraphics[scale=.23]{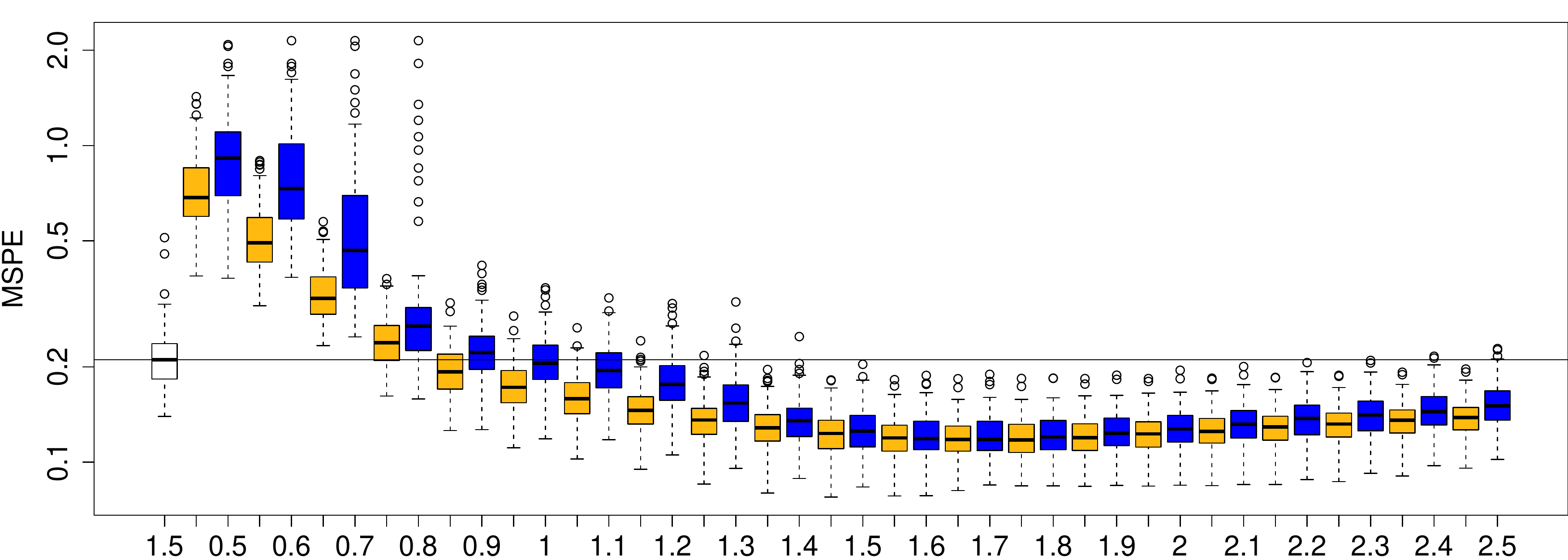}}
		\subfigure[$200 \times 200$ grid with $m=175$]{\includegraphics[scale=.23]{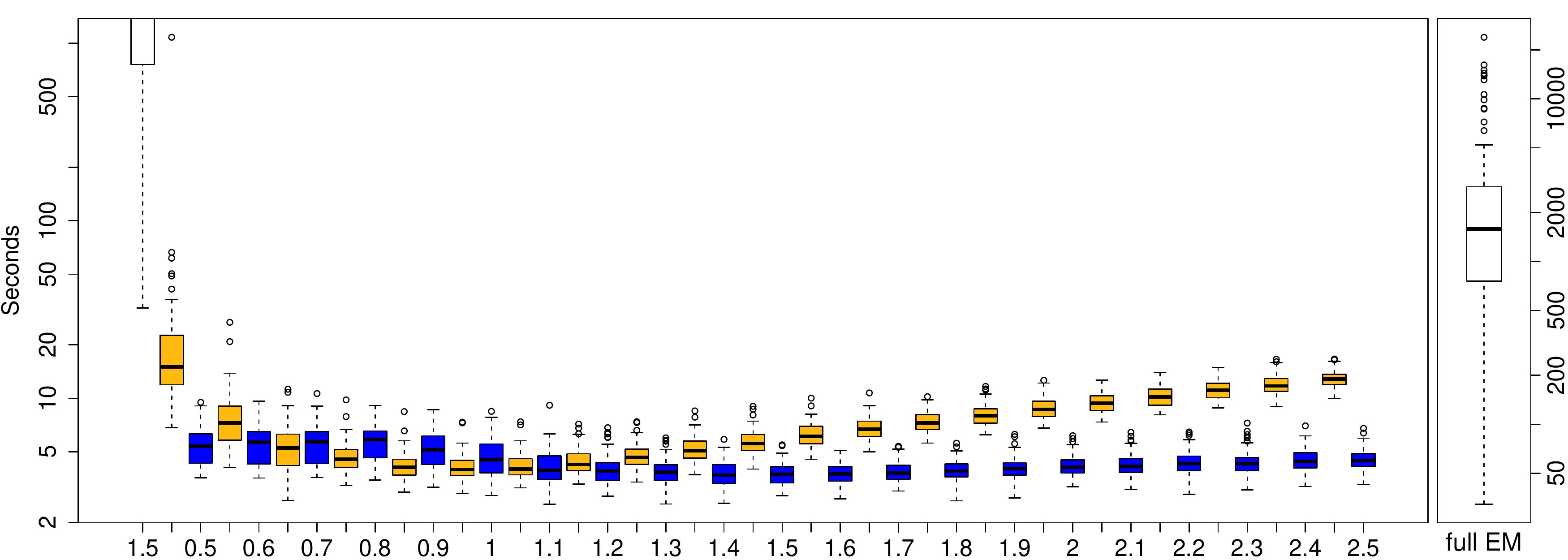}}		
		\subfigure[$200 \times 200$ grid with $m=175$]{\includegraphics[scale=.23]{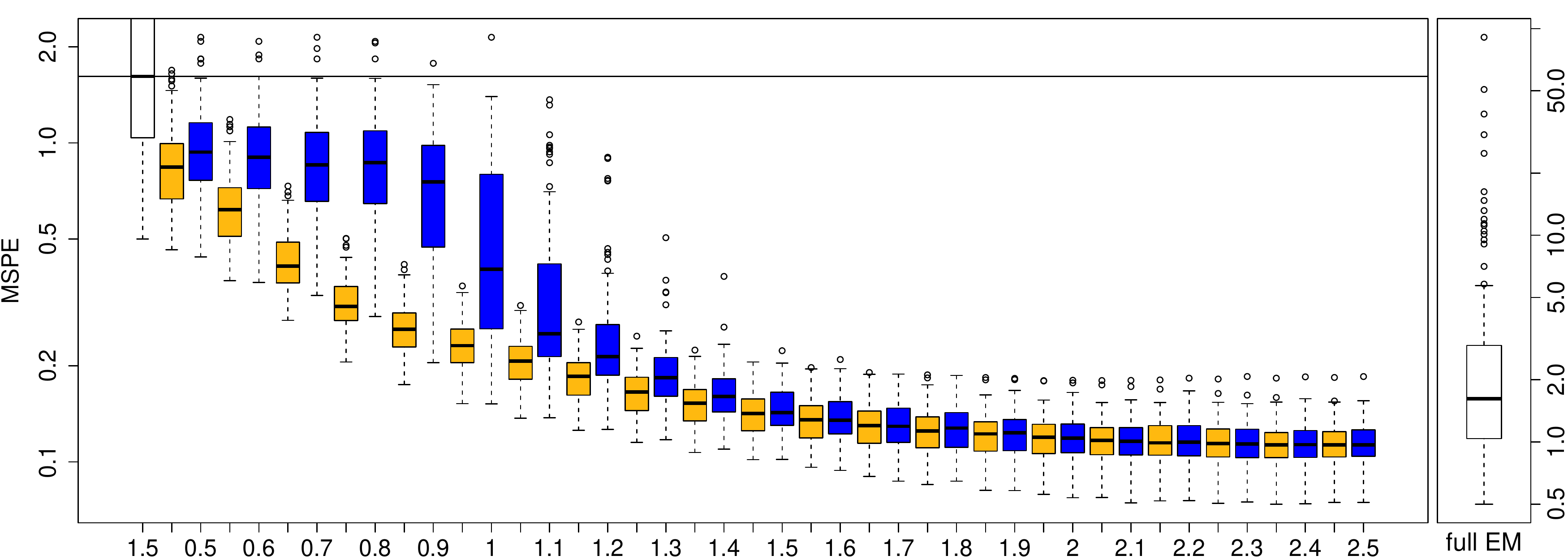}}
	\end{center}
	\caption{Seconds (left panel) and Mean Squared Prediction
          Errors (MSPEs, right panel) for varying bandwidths constants
          (x-axis) and number of knots ($m$) for all three estimation
          methods on both grids when $\nu=1.5$ and $\sigma^2=0.4$.} 
\end{figure}

\begin{figure}[H]
	\begin{center}
		\subfigure[$50 \times 50$ grid with $m=23$]{\includegraphics[scale=.23]{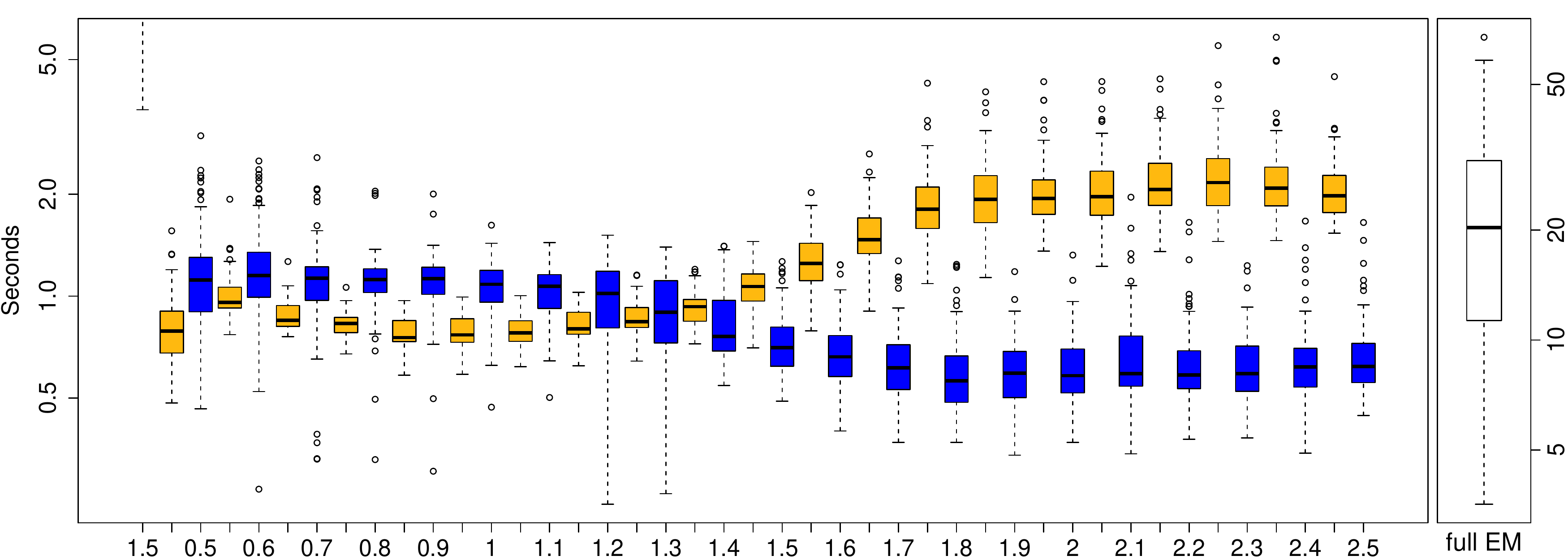}}
		\subfigure[$50 \times 50$ grid with $m=23$]{\includegraphics[scale=.23]{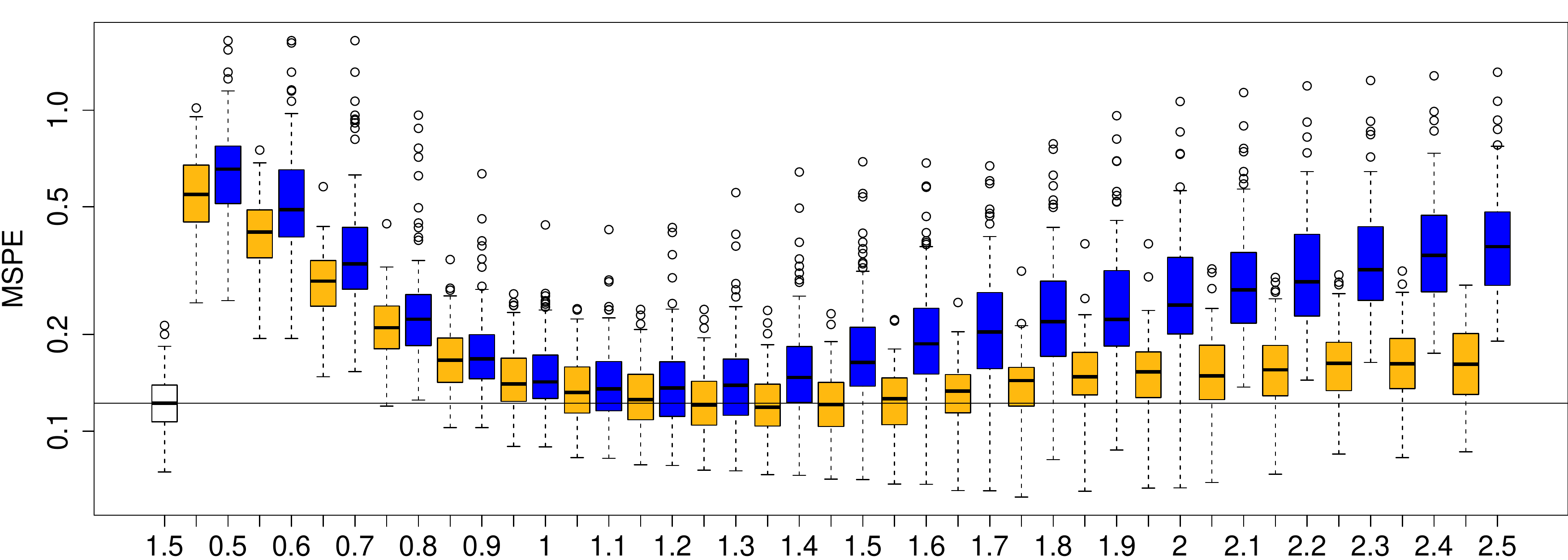}}
		\subfigure[$50 \times 50$ grid with $m=77$]{\includegraphics[scale=.23]{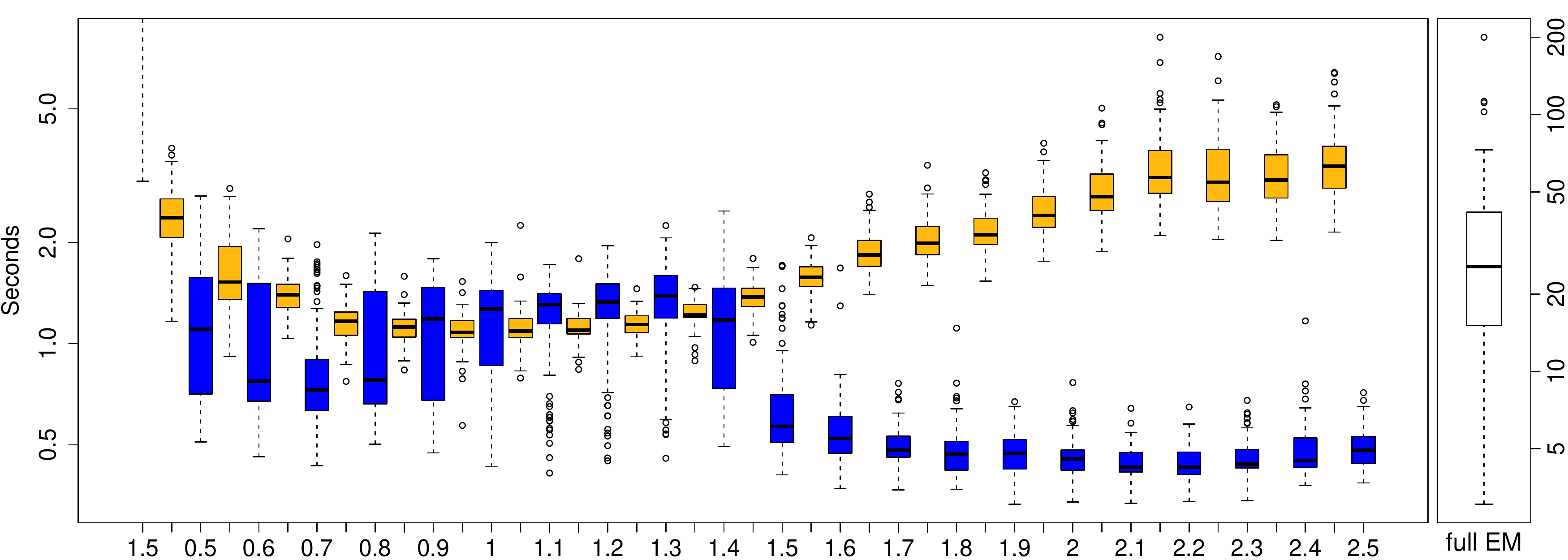}}
		\subfigure[$50 \times 50$ grid with $m=77$]{\includegraphics[scale=.23]{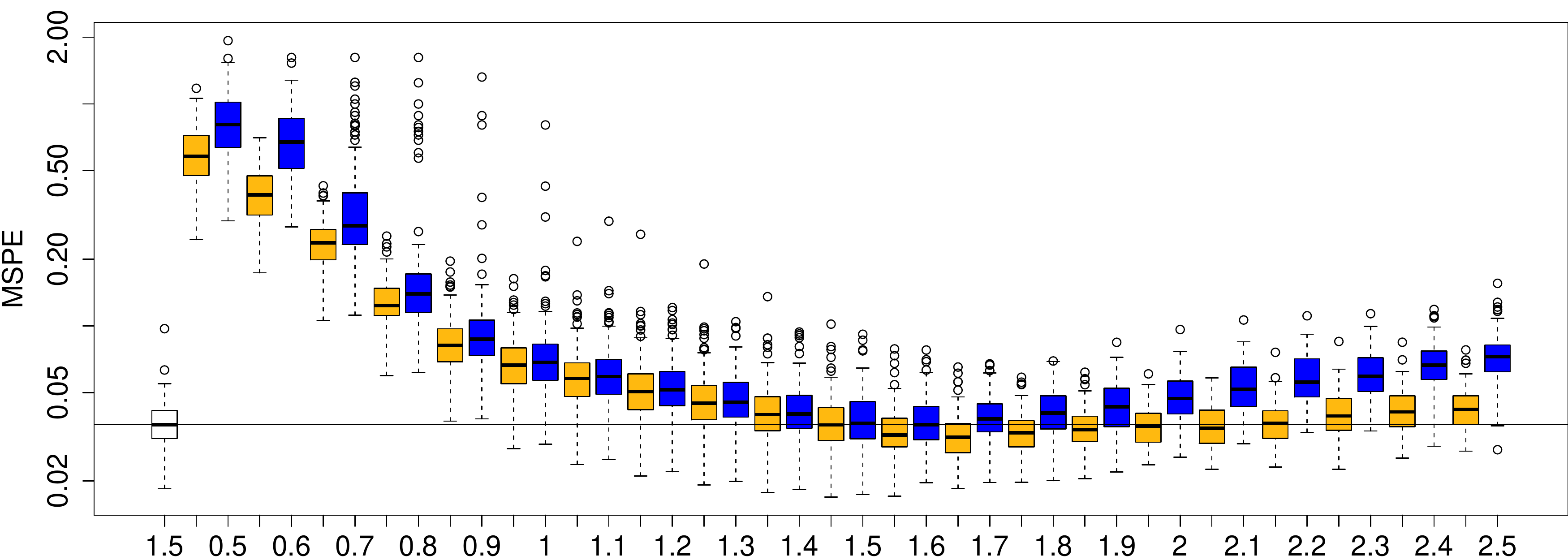}}
		\subfigure[$50 \times 50$ grid with $m=175$]{\includegraphics[scale=.23]{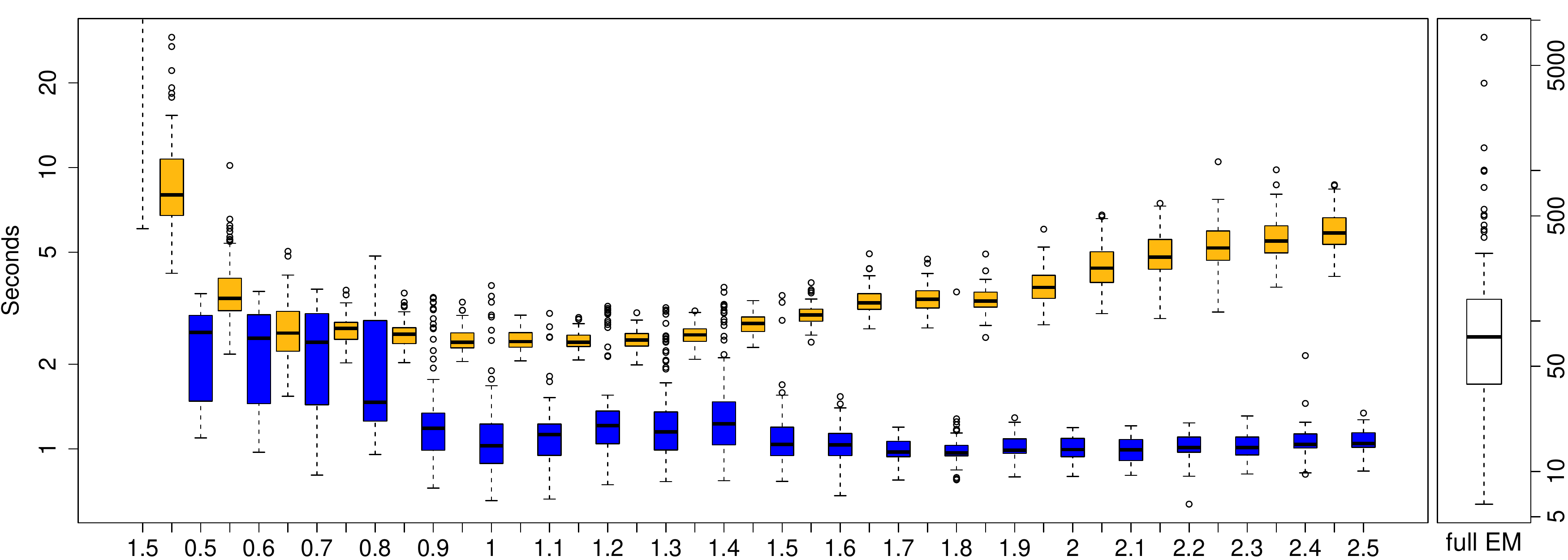}}		
		\subfigure[$50 \times 50$ grid with $m=175$]{\includegraphics[scale=.23]{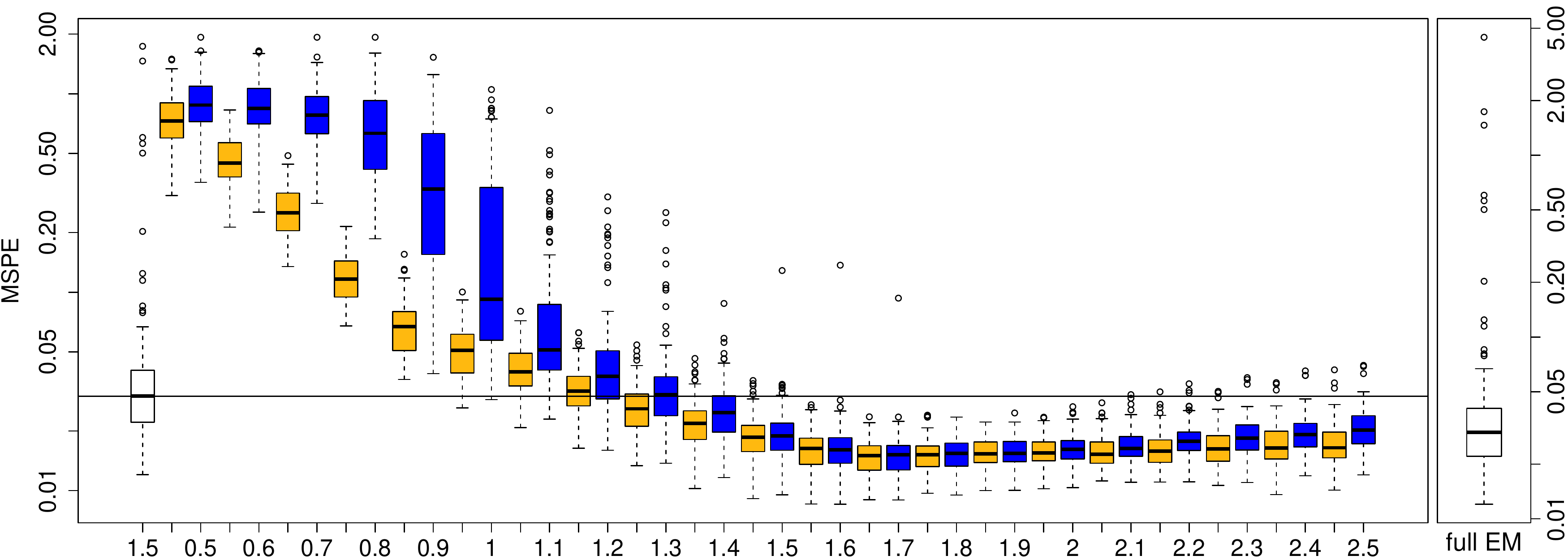}}
		\subfigure[$200 \times 200$ grid with $m=23$]{\includegraphics[scale=.23]{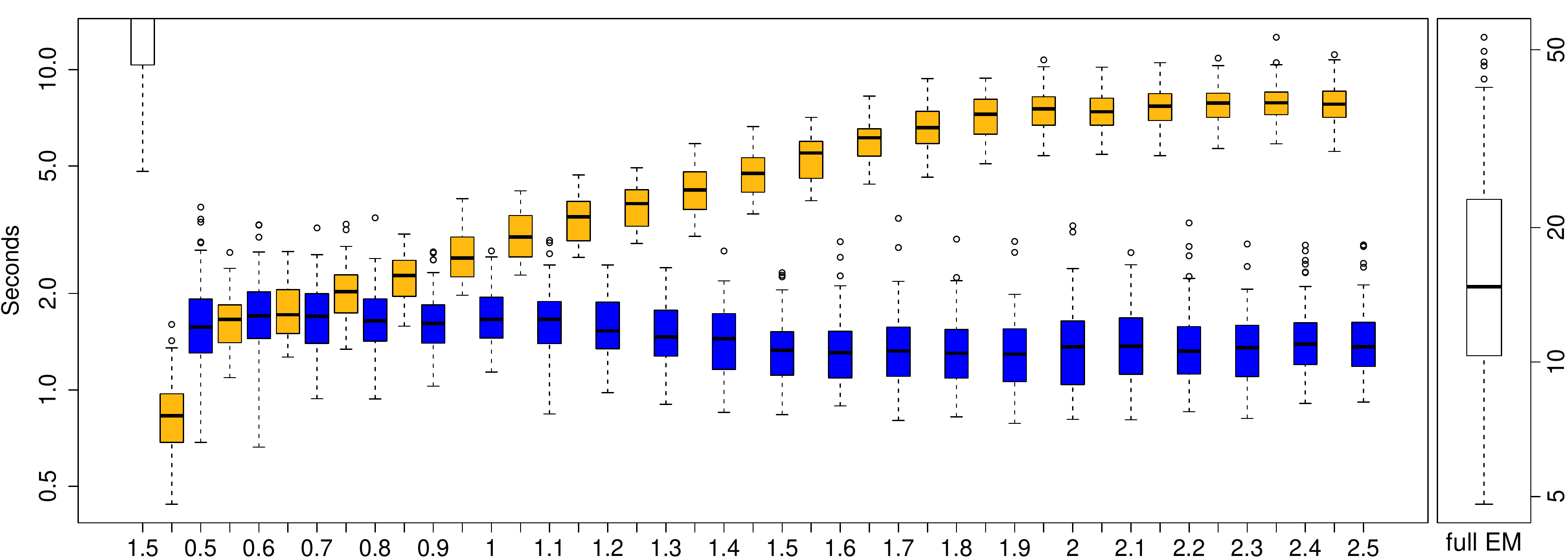}}
		\subfigure[$200 \times 200$ grid with $m=23$]{\includegraphics[scale=.23]{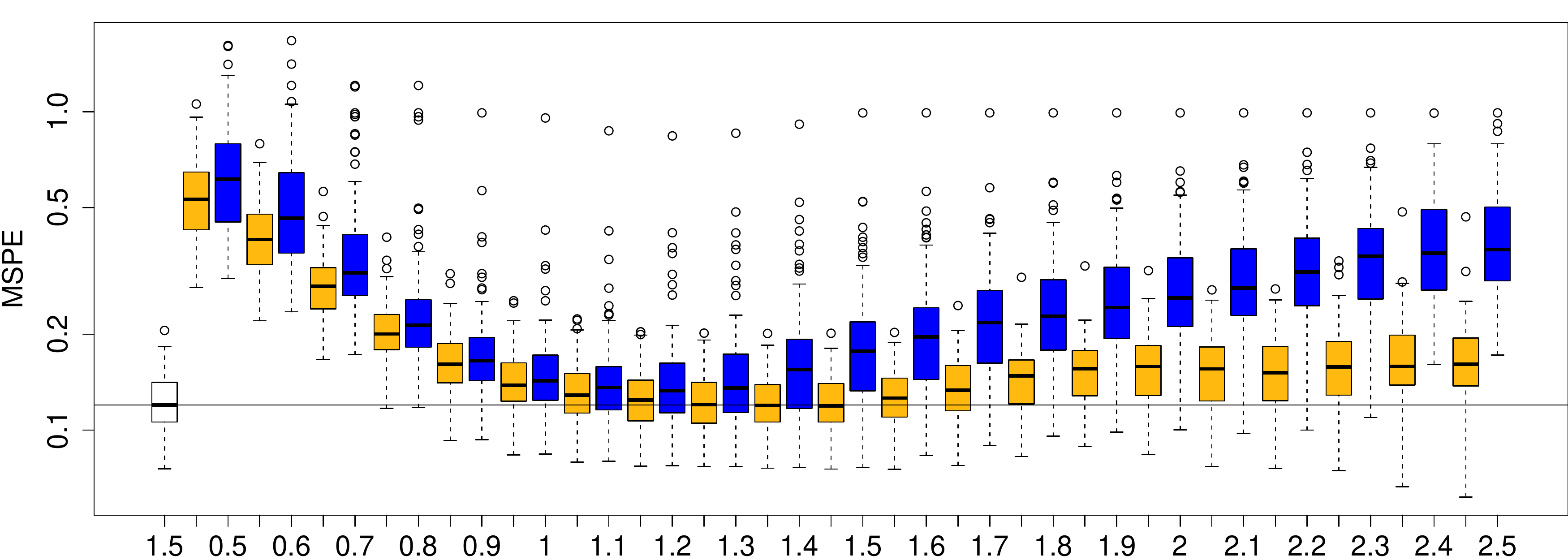}}
		\subfigure[$200 \times 200$ grid with $m=77$]{\includegraphics[scale=.23]{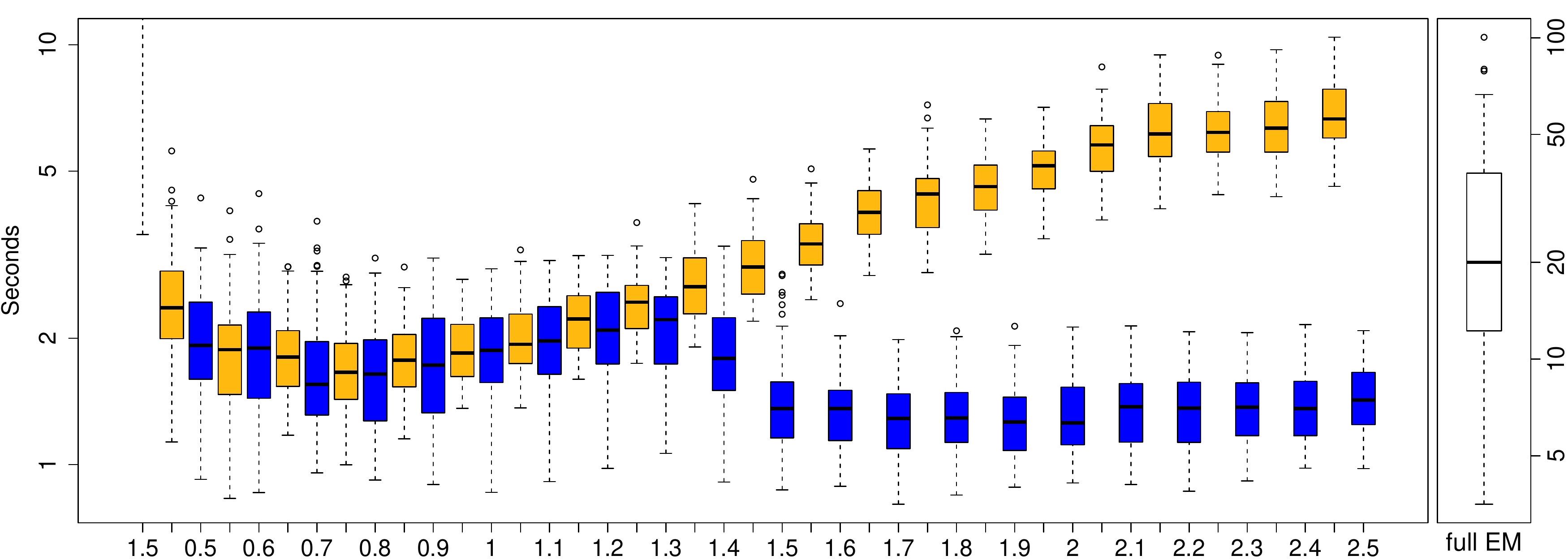}}
		\subfigure[$200 \times 200$ grid with $m=77$]{\includegraphics[scale=.23]{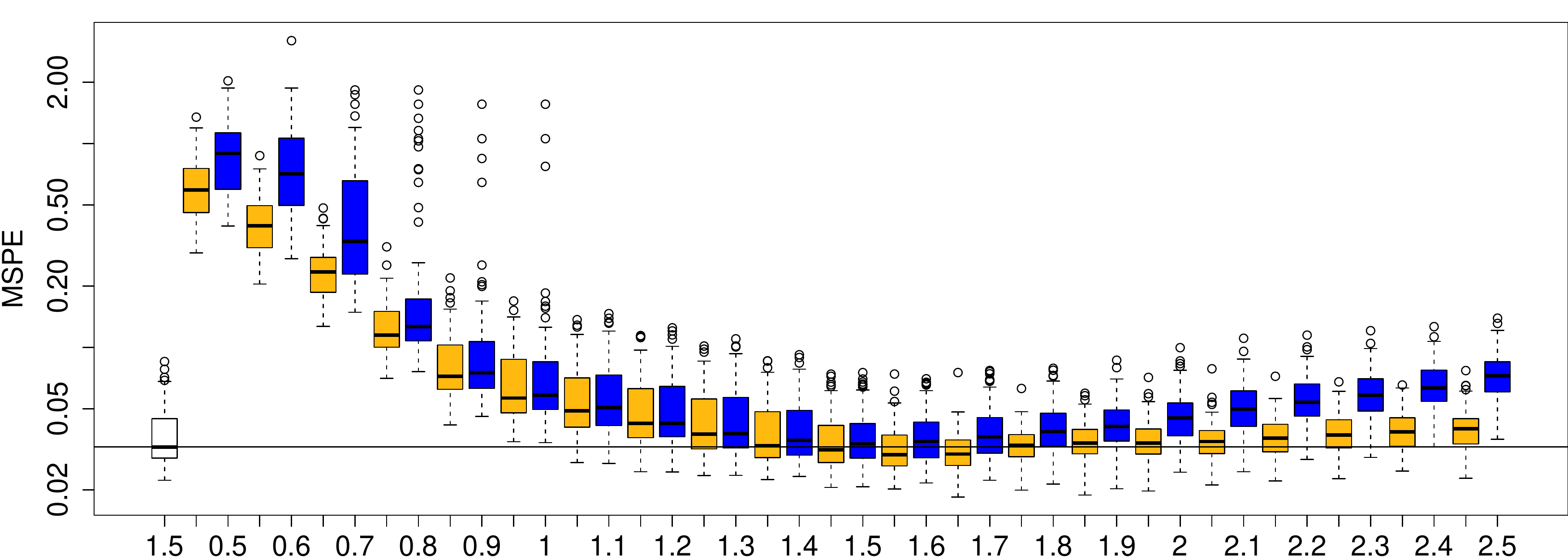}}
		\subfigure[$200 \times 200$ grid with $m=175$]{\includegraphics[scale=.23]{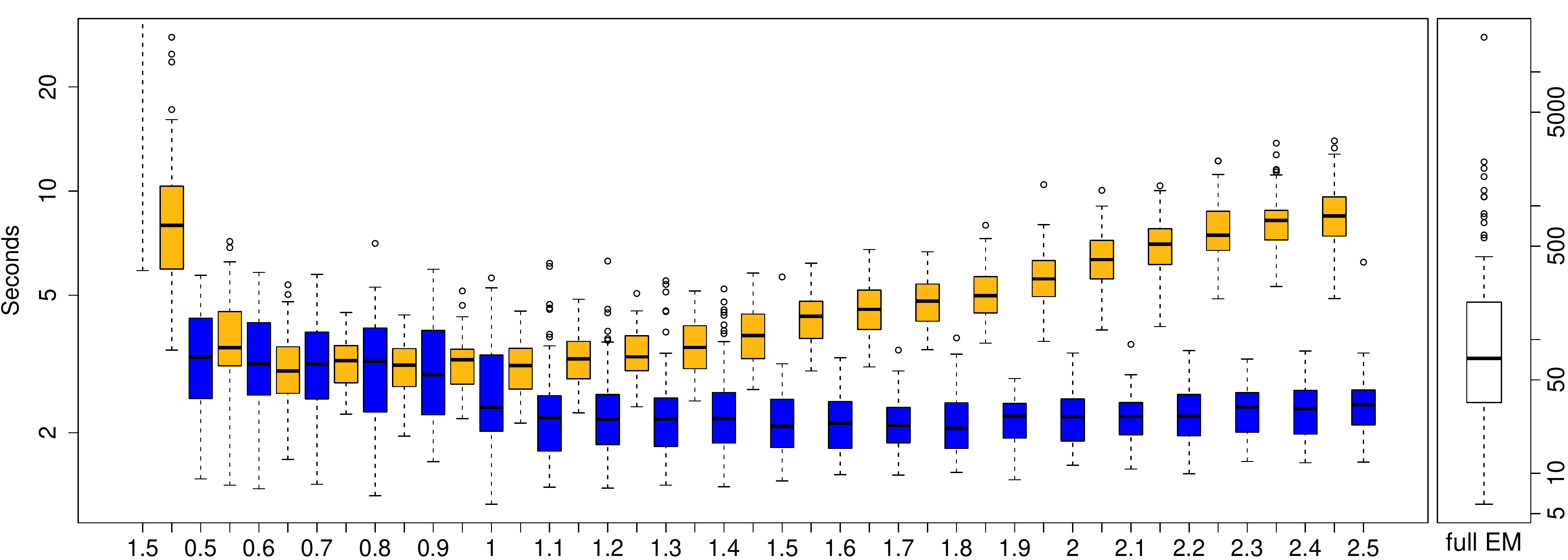}}		
		\subfigure[$200 \times 200$ grid with $m=175$]{ \includegraphics[scale=.23]{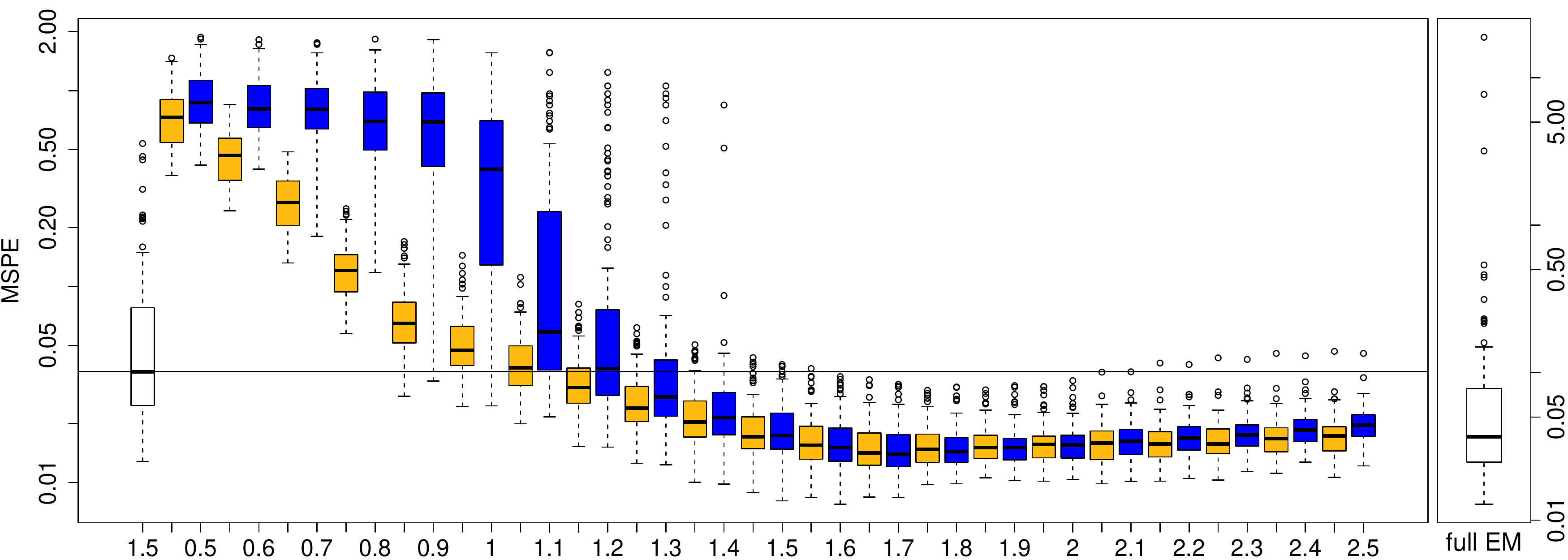}}
	\end{center}
	\caption{Seconds (left panel) and Mean Squared Prediction
          Errors (MSPEs, right panel) for varying bandwidths constants
          (x-axis) and number of knots ($m$) for all three estimation
          methods on both grids when $\nu=2$ and $\sigma^2=0$.}
\end{figure}

\begin{figure}[H]
	\begin{center}
		\subfigure[$50 \times 50$ grid with $m=23$]{\includegraphics[scale=.23]{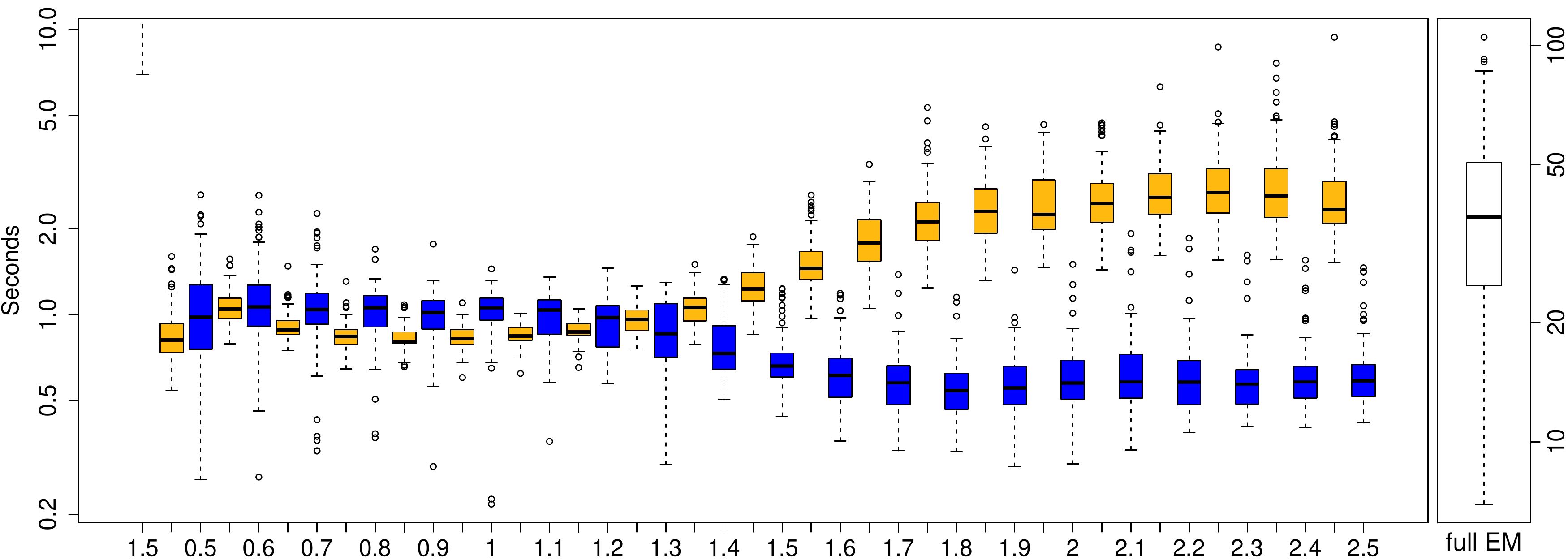}}
		\subfigure[$50 \times 50$ grid with $m=23$]{\includegraphics[scale=.23]{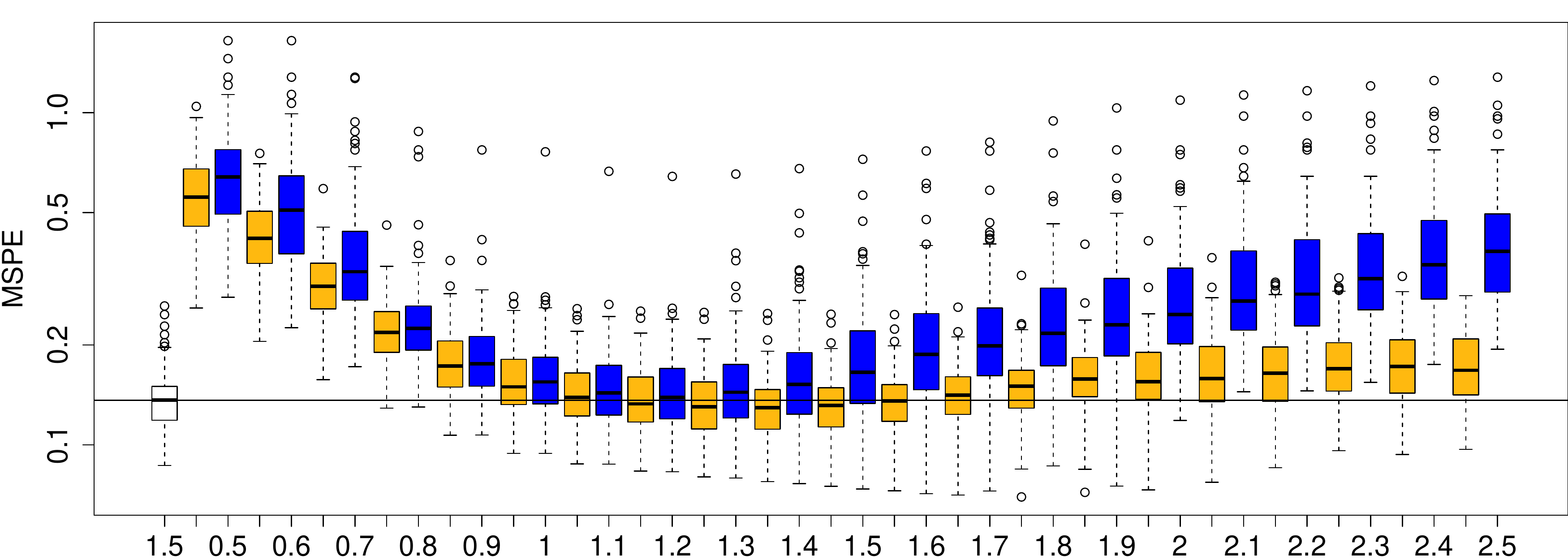}}
		\subfigure[$50 \times 50$ grid with $m=77$]{\includegraphics[scale=.23]{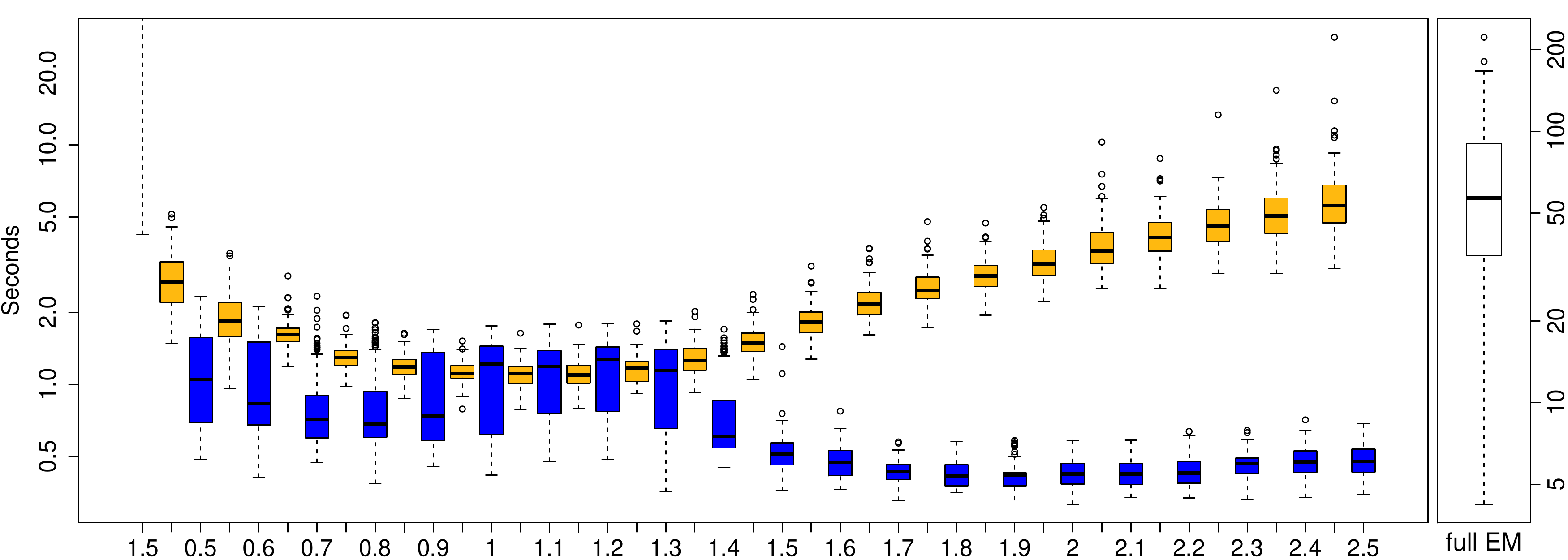}}
		\subfigure[$50 \times 50$ grid with $m=77$]{\includegraphics[scale=.23]{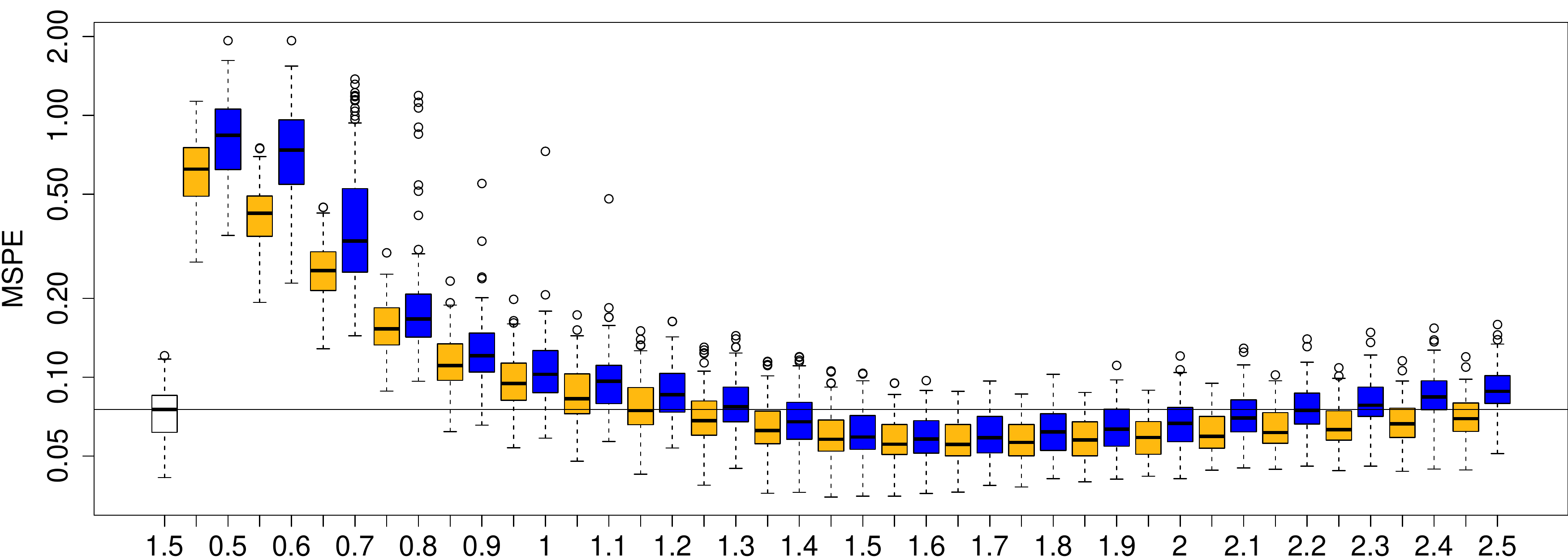}}
		\subfigure[$50 \times 50$ grid with $m=175$]{\includegraphics[scale=.23]{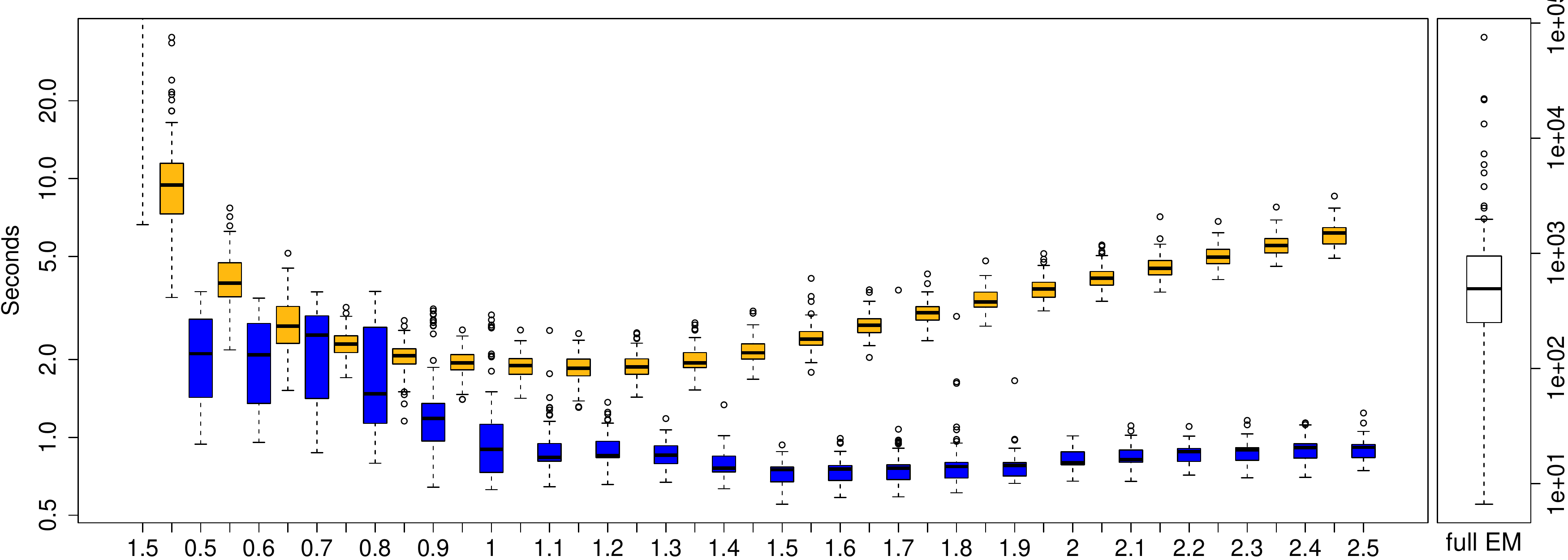}}		
		\subfigure[$50 \times 50$ grid with $m=175$]{\includegraphics[scale=.23]{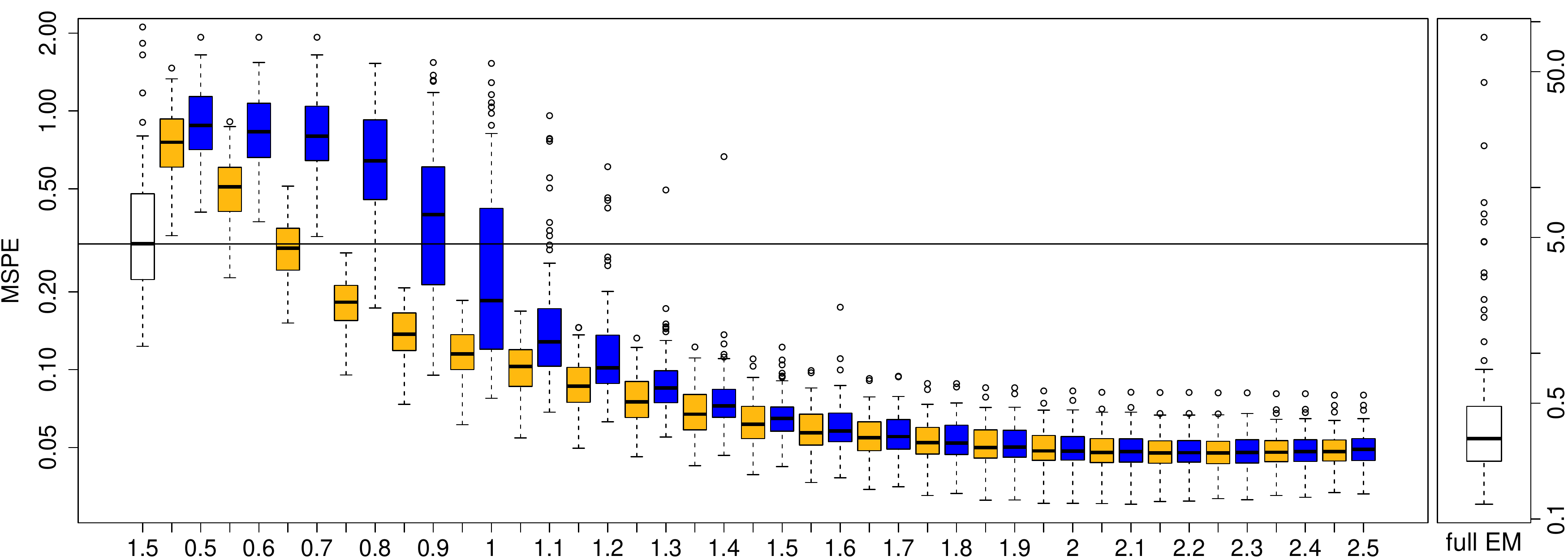}}
		\subfigure[$200 \times 200$ grid with $m=23$]{\includegraphics[scale=.23]{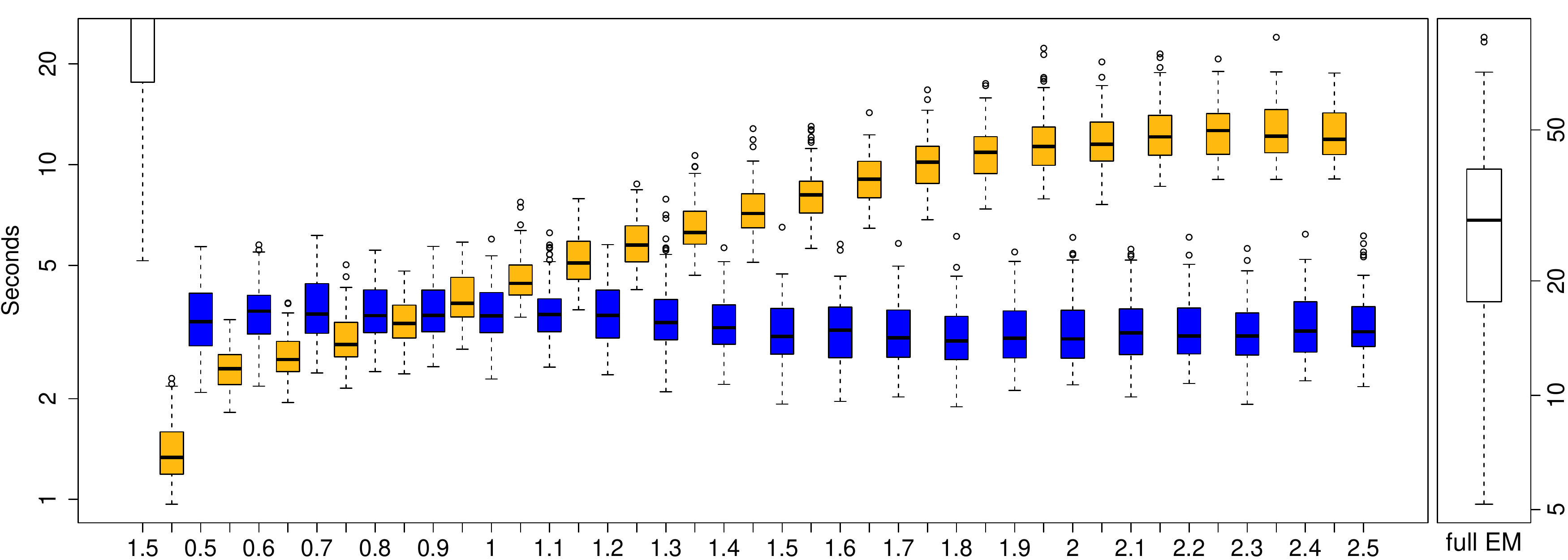}}
		\subfigure[$200 \times 200$ grid with $m=23$]{\includegraphics[scale=.23]{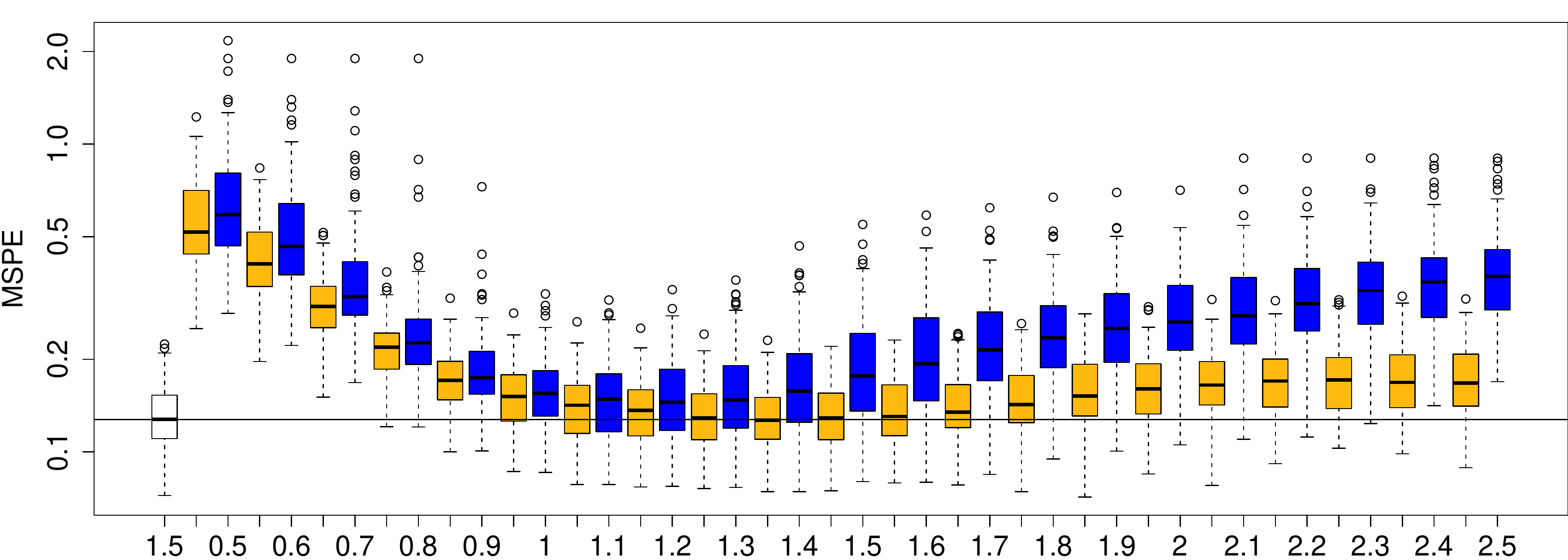}}
		\subfigure[$200 \times 200$ grid with $m=77$]{\includegraphics[scale=.23]{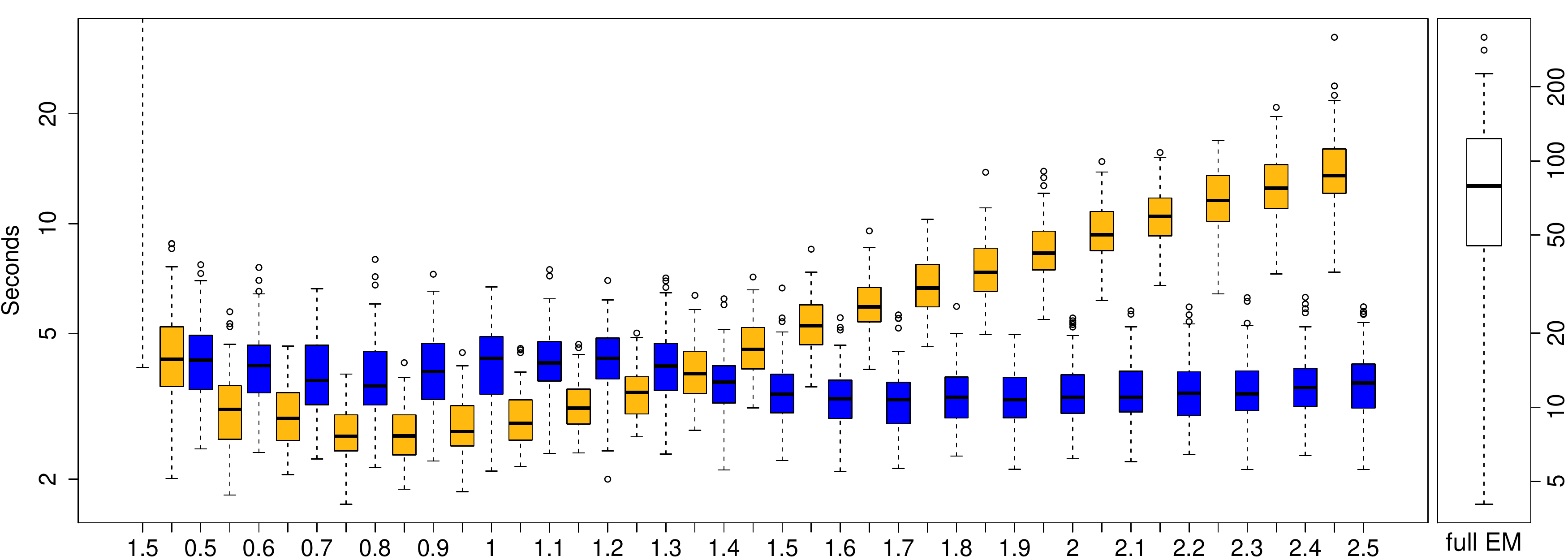}}
		\subfigure[$200 \times 200$ grid with $m=77$]{\includegraphics[scale=.23]{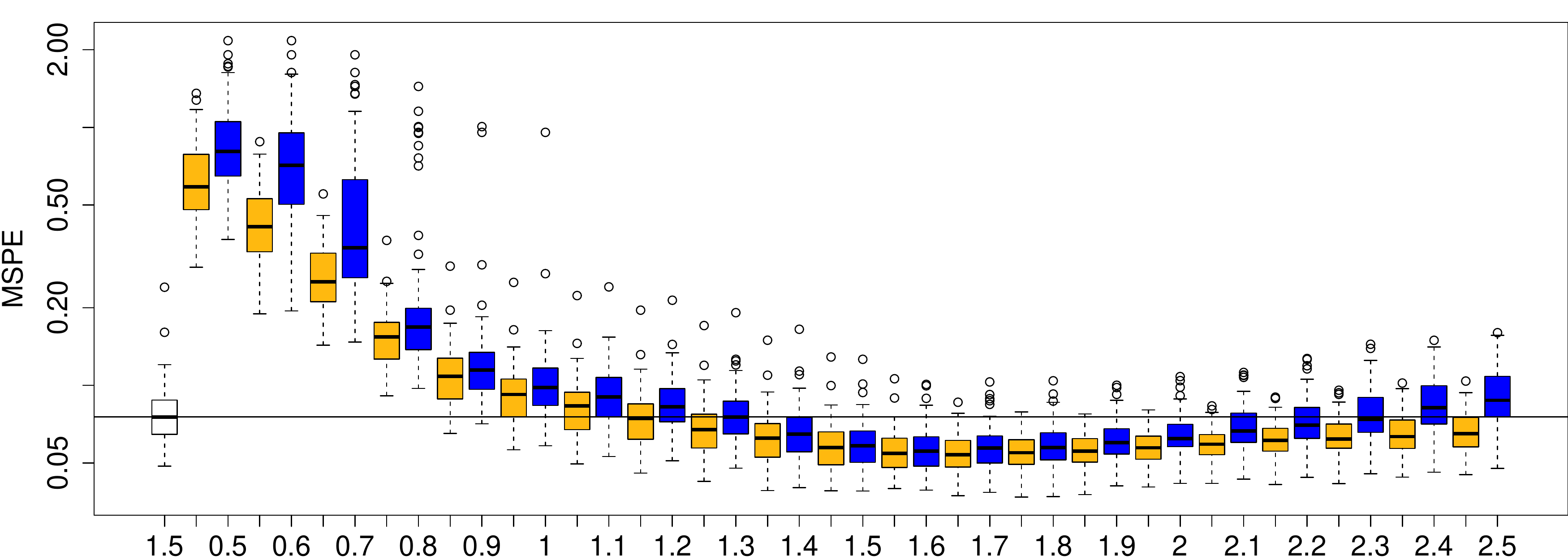}}
		\subfigure[$200 \times 200$ grid with $m=175$]{\includegraphics[scale=.23]{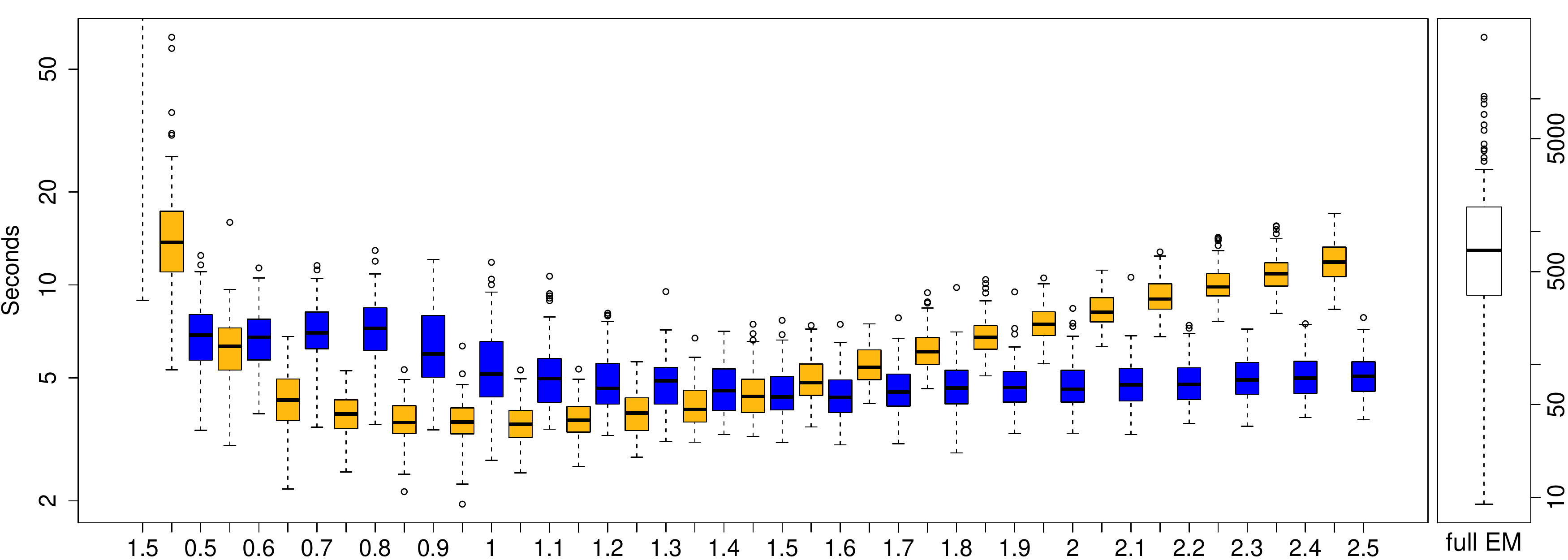}}		
		\subfigure[$200 \times 200$ grid with $m=175$]{\includegraphics[scale=.23]{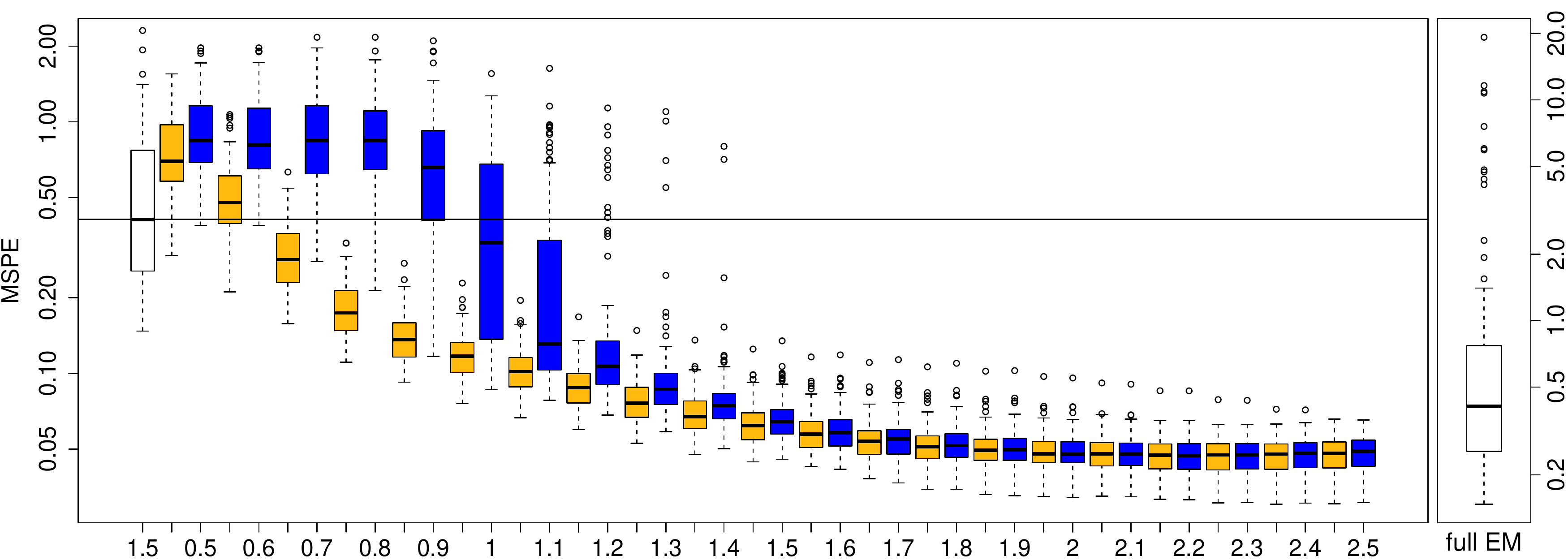}}
	\end{center}
	\caption{Seconds (left panel) and Mean Squared Prediction
          Errors (MSPEs, right panel) for varying bandwidths constants
          (x-axis) and number of knots ($m$) for all three estimation
          methods on both grids when $\nu=2$ and $\sigma^2=0.1$.} 
\end{figure}

\begin{figure}[H]
	\begin{center}
		\subfigure[$50 \times 50$ grid with $m=23$]{\includegraphics[scale=.23]{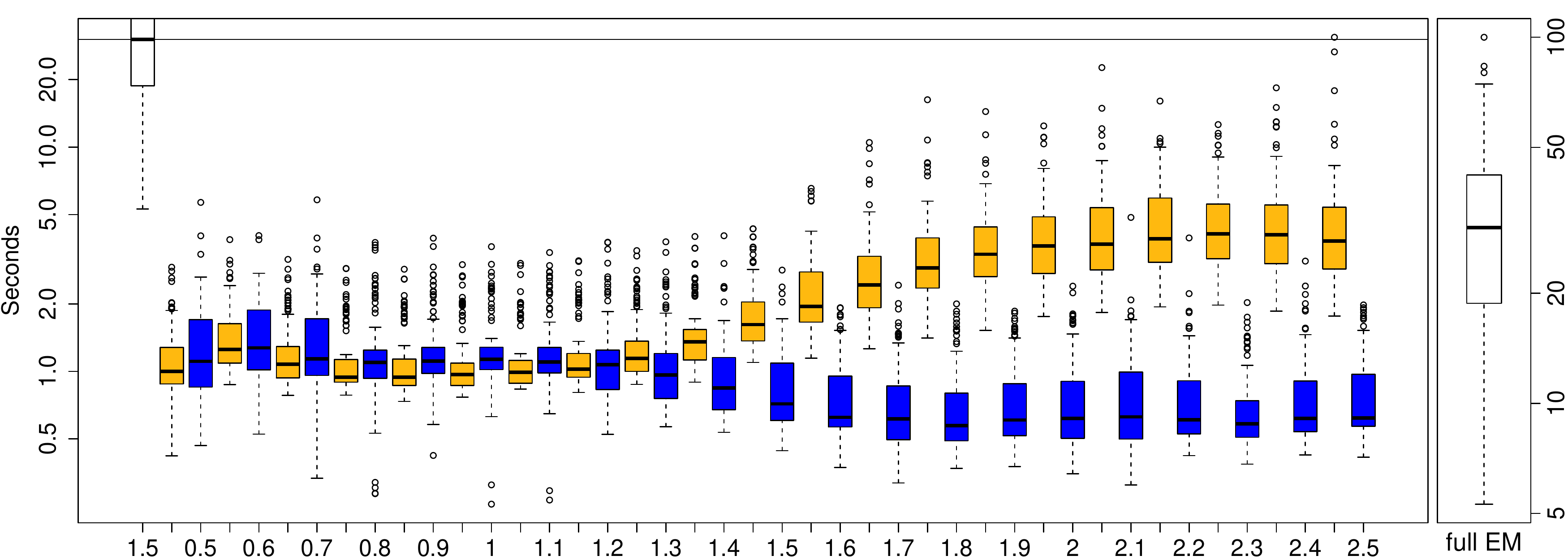}}
		\subfigure[$50 \times 50$ grid with $m=23$]{\includegraphics[scale=.23]{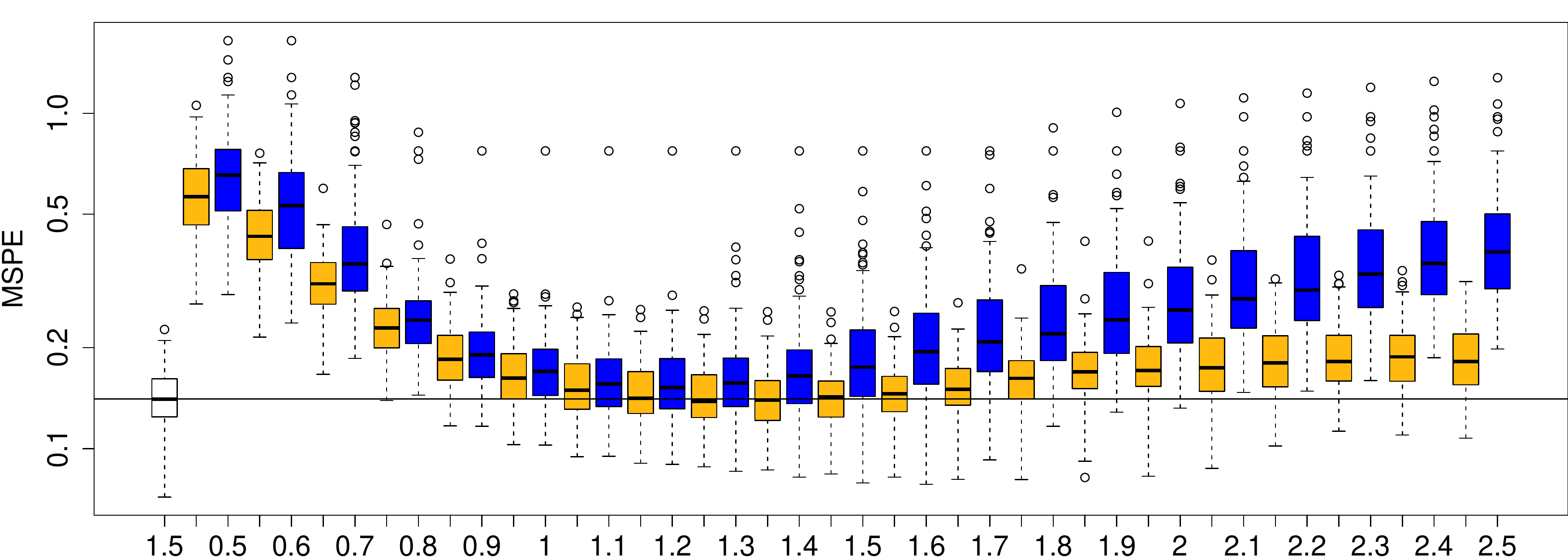}}
		\subfigure[$50 \times 50$ grid with $m=77$]{\includegraphics[scale=.23]{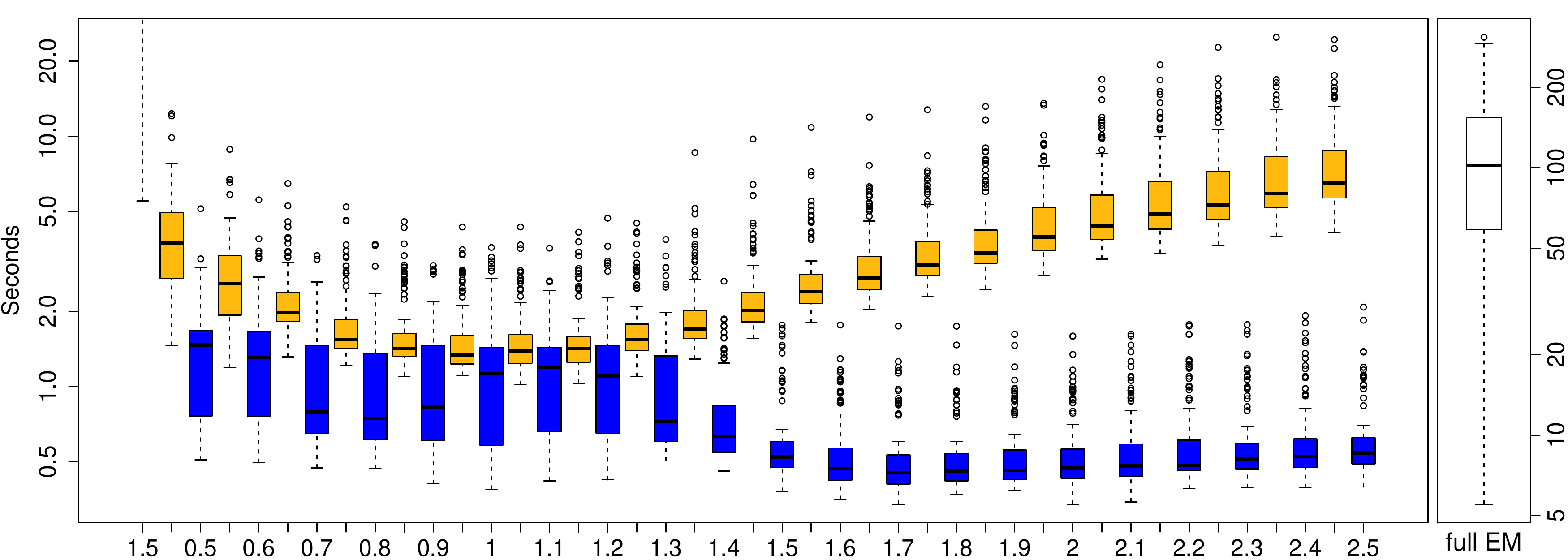}}
		\subfigure[$50 \times 50$ grid with $m=77$]{\includegraphics[scale=.23]{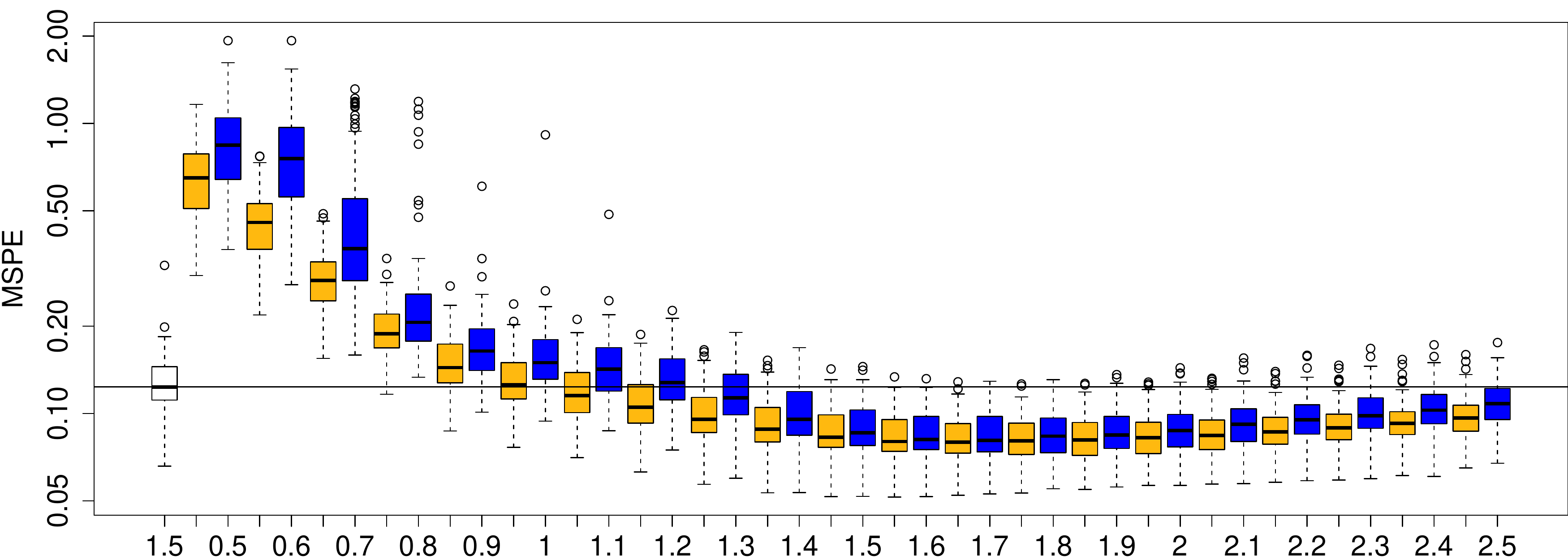}}
		\subfigure[$50 \times 50$ grid with $m=175$]{\includegraphics[scale=.23]{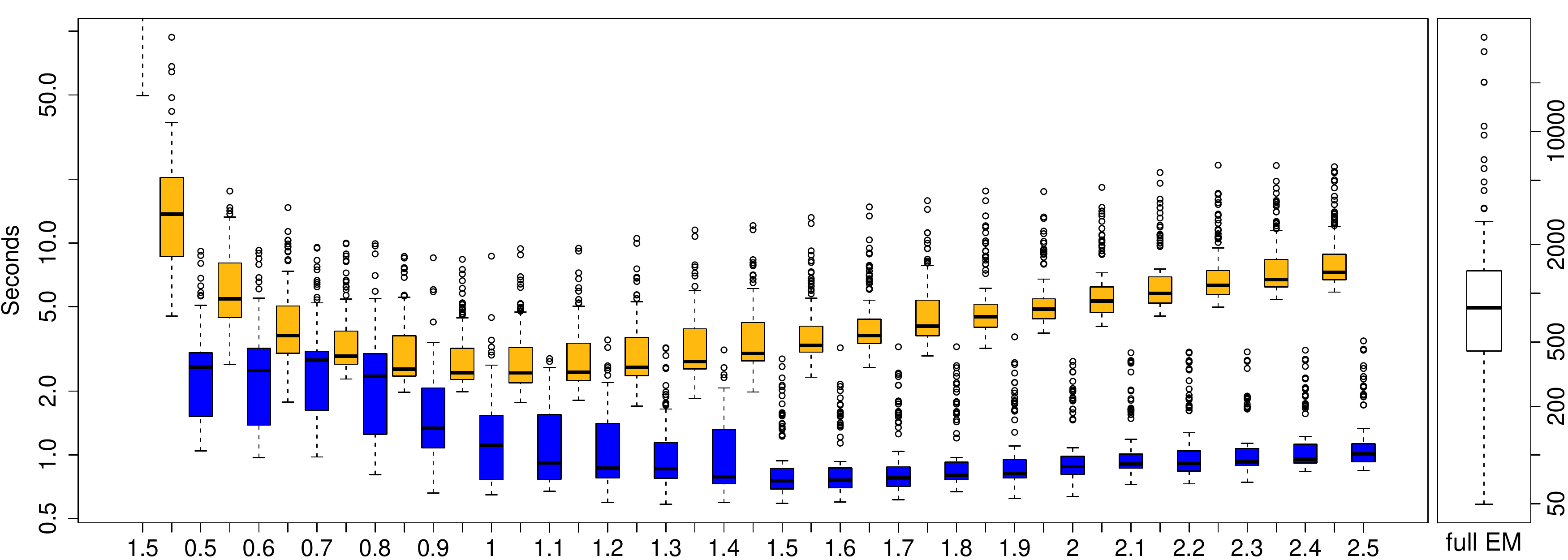}}		
		\subfigure[$50 \times 50$ grid with $m=175$]{\includegraphics[scale=.23]{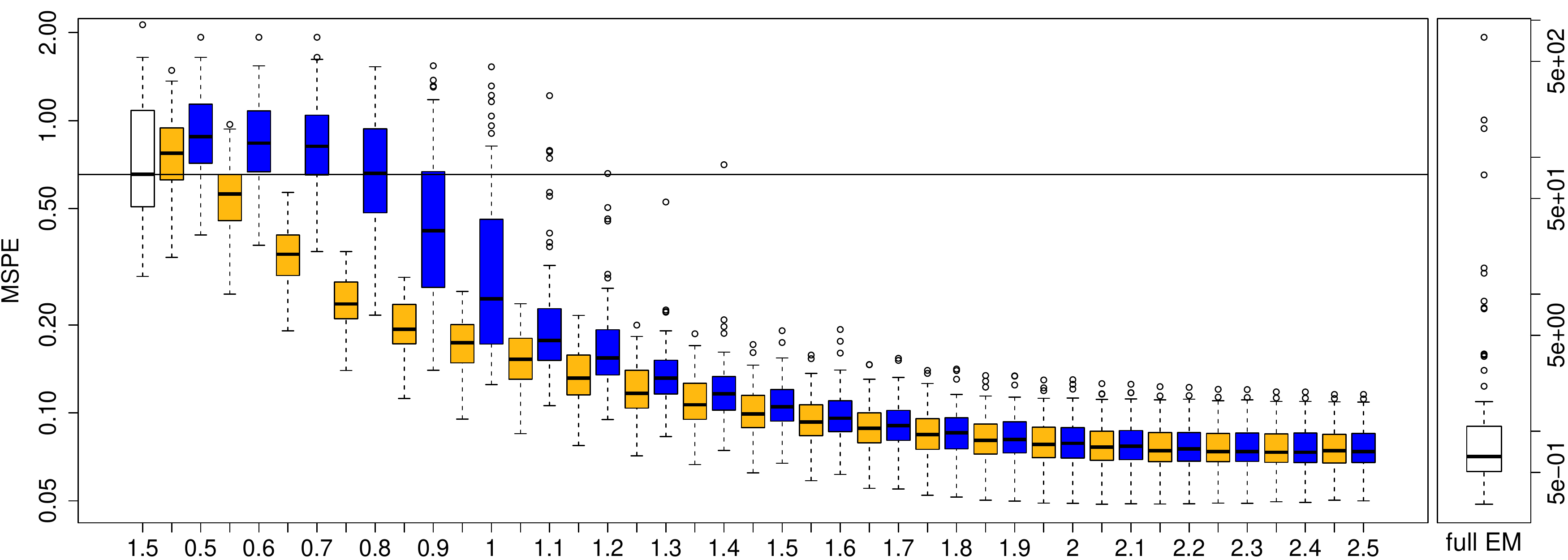}}
		\subfigure[$200 \times 200$ grid with $m=23$]{\includegraphics[scale=.23]{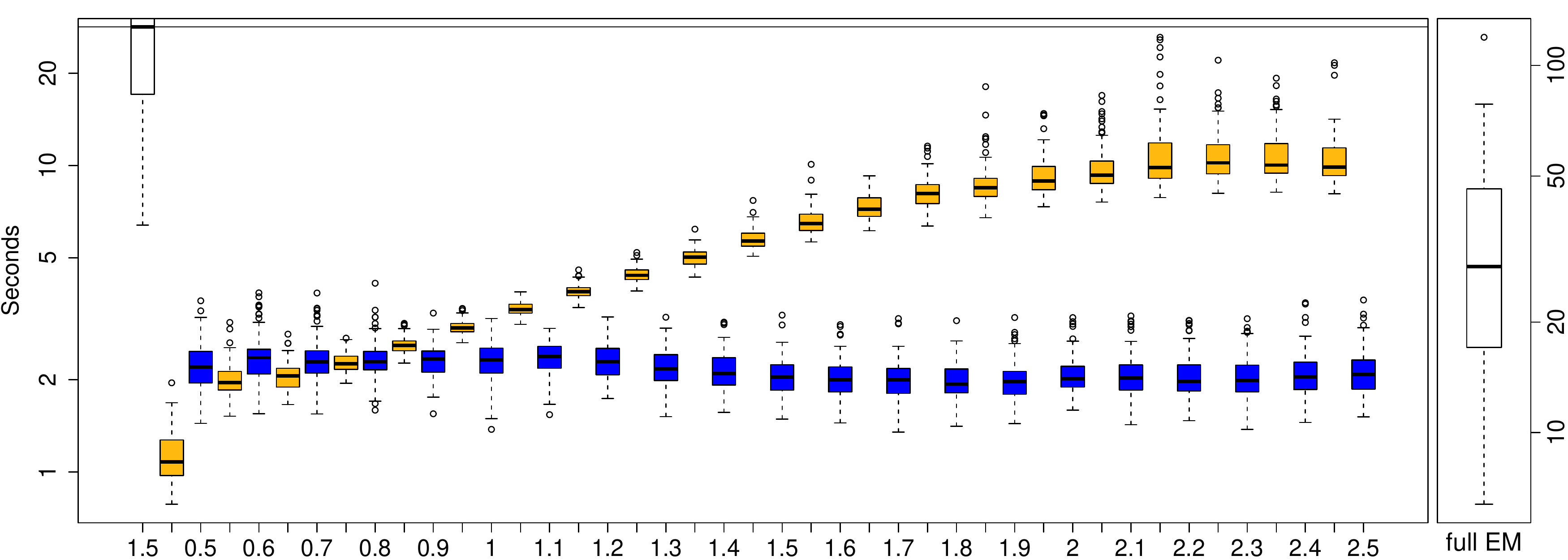}}
		\subfigure[$200 \times 200$ grid with $m=23$]{\includegraphics[scale=.23]{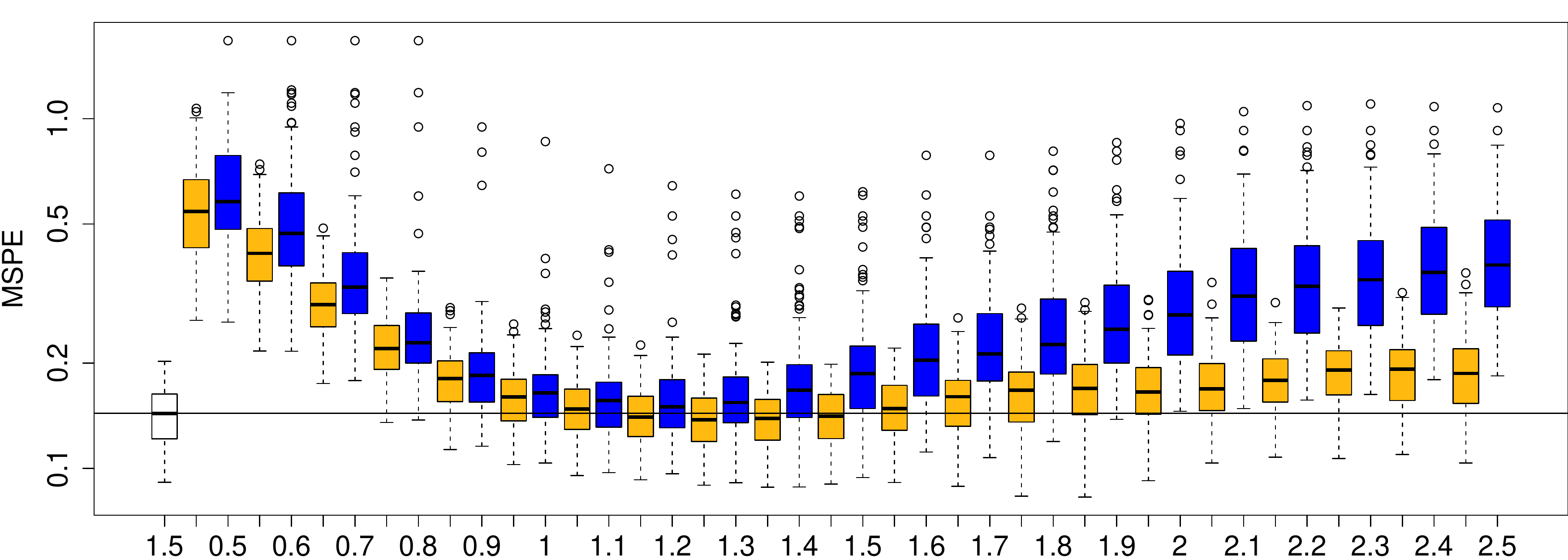}}
		\subfigure[$200 \times 200$ grid with $m=77$]{\includegraphics[scale=.23]{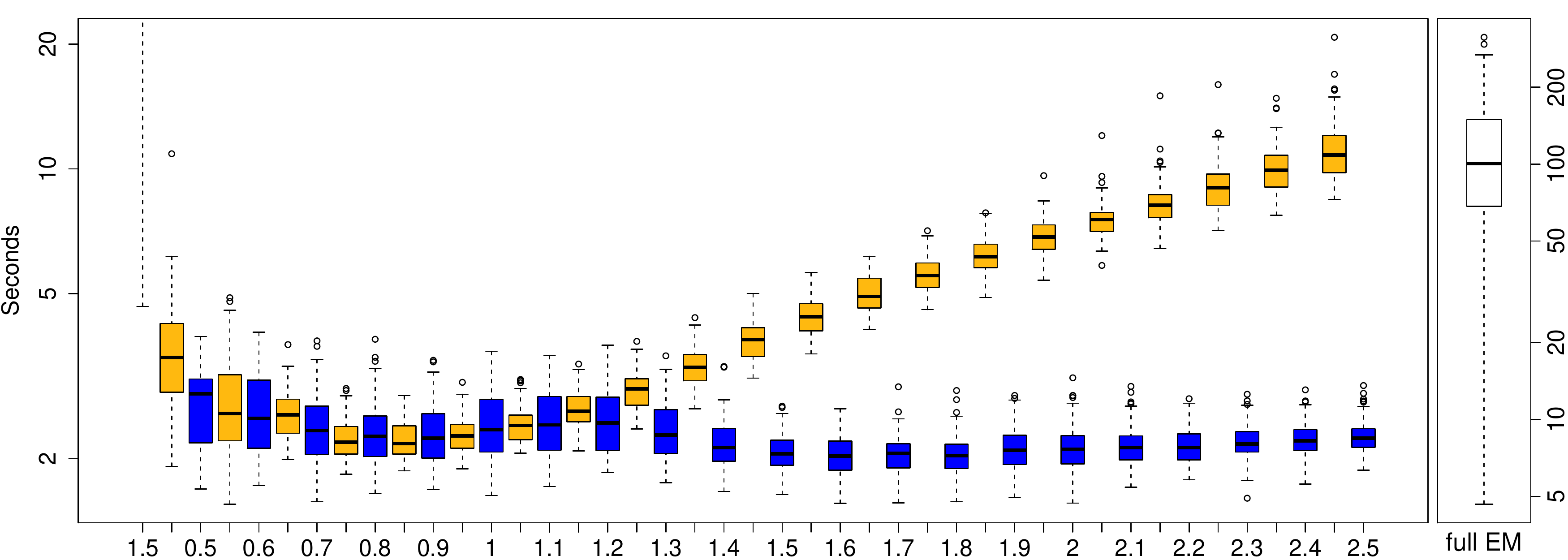}}
		\subfigure[$200 \times 200$ grid with $m=77$]{\includegraphics[scale=.23]{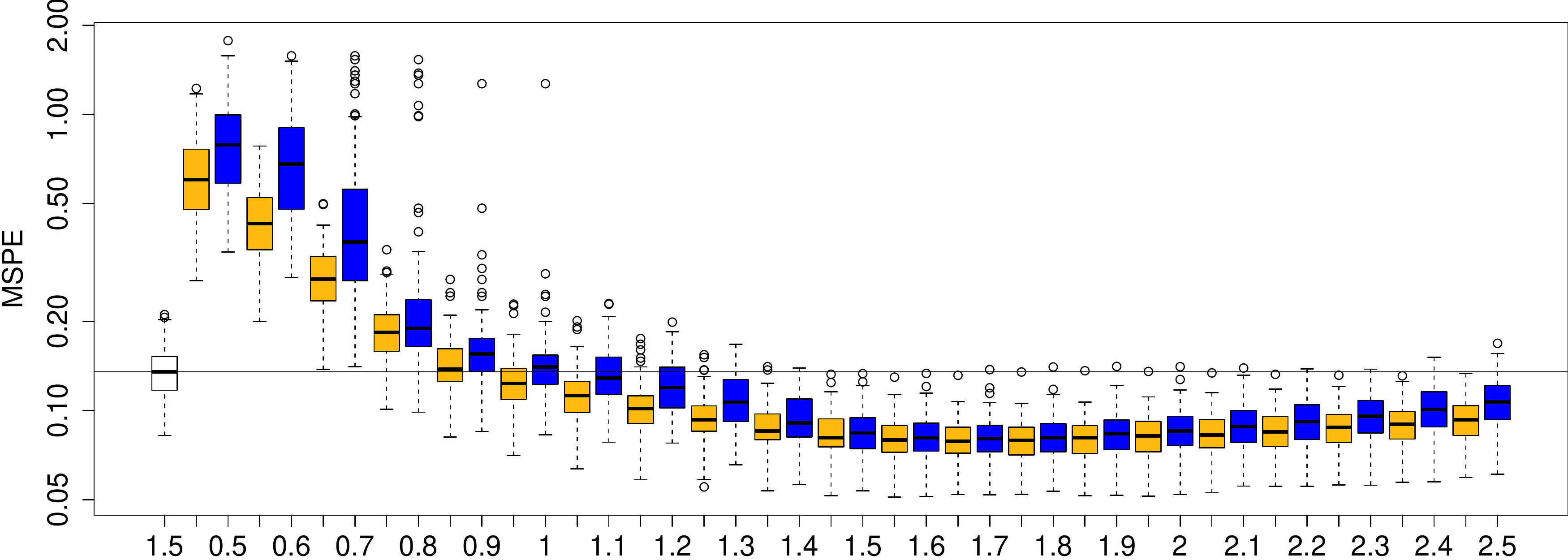}}
		\subfigure[$200 \times 200$ grid with $m=175$]{\includegraphics[scale=.23]{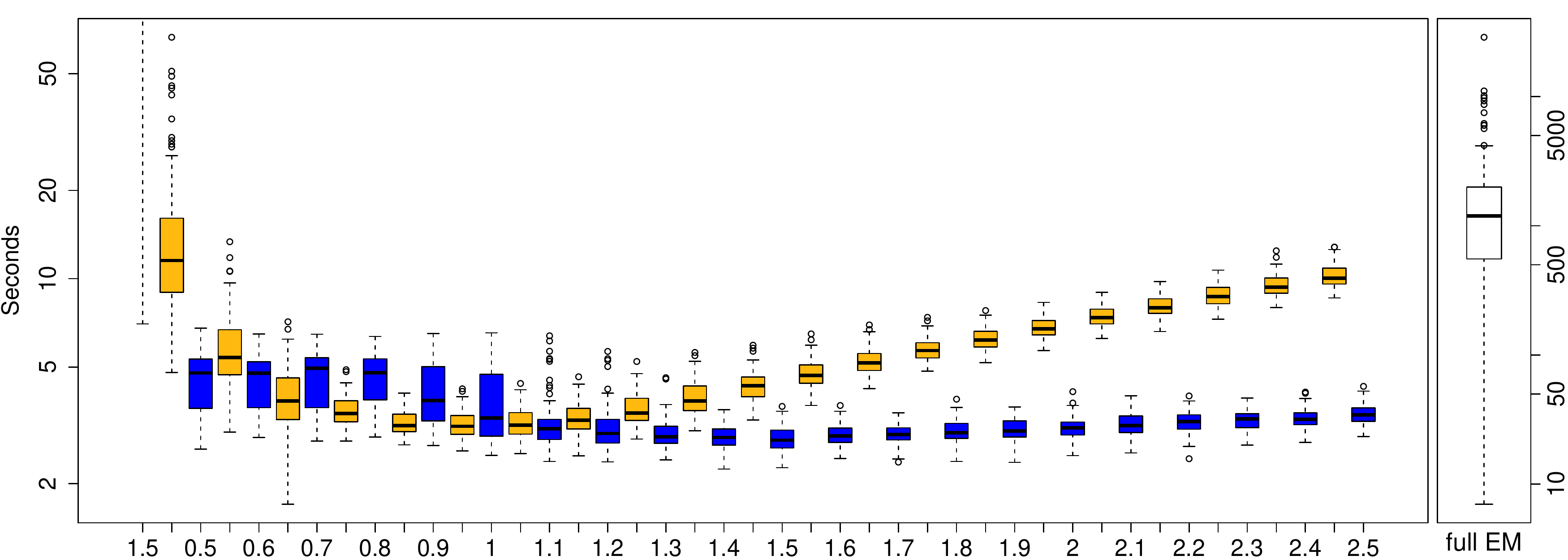}}		
		\subfigure[$200 \times 200$ grid with $m=175$]{\includegraphics[scale=.23]{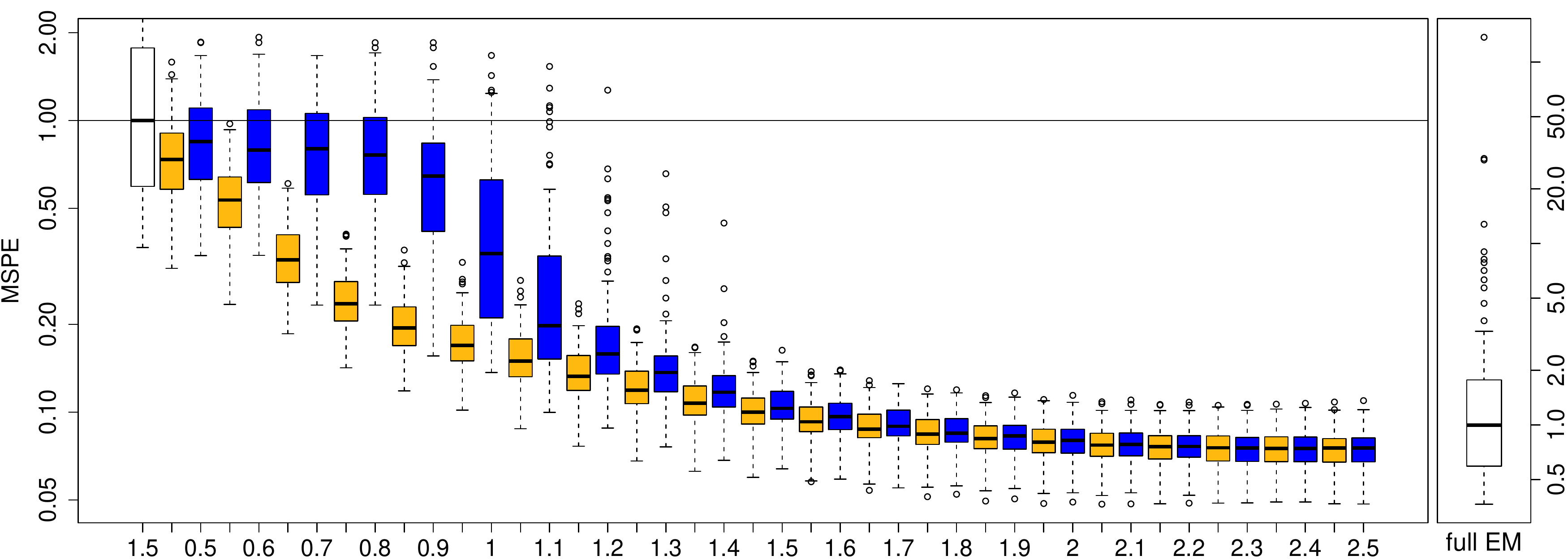}}
	\end{center}
	\caption{Seconds (left panel) and Mean Squared Prediction
          Errors (MSPEs, right panel) for varying bandwidths constants
          (x-axis) and number of knots ($m$) for all three estimation
          methods on both grids when $\nu=2$ and $\sigma^2=0.25$.} 
\end{figure}

\bibliographystyle{apalike}
\bibliography{References}